\documentclass[11pt,a4paper]{article}

\usepackage[margin=1in]{geometry}
\usepackage{amsmath}
\usepackage{amssymb}
\usepackage{amsthm}
\usepackage{booktabs}
\usepackage{placeins}
\usepackage{multirow}
\usepackage{longtable}
\usepackage{array}
\usepackage{url}
\usepackage[numbers]{natbib}
\usepackage[hidelinks,unicode]{hyperref}

\newcommand{\doi}[1]{\href{https://doi.org/#1}{\nolinkurl{doi:#1}}}

\newcommand{\AmpExactFour}{9}
\newcommand{\AmpExactOne}{3}
\newcommand{\AmpExactThree}{7}
\newcommand{\AmpExactThreeSymOne}{3}
\newcommand{\AmpExactThreeSymTwo}{5}
\newcommand{\AmpExactTwo}{5}
\newcommand{\AmpExactUnaryFive}{11}
\newcommand{\AmpFirstOne}{2}
\newcommand{\AmpFirstThree}{10}
\newcommand{\AmpFirstTwo}{11}
\newcommand{\AmpFour}{9}
\newcommand{\AmpOne}{3}
\newcommand{\AmpRunsFirstThree}{31}
\newcommand{\AmpRunsThree}{7}
\newcommand{\AmpThree}{7}
\newcommand{\AmpTwo}{5}
\newcommand{\AmpUnaryFirstFour}{79}
\newcommand{\BeyondProper}{600}
\newcommand{\BeyondUnproper}{600}
\newcommand{\BlockBelow}{6{,}070{,}020}
\newcommand{\BlockEdits}{6{,}541{,}200}
\newcommand{\CapBoth}{632}
\newcommand{\CapDictionaries}{634}
\newcommand{\CapEditFinite}{632}
\newcommand{\CapExactEdits}{361{,}594}
\newcommand{\CapFewer}{267}
\newcommand{\CapFinite}{1{,}898}
\newcommand{\CapGammaBelow}{389}
\newcommand{\CapGammaPairs}{1{,}912}
\newcommand{\CapGammaZero}{322}
\newcommand{\CapIntFewer}{137}
\newcommand{\CapLocalDictionaries}{2}
\newcommand{\CapLocalEdits}{1{,}440}
\newcommand{\CapPrependFinite}{1{,}266}
\newcommand{\CapSeeded}{150}
\newcommand{\CapWeightedBelow}{305}
\newcommand{\CapZeroPairs}{1{,}964}
\newcommand{\CensusPlateaus}{114}
\newcommand{\CensusTotal}{484}
\newcommand{\CensusUnbounded}{370}
\newcommand{\CertCertificates}{466}
\newcommand{\CertEventEdges}{50{,}784}
\newcommand{\CertProducts}{968}

\newcommand{\ChainExactWidthOrdered}{435}

\newcommand{\ChainFrontierStates}{6{,}119}
\newcommand{\ChainGptC}{129}
\newcommand{\ChainGptCstart}{118}
\newcommand{\ChainGptEdges}{1{,}146{,}959}
\newcommand{\ChainGptPruned}{17{,}442}
\newcommand{\ChainGptQD}{167{,}515}
\newcommand{\ChainGptQspan}{129}
\newcommand{\ChainGptQstart}{118}
\newcommand{\ChainGptRows}{41 & 49 & 4 & 2 & 4 & 2 & 3 \\
100 & 120 & 5 & 4 & 5 & 4 & 5 \\
200 & 269 & 8 & 6 & 8 & 5 & 10 \\
400 & 638 & 14 & 10 & 14 & 10 & 11 \\
800 & 1{,}452 & 19 & 17 & 19 & 17 & 13 \\
1{,}600 & 3{,}287 & 28 & 22 & 28 & 22 & 15 \\
3{,}200 & 7{,}644 & 34 & 31 & 34 & 31 & 33 \\
6{,}400 & 17{,}579 & 51 & 44 & 51 & 44 & 64 \\
12{,}800 & 38{,}698 & 68 & 62 & 68 & 62 & 40 \\
25{,}600 & 82{,}483 & 98 & 88 & 98 & 88 & 54 \\
50{,}000 & 167{,}515 & 129 & 118 & 129 & 118 & 71 \\}
\newcommand{\ChainGptSources}{8{,}254}
\newcommand{\ChainGptSpanRules}{67}

\newcommand{\ChainGptStartRules}{70}
\newcommand{\ChainGptStartString}{122}
\newcommand{\ChainGptStates}{71}
\newcommand{\ChainGptTwoHundredCstart}{6}
\newcommand{\ChainGptTwoHundredQstart}{5}
\newcommand{\ChainGraphBelowQ}{1{,}687}
\newcommand{\ChainGraphExactSpan}{2{,}669}
\newcommand{\ChainGraphExactStart}{2{,}604}
\newcommand{\ChainGraphSlackSpan}{7}
\newcommand{\ChainGraphSlackStart}{72}

\newcommand{\ChainLargest}{17}

\newcommand{\ChainNonstatePairs}{10}

\newcommand{\ChainNonstateProductStart}{2}

\newcommand{\ChainNonstateStart}{3}
\newcommand{\ChainOrdered}{2{,}676}

\newcommand{\ChainPrefixClosed}{5{,}980}
\newcommand{\ChainProducts}{3{,}808}

\newcommand{\ChainStrings}{55{,}192}
\newcommand{\ChainSuffixStates}{4{,}430}
\newcommand{\ChainTables}{1{,}904}

\newcommand{\ClAdmissible}{100{,}000}
\newcommand{\ClEffective}{305}
\newcommand{\ClInfiniteHundred}{6}
\newcommand{\ClLeastMerge}{0}
\newcommand{\ClMaxFiniteHundred}{5}
\newcommand{\ClMerges}{100{,}000}

\newcommand{\ClSelfPairs}{305}
\newcommand{\ClSymbolsHundred}{39}

\newcommand{\ClassBytesDigits}{10}
\newcommand{\ClassBytesLetters}{52}
\newcommand{\ClassBytesSymbols}{60}
\newcommand{\ClassBytesWhitespace}{6}
\newcommand{\ClassContractionForms}{7}
\newcommand{\ClassRadiusContractions}{0}
\newcommand{\ClassRadiusDigits}{42}
\newcommand{\ClassRadiusLetters}{129}
\newcommand{\ClassRadiusSymbols}{65}
\newcommand{\ClassRadiusWhitespace}{1}
\newcommand{\ClauseBDicts}{3{,}733}
\newcommand{\CompletionUnordered}{608}
\newcommand{\CppDictionaries}{788}
\newcommand{\CppVerdicts}{1{,}879}
\newcommand{\CrossDefFour}{124{,}123}
\newcommand{\CrossDefThree}{124{,}160}
\newcommand{\CrossDictionaries}{200}
\newcommand{\CrossDiffer}{37}
\newcommand{\CrossImproperInputs}{124{,}160}
\newcommand{\CrossInputs}{64{,}000}
\newcommand{\DensityBRadius}{4}
\newcommand{\DensityBounds}{847}
\newcommand{\DensityBoundsTight}{732}
\newcommand{\DensityConstructive}{25{,}620}
\newcommand{\DensityCuts}{1{,}070{,}499}

\newcommand{\DensityDead}{293}
\newcommand{\DensityDense}{633}

\newcommand{\DensityFinite}{854}
\newcommand{\DensityFourRule}{1{,}500}
\newcommand{\DensityGptFortyOneQs}{2}
\newcommand{\DensityGptFortyOneWidth}{6}
\newcommand{\DensityGptR}{3}
\newcommand{\DensityGptSevenQs}{1}
\newcommand{\DensityGptTwentyQs}{1}

\newcommand{\DensityInert}{14}
\newcommand{\DensityInfinite}{1{,}915}
\newcommand{\DensityJointBelow}{482}
\newcommand{\DensityJointExact}{386}
\newcommand{\DensityLeftInputs}{39{,}160}

\newcommand{\DensityOffsetBelow}{561}
\newcommand{\DensityOffsetExact}{372}
\newcommand{\DensityOrdered}{803}
\newcommand{\DensityPinned}{425{,}557}
\newcommand{\DensityPinnedSampled}{10{,}613}

\newcommand{\DensityPool}{2{,}769}
\newcommand{\DensityPositive}{443}
\newcommand{\DensityPrefixBelow}{326}
\newcommand{\DensityPrefixExact}{397}
\newcommand{\DensityQsZero}{411}

\newcommand{\DensitySeeded}{785}
\newcommand{\DensitySparse}{221}
\newcommand{\DensitySuffixBelow}{344}
\newcommand{\DensitySuffixExact}{386}
\newcommand{\DensityTight}{842}
\newcommand{\DensityTightSpanSlack}{5}
\newcommand{\DensityTrials}{18{,}000}

\newcommand{\DirectBounded}{502}
\newcommand{\DirectBrute}{246{,}840}
\newcommand{\DirectProducts}{968}
\newcommand{\DirectUnbounded}{466}
\newcommand{\EditCensusAbove}{60}
\newcommand{\EditCensusEqual}{424}
\newcommand{\EmittedPrefixes}{10{,}350}
\newcommand{\ExampleFourStates}{49{,}715}
\newcommand{\FiveEditAttaining}{24}
\newcommand{\FiveEditMax}{11}
\newcommand{\FiveEditTables}{308{,}816}
\newcommand{\FourRuleOrdered}{9{,}792}
\newcommand{\FourRuleTables}{9{,}820}
\newcommand{\FourRuleUnbounded}{8{,}030}
\newcommand{\FourRuleUnordered}{28}
\newcommand{\FrozenMismatches}{0}
\newcommand{\GenCanonAgreements}{151{,}224}
\newcommand{\GenCanonAtTwoMOne}{224}
\newcommand{\GenCanonClassPairs}{11{,}217{,}216}
\newcommand{\GenCanonClasses}{4{,}737}

\newcommand{\GenCanonMinimal}{6{,}934}

\newcommand{\GenCanonPairsByPairs}{4{,}289}

\newcommand{\GenCanonSharpBrute}{5}
\newcommand{\GenCanonSharpTop}{1{,}024}
\newcommand{\GenCanonShortestPairs}{5{,}074}
\newcommand{\GenCanonShortestSame}{4{,}003}
\newcommand{\GenCanonShortestWords}{7{,}833}

\newcommand{\GenCanonWitnessTokenizations}{99{,}069}
\newcommand{\GptEffective}{244}

\newcommand{\GptFinitePrefix}{41}

\newcommand{\GptInfiniteEightHundred}{30}

\newcommand{\GptInfiniteFourHundred}{15}
\newcommand{\GptInfiniteHundred}{3}
\newcommand{\GptInfiniteTwoHundred}{6}
\newcommand{\GptMaxFiniteEightHundred}{9}

\newcommand{\GptMaxFiniteFourHundred}{9}
\newcommand{\GptMaxFiniteHundred}{6}
\newcommand{\GptMaxFiniteTwoHundred}{7}
\newcommand{\GptMirroredFortyOne}{3}
\newcommand{\GptNodesEightHundred}{26{,}075}
\newcommand{\GptRadiusEarly}{1}
\newcommand{\GptRadiusEarlyLast}{6}

\newcommand{\GptRadiusLate}{3}
\newcommand{\GptRules}{50{,}000}
\newcommand{\GptSelfPairs}{244}
\newcommand{\GptSymbolsEightHundred}{78}

\newcommand{\GptSymbolsFourHundred}{65}
\newcommand{\GptSymbolsHundred}{32}
\newcommand{\GptSymbolsTwoHundred}{53}
\newcommand{\GrowthBoundOne}{13}
\newcommand{\GrowthBoundThree}{781}
\newcommand{\GrowthBoundTwo}{85}
\newcommand{\GrowthMaxOne}{12}
\newcommand{\GrowthMaxThree}{276}
\newcommand{\GrowthMaxTwo}{53}
\newcommand{\GrowthMedianOne}{11.5}
\newcommand{\GrowthMedianThree}{62.5}
\newcommand{\GrowthMedianTwo}{17.0}
\newcommand{\GrowthRatio}{0.10}
\newcommand{\GrowthSlackThree}{2.8}
\newcommand{\HardBoundFour}{8}
\newcommand{\HardBoundOne}{8}
\newcommand{\HardCases}{4}
\newcommand{\HardCasesPassed}{4}
\newcommand{\LabelCensusProducts}{968}

\newcommand{\LabelFixtureUnbounded}{6}
\newcommand{\LabelFourRuleTables}{9{,}820}
\newcommand{\LabelFourSampleAbove}{16}
\newcommand{\LabelFourSampleEqual}{1{,}984}
\newcommand{\LabelFourSampleTables}{1{,}000}
\newcommand{\LabelPoolFive}{308{,}816}
\newcommand{\LabelPoolFour}{9{,}792}
\newcommand{\LabelPrependMaxOne}{1}
\newcommand{\LabelPrependMaxThree}{4}
\newcommand{\LabelPrependMaxTwo}{3}
\newcommand{\LabelRandomTables}{200}
\newcommand{\LabelRandomUnbounded}{2}
\newcommand{\LassoCertificates}{469}
\newcommand{\LassoNegative}{57}
\newcommand{\LassoNonneg}{412}
\newcommand{\LeastWidthAAttaining}{2}

\newcommand{\LeastWidthAFailNine}{8}

\newcommand{\LeastWidthAL}{6}
\newcommand{\LeastWidthALeast}{10}
\newcommand{\LeastWidthAM}{9}
\newcommand{\LeastWidthAQD}{15}
\newcommand{\LeastWidthAQspan}{6}
\newcommand{\LeastWidthAQstart}{4}
\newcommand{\LeastWidthAR}{6}
\newcommand{\LeastWidthASlack}{1}

\newcommand{\LeastWidthAWindowsNine}{512}

\newcommand{\LeastWidthAWindowsTen}{1{,}024}

\newcommand{\LeastWidthBLeast}{7}

\newcommand{\LeastWidthBQstart}{4}
\newcommand{\LeastWidthBR}{4}

\newcommand{\LeastWidthCLeast}{10}

\newcommand{\LeastWidthCQstart}{4}
\newcommand{\LeastWidthCR}{7}

\newcommand{\LeastWidthDrawn}{100{,}000}
\newcommand{\LeastWidthFails}{2}
\newcommand{\LeastWidthFinite}{72{,}815}
\newcommand{\LeastWidthHard}{42{,}271}

\newcommand{\LeastWidthReplayed}{1{,}168}
\newcommand{\LlamaAdmissible}{127{,}066}
\newcommand{\LlamaClauseA}{104{,}617}
\newcommand{\LlamaEffective}{454}
\newcommand{\LlamaLeastMerge}{0}
\newcommand{\LlamaMerges}{127{,}066}
\newcommand{\LlamaNotWhole}{588}
\newcommand{\LlamaNotWholeLeast}{100{,}769}

\newcommand{\LlamaPublishedFail}{1}
\newcommand{\LlamaPublishedMerges}{280{,}147}
\newcommand{\LlamaPublishedOrdered}{14}
\newcommand{\LlamaPublishedSelfPairs}{3}
\newcommand{\LlamaRankPairDisagree}{110}

\newcommand{\LlamaSelfPairs}{454}
\newcommand{\LlamaSelfSplit}{127{,}156}
\newcommand{\LlamaShared}{99{,}744}
\newcommand{\LlamaUnreachable}{678}
\newcommand{\LlamaVocabTokens}{127{,}744}
\newcommand{\LookAlphabets}{10{,}257}
\newcommand{\LookAlphabetsCovered}{10{,}196}
\newcommand{\LookAlphabetsEqual}{10{,}233}
\newcommand{\LookBelowStart}{1{,}225}
\newcommand{\LookCovered}{2{,}750}
\newcommand{\LookDelimiterProbes}{8{,}040{,}853}
\newcommand{\LookDelimiterSymbols}{1{,}093}
\newcommand{\LookEqual}{2{,}757}
\newcommand{\LookGptAfter}{118}
\newcommand{\LookGptBound}{6{,}400{,}000}
\newcommand{\LookGptConstant}{119}
\newcommand{\LookGptFirst}{2}
\newcommand{\LookGptLongest}{128}
\newcommand{\LookGptString}{120}
\newcommand{\LookHarmlessDicts}{1{,}418}
\newcommand{\LookHarmlessProbes}{378{,}312}
\newcommand{\LookLeftPositive}{3{,}980}
\newcommand{\LookLeftZero}{2{,}048}
\newcommand{\LookNone}{293}
\newcommand{\LookOrdered}{2{,}740}
\newcommand{\LookPairs}{2{,}476}
\newcommand{\LookRevDicts}{2{,}771}
\newcommand{\LookRevReplayed}{4{,}079}
\newcommand{\LookRevSymbols}{6{,}028}
\newcommand{\LookZero}{1{,}949}
\newcommand{\LookaheadCensus}{10{,}304}
\newcommand{\LookaheadMaxRadius}{21}
\newcommand{\LookaheadSeeded}{20{,}000}
\newcommand{\MeasureCases}{1{,}980{,}528}
\newcommand{\MeasureMismatches}{140{,}194}
\newcommand{\MeasureSeededCases}{2{,}000}
\newcommand{\MeasureSeededDicts}{400}
\newcommand{\MeasureSeededUnordered}{169}
\newcommand{\MirroredDictsStrict}{84}
\newcommand{\MirroredStrict}{182}
\newcommand{\MirroredTwoConstant}{1}
\newcommand{\MirroredTwoOrigin}{2}
\newcommand{\MirroredUnbounded}{466}
\newcommand{\MirroredVerdicts}{502}
\newcommand{\NoselfEditFour}{7}
\newcommand{\NoselfEditOne}{2}
\newcommand{\NoselfEditThree}{5}
\newcommand{\NoselfEditTwo}{4}
\newcommand{\NoselfInsFour}{6}
\newcommand{\NoselfInsOne}{1}
\newcommand{\NoselfInsThree}{4}
\newcommand{\NoselfInsTwo}{3}
\newcommand{\NoselfSubFour}{7}
\newcommand{\NoselfSubOne}{2}
\newcommand{\NoselfSubThree}{5}
\newcommand{\NoselfSubTwo}{4}
\newcommand{\OAdmissible}{199{,}742}
\newcommand{\OEffective}{513}
\newcommand{\OInfiniteHundred}{3}
\newcommand{\OLeastMerge}{0}
\newcommand{\OMaxFiniteHundred}{5}
\newcommand{\OMerges}{199{,}742}

\newcommand{\OSelfPairs}{513}
\newcommand{\OSymbolsHundred}{48}

\newcommand{\ObstacleDistances}{\(1, 2, 2, 2, 5\)}
\newcommand{\OverlapAgree}{968}
\newcommand{\OverlapCanonical}{85{,}248}
\newcommand{\OverlapCertificates}{1{,}861}
\newcommand{\OverlapDictionaries}{618}
\newcommand{\OverlapFixtures}{14}
\newcommand{\OverlapInputs}{178{,}147}
\newcommand{\OverlapRandom}{120}
\newcommand{\PathMaxIdentical}{55{,}587}
\newcommand{\PathMaxInputs}{612{,}780}
\newcommand{\PathMaxProducts}{982}
\newcommand{\PathMaxTerminal}{42{,}392}
\newcommand{\PathMaxTerminalNegative}{36{,}916}
\newcommand{\PendingMax}{3}
\newcommand{\PhaseInputs}{247{,}324}
\newcommand{\PlateauFive}{8}
\newcommand{\PlateauFour}{32}
\newcommand{\PlateauOne}{2}
\newcommand{\PlateauSix}{2}
\newcommand{\PlateauThree}{40}
\newcommand{\PlateauTwo}{30}
\newcommand{\PrefixCases}{987{,}360}
\newcommand{\PrependExactFive}{7}
\newcommand{\PrependExactFour}{5}
\newcommand{\PrependExactOne}{1}
\newcommand{\PrependExactThree}{4}
\newcommand{\PrependExactTwo}{3}
\newcommand{\PrependFiveAttaining}{16}
\newcommand{\PrependFiveTables}{308{,}816}
\newcommand{\ProductAgree}{484}
\newcommand{\ProductExact}{114}

\newcommand{\RealAlphabet}{20}
\newcommand{\RealMirroredEight}{2}
\newcommand{\RealMirroredFour}{1}
\newcommand{\RealMirroredTwelve}{2}
\newcommand{\RealRadiusEight}{3}
\newcommand{\RealRadiusFour}{1}
\newcommand{\RealRadiusThirtyTwo}{3}
\newcommand{\RealRadiusTwelve}{3}
\newcommand{\RealStates}{131{,}245}
\newcommand{\RealTail}{3}
\newcommand{\RepairedBounded}{114}
\newcommand{\RepairedUnbounded}{370}
\newcommand{\ReprobeExceeded}{0}
\newcommand{\SelfPairInactive}{2}
\newcommand{\SelfPairSyntactic}{372}
\newcommand{\SlackFamilyTop}{12}
\newcommand{\StaircaseHundred}{1{,}050}
\newcommand{\StaircaseInsertHundred}{1{,}112}
\newcommand{\StaircaseSubEleven}{23}
\newcommand{\StaircaseSubHundred}{1{,}110}
\newcommand{\StaircaseSubNine}{15}
\newcommand{\StaircaseSubTen}{19}
\newcommand{\TailAttaining}{748}
\newcommand{\TailClassPairs}{5{,}268}
\newcommand{\TailClasses}{7{,}320}
\newcommand{\TailComparisons}{2{,}447}
\newcommand{\TailDictionaries}{1{,}096}
\newcommand{\TailFinite}{1{,}431}
\newcommand{\TailInfinite}{1{,}016}
\newcommand{\TailPairs}{195{,}958}
\newcommand{\TailQuotientPairs}{12{,}967}
\newcommand{\TailRefinedBelow}{342}
\newcommand{\TailReset}{290}
\newcommand{\TailRounds}{9}
\newcommand{\TailStringBrute}{1{,}308}
\newcommand{\TailStringFinite}{4{,}030}
\newcommand{\TailStringTargets}{5{,}659}
\newcommand{\TailSuffixes}{102{,}997}
\newcommand{\TailUnordered}{608}
\newcommand{\TailUnorderedStates}{9{,}598}
\newcommand{\UnaryRightTables}{2{,}036}
\newcommand{\VocabSeededStrings}{1{,}000}
\newcommand{\WholeByB}{354}
\newcommand{\WholeBySpan}{2{,}136}
\newcommand{\WholeDictionaries}{2{,}490}
\newcommand{\WholeInTrie}{1{,}803}
\newcommand{\WholeInputs}{1{,}492}
\newcommand{\WholePrefixes}{21{,}245}
\newcommand{\WholeWithB}{373}
\newcommand{\WitnessTables}{1{,}904}
\newcommand{\WitnessTight}{272}
\newcommand{\ZeroDelayBelowQ}{1{,}369}
\newcommand{\ZeroDelayBrute}{35{,}644}
\newcommand{\ZeroDelayDictionaries}{2{,}490}
\newcommand{\ZeroDelayInputs}{9{,}960}
\newcommand{\ZeroDelayPrefixes}{134{,}084}
\newcommand{\ZeroDelayUnordered}{924}

\hypersetup{
  pdftitle={The Left Ripple of Byte-Pair Encoding: Certified Lookbehind for Every Dictionary},
  pdfauthor={Nicklas Nidhögg},
  pdfsubject={Byte-pair encoding; tokenization; incremental retokenization; finite automata},
  pdfkeywords={byte-pair encoding, tokenization, lookbehind, retokenization, certificates}
}

\theoremstyle{definition}
\newtheorem{definition}{Definition}

\theoremstyle{plain}
\newtheorem{lemma}{Lemma}
\newtheorem{theorem}{Theorem}
\newtheorem{proposition}{Proposition}
\newtheorem{corollary}{Corollary}

\newcommand{\code}[1]{\texttt{#1}}
\newcommand{\T}{\mathbb{T}}
\newcommand{\ripple}{R_{L}}
\newcommand{\yield}{\mathrm{yield}}
\newcommand{\Stable}{\mathrm{Stable}}
\newcommand{\Csym}{C^{\mathrm{sym}}}
\newcommand{\Eedit}{E^{\mathrm{edit}}}

\title{The Left Ripple of Byte-Pair Encoding:\\Certified Lookbehind for Every Dictionary}
\author{Nicklas Nidh\"ogg\\[0.4em] \normalsize Independent Researcher\\ \normalsize
\texttt{nicklas.nidhogg@gmail.com}\\ \normalsize ORCID: 0009-0006-6161-6150}
\date{November 9, 2026}

\begin{document}

\maketitle

\begin{abstract}
Berglund and van der Merwe formalized byte-pair encoding and bounded how far an appended string can change a
tokenization, leaving the converse open: how far can a string prepended on the left change it, which dictionaries make
that ripple infinite, and can it be computed efficiently when finite? We answer all three for every dictionary. The
origin radius, how far a prepend can change covering spans, is decided with radius certificates by a product of two runs
of the automaton that guesses the final tokens, in time polynomial in the dictionary written out, and it is infinite
exactly when one periodic word has two canonical tilings with no common boundary. Under their ordering condition it is
infinite exactly when some rule \((t, t)\) finds its two copies adjacent at its own stage, and such a rule makes any
dictionary infinite. A finite radius makes every window of a width read off the radii have a cut, and the radius is
finite exactly when such cuts are dense. The token machine completes for every dictionary, and the tokenization is
subsequential, extending what Berglund, Martens and van der Merwe proved for proper ones under one pass per rule. A
streaming tokenizer emits after every symbol exactly the tokens that no continuation changes, in polynomial time,
answering a question of Dalirrooyfard and Ye. Boundedness of the label measure, the edit distance of the token
sequences, is decided for every dictionary in polynomial time as well, and under the ordering condition the largest
distance one edit causes over dictionaries of \(m\) rules lies asymptotically between \(m^2/10\) and \(m^2/2\). GPT-2's
merge table carries \GptSelfPairs{} effective self-pairs, so its origin radius is infinite from the forty-second merge
on, and that of cl100k, o200k and Llama 3 from their first. Every measured number is its program's output, and what is
not executed is named.
\end{abstract}

\section{Introduction}

A byte-pair tokenizer applies a dictionary of merge rules in priority order until none applies. Its output on a string
depends on the whole string. Incremental use asks how much of a tokenization survives an edit and how much must be
recomputed, and both questions reduce to how far a local change can propagate. For appending,
\citet{berglund2023formalizing} proved the propagation bounded by \(|D|\) tokens and defined a lookahead constant. For
strings appended on the left they gave an example with infinite propagation. Their closing section asks whether the
dictionaries exhibiting these infinite ripples do so due to some easily decidable property, and whether the lookbehind
is efficiently computable when finite. They name the lookbehind in quotation marks, by analogy with their Definition~6,
and leave its definition open. From random testing they recorded that the ripple seems to often be finite. An answer
therefore needs three things: a definition of how far a prepend reaches, a property that decides which dictionaries make
that reach infinite, and an efficient computation of it when finite.

This paper supplies all three, with two definitions for the first. For every dictionary the origin radius is decidable
and exactly computable with executable certificates, in polynomial time. Under the ordering condition the dictionaries
with infinite radius are exactly those with an effective self-pair rule, a property read off the rules. A streaming
tokenizer emits after every symbol exactly the tokens that no continuation changes, in polynomial time, which answers
the question \citet{dalirrooyfard2026online} leave open, whether byte-pair encoding alone admits zero delay. The
constants are computed on a census of \CensusTotal{} small dictionaries and on GPT-2's merge table, and the self-pair
threshold on three more production tables. All costs here are in the dictionary written out, where token lengths can be
exponential in the number of rules. The compact companion~\citep{nidhogg2026compact} measures costs for the same
questions by the size of a straight-line program over the dictionary's operands. Nothing here depends on it.

Throughout, results of \citet{berglund2023formalizing} are cited as the source paper's Definition~3, Lemma~1 and so on,
by the numbering of that text. Where the revised paper \citep{berglund2024formalizing} is cited, its own numbering is
named, as in the revised Definition~5(b). Unqualified numbers refer to this paper's own statements.

The \emph{origin ripple} counts the positions up to and including the furthest one whose covering span changes, which is
also the furthest one whose token origin, its covering token's start, moves (Proposition~\ref{prop:sides}), and is zero
when none changes. Its supremum is the origin radius. The \emph{label distance} is the edit distance between the two
token sequences. The former measures the region an implementation that maintains spans may need to revisit, and the
latter the change seen by a consumer of the token stream. A bounded origin ripple bounds label distance
(Proposition~\ref{prop:labels}), but not conversely: the one-rule dictionary \((a, a)\) is bounded in labels and
unbounded in origins. Under the ordering condition the label distance is bounded per dictionary, while the origin radius
is infinite on \CensusUnbounded{} of \CensusTotal{} census dictionaries. On that census an infinite radius is the common
case, which reads against the source paper's impression from random testing. No claim about a distribution on
dictionaries is made from the census. That a proper, ordered dictionary can have infinite reach comes from the source
paper. Example~3 of the revised paper is \((a, b), (ab, ab)\), and prepending \(a\) to \(b(ab)^k\) changes the covering
span of every position. The census counts infinite reach on the smallest dictionaries.

The contributions are:

\begin{itemize}
\item The lookbehind is defined twice, as the origin ripple and the label distance, which are not equivalent, the second
being at most one above the first (Proposition~\ref{prop:labels}), and the mirror of the source paper's Definition~6
lies at most \(M - 1\) below the first and never above it, \(M\) the longest token (Proposition~\ref{prop:mirror}).
Canonical adjacency (Theorem~\ref{thm:bigram}) is the bigram characterization of \citet{vieira2025canonical}, stated
before for ranked token lists in the documentation of GitHub's \code{bpe} crate \citep{vanantwerpen2024bpe}, and the
proof here makes its projection lemma explicit. For every dictionary an appended string changes the tokens of at most
\(Q(D)\) trailing symbols (Theorem~\ref{thm:lookahead}), the delay bound of \citet{mamouras2026bpe} for trained
rulesets, so the tokenization is subsequential (Corollary~\ref{cor:subsequential}), which
\citet[Theorem~4]{berglund2024dfa} proved for proper dictionaries under one pass per rule, the tokenizer of their
Algorithm~1. The tokens that no continuation changes are the common prefix over finitely many continuations, and a
streaming tokenizer emits exactly those after every symbol in polynomial time (Theorems~\ref{thm:zerodelay}
and~\ref{thm:zerodelayauto}), and so does a tokenizer that looks a whole piece up in its vocabulary before merging
(Corollary~\ref{cor:wholepiece}). Emitting that prefix from a compiled subsequential transducer is classical
\citep{mohri1994compact}, and the polynomial procedure answers the question \citet{dalirrooyfard2026online} leave open
for byte-pair encoding.
\item The overlap graph (Theorem~\ref{thm:overlap}) decides the origin radius for every dictionary and computes it
exactly from a product of two runs of the automaton that guesses the final tokens, with radius certificates in both
directions, in polynomial time. The radius is infinite exactly when one periodic word has two canonical tilings with no
common boundary (Theorem~\ref{thm:circular}). Under the ordering condition it is infinite exactly when some rule
\((t, t)\) is effective at its own stage (Theorem~\ref{thm:selfpair}), for every dictionary such a rule makes it
infinite (Theorem~\ref{thm:selfgeneral}), and a six-rule dictionary with clause (a) alone shows the condition is needed
(Proposition~\ref{prop:noselfpair}). The exploration of the token machine built by the source paper's Algorithm~1
completes for every dictionary (Corollary~\ref{cor:completion}), and the machine's quotient by future last-token
behavior gives the radius under every prepended string from its transitions alone, by the contraction of definite
automata \citep{steinby1969definite} (Corollary~\ref{cor:tail}).
\item With the right start radius \(Q_{\mathrm{start}}(D)\), at most \(\max(0, Q(D) - 1)\), every window of width
\(\ripple(D) + Q_{\mathrm{start}}(D) + 1\) pins its byte \(\ripple(D)\) in every context and so has a cut
(Theorem~\ref{thm:pinned}), and the radius is finite exactly when cuts are dense at some width
(Theorem~\ref{thm:density}). Under the ordering condition the chain search computes both right radii exactly from the
rules (Theorem~\ref{thm:chainsearch}). On GPT-2's whole table the right span radius, the streaming delay of
\citet{mamouras2026bpe}, is \ChainGptQspan{}, where their Theorem~10, the sum of the left operands, gives \ChainGptQD{}.
The lookahead constant of the source paper's Definition~6 is computed for every dictionary with an attaining input, and
it is \LookGptConstant{} on GPT-2's table (Corollary~\ref{cor:lookconst}).
\item Boundedness of the label distance under a prepended symbol, and under one symbol edit anywhere, is decided for
every dictionary in time polynomial in the overlap graph, with a pumpable certificate when unbounded
(Theorem~\ref{thm:labelgraph}). When it is bounded the exact supremum is computed with an attaining input, by profiles
exponential in the value and the longest token. That it is decidable at all follows from Corollary~\ref{cor:rational}
and the edit distance of functional transducers of \citet{aiswarya2024edit}. A bound under every prepended symbol gives
one under every edit (Corollary~\ref{cor:prependedit}). Under the ordering condition no single symbol edit moves the
tokenization by more than the Fibonacci number \(F_{2|D|+2}\) of tokens (Theorem~\ref{thm:label}), a bound nowhere
sharper than the quadratic one that follows, and the largest label distance one edit causes over ordered dictionaries of
\(m\) rules is \(\Theta(m^2)\), at least \(\lfloor (m^2 + 11m + 20)/10 \rfloor\) and at most \((m + 1)(m + 2)/2\) for
\(m \ge 1\) (Theorems~\ref{thm:amp} and~\ref{thm:quadratic}, Proposition~\ref{prop:staircaseinsert}), with the exact
constant computed over finite universes (Section~\ref{sec:ampcomputed}).
\item GPT-2's whole merge table, read as a dictionary with the pre-tokenizer not modeled, is ordered, with
\GptSelfPairs{} effective self-pairs and the first at merge \GptFinitePrefix{}, counted from zero, so its origin radius
is infinite from the forty-second merge on, where the source paper's random testing suggested finiteness is common. The
merge tables of cl100k\_base, o200k\_base and Llama~3 begin with the rule that joins two spaces, effective at its stage,
so their origin radius is infinite from the first merge on (Section~\ref{sec:production}). The \code{tokenizers} library
implements the semantics the source paper calls SentencePiece, not the one it names after HuggingFace, and the ordering
condition this paper works under is theirs, clause (b) of their revised Definition~5 (Section~\ref{sec:semantics}).
\end{itemize}

Table~\ref{tab:results} lists the main results with their hypotheses and costs. Canonical adjacency, the decidability of
the label distance, the subsequentiality and the equivalence test of proper dictionaries under one pass per rule, the
producers of Lemma~\ref{lem:gtrees}, the product of two runs behind the overlap graph
\citep[Section~1]{beal1998synchronization}, the cycle test of the second form of Theorem~\ref{thm:circular}, the
contraction behind Corollary~\ref{cor:tail} and the emission of the stable output from a compiled transducer are prior
results, credited where they are used. The exact origin radius with its certificates, the effective self-pair criterion,
the width at which every window has a cut and the equivalence of a finite radius with dense cuts, the chain search, the
exact lookahead constant, the label decision in polynomial time and the Fibonacci and quadratic bounds of the label
measure were not found in the literature that Section~\ref{sec:related} sets out, and are what the paper claims as its
own. Appendices~\ref{app:machine} to~\ref{app:labels} hold the longer proofs, the auxiliary lemmas and the sharpness
constructions, and Appendix~\ref{app:exec} lists the executions that check the results.

\begin{table}[!t]
\centering
\small
\begin{tabular}{@{}>{\raggedright\arraybackslash}p{0.34\textwidth}>{\raggedright\arraybackslash}p{0.17\textwidth}
  >{\raggedright\arraybackslash}p{0.16\textwidth}>{\raggedright\arraybackslash}p{0.23\textwidth}@{}}
\toprule
result & hypothesis & cost & statement \\
\midrule
origin radius decided and computed, with certificates & every dictionary & polynomial &
Theorem~\ref{thm:overlap} \\
\addlinespace
tail product of the token machine & every dictionary & polynomial in the machine & Corollary~\ref{cor:tail} \\
\addlinespace
canonical adjacency, rationality & every dictionary & & Theorem~\ref{thm:bigram} \\
\addlinespace
right lookahead of \(Q(D) \le 2^s - 1\) symbols, subsequentiality & every dictionary & &
Theorem~\ref{thm:lookahead} \\
\addlinespace
streaming with zero delay & every dictionary & polynomial & Theorem~\ref{thm:zerodelay} \\
\addlinespace
pinned bytes & finite radius & & Theorem~\ref{thm:pinned} \\
\addlinespace
a finite radius exactly when cuts are dense & every dictionary & & Theorem~\ref{thm:density} \\
\addlinespace
both right radii by the chain search & ordering condition
  & one test per candidate extension, exponentially many states in the span
radius & Theorem~\ref{thm:chainsearch} \\
\addlinespace
lookahead constant \(l(D)\) & every dictionary & polynomial & Corollary~\ref{cor:lookconst} \\
\addlinespace
circularity certificates & every dictionary & & Theorem~\ref{thm:circular} \\
\addlinespace
effective self-pair criterion & ordering condition & & Theorem~\ref{thm:selfpair} \\
\addlinespace
an effective self-pair makes the radius infinite & every dictionary & & Theorem~\ref{thm:selfgeneral} \\
\addlinespace
canonical words and pairs, equal tokenizers & every dictionary & polynomial &
Corollary~\ref{cor:equivalence} \\
\addlinespace
boundedness of the label measure, with pumpable certificates & every dictionary & polynomial &
Theorem~\ref{thm:labelgraph} \\
\addlinespace
the exact supremum of the label measure, with an attaining input & bounded label measure & by profiles,
exponential in the value and the longest token & Theorem~\ref{thm:labelgraph},
Lemma~\ref{lem:profiles} \\
\addlinespace
one edit moves at most \(F_{2m+2}\) and \((m + 1)(m + 2)/2\) tokens & ordering condition & &
Theorems~\ref{thm:label}, \ref{thm:quadratic} \\
\addlinespace
edit bounds read off prepend bounds & every dictionary & & Corollary~\ref{cor:prependedit} \\
\bottomrule
\end{tabular}
\caption{The main results, each with its hypothesis, the cost of its procedure where it has one, and its statement,
with \(m\) the number of rules and \(s\) the number that fire. Unless another parameter is named, a polynomial cost is
in the size of the dictionary written out.}
\label{tab:results}
\end{table}

\FloatBarrier
\section{Semantics}\label{sec:semantics}

We follow \citet{berglund2023formalizing}, Section~3. A dictionary \(D\) over an alphabet \(\Sigma\) of symbols is an
ordered list of rules \((u_i, v_i)\), higher priority first. A rule's \emph{rank} is its place in \(D\), counted from
one. Written out, a dictionary lists every rule's strings in full, and its size is the alphabet's size plus their total
length. Polynomial bounds below are in that size, unless another parameter is named. A tokenization \(\tau\) is a list
of strings, and its yield \(\yield(\tau)\) is their concatenation. One step replaces an adjacent pair \(u, v\) by
\(uv\).


\begin{definition}[The source paper's Definition 3]\label{def:sp}
\(\T^D(w)\) applies at every step the highest-priority applicable rule at its leftmost occurrence, one merge per step,
until none applies. The source paper names this after SentencePiece.
\end{definition}

\begin{definition}[The source paper's Definition 4]\label{def:hf}
At every step the highest-priority applicable rule is applied left to right until it no longer applies, then priorities
are reconsidered. The source paper names this after HuggingFace.
\end{definition}


\begin{definition}[The ordering condition]\label{def:ordered}
Every rule producing a token precedes every rule consuming it, that is, \(u_i v_i \in \{u_j, v_j\}\) implies \(i < j\).
\end{definition}

The ordering condition is not ours. The source paper's Definition~5 of the conference text calls a dictionary
\emph{proper} when some earlier rule produces each multi-symbol operand, and admits a later producer too. Their revised
Definition~5 \citep{berglund2024formalizing} asks for that as its clause (a), and as its clause (b) that no later rule
produce an operand. Clause (b) is the condition above. Call a rule \emph{useful} when it fires on some input. Their
Corollary~2 there derives the second clause from the first for dictionaries all of whose rules are useful. We keep the
name for the clause we work under and cite their Definition~5(b) for it. The two are the same condition, since no rule
can produce its own operand, so the execution is a transcription check, which adds no evidence. The clause taken from
its own sentence and the test of it agree on all \ClauseBDicts{} proper dictionaries of one to three rules on two and
three symbols.

Clause (a) alone is not enough for their Lemma~1. The dictionary \((a,a)\), \((a,b)\), \((a,ab)\), \((aab,aa)\),
\((aa,b)\) satisfies it and tokenizes \(aabaab\) as \(aabaa \mid b\) under Definition~\ref{def:sp} and as
\(aab \mid aab\) under Definition~\ref{def:hf}. Its lowest rule creates \(aab\), an operand of a higher one. One pass
per rule in priority order, the tokenizer of Algorithm~1 of \citet{berglund2024dfa}, gives \(aab \mid aab\) on it as
well, and \(aab \mid aa\) on \(aabaa\), where both definitions give the one token \(aabaa\); under the ordering
condition it is both definitions, by Lemma~\ref{lem:stab} below. The dictionary fails clause (b), so it is not proper in
the revised sense and contradicts nothing that edition states. It shows why the hypothesis used here is clause (b) and
not clause (a). The source paper's argument for its Lemma~3, that a lower-priority rule cannot create the pair of a
higher-priority one, is exactly clause (b), and under it that paper's results hold with their existing proofs.
Properness is not assumed beside it. A rule that never fires can be dropped, such as one with a composite operand that
no rule produces. Since the run with the first \(r\) rules is the beginning of the full run, no prefix run changes
either, so the condition need hold only on the rules that fire. The dictionary \((b, a)\), \((ab, a)\), \((a, b)\) fails
it as written and satisfies it on the two rules that fire. The source paper's arguments used below go through under the
ordering condition alone. The condition holds on all \CensusTotal{} census dictionaries, the \HardCases{} dictionaries
of Table~\ref{tab:exec-product} that lie outside the census, and GPT-2's whole merge table
(Section~\ref{sec:production}). The five-rule dictionary above and the six-rule dictionary of
Proposition~\ref{prop:sixrule} satisfy clause (a) and fail clause (b).


\begin{lemma}[Passes and stability under the ordering condition]\label{lem:stab}
Under the ordering condition, Definitions~\ref{def:sp} and~\ref{def:hf} coincide and the tokenization is the composition
of one pass per rule in priority order, each applied left to right to exhaustion. If \(\T^D(p) = u_1 \cdots u_n\) with
\(n > |D|\), then for every continuation \(z\), \(\T^D(pz)\) begins with \(u_1 \cdots u_{n - |D|}\).
\end{lemma}

\begin{proof}
Applying a rule can only enable rules of lower priority, so along any derivation under Definition~\ref{def:hf} the rule
indices never decrease. This is the source paper's Lemma~3 under clause (b), and the two definitions coincide, which is
the source paper's Lemma~1. A merge creates no new pair for its own rule, since \(uv\) is neither \(u\) nor \(v\). A
rule never becomes applicable again after its pass, because a merge's only new adjacencies involve the new token and
every producer of \(u_i\) or \(v_i\) precedes rule \(i\). The run is thus one pass per rule in order. The stability
statement is the ordered-pass argument of the source paper's Theorem~2.
\end{proof}

The implementation of both definitions is asserted on the source paper's three examples and cross-checked against the
\code{tokenizers} library. On \CrossDictionaries{} proper dictionaries and \CrossInputs{} inputs the two definitions and
the library agree. On \CrossImproperInputs{} inputs to improper dictionaries the library matches Definition~\ref{def:sp}
on all \CrossDefThree{} and Definition~\ref{def:hf} on \CrossDefFour{}, failing exactly on the \CrossDiffer{} where the
two definitions differ. The library thus implements the semantics the source paper calls SentencePiece and not the one
they name after HuggingFace, whose Definition~4 models the older Python GPT-2 implementation they cite. Every theorem
below names a definition and never a library.

\section{Two measures of ripple}\label{sec:measures}


\begin{definition}[Origin ripple]\label{def:origin}
For a symbol \(a\) and input \(x\), the origin ripple \(\rho(a, x)\) is one plus the largest position \(p\) of \(x\)
whose covering span in \(\T^D(x)\) differs, as an interval, from the covering span of the same position in \(\T^D(ax)\),
the latter's spans shifted left by one; zero when none differs. A token of the second run that also covers the prepended
symbol has a shifted span beginning at minus one and so differs wherever it reaches. For a nonempty string \(w\),
\(\rho(w, x)\) is the same measure with \(w\) prepended and the spans of \(\T^D(wx)\) shifted left by \(|w|\). The
origin radius \(\ripple(D)\) is the supremum over \(a\) and \(x\).
\end{definition}

On the one-rule dictionary \((a,a)\), \(\T^D(a^{2n})\) is \(n\) tokens \(aa\) and \(\T^D(a^{2n+1})\) is \(n\) tokens
\(aa\) followed by one \(a\). After the shift the first \(aa\) begins at minus one, so every span of the original moves
by one, the origin ripple is \(2n\), and the two label sequences differ by one token.

\begin{definition}[Label distance]\label{def:label}
Write \(d\) for edit distance on token sequences, one insertion, deletion or substitution of a token per unit. The label
distance \(\lambda(a, x)\) is \(d(\T^D(x), \T^D(ax))\).
\end{definition}


\begin{proposition}[Labels are bounded by origins]\label{prop:labels}
For every dictionary, symbol \(a\) and input \(x\), \(\lambda(a, x) \le \rho(a, x) + 1\).
\end{proposition}

\begin{proof}
Write \(r = \rho(a, x)\). Every position of \(x\) from \(r\) on has the same covering span in both runs, the second
run's shifted. That span is an interval of positions of \(x\), so the token covering it is the same substring of \(x\)
in both runs. If \(r < |x|\), the span of position \(r\) does not reach back to position \(r - 1\). When \(r \ge 1\),
that position has different spans in the two runs by the maximality of \(r\), and when \(r = 0\) the first run's spans
begin at position \(0\) or later. Then \(r\) begins a token in both runs, and the tokens from that one on are the same
in both, a shared suffix. When \(r = |x|\) the shared suffix is empty. The part before it covers \([0, r)\) in the first
run and \([-1, r)\) in the second, with at most \(r\) and \(r + 1\) tokens. Two sequences of at most \(r\) and \(r + 1\)
tokens are at edit distance at most \(r + 1\), substituting along the shorter and inserting or deleting the rest. The
shared suffix costs nothing, so \(\lambda(a, x) \le r + 1\).
\end{proof}


\begin{proposition}[Unbounded without the ordering condition]\label{prop:sixrule}
For the proper dictionary \((a,a)\), \((a,b)\), \((a,ab)\), \((b,aab)\), \((aab,b)\), \((aa,b)\), which fails the
ordering condition, and \(x = (aabb)^n\), \(\T^D(x)\) is \(n\) copies of \(aabb\) and \(\T^D(bx)\) is \(n\) copies of
\(baab\) followed by \(b\), giving label distance \(n + 1\) and origin ripple \(4n\).
\end{proposition}

\begin{proof}
Rule one merges every \(aa\), leaving blocks \(aa \mid b \mid b\), to which no rule of priority two, three or four
applies. From then on rule six creates one \(aab\) at the leftmost block and a higher rule consumes it before the next
block is touched, so the run proceeds block by block. In the base run, before block \(k\) the tokens are \(k - 1\)
copies of \(aabb\), ending in a \(b\) already absorbed, so rule four cannot fire and rule five merges \(aab\) with its
following \(b\). In the prepended run, before block \(k\) the tokens are \(k - 1\) copies of \(baab\) and one free
\(b\), the prepended one when \(k = 1\). Rule four merges that free \(b\) with the block's \(aab\), leaving the block's
own \(b\) free for the next. No target label equals \(aabb\), so the distance is at least \(n + 1\), and \(n\)
substitutions with one insertion attain it.
\end{proof}


\begin{lemma}[Longer prefixes add nothing]\label{lem:D}
The supremum of the origin ripple over arbitrary prepended strings equals the supremum over single symbols.
\end{lemma}

\begin{proof}
Prepend \(w\) symbol by symbol from the right. At stage \(i\) an original position \(p \ge \ripple\) sits at coordinate
\(p + i - 1 \ge \ripple\) of the string that stage prepends to, so the single-symbol prepend preserves its covering
span. Induction over the stages gives the claim.
\end{proof}

\section{The token machine}\label{sec:machine}

Fix a tail length \(B\). The machine's state after reading \(p\) is a retained suffix \(S(p)\) of at most \(B\) tokens,
with the invariant \(\T^D(p) = E(p)\,S(p)\), where \(E(p)\) is the emitted prefix. \(S(p)\) is not in general the last
\(B\) tokens of \(\T^D(p)\). On the dictionary \((b,c), (a,bc)\) with \(B = 3\), the state after \(aaaaabc\) is
\(a \mid abc\), while the last three tokens are \(a \mid a \mid abc\). They differ because a merge shortened a suffix
whose earlier tokens had already been emitted. The invariant is checked by execution on \EmittedPrefixes{} prefixes.

The transition on a byte \(c\) is the source paper's Algorithm~1, specialized to one appended byte and run on the
retained suffix alone. For \(j = 1, 2, \dots\), it tokenizes the last \(j\) tokens of the suffix followed by \(c\) in
isolation and halts at the first \(j\) for which the result begins with the \(j\)th token from the end. The new suffix
is the retained suffix less its last \(j\) tokens, followed by that result. This is the object the proof of the source
paper's Theorem~1 establishes. Its printed step~3 would drop the result's last token and append \(c\), which is wrong
whenever \(c\) was merged. Whether a suffix was truncated is a property of the path to it. A backup that would step past
the first kept token of a truncated suffix raises instead of retokenizing a partial string. The flag decides only
whether a backup raises, and the successor of a suffix on a byte does not depend on it, so once the exploration has
completed the transition is a function of the suffix and the byte alone. The exploration of the machine from the empty
suffix over all bytes either completes or raises, and the tail length \(B\) is searched upward from one. The proofs of
Lemmas~\ref{lem:A} and~\ref{lem:B} below are in Appendix~\ref{app:machine}.


\begin{lemma}[Correctness when exploration completes]\label{lem:A}
If the exploration at \(B\) completes, every emitted token is final, that is, for every continuation \(z\), \(\T^D(pz)\)
begins with \(E(p)\).
\end{lemma}


\begin{lemma}[The backup halts inside the retained suffix, and the exploration completes]\label{lem:B}
Let \(D\) satisfy the ordering condition, \(m = |D|\), \(B = m + 1\). Then on every transition the backup halts inside
the retained suffix, and the exploration at \(B\) completes.
\end{lemma}

\section{The frozen-token product and the origin theorem}\label{sec:product}

Run the machine on \(x\) from the empty suffix and on \(ax\) from the suffix after \(a\), stepping both on each byte of
\(x\). By Lemma~\ref{lem:A} a token that has left the retained suffix is final, so the two runs can be compared token by
token once both have frozen the tokens covering a position.


\begin{definition}[The product]\label{def:product}
A node of the product is a pair of retained suffixes together with each run's frozen tokens not yet compared with the
other's, the pending tokens, and each run's \emph{freeze point}, the position up to which it has frozen. An edge reads
one byte, steps both runs, compares the covering tokens on the positions the lesser freeze point passes at that step,
and carries the position of the last disagreement it reveals, if any, an event. A node carries as its terminal
contribution the position of the last disagreement visible if the input ended there, if any. A contribution at position
\(p\) counts \(p + 1\), Definition~\ref{def:origin}'s convention, and is stored as an offset relative to the node's own
position, so that a contribution read at a node reached after \(k\) bytes has value \(k\) plus the offset.
\end{definition}

The \emph{start node} of the product is its node before any byte of \(x\) is read, pairing the empty suffix with the
suffix after \(a\).


\begin{lemma}[The product is finite]\label{lem:finite}
The product has finitely many nodes.
\end{lemma}


\begin{lemma}[The ripple is the path maximum]\label{lem:max}
At any tail length at which the exploration completes, for every prepended symbol and every input the origin ripple is
the largest of the events the path of the input \(x\) records, the terminal contribution of the node it ends in, and
zero. When that node carries a terminal contribution of offset \(e\), the ripple is \(|x| + e\), whatever the sign of
\(e\).
\end{lemma}

\begin{lemma}[Identical runs are a dead end]\label{lem:dead}
If at a node the two runs have the same retained suffix, the same pending tokens and the same freeze point, then no
later edge records an event and no later node carries a terminal contribution, so giving the node no successors changes
no value Lemma~\ref{lem:max} reads.
\end{lemma}

The proofs of Lemmas~\ref{lem:finite}, \ref{lem:max} and~\ref{lem:dead} are in Appendix~\ref{app:machine}. The
construction therefore gives no successors to a node whose two runs have the same suffixes, pending tokens and freeze
points. Such a node reaches no contribution, so the subgraph of nodes from which a contribution is reachable is
unchanged, and with it the cycle verdict and every value the theorem reads.


Neither part of the maximum alone is the ripple. Take \((b,c), (a,bc)\) with \(a\) prepended. The input \(b\) has no
disagreement at any of its positions, while \(bc\) ripples by two. The prepended \(a\) sits in the second run's retained
suffix until the rule that consumes \(bc\) reaches it, so the reading at a node bounds no later one. In the other
direction an event already frozen behind a node can exceed that node's own reading. Table~\ref{tab:exec-product} lists
the executions of the results of this section.

The comparison of final covering tokens is exact per input. On every census dictionary, both prepends and every input to
length ten, \MeasureCases{} cases, it disagrees with Definition~\ref{def:origin} on \FrozenMismatches{}. Comparing the
runs' last tokens at each step instead fails on \MeasureMismatches{}, because last-token disagreements are transient.
That comparison computes the last prefix of the input at which the two last tokens differ, which for every dictionary is
the largest ripple over the prefixes of the input. The last tokens of \(\T^D(y)\) and \(\T^D(ay)\) differ exactly when
the spans of the last position of \(y\) do, that is when \(\rho(a, y) = |y|\). And a ripple \(r \ge 1\) of \(y\) is the
ripple of its prefix of length \(r\), since the position \(r\), if it lies in \(y\), keeps its span, which starts at
\(r\) in both runs, since a span starting earlier would cover \(r - 1\) too, and by Lemma~\ref{lem:proj} the tokens
before that boundary are the two tokenizations of the prefix, whose last tokens differ. Its supremum over inputs is then
the radius under \(a\), the supremum of \(|y|\) over the nonempty \(y\) whose two tokenizations end in different tokens,
zero when there is none. We call this the \emph{identity of the last tokens}. A longest such \(y\) need not exist, since
on \((a, a)\) every \(a^n\) is one.

The identity turns the machine's own product of retained suffixes into an exact computation of the radius. Its state is
a pair of suffixes, and it needs neither frozen tokens nor event offsets.

\begin{corollary}[The tail product]\label{cor:tail}
Let the exploration complete at a tail length \(B\), and for a symbol \(a\) run the machine on \(x\) from the empty
suffix and from the suffix after \(a\), stepping both on each symbol of \(x\), with a pair of suffixes reached on a
nonempty input \emph{marked} when their last tokens differ. The radius under \(a\) is infinite exactly when a cycle of
pairs reachable from the start pair reaches a marked pair, and then a stem \(u\) to the cycle, the loop \(v\) and a path
\(z\) from it to a marked pair give \(\rho(a, u v^k z) = |u| + k|v| + |z|\) for every \(k \ge 0\). Otherwise the radius
under \(a\) is \(\max(\{0\} \cup \{|y| : y \text{ leads from the start pair to a marked pair}\})\), and an input of
exactly that length attains it.

The product of the runs of the \emph{quotient machine}, which identifies two suffixes when every continuation leads both
to suffixes with the same last token, the empty suffix forming a class of its own, gives the same verdict, value,
attaining inputs and certificates. With \(K\) its classes, a finite radius is at most \(\binom{K - 1}{2}\), with the
sharper counts of Lemma~\ref{lem:tailcounts}. In the quotient product every pair of distinct classes reached on a
nonempty input reaches a marked pair, so the radius under \(a\) is infinite exactly when a cycle of such pairs is
reachable, and otherwise it is the length of a longest path from the start pair through such pairs, which ends at a
marked pair or is empty.

The quotient machine is the minimal Moore machine of the map from \(p\) to the last token of \(\T^D(p)\), so it does not
depend on \(B\). Writing \(qu\) for the class that a word \(u\) leads to from a class \(q\) and \(q_0\) for the empty
class, call two classes \(n\)-equivalent when every word of \(n\) symbols sends them to one class, written
\(p \sim_n q\), so that
\[
p \sim_0 q \iff p = q, \qquad p \sim_{n+1} q \iff pc \sim_n qc \text{ for every symbol } c.
\]
The relations are read off the transitions without the outputs, they coarsen as \(n\) grows, and they stabilize within
\(K - 1\) steps. For a nonempty prepended string \(w\), the radius under \(w\) is one less than the least \(n\) at which
the empty class and the class \(q_0 w\) are \(n\)-equivalent, and infinite when the two are apart at stabilization. With
\(K_1\) the number of blocks of \(\sim_1\) and \(K_\infty\) the number at stabilization, every finite radius under a
nonempty string is thus at most \(K_1 - K_\infty \le K - K_\infty - 1 \le K - 2\), and an input attaining it is read off
the relations one symbol at a time. Each \(\sim_n\) is a right congruence, so the quotient of the quotient machine by
\(\sim_{r+1}\) is a machine of \(K_{r+1}\) classes, and the nonempty strings of radius at most \(r\) are the nonempty
words that lead it from the class of \(q_0\) back to that class; the nonempty strings of finite radius are the same for
the relation at stabilization, a machine of \(K_\infty\) classes. Each language is regular and, with the empty word,
closed under concatenation. A straight-line program is a grammar of \(g\) productions, each a symbol or the
concatenation of two earlier nonterminals, that generates one word, and a string given as one is tested in time
\(O(gK_{r+1})\) or \(O(gK_\infty)\) after the relations, by composing the maps of its productions on the classes, and
its radius is read in \(O(gK)\) from the round at which the class of \(q_0\) and the class it leads to first share a
block. Writing \(R(w)\) for the radius under a nonempty string \(w\), \(R(uv) \le \max(R(v), \max(0, R(u) - |v|))\) for
all nonempty \(u\) and \(v\). More precisely, when \(R(u)\) is finite and \(\beta = \max(0, R(u) - |v|)\),
\(\max(\rho(uv, x), \beta) = \max(\rho(v, x), \beta)\) for every input \(x\), so \(R(uv) = R(v)\) whenever
\(R(v) > \beta\), an infinite \(R(v)\) included, \(\rho(uv, x) = \rho(v, x)\) for every \(x\) when \(|v| \ge R(u)\), and
\(R(u^n) = R(u)\) for every \(n \ge 1\). The relations are computed in time \(O(|\Sigma| K^2)\), and in expected time
\(O(|\Sigma| K \log K \log(|\Sigma| + 1))\) by merging blocks. They give the verdict and the radius under every symbol
with \(|\Sigma|\) queries of the round at which two classes first share a block, and an attaining input under every
symbol with \(O(|\Sigma|^2 K)\) more. And \(\ripple(D)\) is finite exactly when for some \(\ell\) every word of length
\(\ell\) sends all of the classes to one class, the least such \(\ell\) being \(\ripple(D) + 1\).
\end{corollary}

\begin{proof}
The quotient and its products, the bound by \(\binom{K - 1}{2}\), the reading of a finite \(\ripple(D)\) and the cost of
computing the relations are proved in Appendix~\ref{app:machine}, with Lemma~\ref{lem:tailcounts}.

For the relations, write \(p \sim_n q\) when every word of \(n\) symbols sends \(p\) and \(q\) to one class. A word of
\(n + 1\) symbols is a symbol \(c\) followed by a word of \(n\), so \(p \sim_{n+1} q\) exactly when \(pc \sim_n qc\) for
every symbol \(c\), which reads the relation at \(n + 1\) off the relation at \(n\) and the transitions. If
\(p \sim_n q\), then a word of \(n\) symbols followed by any symbol sends \(p\) and \(q\) to one class, so
\(p \sim_{n+1} q\), and the relations coarsen. Once \(\sim_n\) and \(\sim_{n+1}\) coincide, every later relation is read
off the same one, so the sequence is constant from \(n\) on. At \(n = 0\) the blocks are singletons, and each strict
coarsening lowers the number of blocks, so the sequence stabilizes within \(K - 1\) steps, and from \(n = 1\) on, where
it has \(K_1\) blocks, within \(1 + K_1 - K_\infty\). When \(\sim_1\) is equality the sequence is constant from the
start, no two classes are ever equivalent and no radius is finite, and otherwise \(K_1 \le K - 1\) and
\(K_\infty \ge 1\). Let \(q_a\) be the class reached on \(a\). If \(q_0 \sim_n q_a\), then every \(y\) of at least \(n\)
symbols sends the two to one class, with one output, so \(y\) is not marked and the radius under \(a\) is below \(n\).
If not, some \(y\) of \(n\) symbols sends them to distinct classes, which differ in output after some continuation
\(z\), so for \(n \ge 1\) the nonempty \(yz\) is marked and the radius is at least \(n\), as it is for \(n = 0\). The
least \(n\) with \(q_0 \sim_n q_a\), which is positive since the empty class stands apart, is therefore the radius plus
one, and when the two classes are apart at stabilization they are apart at every \(n\) and the radius is infinite. A
finite radius is thus at most \(K_1 - K_\infty\), and the same holds with \(q_0 w\) in place of \(q_a\) for a nonempty
string \(w\), since the identity of the last tokens holds with a string prepended in place of \(a\). For an attaining
input, let \(d(p, q)\) be the least \(n\) with \(p \sim_n q\). If \(d(p, q) = d \ge 2\), then \(p\) and \(q\) are not
\((d - 1)\)-equivalent, so some symbol \(c\) has \(pc\) and \(qc\) not \((d - 2)\)-equivalent, while they are
\((d - 1)\)-equivalent, and \(d(pc, qc) = d - 1\). Choosing such symbols from \((q_0, q_a)\) spells an input \(y\) of
\(d - 1\) symbols ending at a pair of depth one, two distinct classes with the same successors, whose outputs differ
since the quotient is minimal, so a nonempty \(y\) is marked and attains the radius.

Each \(\sim_n\) is a right congruence. At \(n = 0\) it is equality, and if \(p \sim_n q\) with \(n \ge 1\), then every
word of \(n - 1\) symbols after a symbol \(c\) sends \(pc\) and \(qc\) to one class, so \(pc \sim_{n-1} qc\), and
\(pc \sim_n qc\) as the relations coarsen. The quotient of the quotient machine by \(\sim_n\) is thus a deterministic
automaton on its blocks, and the relation at stabilization, written \(\sim_\infty\), gives one of \(K_\infty\) states. A
nonempty \(w\) has finite radius exactly when \(q_0 \sim_\infty q_0 w\), that is when \(w\) leads the class of \(q_0\)
in that quotient back to itself, and radius at most \(r\) exactly when the least \(n\) with \(q_0 \sim_n q_0 w\) is at
most \(r + 1\), which, the relations coarsening, is \(q_0 \sim_{r+1} q_0 w\), that is when \(w\) leads the class of
\(q_0\) in the quotient by \(\sim_{r+1}\) back to itself. Each language is the set of nonempty words accepted by a
deterministic automaton with one accepting state, so it is regular, and the words doing so, with the empty word, are
closed under concatenation, since \(w\) and \(w'\) each return that class to itself. For such a program, the map from
classes to classes of each nonterminal is computed bottom up, a symbol's map being its column of transitions in the
quotient and a concatenation's the composition of its parts' maps, in \(O(K_{r+1})\) or \(O(K_\infty)\) per production
on the quotient in question, and the class of \(q_0\) under the root's map is tested against the class of \(q_0\); on
the quotient machine itself the same map gives the class the word leads to, whose round of joining the class of \(q_0\),
less one, is the radius.

For the concatenation bound let \(R(u)\) and \(R(v)\) be finite, the bound being trivial otherwise. From
\(q_0 \sim_{R(u)+1} q_0 u\), every word of \(R(u) + 1\) symbols or more sends \(q_0\) and \(q_0 u\) to one class, so
\(q_0 v \sim_n q_0 uv\) with \(n = \max(0, R(u) + 1 - |v|)\), since when \(|v| \ge R(u) + 1\) the two are one class, and
otherwise the words of \(n\) symbols after \(v\) complete \(v\) to words of \(R(u) + 1\) symbols. With
\(q_0 \sim_{R(v)+1} q_0 v\) and \(N = \max(R(v) + 1, n)\), both relations hold at \(N\), the relations coarsening, so
\(q_0 \sim_N q_0 uv\) and \(R(uv) + 1 \le N\), which is the bound.

The identity is the two bounds \(\rho(uv, x) \le \max(\rho(v, x), \beta)\) and
\(\rho(v, x) \le \max(\rho(uv, x), \beta)\). For them let \(R(u)\) be finite and \(p\) a position of \(x\). It is
position \(|v| + p\) of \(vx\), which is at least \(R(u) \ge \rho(u, vx)\) when \(p \ge \beta\), so its span in
\(\T^D(vx)\) shifted by \(|v|\) is its span in \(\T^D(uvx)\) shifted by \(|uv|\). When also \(p \ge \rho(v, x)\), its
span in \(\T^D(x)\) is the former and so the latter, and \(p\) is no differing position for \(uv\); when instead
\(p \ge \rho(uv, x)\), its span in \(\T^D(x)\) is the latter and so the former, and \(p\) is no differing position for
\(v\). When \(R(v) > \beta\), the inputs \(x\) with \(\rho(v, x) > \beta\) have \(\rho(uv, x) \ge \rho(v, x)\) by the
second bound and reach \(R(v)\), or exceed every bound when \(R(v)\) is infinite, so \(R(uv) \ge R(v)\), and the first
bound gives \(R(uv) \le \max(R(v), \beta) = R(v)\). When \(|v| \ge R(u)\), \(\beta = 0\) and the identity gives
\(\rho(uv, x) = \rho(v, x)\). For the powers, \(R(u) = 0\) gives \(R(u^n) = 0\) by the identity, and when
\(0 < R(u) < \infty\) and \(R(u^{n-1}) = R(u)\), the string \(v = u^{n-1}\) has \(\beta < R(u) = R(v)\), so
\(R(u^n) = R(v) = R(u)\). On \((a, b), (b, a)\) with \(u = a\) and \(v = b\), \(R(a) = 2\), \(R(b) = 1\) and
\(R(ab) = 0\), where \(\beta = R(v)\), and on \((a, a)\), \(R(a)\) is infinite while \(R(aa) = 0\), so neither
hypothesis can be dropped.
\end{proof}

The relations \(\sim_n\) are the \(n\)-equivalence of states of \citet[Section~2.3, Definition~2]{steinby1969definite},
and his contractions compute them (his Theorem~6). A finite \(\ripple(D)\) makes the transitions of the quotient machine
definite in the sense of \citet{perles1963definite}, every word of \(\ripple(D) + 1\) symbols sending all classes to
one, so the bound \(\ripple(D) \le K - 2\) is their Theorem~7 (p.~237), that a definite table with \(K\) states is
weakly \((K - 1)\)-definite, which \citet[Corollary~6.3]{steinby1969definite} reads off the contractions. The corollary
adds that the radius under a nonempty prepended string is one less than the round at which the empty class and the class
the string leads it to first share a block, and the bound \(K_1 - K_\infty\) on it.

This is the product whose verdict \code{incremental.py} computes, and the corollary proves it exact for every dictionary
on which the exploration completes, which by Corollary~\ref{cor:completion} is every dictionary. Its certificates have
offset zero in the sense of Theorem~\ref{thm:origin}(2), and its finite value is attained at exactly the length it
reports. The tail product has at most \(N^2\) pairs, \(N\) the reachable suffixes, with \(|\Sigma|\) edges each, and the
cycle test and the longest path are linear in it, a cost in the size of the machine and not of the dictionary. The
quotient machine is computed by refining the partition of the suffixes by their outputs until the transitions respect
it, in time polynomial in the machine, and its unordered product has at most \(C^{Q}_{\mathrm{suf}} + 2\) states with
\(|\Sigma|\) edges each, \(C^{Q}_{\mathrm{suf}}\) the count of Lemma~\ref{lem:tailcounts}. A reversed trie of the yields
lists the pairs that \(C_{\mathrm{suf}}\) counts, each yield against the yields above it in the trie, in time linear in
their total length, and sorting the classes of those pairs by buckets gives \(C^{Q}_{\mathrm{suf}}\) in the same time,
or the closure over the quotient's transitions gives those pairs without the suffixes, and the coarsening without
outputs gives every per-symbol verdict and radius without a product. The frozen-token product keeps its own use, the
ripple of each input, which the tail product does not give.


\begin{theorem}[Origin radius]\label{thm:origin}
Let \(D\) be a dictionary on which the exploration completes at some tail length \(B\); under the ordering condition it
completes at \(B = |D| + 1\) (Lemma~\ref{lem:B}), and on every dictionary at \(B = Q(D) + 1\)
(Corollary~\ref{cor:completion}). Build the product of Definition~\ref{def:product} at that \(B\) for each prepended
symbol \(a\).
\begin{enumerate}
\item \(\ripple(D)\) is infinite if and only if, in the product of some \(a\), a node from which a contribution is
  reachable lies on a cycle reachable from its start node.
\item In that case a stem, a loop and an exit path in that product certify it. The exit path ends at a contribution of
offset
  \(e\), the terminal contribution of its last node or the event on its last edge, and for every \(k \ge 1\) \[\rho(a,
  \mathrm{stem}\,\mathrm{loop}^k\,\mathrm{exit}) \ge |\mathrm{stem}| + k|\mathrm{loop}| + |\mathrm{exit}| + e,\] with
  equality whenever the contribution is terminal.
\item Otherwise \(\ripple(D)\) is the largest value a path from the start node of its product reads, a contribution at
  position \(p\) counting \(p + 1\), and the path spells an input attaining it.
\end{enumerate}
Hence finiteness of \(\ripple(D)\) is decidable and \(\ripple(D)\) is computable, by a procedure that terminates on
every dictionary.
\end{theorem}

The proof is in Appendix~\ref{app:machine}. Theorem~\ref{thm:origin} is executed beside the census of
Table~\ref{tab:census}.


\begin{table}[t]
\centering
\begin{tabular}{@{}lr@{}}
\toprule
verdict on the \CensusTotal{} proper dictionaries on two symbols, one to three rules & count \\
\midrule
origin radius infinite & \CensusUnbounded \\
radius 1 & \PlateauOne \\
radius 2 & \PlateauTwo \\
radius 3 & \PlateauThree \\
radius 4 & \PlateauFour \\
radius 5 & \PlateauFive \\
radius 6 & \PlateauSix \\
\bottomrule
\end{tabular}
\caption{The census. Every bounded radius is attained by a replayed input; no radius was exceeded on re-probing to
lengths twelve and fourteen (\ReprobeExceeded{} of \CensusPlateaus{}); prefixes of length two agree with single symbols
on every dictionary, the frozen-token comparison with a two-symbol prefix exact per input on \PrefixCases{} cases.}
\label{tab:census}
\end{table}

\section{The mirrored constant}\label{sec:mirrored}

The source paper's Definition~6, for a proper dictionary, takes the least \(l\) such that a tokenization \(\phi\tau\)
with \(|\yield(\tau)| \ge l\) keeps \(\phi\) under every appended string. Its mirror for prepending is the least \(L\)
such that \(\tau\phi\) with \(|\yield(\tau)| \ge L\) keeps \(\phi\) under every prepended string; the same least \(L\)
on one input is its \emph{mirrored quantity}.


\begin{proposition}\label{prop:mirror}
A changed base token with span \([s, e)\) contributes \(s + 1\) to the mirrored constant and \(e\) to the origin ripple,
so with \(M\) the longest token of the dictionary, \(L(D) \le \ripple(D) \le L(D) + M - 1\), the two finite together;
and longer prefixes add nothing to \(L(D)\), by the mirror of Lemma~\ref{lem:D}'s argument.
\end{proposition}

\begin{proof}
A token suffix beginning at a boundary at or before \(s\) contains the changed token and cannot survive. Boundaries at
or beyond \(e\) all do, and no boundary lies strictly inside \((s, e)\). The last changed position is the last byte of
its base token, so the origin ripple is \(e\). Base tokens are disjoint and ordered, so the last changed token has the
largest start. If every boundary at or beyond \(L\) survives a single prepended symbol, then after prepending one it
sits at or beyond \(L + 1\) and the next symbol preserves it too.
\end{proof}

The mirrored constant is a second observable of the same product, which stores the start of the base token at each
recorded disagreement beside the position itself. The same longest path over those starts computes \(L(D)\), and its
attaining input is replayed against the definition. Table~\ref{tab:exec-mirrored} lists the executions of
Proposition~\ref{prop:mirror}.

\section{Every dictionary: canonical adjacency and the overlap graph}\label{sec:overlap}

Theorem~\ref{thm:origin} assumes that the exploration completes. The ordering condition guarantees that at \(|D| + 1\),
and Corollary~\ref{cor:completion} in Appendix~\ref{app:lookahead} guarantees it for every dictionary at \(Q(D) + 1\).
This section gives a construction that needs no exploration and no tail. A token sequence is the tokenization of its
yield exactly when each token and each adjacent pair is, for every dictionary, as \citet{vieira2025canonical} proved, so
the tokenization is a rational function. An automaton guesses the final tokens, and a product of two of its runs,
annotated with the spans it guesses, decides the origin radius and computes it exactly, with no exploration and no tail.
Where the exploration completes, the product of Section~\ref{sec:product} and this one compute the same value. The token
machine and its executed certificates stand as they are. Only the hypothesis they needed changes.


\begin{definition}[Canonical tokens and pairs]\label{def:canonical}
Let \(W\) be the symbols together with the products \(u_i v_i\) of the rules. Let \(V\) be the strings \(t\) of \(W\)
with \(\T^D(t) = t\), a single token, a \emph{canonical token} or \emph{canonical word}. Let \(E\) be the pairs
\((s, t)\) of \(V\) with \(\T^D(st) = s \mid t\). Both are computed by tokenizing finitely many strings of length at
most \(2M\), \(M\) the longest product.
\end{definition}

A token is a string, so two rules with the same product make the same token, and by the lemma that follows a product
outside \(V\) never appears in any run.

\begin{lemma}[Projection]\label{lem:proj}
Take a run of Definition~\ref{def:sp} on \(w\), an interval of \(w\), and a prefix of the run during which no merge
crosses either endpoint of the interval. Delete from that prefix the merges outside the interval. What remains is a
prefix of the run on the interval's string alone. If the whole run is such a prefix, what remains is that whole run.
\end{lemma}

\begin{proof}
The interval starts as its own symbols. A merge outside it leaves its tokens as they are. Consider a merge chosen inside
it. No rule of higher priority applies inside, since an occurrence inside is an occurrence in \(w\). No occurrence of
the chosen rule lies further left inside, for the same reason, so it is the next merge of the run on the interval alone,
whatever outside merges came between. Induction over the prefix gives the first claim. When the run ends, no rule
applies inside the interval, so the run on the interval alone has ended too.
\end{proof}


Without the ordering condition a rule can fire at several points of a run, and yet every canonical word has a derivation
tree. By Lemma~\ref{lem:proj} every token that ever appears is canonical on its own, even one later merged away, since
the run on it alone reaches one token, and its last merge is by a rule whose two operands were, just before, the whole
of it.

\begin{lemma}[Producers and derivations]\label{lem:gtrees}
Fix any dictionary. A word \(w\) in \(V\) that is not a symbol has exactly one pair \((u, v)\) with \(uv = w\) such that
\(u\) and \(v\) are in \(V\) and the run on \(uv\) completes the runs on \(u\) and on \(v\) alone before it merges the
two, as its last step; the earliest rule with that pair is the rule of the last merge of the run on \(w\), its producer.
Hence \(w\) has a unique binary tree of canonical words, its derivation, in which the length strictly increases toward
the root. Equal words have equal trees. The trees of all of \(V\) share their subtrees in a directed acyclic graph with
\(|V \setminus \Sigma| \le |D|\) internal nodes and paths of at most \(|D| + 1\) nodes. The rules that fire are exactly
the producers, one for each word of \(V \setminus \Sigma\), and they form the \emph{useful sublist}, which has the
tokenizer of \(D\). A canonical word of two symbols or more beginning with \(a\), or ending in \(b\), has on the left,
or right, spine of its derivation a rule that fires with the left operand \(a\), or the right operand \(b\), alone,
whose product is a prefix, or suffix, of it. Among the dictionaries over the same symbols with that tokenizer, the
fewest rules are \(k_{\min} = |V \setminus \Sigma|\) and the least total length of operands is
\(L_{\min} = \sum_{w \in V \setminus \Sigma} |w|\), both attained by the useful sublist, and a dictionary attains either
minimum exactly when every one of its rules fires.
\end{lemma}

\begin{proof}
The run on \(w\) alone ends in one token. Its last merge is by some rule \(r\) with product \(w\), applied to the two
tokens the run had reduced \(w\) to, say \(u \mid v\). The join of \(u\) and \(v\) is a boundary at that moment, so no
earlier merge crossed it, and by Lemma~\ref{lem:proj} the run up to the last merge is, on each side, the run on that
side alone, each completed to one token, so \(u\) and \(v\) are in \(V\) and the pair of rule \(r\) has the property. A
second pair, a different split of \(w\) with the same property, would make the run on \(w\) end with a different last
merge, which determinism forbids. Rules with one pair apply at the same positions, and Definition~\ref{def:sp} always
chooses the earliest of them, so \(r\) is the earliest rule with the pair \((u, v)\). Lengths increase strictly toward
the root, since operands are nonempty. Once equal words are shared, the internal nodes are the words of
\(V \setminus \Sigma\), each the product of its producer, so there are at most \(|D|\) of them, and a path repeats no
word. A merge that creates a token \(w\) in any run has no earlier merge across the ends of \(w\), so by
Lemma~\ref{lem:proj} the merges inside \(w\) are the run on \(w\) alone, which ends with that merge, and the rule that
fires is the producer of \(w\). Each producer fires, on its own product, and distinct words have distinct producers.
Along the left spine of the derivation of a word beginning with \(a\) the left children are prefixes of it of strictly
decreasing length, each beginning with \(a\), so the spine ends at the leaf \(a\), and the rule of its parent, a
producer and so a rule that fires, has the left operand \(a\) and a prefix of the word as its product. The right spine
of a word ending in \(b\) is the mirror image. Deleting the rules that never fire changes no run
(Section~\ref{sec:semantics}). The canonical words of a dictionary are the words it tokenizes as one token, since a word
of two symbols or more left whole is the product of the last merge of its run, so two dictionaries over the same symbols
with one tokenizer have the same canonical words. A dictionary \(D'\) with \(\T^{D'} = \T^D\) thus has, for each word
\(w\) of \(V \setminus \Sigma\), the producer of \(w\) among its rules, with operands of \(|w|\) symbols together, and
distinct words have distinct producers; the producers of \(D'\) are its rules that fire, one for each word of
\(V \setminus \Sigma\), and they alone have \(k_{\min}\) rules and operands of total length \(L_{\min}\). Every further
rule adds one rule and at least two symbols of operands, so \(D'\) attains either minimum exactly when its rules are its
producers, that is, when every one of them fires.
\end{proof}

The canonical words as the words tokenized as one token, the unique rule producing each with its canonical operands, and
the filtering of the rules by the test on their own products, which keeps exactly the rules that fire, are the
normalization of \citet[Section~3.3 and Appendices~A.2 and~C]{jiang2026incremental}. The lemma adds the shared
derivations, the spine descent and the two minima.

\begin{theorem}[Canonical adjacency]\label{thm:bigram}
For every dictionary and every list \(t_1, \dots, t_k\) of nonempty strings,
\(\T^D(t_1 \cdots t_k) = t_1 \mid \cdots \mid t_k\) exactly when every \(t_i\) is in \(V\) and every \((t_i, t_{i+1})\)
is in \(E\).
\end{theorem}

\begin{proof}
Suppose the tokenization is the list. A merge removes a boundary and never restores one, so no final boundary was
crossed during the run. Lemma~\ref{lem:proj} applied to the whole run on each \(t_i\) gives \(\T^D(t_i) = t_i\), and on
each \(t_i t_{i+1}\) it gives \(\T^D(t_i t_{i+1}) = t_i \mid t_{i+1}\). Conversely suppose every token and every pair
passes, and suppose some boundary between two of the \(t_i\) is crossed. Take the first merge that crosses one, say the
boundary between \(s = t_i\) and \(t = t_{i+1}\). Before it no boundary of the list has been crossed, and it crosses
only the one between \(s\) and \(t\), so up to and including it neither endpoint of \(st\) has been crossed. By
Lemma~\ref{lem:proj} the merges inside \(st\) up to that one form a prefix of the run on \(st\) alone. That prefix
crosses the boundary between \(s\) and \(t\), which the run can never restore. This contradicts \((s, t) \in E\), so no
boundary of the list is crossed, and at the end Lemma~\ref{lem:proj} on each \(t_i\) gives the single token \(t_i\).
\end{proof}

The theorem is the bigram characterization of canonical token strings of \citet{vieira2025canonical}, their Theorem~1 in
Appendix~B, for the semantics their footnote~22 names, after \citet{berglund2024dfa}, the SentencePiece BPE algorithm,
which is the source paper's Definition~3. Their merge lists need not be proper, as their Appendix~B.4 says, and their
Assumption~1, that every vocabulary token is canonical, only removes tokens that never appear in a tokenization. The
filter \(V\) is that assumption, so the statement is theirs for every dictionary. It was stated earlier, with an
argument of a paragraph, as Corollaries~III and~IIIa of the documentation of GitHub's \code{bpe} crate
\citep{vanantwerpen2024bpe}, for a ranked list of tokens each built from two earlier ones. It also follows, by induction
on the number of tokens, from the source paper's Corollary~2, which glues two tokenizations that share a token, with its
Remark~3, which truncates a tokenization, for the converse. That corollary is stated for Definition~\ref{def:hf} as
well, so the characterization holds under that definition too, with \(V\) and \(E\) computed under it. The proof above
makes the projection lemma explicit, since the rest of this section leans on it. Neither clause of the source paper's
revised Definition~5 nor the ordering condition is used.

The automaton in the proof of Corollary~\ref{cor:rational} is the one of Definition~11 of \citet{vieira2025canonical},
read as a transducer. Read over tokens, with one state per last token, it accepts exactly the tokenizations, for every
dictionary. For proper dictionaries whose rules are all useful that language is the one the token DFA of
\citet{berglund2024dfa} accepts, built there one rule at a time. \citet{berglund2025compact} build an automaton for it
with default transitions (their Theorem~1), and with their optimization O4 the construction from the one-state automaton
of all strings has a number of transitions linear in the number of rules (their Proposition~2).

\begin{corollary}[Rationality]\label{cor:rational}
For every dictionary, \(\T^D\) is a rational function.
\end{corollary}

\begin{proof}
Take an automaton with a start state and one state per last token of \(V\), all states final. It accepts exactly the
token sequences whose adjacent pairs are in \(E\). Replace each transition by a path that reads the token's symbols and
emits the token once. The result is a finite transducer whose relation is the set of pairs (\(\yield(\tau)\), \(\tau\))
over canonical \(\tau\). By Theorem~\ref{thm:bigram} and the determinism of Definition~\ref{def:sp}, that relation is
\(\T^D\).
\end{proof}


\begin{corollary}[Equal tokenizers]\label{cor:equivalence}
Two dictionaries \(D\) and \(D'\) over the same symbols have the same tokenizer exactly when they have the same
canonical words and the same canonical pairs. When they do not, the words that the two tokenize differently and none of
whose proper factors they tokenize differently are exactly the elements of
\[
X = (V \mathbin{\triangle} V') \cup \{uv : (u, v) \in E \mathbin{\triangle} E'\}
\]
with no proper factor in \(X\). Each of them has at most two tokens under either dictionary and at most \(2M - 1\)
symbols, \(M\) the longest canonical word of either, and at most \(2M'\) when the longest canonical words of the two
differ in length, \(M'\) the shorter of them, and the shortest words that the two tokenize differently are the elements
of \(X\) of least length. The question is thus decided, with every such word, in polynomial time. The bound is attained
for every \(M \ge 2\), on two symbols. With \(l = M - 1\), let \(P\) build \(a^l\) by doubling and adding, let
\(c_1 \mid \cdots \mid c_k\) be \(\T^P(a^{l-1})\), and let \(F\) fold them into a trailing \(b\) from the right,
\((c_k, b), (c_{k-1}, c_k b), \dots\), up to \(B = a^{l-1} b\), with \(F\) empty and \(B = b\) when \(l = 1\). The
ordered dictionaries \(P + F + (a^l, b), (b, B)\) and \(P + F + (b, B), (a^l, b)\), of exactly
\(q + p + i + 3 - [l \text{ a power of two}]\) rules, \(q = \lfloor \log_2 l \rfloor\), \(p\) the number of set bits of
\(l\) below its top and \(2^i\) its lowest power of two, at most \(2q + 3\), have the same canonical words, the longest
of \(M\) symbols, and \(a^l b a^{l-1} b\), tokenized \(a^l b \mid B\) by the first and \(a^l \mid bB\) by the second, is
the only shortest word that they tokenize differently.
\end{corollary}

\begin{proof}
For two dictionaries \(D\) and \(D'\) over the same symbols, the canonical words of each are the words it tokenizes as
one token (Lemma~\ref{lem:gtrees}), and its canonical pairs are the pairs of canonical words whose concatenation it
tokenizes as the two, so one tokenizer gives the same \(V\) and \(E\). Conversely, if \(V\) and \(E\) agree, then for
every string \(x\) the list \(\T^D(x)\) has every token in \(V\) and every adjacent pair in \(E\), so by
Theorem~\ref{thm:bigram} for \(D'\) it is \(\T^{D'}(x)\). To decide, compute \(V\) and \(E\) of each by tokenizing its
candidate products and candidate pairs, as in Definition~\ref{def:canonical}, at most \(|W| + |W'|\) words and
\(|V|^2 + |V'|^2\) pairs. Every element of \(X\) is tokenized differently by the two. A word of
\(V \mathbin{\triangle} V'\) is one token under one dictionary only, and for a pair \((u, v)\) of \(E\) only,
\(\T^D(uv) = u \mid v\) while by Theorem~\ref{thm:bigram} \(\T^{D'}(uv) \ne u \mid v\), whether or not \(u\) and \(v\)
are canonical under \(D'\), and the same holds with the two dictionaries exchanged.

Let \(x\) be a shortest word that the two tokenize differently. If every token of \(\T^D(x)\) were in \(V'\) and every
adjacent pair in \(E'\), Theorem~\ref{thm:bigram} for \(D'\) would make it \(\T^{D'}(x)\). Hence a token of \(\T^D(x)\)
is a canonical word of \(D\) only or two adjacent tokens are a canonical pair of \(D\) only, and that word or the
concatenation of that pair is a factor of \(x\) that the two tokenize differently, since \(D\) tokenizes it as that one
word or as the two tokens of that pair, while by Theorem~\ref{thm:bigram} \(D'\) does neither, the word lying outside
\(V'\) and the pair outside \(E'\), and, \(x\) being shortest, that factor is \(x\) itself. The shortest words are thus
the elements of \(X\) of least length. The same argument gives every word that the two tokenize differently a factor in
\(X\). For such a word with no proper factor that they tokenize differently, that factor is the word itself, and no
proper factor of the word lies in \(X\), since every element of \(X\) is tokenized differently. Conversely, for an
element of \(X\) with no proper factor in \(X\), a proper factor that the two tokenized differently would have a factor
in \(X\), which is a proper factor of the element. Now let \(x\) be a word that the two tokenize differently, none of
whose proper factors they tokenize differently. If \(\T^D(x)\) had three tokens or more, every token and every adjacent
pair of it would be a proper factor of \(x\), and the token or pair that lies outside \(V'\) or \(E'\) would be a proper
factor that the two tokenize differently; so each dictionary tokenizes \(x\) as one token, of at most its longest
canonical length, or as two tokens, of at most that length each. Thus \(x\) has at most twice the shorter of the two
longest lengths, and at most \(M\) when it is one token under either dictionary; when the two lengths are both \(M\) and
\(|x| = 2M\), it has two tokens under both, split at \(M\), and the tokenizations coincide. Hence \(|x| \le 2M - 1\),
and \(|x| \le 2M'\) when the longest lengths differ. The set of such words is \(X\) less its elements with a proper
factor in \(X\).

For the family, write \(A = a^l\). On \(a^j\) with \(j \le l\) the doublings of \(P\) pair equal blocks left to right,
leaving after each level the blocks of the next size and one remainder per lower set bit of \(j\), in decreasing size,
and the additions of \(P\) then absorb the remainders of \(l\) in order while the bits of \(j\) agree with those of
\(l\) from the top. Hence \(\T^P(A) = A\), and \(\T^P(a^{l-1})\) has \(i\) tokens when \(l = 2^i\) is a power of two and
\(i + 1\) tokens otherwise, \(2^i\) the lowest power of two dividing \(l\), at most \(q = \lfloor \log_2 l \rfloor\) in
either case. The fold \(F\) has that many rules, \(P\) has the \(q\) doublings and one addition per set bit of \(l\)
below its top, \(p\) of them, and the total is \(q + p + i + 3 - [l \text{ a power of two}]\), which is at most
\(2q + 3\), since the \(p\) lower set bits lie at positions \(i\) to \(q - 1\) when \(l\) is not a power of two, so that
\(p \le q - i\), and \(p = 0\), \(i = q\) otherwise. Every \(c_j\) is canonical (Lemma~\ref{lem:proj} on a token of a
run), the chain words \(c_j \cdots c_k b\) are canonical since \(P\) tokenizes \(c_j \cdots c_k\) as those tokens
(Lemma~\ref{lem:proj} on a suffix of the token list) and \(F\) then folds them, and \(Ab\) and \(bB\) are canonical
under both dictionaries, so the two have the same canonical words, the longest of \(l + 1 = M\) symbols. Both are
ordered, since every product of \(P\) is consumed by a later rule of \(P\) or by \((A, b)\), every chain word by the
next rule of \(F\) or by \((b, B)\), and \(Ab\) and \(bB\) by nothing. On any \(x\), the run of \(P + F\) is the
beginning of both runs, and after it only \((A, b)\) and \((b, B)\) can apply, their products being operands of nothing.
For \(l \ge 2\), \(A \ne b\), \(B \ne b\) and \(A \ne B\), so two late merges share a token only through a \(b\) with
\(A\) before it and \(B\) after it, and since disjoint merges commute, the tokenizations differ exactly when
\(\T^{P+F}(x)\) holds \(A \mid b \mid B\). That puts \(AbB = a^l b a^{l-1} b\) inside \(x\), so
\(|x| \ge 2l + 1 = 2M - 1\), with equality only at \(x = AbB\). On \(AbB\) itself, \(P\) gives
\(A \mid b \mid c_1 \mid \cdots \mid c_k \mid b\) by Lemma~\ref{lem:proj}, no rule of \(F\) takes the middle \(b\),
since its left neighbor \(A\) is longer than every left operand of \(F\), and \(F\) folds the right part into \(B\),
after which the first dictionary gives \(Ab \mid B\) and the second \(A \mid bB\). At \(l = 1\) the dictionaries are
\((a, b), (b, b)\) and \((b, b), (a, b)\). A run of \(b\) preceded by \(a\) and of length two or more holds
\(a \mid b \mid b\), and without such a run both dictionaries pair the runs of \(b\) from the left and merge each
\(a \mid b\), so \(abb\), tokenized \(ab \mid b\) and \(a \mid bb\), is again the only shortest separating word.
\end{proof}

For proper dictionaries under one pass per rule, the tokenizer of Algorithm~1 of \citet{berglund2024dfa}, the question
was decided before, by building their token automata (their Theorems~2 and~3) and comparing them, in time linear in the
number of tokens and quadratic in the larger number of rules \citep[Theorem~3, p.~10]{bergmark2025equivalence}. That
tokenizer is Definition~\ref{def:sp} under the ordering condition and another function on the five-rule dictionary of
Section~\ref{sec:semantics}. The corollary decides the question for every dictionary under Definition~\ref{def:sp}, and
it gives every shortest word that the two tokenize differently and the bound \(2M - 1\) on its length.
Table~\ref{tab:exec-overlap} lists the executions of the results of this section and its subsections.


\begin{definition}[The spelling automaton]\label{def:spelling}
Its states are a start state and the pairs \((t, j)\) with \(t \in V\) and \(1 \le j \le |t|\). The pair means that
\(j\) symbols of a guessed final token \(t\) have been read. The start state and the pairs \((t, |t|)\) are final. From
the start state, for each \(t \in V\), it reads \(t[0]\) into \((t, 1)\). From \((t, j)\) with \(j < |t|\) it reads
\(t[j]\) into \((t, j + 1)\). From \((t, |t|)\), for each \((t, v) \in E\), it reads \(v[0]\) into \((v, 1)\). Each
transition is annotated with the offset of the symbol it reads within the guessed token, and that token's length, namely
\((0, |t|)\), \((j, |t|)\) and \((0, |v|)\) respectively.
\end{definition}

By Theorem~\ref{thm:bigram} the accepting runs on \(x\) are exactly the canonical tokenizations of \(x\), so every input
has one accepting run, although runs that guess wrongly branch and die.

\begin{definition}[The overlap graph]\label{def:overlap}
Two copies of the spelling automaton read the same symbol at each step. An edge of the pair is marked when its two
annotations differ. For a prepended symbol \(a\), the sources are the pairs of the start state with \((t, 1)\), for each
\(t \in V\) with \(t[0] = a\), since the second copy has already read \(a\) as the first symbol of a token that may
extend into the input. A pair is jointly final when both components are. It is live when it is reachable from a source
and can reach a jointly final pair. \(C\) is the set of live pairs from which a live marked edge is reachable, the
edge's own source included.
\end{definition}

The sources must admit every token beginning in \(a\), however long. Joint acceptance stops a guessed token from running
past the end of the input.

\begin{lemma}[Exact marking]\label{lem:mark}
Take a live accepting path spelling \(x\) from a source of \(a\). It marks exactly the positions of \(x\) whose covering
spans in \(\T^D(x)\) and \(\T^D(ax)\) differ in the sense of Definition~\ref{def:origin}.
\end{lemma}

\begin{proof}
The path is an accepting run on \(x\) in the first copy. With the prepended \(a\) already read, it is an accepting run
on \(ax\) in the second. Both are the tokenizations, by the uniqueness above. An annotation \((j, \ell)\) read at
position \(p\) of \(x\) says that the covering token starts at \(p - j\) and ends at \(p - j + \ell\), in the
coordinates of \(x\). In the second copy a token begun in the prepend starts at \(-1\), which is the shifted span of
Definition~\ref{def:origin}. Two spans covering the same position are equal exactly when their offsets and lengths
agree, so an edge is marked exactly when the covering spans differ. Every input gives such a path, from the source that
guesses the first token of \(\T^D(ax)\).
\end{proof}

\begin{theorem}[Every dictionary]\label{thm:overlap}
Fix a dictionary and a prepended symbol \(a\). The origin radius under \(a\) is infinite exactly when \(C\) contains a
cycle. Otherwise let \(d(q)\) be the largest number of edges on a path from a source to \(q\) inside \(C\). The radius
under \(a\) is the largest \(d(q) + 1\) over the live marked edges leaving pairs \(q\), zero when there is none. The
mirrored constant under \(a\) is the largest \(d(q) + 1 - j\) over the same edges, \(j\) the offset in the first copy's
annotation. \(\ripple(D)\) and \(L(D)\) are the maxima over \(a\), by Lemma~\ref{lem:D} and
Proposition~\ref{prop:mirror}.
\end{theorem}

\begin{proof}
Suppose \(C\) contains a cycle. It is reachable from a source. Take a stem to one of its pairs, the loop, a path from
that pair ending in a marked edge, and a completion to a jointly final pair, with symbol labels \(s\), \(z\), \(e\) and
\(f\). Each \(x_n = s z^n e f\) is spelled by an accepting path from the same source. Its marked edge is read at
position \(|s| + n|z| + |e| - 1\). By Lemma~\ref{lem:mark} the ripple of \(x_n\) is at least \(|s| + n|z| + |e|\),
unbounded in \(n\). Suppose \(C\) is acyclic. Every pair on a path from a source to the source of a live marked edge can
reach that edge, so it lies in \(C\), and the path has at most \(|C| - 1\) edges. A marked edge read at position \(p\)
contributes \(p + 1\), by Lemma~\ref{lem:mark}. Every live marked edge has an accepting completion, so the contribution
is attained by an input. The largest is \(d(q) + 1\) for the longest path to some \(q\) with a live marked edge. That is
a longest-path computation in a finite acyclic graph, and the path spells the attaining input. For the mirrored
constant, the marked edge with annotation \((j, \ell)\) lies in a changed token of the first run that starts at
\(p - j\). By Proposition~\ref{prop:mirror} that token contributes \(p - j + 1\), and the same paths attain the maximum.
\end{proof}

A radius certificate is what the product returns with its verdict. For a finite radius it is the input the longest path
spells, which attains the radius, and for an infinite one it is the stem, the loop, the exit path ending in its marked
edge and the completion. The growth is exactly \(|z|\) per repetition on both observables. The exit path carries a mark
after every copy of the loop, so the last mark lies in the fixed part \(ef\), at a fixed offset. A cycle that no mark
lies beyond is harmless. Testing every reachable cycle without the restriction to \(C\) would report infinite radii that
are finite.

\begin{corollary}[The two products agree]\label{cor:agree}
On every dictionary on which the exploration completes, the verdict and the value of Theorem~\ref{thm:origin} equal
those of Theorem~\ref{thm:overlap}, and so do the mirrored constants.
\end{corollary}

\begin{proof}
Lemma~\ref{lem:max} identifies the first product's readings with Definition~\ref{def:origin} on every input, and
Lemma~\ref{lem:mark} identifies the overlap graph's with the same definition. The suprema agree. For the mirrored
constant, the first product stores at each recorded disagreement the start of the base token covering it, so by the
argument of Lemma~\ref{lem:max} its longest path over those starts is one plus the largest start of a changed base token
per input, or zero when none changes, which Proposition~\ref{prop:mirror} identifies with the mirrored quantity.
Lemma~\ref{lem:mark} and Theorem~\ref{thm:overlap} identify the overlap graph's with the same. This is an equivalence of
two computations of one definition. The census agreement of Table~\ref{tab:exec-overlap} is its transcription check.
\end{proof}
The construction is bounded in the size of the dictionary. Let \(n\) be the number of symbols plus the total length of
the rules' operands. \(V\) and \(E\) cost at most \(|W| + |V|^2\) tokenizations of strings of length at most \(2M\),
each polynomial in \(n\). The spelling automaton has \(1 + \sum_{t \in V} |t|\) states, and the overlap graph at most
the square of that. Cycle detection and the longest path are linear in the graph, so the origin radius and the mirrored
constant are computed in time polynomial in \(n\).

\paragraph{The number of rules.} That is efficiency in the size of the dictionary as written, though not in the number
of rules, since under clause (a) a product can have length \(2^i\) at rank \(i\). Measured by the size of a dictionary
given as a straight-line program over its operands, the decision, the exact radius and the canonical words become
questions about compressed words, which the compact companion~\citep{nidhogg2026compact} takes up, citing this one for
the semantics, canonical adjacency, the overlap graph and the criterion.

The graph decides the origin measure and the mirrored constant, not the label measure. On the one-rule dictionary
\((a, a)\) the cycle condition holds and the label distance is one on every input, so a cycle in \(C\) is no label
criterion. A finite origin radius bounds the label distance by Proposition~\ref{prop:labels}, and the ordering condition
by Theorem~\ref{thm:label}. Section~\ref{sec:labelgraph} decides it for every dictionary.

\begin{definition}[Reversed overlap graph]\label{def:revoverlap}
The \emph{reversed overlap graph} is the overlap graph of Definition~\ref{def:overlap} built from the reversed canonical
tokens, with the pair \((t^{\mathrm{rev}}, s^{\mathrm{rev}})\) for each pair \((s, t)\) of \(E\).
\end{definition}

By Theorem~\ref{thm:bigram} read from the right, its spelling automaton, the \emph{reversed spelling automaton}, accepts
exactly the reversed tokenizations, so Lemma~\ref{lem:mark} and Theorem~\ref{thm:overlap} hold for it with a symbol
\(b\) appended to \(x\) in place of the symbol prepended to the reversed string.

\begin{lemma}[Reset]\label{lem:reset}
For every dictionary, symbol \(a\) and input \(x\), the origin ripple \(\rho(a, x)\) is the first position at or after
zero that is a boundary of both \(\T^D(x)\) and \(\T^D(ax)\) shifted, the end of \(x\) counting as a boundary. Every
position before it has different covering spans, and every position from it on the same.
\end{lemma}

\begin{proof}
At a common boundary, Lemma~\ref{lem:proj} applied to the suffix from it gives the same tokenization of the same string
in both runs, so every later span agrees. Suppose a position before the first common boundary had equal spans. Their
common start would be an earlier common boundary, unless it is \(-1\), which is no boundary of the base run.
\end{proof}

\subsection{Finite right lookahead for every dictionary}\label{sec:lookahead}

Rationality says nothing about delay. How far an appended string reaches back into a tokenization is known in two
settings. The source paper's Theorem~2 bounds it by \(|D|\) tokens for proper dictionaries, and Lemma~\ref{lem:stab}
restates that under the ordering condition. From that bound \citet{berglund2024dfa} derive bounded variation and
conclude that the tokenization of a proper dictionary by one pass per rule, their Algorithm~1, is subsequential, their
Lemma~7 and Theorem~4, with no bound on the buffer. \citet{mamouras2026bpe} define the delay of a ruleset and bound it
by the sum of the lengths of the rules' left operands, their Theorem~10, and prove that the shortest prefix of a
tokenization that leaves at most that many symbols is a prefix of the tokenization of every extension, their Theorem~11,
the property a streaming tokenizer needs. Their rulesets are trained ones, in which every rule produces a fresh token,
and their proof composes one transducer per rule, the tokenizer being one pass per rule in that setting. That is the
ordering condition, in this paper's terms. This subsection proves their bound for every dictionary under
Definition~\ref{def:sp}, where a run is not a sequence of passes, sharpens it to the rules that can fire, and shows the
bound \(2^{s} - 1\) in the number \(s\) of those rules to be attained.

\begin{definition}[Right ripples]\label{def:rightripple}
For a string \(x\) and a symbol \(b\), the \emph{right span ripple} is the largest \(|x| - p\) over positions \(p\) of
\(x\) whose covering span in \(\T^D(xb)\) differs from its covering span in \(\T^D(x)\), zero when none differs, and the
\emph{right start ripple} is the same with the start of the span in place of the span. Their suprema over \(x\) and
\(b\) are the \emph{right span radius} \(Q_{\mathrm{span}}(D)\) and the \emph{right start radius}
\(Q_{\mathrm{start}}(D)\). Mirroring Definition~\ref{def:origin}, the \emph{left start ripple} of \(a\) on \(x\) is one
plus the largest position of \(x\) whose start differs between the two runs compared there, zero when none differs, and
the left start radius is its supremum.
\end{definition}

The right span radius is the value the reversed overlap graph of Definition~\ref{def:revoverlap} computes, and the
\emph{append product}, the pair graph of the spelling runs on \(x\) and \(xb\) read left to right, computes both right
radii.
Write \(N_r(w)\) for the result of Definition~\ref{def:sp} run on \(w\) with the first \(r\) rules alone, \(N_0(w)\)
being the symbols of \(w\), and call a token sequence \emph{\(r\)-normal} when it is \(N_r(w)\) for some \(w\). None of
the first \(r\) rules applies to it. For an \((r-1)\)-normal sequence \(U\) let \(A_r(U)\) be the result of running
Definition~\ref{def:sp} with the first \(r\) rules from \(U\). Then \(N_r(w) = A_r(N_{r-1}(w))\), since rule \(r\) is
first used only when no rule before it applies, so the run with \(r\) rules begins with the run with \(r - 1\) rules, to
its end. A word \(t\) is \emph{born} at the least \(r\) with \(N_r(t) = t\), a single token. Symbols are born at zero,
and a word canonical for the first \(r\) rules stays canonical for more, since \(A_{r+1}\) has nothing to merge in a
single token. By Lemma~\ref{lem:proj} a rule that fires is the last merge of the run on its own product, so each
canonical word has one useful producer, as in Lemma~\ref{lem:gtrees}. Write \(Q(D)\) for the sum of \(|u_j|\) over the
useful rules.

Since every token made at stage \(r\) is born at \(r\), a useful rule fires in one stage only, the birth stage of its
product. Write \(Q_r\) for the sum of \(|u_j|\) over the useful rules whose product is born at \(r\), so that
\(Q(D) = \sum_r Q_r\). Lemmas~\ref{lem:episodes} and~\ref{lem:onestage} in Appendix~\ref{app:lookahead} follow the run
of one stage \(r\) as a sequence of episodes, and show that two \((r-1)\)-normal sequences sharing a complete token
prefix of \(c\) symbols keep one of at least \(c - Q_r\) symbols through it.

\begin{theorem}[Right lookahead]\label{thm:lookahead}
For every dictionary and all strings \(w\) and \(z\), \(\T^D(w)\) and \(\T^D(wz)\) share a complete token prefix of at
least \(|w| - Q(D)\) symbols. With \(s\) the number of useful rules, \(Q(D) \le 2^{s} - 1\), and the dictionary
\((a, a), (a^2, a^2), \dots, (a^{2^{m-1}}, a^{2^{m-1}})\) attains \(2^m - 1\) on \(w = a^{2^m - 1}\) and \(z = a\).
\end{theorem}

\begin{proof}
As symbols, \(w\) and \(wz\) share the prefix \(w\). Apply Lemma~\ref{lem:onestage} for \(r = 1, \dots, |D|\) to the
\((r-1)\)-normal sequences \(N_{r-1}(w)\) and \(N_{r-1}(wz)\), and the shared prefix loses at most \(Q_r\) symbols at
stage \(r\). For the second claim, list the useful products in order of length. The operands of a useful rule are
canonical, so each is a symbol or an earlier, shorter useful product. By induction the \(i\)th product has length at
most \(2^i\) and its left operand at most \(2^{i-1}\), and the sum is at most \(2^s - 1\). In the doubling dictionary
every rule is useful, \(\T^D(a^{2^m})\) is one token, and no token of \(\T^D(a^{2^m - 1})\) is that long, so the two
share no token.
\end{proof}

\begin{corollary}[Subsequentiality]\label{cor:subsequential}
For every dictionary, \(\T^D\) is a subsequential function, computed by a deterministic left-to-right transducer whose
state is a buffer of at most \(Q(D)\) symbols, with a final output per state.
\end{corollary}

\begin{proof}
Keep a buffer \(u\), initially empty, and the invariant that with \(P\) the tokens emitted after reading \(w\),
\(\T^D(wz) = P\,\T^D(uz)\) for every \(z\). On a symbol \(a\) compute \(\T^D(ua)\). If \(|ua| \le Q(D)\), emit nothing
and keep \(ua\). Otherwise emit the tokens of \(\T^D(ua)\) up to its first boundary at or after \(|ua| - Q(D)\), and
keep the symbols after that boundary. By Theorem~\ref{thm:lookahead} the emitted tokens are a prefix of \(\T^D(uaz)\)
for every \(z\). By Lemma~\ref{lem:proj} the rest of \(\T^D(uaz)\) is the tokenization of the kept symbols followed by
\(z\), so the invariant holds, and the kept buffer has at most \(Q(D)\) symbols. At the end of the input emit
\(\T^D(u)\). The buffers are finitely many, and each transition and each final output is a tokenization of at most
\(Q(D) + 1\) symbols.
\end{proof}

On a trained ruleset every rule is useful or can be deleted, and \(Q(D)\) is then the bound of \citet{mamouras2026bpe},
whose delay is the loss measured here. Their batch bounds, which merge rules that cannot interfere into one transducer,
are a different refinement and are not compared. For proper dictionaries under the pass-based semantics of that paper,
one pass per rule in priority order, the corollary is Theorem~4 of \citet{berglund2024dfa}, obtained there through
bounded variation and stated with no buffer. Under the ordering condition that tokenizer is \(\T^D\)
(Lemma~\ref{lem:stab}); without it, one pass per rule need not compute \(\T^D\), as the five-rule dictionary of
Section~\ref{sec:semantics} shows on \(aabaab\). The corollary adds the buffer, the dictionaries that are not proper,
and the dictionaries whose priorities invert, under Definition~\ref{def:sp}. Without the ordering condition a run can
revisit rules of higher priority, as that run does, and alternate between rules without bound. The episodes are what
remains of the passes. The bound is in symbols, where the source paper's is in tokens. It is attained by the doubling
dictionary and lies far above the right span radius \(Q_{\mathrm{span}}(D)\) of Definition~\ref{def:rightripple} on
GPT-2's table, \ChainGptQspan{} against \ChainGptQD{}. On \LookaheadSeeded{} seeded dictionaries and the
\LookaheadCensus{} proper ones of one to four rules on two symbols, the right span radius never exceeds \(Q(D)\) and is
at most \LookaheadMaxRadius{}. It can be exponential in the number of rules, and no transducer polynomial in that
parameter is claimed. The left ripple can be infinite and the right one never is, for any dictionary. Reading left to
right, a byte-pair tokenizer needs a bounded window, and reading right to left it may need the whole input.
\citet[Example~8]{berglund2024dfa} make the mirror observation that with rules applied right-most, the dictionary
\((a, a)\) has no subsequential transducer.

The right span radius is also the delay of \citet{mamouras2026bpe}, who define the delay of a ruleset as that of its
tokenizer's normalized transducer, which emits output as early as possible, so that after reading \(x\) it has emitted
the longest prefix of \(\T^D(x)\) that every \(\T^D(xy)\) keeps and withholds the rest of \(x\).

\begin{lemma}[Delay]\label{lem:delay}
For every dictionary, \(Q_{\mathrm{span}}(D)\) is the supremum over strings \(x\) of \(|x|\) less the symbols of the
longest prefix of \(\T^D(x)\) that is a prefix of \(\T^D(xy)\) for every string \(y\). More generally, with a
nonnegative weight on each symbol, the largest weight, over the strings \(x\), of the symbols of \(x\) after that prefix
equals the largest weight, over the strings \(x\) and symbols \(b\), of the symbols of \(x\) after the longest common
prefix of complete tokens of \(\T^D(x)\) and \(\T^D(xb)\).
\end{lemma}

\begin{proof} For strings \(x\) and \(y\) let \(c(x, y)\) be the symbols of the longest common prefix of complete tokens
of \(\T^D(x)\) and \(\T^D(xy)\). The prefix of the lemma is the shortest of these over \(y\), so the supremum of the
lemma is the supremum of \(|x| - c(x, y)\) over all \(x\) and \(y\), and \(Q_{\mathrm{span}}(D)\) is the same supremum
over single symbols \(y\), since the right span ripple of \((x, b)\) is \(|x| - c(x, b)\), as the positions before
\(c(x, b)\) keep their spans, and when \(c(x, b) < |x|\) the token of \(\T^D(x)\) beginning at \(c(x, b)\) is not a
token of \(\T^D(xb)\), by maximality. This makes the first supremum at least the second. For the converse take any
positive \(d\) attained by some pair \((x, y)\) with \(|x| - c(x, y) \ge d\), and among such pairs one with \(y\)
shortest; \(y\) is nonempty, so write \(y = by'\). The tokens of \(\T^D(xb)\) ending at or before
\(\min(c(x, b), c(xb, y'))\) are tokens of \(\T^D(x)\) and of \(\T^D(xby')\), so \(c(x, y)\) is at least that minimum.
If \(c(x, b) \le c(x, y)\), the pair \((x, b)\) reaches \(d\), so \(y = b\). Otherwise \(c(xb, y') \le c(x, y)\), and
\(|xb| - c(xb, y') \ge d + 1\) with a shorter right part, against the choice of \(y\). The supremum over all \(y\) is
attained over single symbols. With weights the same argument holds with the weight of the symbols after the common
prefix in place of their number, since a longer suffix weighs at least as much and the appended symbol adds a
nonnegative weight.
\end{proof}


\begin{lemma}[Right start stability]\label{lem:rightstart}
For every dictionary: (i) \(Q_{\mathrm{start}}(D) \le Q_{\mathrm{span}}(D) \le Q(D)\), so both are finite; (ii) for all
strings \(x\) and \(z\) and every position \(p < |x| - Q_{\mathrm{start}}(D)\) of \(x\), the token covering \(p\) in
\(\T^D(xz)\) begins where it begins in \(\T^D(x)\); (iii) when \(\ripple(D)\) is finite, for all strings \(w\) and \(y\)
and every position \(p \ge \ripple(D)\) of \(y\), the covering span of \(p\) in \(\T^D(wy)\), shifted left by \(|w|\),
is its covering span in \(\T^D(y)\).
\end{lemma}

\begin{proof}
(i) A changed start is a changed span. By Theorem~\ref{thm:lookahead}, \(\T^D(x)\) and \(\T^D(xb)\) share a complete
token prefix of at least \(|x| - Q(D)\) symbols, and a position \(p < |x| - Q(D)\) lies in a token of that prefix, whose
span is then a span of both runs. Every changed span then lies at a position \(p \ge |x| - Q(D)\). (ii) For one appended
symbol this is the definition of \(Q_{\mathrm{start}}(D)\). Append \(z\) symbol by symbol. After the first symbol \(p\)
keeps its start, and \(p < |x| - Q_{\mathrm{start}}(D) \le |xb| - Q_{\mathrm{start}}(D)\), so the next symbol leaves it
too, and induction on \(|z|\) gives the claim. (iii) By Lemma~\ref{lem:D} the origin ripple of a prepended string is at
most \(\ripple(D)\), which is the claim.
\end{proof}

The induction in (ii) composes because the condition is on starts alone. The same argument with spans gives the mirror
of (iii) on the right, with \(Q_{\mathrm{span}}(D)\) in place of \(\ripple(D)\).

A streaming tokenizer can emit more than the buffer of Corollary~\ref{cor:subsequential} releases, namely every token
that no continuation of the input read so far can change, and nothing else. That is the zero delay of
\citet[Definition~4.9]{dalirrooyfard2026online}, and the list it emits is the greatest stable prefix of their
Definition~4.7. They leave open whether byte-pair encoding alone admits it efficiently (their Remark~1.4), having shown
that with a pre-tokenizer in front no polynomial algorithm achieves it unless \(\mathrm{P} = \mathrm{NP}\) (their
Theorem~1.1). The quantifier over all continuations reduces to finitely many of them.

\begin{definition}[Stable output]\label{def:stable}
For a string \(p\), \(\Stable(p)\) is the longest common prefix, as a list of tokens, of \(\T^D(pq)\) over all strings
\(q\). Let \(S\) be the empty string together with the proper suffixes \(t[i{:}]\), \(1 \le i < |t|\), of the canonical
tokens \(t \in V\).
\end{definition}

\begin{lemma}[Finitely many continuations]\label{lem:finitecont}
For every dictionary and every string \(p\), \(\Stable(p)\) is the longest common prefix of \(\T^D(ps)\) over
\(s \in S\). The same holds with \(S\) built from \(W\) in place of \(V\).
\end{lemma}

\begin{proof}
Write \(H\) for the longest common prefix over \(S\). Every \(s \in S\) is a continuation, so \(\Stable(p)\) is a prefix
of \(H\). Conversely fix \(q\), and take the first token boundary of \(\T^D(pq)\) at or after \(|p|\), which exists
since \(|pq|\) is one. Let \(r\) be the part of \(q\) before it. If the boundary is at \(|p|\), then \(r\) is empty.
Otherwise \(p\) ends strictly inside a token \(t\) of \(\T^D(pq)\), and \(r\) is a proper suffix of \(t\). Every token
of a run is a symbol or a product, and by Lemma~\ref{lem:proj} on the interval it spans \(\T^D(t) = t\), so \(t \in V\).
In both cases \(r \in S\). No merge of the run on \(pq\) crosses the chosen boundary, since it is a boundary of the
result, so by Lemma~\ref{lem:proj} the tokens of \(\T^D(pq)\) up to it are \(\T^D(pr)\). Hence \(H\) is a prefix of
\(\T^D(pr)\) and hence of \(\T^D(pq)\), for every \(q\), and \(H\) is a prefix of \(\Stable(p)\). With \(S\) built from
\(W\) the added suffixes are still continuations, so the common prefix lies between \(\Stable(p)\) and \(H\).
\end{proof}

The empty string in \(S\) is needed. On \((a, b)\) the suffixes alone give the single token \(ab\), while \(\T^D(a)\) is
the token \(a\), so \(\Stable(a)\) is empty. Tokens are compared as labels, not by the common prefix of their spellings,
since \(ab\) and \(a \mid b\) share no token.

\begin{theorem}[Zero delay]\label{thm:zerodelay}
For every dictionary there is a streaming tokenizer that, after reading each symbol of \(w\), has emitted exactly
\(\Stable\) of the input read, keeps a held suffix of at most \(Q_{\mathrm{span}}(D) \le Q(D)\) symbols, and at the end
of the input has emitted \(\T^D(w)\). After polynomial preprocessing, each symbol costs polynomial time, and the working
space is polynomial and independent of \(|w|\).
\end{theorem}

\begin{proof}
Let \(P\) denote the output already written, and keep a held suffix \(u\), both initially empty, with the invariant that
\(\T^D(wz) = P\,\T^D(uz)\) for every \(z\). On a symbol \(a\) replace \(u\) by \(ua\), compute \(H = \Stable(u)\) by
Lemma~\ref{lem:finitecont}, emit \(H\), and remove the yield of \(H\) from the front of \(u\), leaving \(u'\). Since
\(H\) is a prefix of every \(\T^D(uz)\), the end of its yield is a boundary of \(\T^D(uz)\), and Lemma~\ref{lem:proj} on
the interval from that boundary to the end makes the remaining tokens \(\T^D(u'z)\). The invariant persists with \(PH\)
and \(u'\). The longest common prefix of \(P\,\T^D(uz)\) over \(z\) is \(P\) followed by that of \(\T^D(uz)\), so after
the step the emitted list is \(\Stable\) of the input read. At the end, \(z\) empty gives \(\T^D(w) = P\,\T^D(u)\), and
\(\T^D(u)\) is flushed. The empty string in \(S\) keeps \(H\) a prefix of \(\T^D(u)\), so \(H\) spells only symbols
already read.

For the held suffix, the emitted list after the input \(x\) read so far is \(\Stable(x)\), the longest prefix of
\(\T^D(x)\) that is a prefix of \(\T^D(xy)\) for every \(y\), and the held suffix is the rest of \(x\). By
Lemma~\ref{lem:delay} its length is at most \(Q_{\mathrm{span}}(D)\), the right span radius of
Definition~\ref{def:rightripple}, which by Lemma~\ref{lem:rightstart}(i) is at most \(Q(D)\), so finite and attained.
Before a step the held suffix therefore has at most \(Q_{\mathrm{span}}(D) + 1\) symbols. The operands of a useful rule
are canonical, so \(Q(D) \le |D|\,M\), \(M\) the longest product. A step tokenizes at most
\(1 + \sum_{t \in V} (|t| - 1)\) words, each of at most \(Q_{\mathrm{span}}(D) + M\) symbols, and a run of
Definition~\ref{def:sp} on a word of length \(\ell\) makes at most \(\ell - 1\) merges, each found by scanning the rules
over the adjacent pairs. The preprocessing is \(V\), or just \(W\), and \(S\). The working space is the held suffix, one
run's workspace, the running common prefix and the dictionary.
\end{proof}

The spelling automaton of Definition~\ref{def:spelling} maintains the same stable output with no tokenization at run
time. Label each of its edges that reads the last symbol of its guessed token with that token, and the other edges with
nothing, so that a run outputs the tokens it completes, each on its last symbol, and an accepting run on \(x\) outputs
\(\T^D(x)\). Write \(n = 1 + \sum_{t \in V} |t|\) for its states and \(E_s\) for its edges.

\begin{theorem}[Zero delay by the spelling automaton]\label{thm:zerodelayauto}
For every dictionary, after computing \(V\), \(E\) and the spelling automaton, a streaming tokenizer that keeps one
partial run of the spelling automaton per active state, each with its pending output, emits exactly \(\Stable\) of the
input read after each symbol of \(w\) and \(\T^D(w)\) at the end, in \(O(E_s + n\,(Q_{\mathrm{span}}(D) + 1))\)
operations on token identifiers per symbol and that much working space, independent of \(|w|\).
\end{theorem}

The proof is in Appendix~\ref{app:lookahead}. Dalirrooyfard and Ye's \(\mathrm{Merge}\) of their Definition~4.4 makes
one left-to-right pass per merge in list order on a canonical trained list, whose operands precede their product. That
list satisfies the ordering condition, and by Lemma~\ref{lem:stab} its passes are Definition~\ref{def:sp}. Their online
byte-pair encoding of Definition~4.15 reads a list of segments with hard boundaries, which a flush at each boundary
handles. On their lists every product is a token, of at most \(t\) symbols by their Definition~4.5, so the list written
out has size \(O(|\Sigma| + mt)\) in their \(m\) merges, and polynomial in it is polynomial in their parameters. The two
theorems thus answer their Remark~1.4 for the byte-pair stage, for their lists and for every dictionary, with
preprocessing and total time polynomial in the input length and the merge list written out, the efficiency their
Theorem~1.1 rules out for the composed problem unless P equals NP. Neither reaches their near-linear dependence on the
dictionary's parameters, and the composed problem with special tokens and a pre-tokenizer stays NP-hard by that theorem.

Both are polynomial in the dictionary written out and not in the number of rules, since \(Q_{\mathrm{span}}(D)\) and
\(n\) can be exponential in it under clause (a). The first is the shorter proof, and the second is the implementation.
Indexing the automaton's edges by the symbol they read, a step on \(a\) scans only the active states' edges that read
\(a\), at most the \(E_a\) edges of the automaton that read \(a\), for \(O(E_a + n\,(Q_{\mathrm{span}}(D) + 1))\)
operations. The bound on the held suffix can be far below \(Q(D)\). On \((a_1, b_1), \dots, (a_m, b_m)\) over \(2m\)
distinct symbols \(Q(D) = m\) while \(Q_{\mathrm{span}}(D) = 1\). Two symbols suffice for an unbounded ratio. On
\((a, b), (ab, b), \dots, (ab^{m-1}, b)\) every rule fires, so \(Q(D) = 1 + 2 + \dots + m = m(m + 1)/2\), while the run
turns each \(a\) and the run of \(b\) after it into \(ab^j\), \(j\) at most \(m\), and single \(b\), so an appended
symbol changes only a last token \(ab^j\) with \(j < m\), of at most \(m\) symbols, and \(Q_{\mathrm{span}}(D) = m\),
attained by \(ab^{m-1}\) with \(b\) appended. By Corollary~\ref{cor:completion} in Appendix~\ref{app:lookahead}, the
exploration of Section~\ref{sec:machine} completes on every dictionary at \(B = Q(D) + 1\), so Theorem~\ref{thm:origin}
holds for every dictionary, at a tail that can be exponential in the number of rules.

Without a bound on the cost, emitting the stable output is classical for any function that a subsequential transducer
computes. Pushing onto each transition the longest common prefix of the outputs that can follow makes the transducer
emit, on every input, the longest common prefix of the outputs of its continuations \citep[Sections~2 and~3, pp.~205
and~207]{mohri1994compact}, and \citet[Theorem~4]{berglund2024dfa} build a subsequential transducer for every proper
dictionary under one pass per rule, leaving the state complexity of building it open (their Section~7).
Lemma~\ref{lem:finitecont} and the two theorems add the finite set of continuations for every dictionary and a
polynomial procedure, with no transducer compiled. Zero delay concerns what is emitted, and the held suffix of up to
\(Q_{\mathrm{span}}(D)\) symbols is still held. Two implementations stream byte-pair tokens in a related way. The
\code{encode\_with\_unstable} of \code{tiktoken} \citep{openai2023tiktoken} keeps back the tokens of at least the last
piece of its pattern, earlier tokens too when that piece begins with a whitespace token and they are whitespace tokens,
and none when the input ends in an allowed special token, and lists the token sequences that can complete the bytes it
keeps back, a finite completion whose stable part ends before that piece. The streaming tokenizer of \citet[Section~3.5
and Algorithm~2]{hayase2025bytesampler} keeps the unwritten part of the valid covering tree of the input read and writes
out its root once that node has a single child, so a token that no continuation changes can wait more than one byte, as
\(a\) does on \(abb\) under the one rule \((b, b)\). Cut each sequence of their Definition~3.1 after the first token
that reaches the end of the input. The cut sequences are the tokenizations of the continuations of
Lemma~\ref{lem:finitecont} so cut, the tokenization of the input read among them, and by that lemma their longest common
prefix is \(\Stable\) of the input read, for every dictionary. Writing the common prefix of the kept paths in place of
the root alone reaches zero delay on their lists at the cost of the tree. Lemma~\ref{lem:delay} bounds the symbols of
the input read after \(\Stable\) by \(Q_{\mathrm{span}}(D) \le |D|M\) for every dictionary, \(M\) the longest product,
while the last token of a cut sequence can reach \(M - 1\) symbols past the input, as \(ab\) does on \((a, b)\) after
\(a\), where \(Q_{\mathrm{span}}(D) = 1\); their Proposition~3.2, for merge lists in the normal form of their
Section~C.5, bounds the symbols of the input outside a shared trunk by the lookahead constant of
\citet{berglund2023formalizing}, and the two theorems emit \(\Stable\) with no tree.


On one symbol the right span radius is exact and cheap. A one-symbol dictionary's canonical tokens are lengths, and its
adjacency graph has exactly one token \(T\) with an edge to itself, no other cycle, and no edge into \(T\) from another
token. A longer cycle or a second loop would tokenize one length in two ways, and an edge into \(T\) would let a
tokenization begin with another token before arbitrarily many \(T\), against Theorem~\ref{thm:lookahead}. Every
tokenization is thus copies of \(T\) followed by a path that avoids \(T\), and the right span radius is the largest
total length \(H\) of such a path, since two tokenizations differ only in those tails, the input \(a^H\) is such a path,
and \(a^{H+1}\) must begin with \(T\). Executed, it equals the reversed overlap graph's radius on \UnaryRightTables{}
one-symbol dictionaries.

\subsection{Pinned bytes and cuts from a finite radius}\label{sec:density}

A finite origin radius and the right start radius of Definition~\ref{def:rightripple} together localize token starts.
Every window of width at least \(\ripple(D) + Q_{\mathrm{start}}(D) + 1\) pins each byte between offsets \(\ripple(D)\)
and \(|W| - 1 - Q_{\mathrm{start}}(D)\) at the start that byte has when the window is tokenized alone, in every context.
In particular, every window of width \(\ripple(D) + Q_{\mathrm{start}}(D) + 1\) has a cut, at the start of its byte
\(\ripple(D)\), and when the cut must lie at or after that byte, \(L - 1\) more symbols suffice, \(L\) the longest
canonical token. This is a uniform ceiling, and a planner need not enumerate windows, since it can decide the narrower
windows an input actually contains. A finite radius alone does not guarantee window certificates, which pin a window's
last byte, as the two-rule dictionary \((a, b), (a, ab)\) shows. The radius is finite exactly when cuts are dense at
some width. This subsection, with Appendix~\ref{app:windows}, proves these statements, with a bound on each radius read
off the rules, computes both right radii exactly from the rules under the ordering condition, and settles the least
width.

A window certificate asks the question of this paper at a fixed window instead of a fixed input. For a window \(W\), a
nonempty string over the alphabet, and an origin \(0 \le o < |W|\), \(D\) \emph{certifies} \((W, o)\) when at every
occurrence of \(W\) at a position \(p\) of every string, the token covering position \(p + |W| - 1\) begins at
\(p + o\). That is the certification of \citet[Definition~2]{nidhogg2026finitelyverifiable}, stated there for a
tokenizer presented by a finite-state verifier and decided over the verifier. A position of a window is called a byte,
as there, over any alphabet. More generally \(W\) \emph{pins} its byte \(j\) at \(o\), written \((W, j, o)\) with
\(0 \le o \le j < |W|\), when at every occurrence the token covering \(p + j\) begins at \(p + o\). Certification is the
case \(j = |W| - 1\), and a \emph{cut} at \(a\) is the case \((W, a, a)\), a token boundary at \(p + a\) at every
occurrence, the byte-pair form of a forced gap in the boundary profile of \citet{nidhogg2026splitwindows}. Since \(W\)
occurs in itself, a byte is pinned at its start in \(\T^D(W)\) or not at all, and since the token covering \(p + j\)
begins at \(p + o\), a pinned byte \((W, j, o)\) gives the cut \((W, o, o)\). A window thus pins a byte exactly when it
has a cut. A set of windows is \emph{dense} when only finitely many strings contain no occurrence of any window in the
set, and its \emph{density bound} \(N\) is one more than the length of the longest such string, so that every string of
length at least \(N\) contains one. Density here is the byte-pair counterpart of a bounded anchor-free span in
\citet{nidhogg2026splitting}, with every string in place of the completely tokenizable inputs, since a dictionary
tokenizes them all, and a different notion from the anchor density of \citet{nidhogg2026splitwindows}, which counts
anchors per kibibyte of input. The simplest cut needs no radius at all. When no token of \(D\) contains \(xy\), no token
can cover both symbols of an occurrence, so \((xy, 1, 1)\) is a cut; that sufficient condition is the test three recent
tokenizer implementations use to find cuts, Gigatoken only before selected punctuation
\citep{rod2026gigatoken, sentencepiecelite2026, xu2026parallel}, and it supplies cuts only where such pairs occur.


\begin{lemma}[Local refutation]\label{lem:localref}
\((W, j, o)\) fails exactly when some list of tokens of \(V\) with adjacent pairs in \(E\) spells a string that contains
\(W\) at a position \(q\), its first token covering \(q\), its last covering \(q + |W| - 1\), and the token covering
\(q + j\) not beginning at \(q + o\). Hence a window that fails has a counterexample of at most \(2L + |W| - 2\)
symbols, and whether \(W\) pins a byte is decided by tokenizing the strings \(uWv\) with \(|u|, |v| < L\).
\end{lemma}

\begin{proof}
Such a list is the tokenization of its spelling by Theorem~\ref{thm:bigram}, and the spelling refutes. Conversely take a
counterexample and an occurrence it covers wrongly. The tokens of its tokenization that overlap the occurrence are in
\(V\) with adjacent pairs in \(E\), by the theorem, and by the theorem again they are the tokenization of their own
spelling, which contains the occurrence with the same covering. Each of the two end tokens overhangs the occurrence by
fewer than \(L\) symbols.
\end{proof}


\begin{theorem}[Pinned byte]\label{thm:pinned}
Let \(R = \ripple(D)\) be finite and \(q = Q_{\mathrm{start}}(D)\). Every window \(W\) of width at least \(R + q + 1\)
pins each byte \(j\) with \(R \le j \le |W| - 1 - q\) at its start in \(\T^D(W)\). In particular every window of width
\(R + q + 1\) pins its byte \(R\) and has a cut at the start \(s\) of that byte, with
\(\max(0, R - L + 1) \le s \le R\), \(L\) the longest canonical token.
\end{theorem}

\begin{proof}
Let \(W\) occur at \(p\) in \(x = uWv\), and let \(s\) be the start of position \(j\) in \(\T^D(W)\). By
Lemma~\ref{lem:rightstart}(ii) on \(W\) and \(v\), the position \(j < |W| - q\) has the start \(s\) in \(\T^D(Wv)\) too.
By Lemma~\ref{lem:rightstart}(iii) on \(u\) and \(Wv\), the position \(j \ge R\) of \(Wv\) keeps its covering span in
\(\T^D(uWv)\), shifted by \(|u|\), so the token covering \(p + j\) in \(\T^D(x)\) begins at \(p + s\). That token begins
at \(p + s\) at every occurrence, so \((W, s, s)\) is a cut, and the token covering position \(R\) has at most \(L\)
symbols, so its start is at least \(R - L + 1\).
\end{proof}

The theorem has an operational form. For an input \(x\) and a byte \(i\), let \(h(x, i)\) be the least width of a window
of \(x\) around \(i\) that pins the byte \(i\), infinite when there is none. Extending a window keeps its pinned bytes,
with the offsets shifted, since every occurrence of the extension is an occurrence of the window. A planner can then try
short and lopsided windows around a wanted cut and widen only those that fail, deciding each by Lemma~\ref{lem:localref}
with no radius in hand. Theorem~\ref{thm:pinned} says that \(h(x, i) \le R + q + 1\) whenever \(i \ge R\) and
\(i + q < |x|\), and it says which window. Tokenize \(x[i - R{:}i + q + 1]\) alone and read the start \(s\) of its byte
\(R\). Then \(i - R + s\) is a boundary of \(\T^D(x)\), between \(i - L + 1\) and \(i\), and by Lemma~\ref{lem:proj} the
two pieces of \(x\) on either side of it tokenize independently. A byte that is interior to a token of \(\T^D(x)\) never
becomes a cut by more context, since \(x\) refutes it, and the pinned start is the nearby cut that context does give. A
finite radius bounds the left. On the right the cost is the right start radius, which is cheaper than the span radius by
at least one whenever the span radius is positive.

For every dictionary the left start radius is \(\ripple(D)\), and the right start radius is at most the right span
radius less one when that is positive, by Proposition~\ref{prop:sides} in Appendix~\ref{app:windows}, which also shows
that the literal converse of the density bound, that a finite radius makes the certified windows of some width dense, is
false, Proposition~\ref{prop:converse}. A finite radius does give pinned bytes, and the next theorem shows that the
radius is finite exactly when, for some \(h\), the windows of width at most \(h\) that pin a byte are dense.


\begin{theorem}[Density]\label{thm:density}
For every dictionary, \(\ripple(D)\) is finite exactly when for some width \(h\) the windows of width at most \(h\) that
have a cut, equivalently that pin a byte, are dense. When \(R = \ripple(D)\) is finite and
\(q = Q_{\mathrm{start}}(D)\), every window of width \(R + q + L\) has a cut at an offset in \([R, R + L - 1]\), and the
density bound \(N\) of the windows of width at most \(R + q + 1\) that have a cut satisfies \(R \le N - 1 \le R + q\).
\end{theorem}

\begin{proof}
Let \(R\) be finite. By Theorem~\ref{thm:pinned} every window of width \(R + q + 1\) has a cut, and every string of that
length or more contains its own prefix of that width, so the cuts are dense at that width and \(N \le R + q + 1\). For
the cut at or beyond \(R\), let \(|W| = R + q + L\). The positions \(R, \dots, R + L - 1\) are at most \(|W| - 1 - q\),
so Theorem~\ref{thm:pinned} pins each at its start in \(\T^D(W)\). The token covering position \(R\) begins at some
\(s \le R\) and has at most \(L\) symbols, so either \(s = R\) or the next token begins at a position \(a\) with
\(R < a \le R + L - 1\); in both cases a position \(a \in [R, R + L - 1]\) begins a token of \(\T^D(W)\), and
\((W, a, a)\) is a cut.

Conversely let \((W, j, o)\) pin a byte, let \(W\) occur in \(x\) at \(p\), and let \(a = p + o\). Then \(x\) has a
token boundary at \(a\). The tokens before \(a\) and the tokens from \(a\) on are, by Lemma~\ref{lem:proj}, since no
merge crosses \(a\), \(\T^D(x[0{:}a])\) and \(\T^D(x[a{:}])\). For every symbol \(c\) the string \(cx\) contains \(W\)
at \(p + 1\), so it has a boundary at \(a + 1\) and its tokens from there are again \(\T^D(x[a{:}])\). Every position of
\(x\) at or beyond \(a\) keeps its span, and \(\rho(c, x) \le a\). Now suppose every string of length \(N\) contains a
window of width at most \(h\) that pins a byte. A string \(x\) of length at least \(N\) contains such an occurrence
inside its first \(N\) symbols, with \(p + |W| \le N\), so \(a \le p + j \le N - 1\) and \(\rho(c, x) \le N - 1\); a
shorter string has ripple at most its length. This gives \(R \le N - 1\), which is the lower bound on \(N\) and the
converse direction at once.
\end{proof}

With an infinite radius the bounded distance between cuts is lost, while the windows with a cut remain. A window can
still pin a byte, and only a width at which every string is guaranteed one is gone. Certification, the last byte pinned,
asks more than Theorem~\ref{thm:pinned} gives.


\begin{proposition}[The last byte]\label{prop:lastbyte}
Let \(R = \ripple(D)\) be finite and \(|W| \ge R + 1\). Then \(D\) certifies \((W, o)\) exactly when for every string
\(v\) the token covering the last byte of \(W\) in \(\T^D(Wv)\) begins at \(o\). For every dictionary, whether the
certified windows of some width are dense is decidable.
\end{proposition}

\begin{proof}
At an occurrence of \(W\) at \(p\) in \(uWv\), Lemma~\ref{lem:rightstart}(iii) on \(u\) and \(Wv\) at the position
\(|W| - 1 \ge R\) says the token covering the last byte of \(W\) has the span it has in \(\T^D(Wv)\), shifted by
\(|u|\). The left context thus drops out, and \((W, o)\) is certified exactly when that token begins at \(o\) in
\(\T^D(Wv)\) for every \(v\).

For the decision, Theorem~\ref{thm:overlap} decides whether \(R\) is finite. If it is not, the certified windows of no
width are dense by Theorem~\ref{thm:density}, since a certified window pins a byte. Let \(R\) be finite. If \((W, o)\)
is certified with \(|W| = h > R + 1\), let \(W'\) be the suffix of \(W\) of width \(R + 1\) and
\(o' = o - (h - R - 1)\). For every \(v\), the start of the last byte of \(W'\) in \(\T^D(W'v)\) equals, by
Lemma~\ref{lem:rightstart}(iii) on the prefix \(W[{:}h - R - 1]\) of \(Wv\) and the position \(R\) of \(W'v\), the start
of the last byte of \(W\) in \(\T^D(Wv)\) less \(h - R - 1\), which is \(o'\). Starts in \(\T^D(W'v)\) are at least
\(0\), so \(o' \ge 0\), and by the first claim \((W', o')\) is certified. Every string containing a certified window
then contains one of width at most \(R + 1\), and the certified windows of some width are dense exactly when those of
width at most \(R + 1\) are. Those are finitely many, each decided by Lemma~\ref{lem:localref}, and the strings avoiding
a finite set of windows form a regular language, whose finiteness is decidable.
\end{proof}

\begin{corollary}[Zero start radius]\label{cor:lastbyte}
If \(R = \ripple(D)\) is finite and \(Q_{\mathrm{start}}(D) = 0\), every window of width \(R + 1\) is certified at the
start of its last byte in its own tokenization, and the certified windows of width at most \(R + 1\) are dense with
density bound \(R + 1\).
\end{corollary}

\begin{proof}
Theorem~\ref{thm:pinned} at width \(R + 1\) pins the byte \(R\), the last. Every string of length \(R + 1\) is then
certified, so the density bound is at most \(R + 1\), and Theorem~\ref{thm:density} gives \(R \le N - 1\), since
certified windows have cuts and so their density bound is at least that of the windows with a cut.
\end{proof}


\begin{proposition}[Bounds in the rules]\label{prop:radiibounds}
Let \(S\) be the sum of the lengths of the left operands over all rules and \(n_0\) one more than the number of symbols
plus the lengths of the distinct products summed. For every dictionary, with \(L\) the longest canonical token: (i)
\(\ripple(D) \ge L - 1\); (ii) call a \emph{pair set} a set of two states of the spelling automaton, other than its
start state, entered by one symbol at different offsets, the automaton having \(n = 1 + \sum_{t \in V} |t| \le n_0\)
states; call a pair set of states \((t, j)\) and \((t', j')\) \emph{comparable} when its unread suffixes \(t[j{:}]\) and
\(t'[j'{:}]\) are one a prefix of the other, and \emph{doubly comparable} when its consumed prefixes \(t[{:}j]\) and
\(t'[{:}j']\) are also one a suffix of the other. Let \(B\), \(B'\) and \(B''\) be the numbers of pair sets, of
comparable ones and of doubly comparable ones, and \(F\) and \(F''\) the numbers of pair sets and of doubly comparable
ones with both states ending their tokens. Then
\(B = \sum_{c} \bigl(\binom{n_c}{2} - \sum_{j} \binom{n_{c,j}}{2}\bigr)\), with \(n_c\) the occurrences of the symbol
\(c\) in the canonical tokens, the states entered by \(c\), and \(n_{c,j}\) the canonical tokens whose symbol at offset
\(j\) is \(c\), and \(F\) is the same sum with the canonical tokens ending in \(c\) in place of \(n_c\) and those of
length \(j + 1\) among them in place of \(n_{c,j}\). If \(\ripple(D)\) is finite, then
\(\ripple(D) \le B'' - \max(0, F'' - 1) \le B' - \max(0, F - 1) \le B - \max(0, F - 1) \le B \le \binom{n - 1}{2}\);
(iii) \(Q_{\mathrm{start}}(D) \le \max(0, S - 1)\); (iv) under the ordering condition with \(\ripple(D)\) finite,
\(\ripple(D) + Q_{\mathrm{start}}(D) + 1\) is at most the sum of the lengths of the useful products, and is \(1\) when
no rule is useful. (v) For \(m \ge 1\) and distinct symbols \(a_0, \dots, a_m\), the chain dictionary
\(K_m = (a_{m-1}, a_m), (a_{m-2}, a_{m-1}), \dots, (a_0, a_1)\), in that priority order, is ordered with
\(\ripple(K_m) = 1\), \(L = 2\), \(Q_{\mathrm{span}}(K_m) = m\) and \(Q_{\mathrm{start}}(K_m) = m - 1 = S - 1\).
\end{proposition}

The proof is in Appendix~\ref{app:windows}. The bound (iii) needs no graph and only the list of rules, and (v) says it
can be tight even with tokens of two symbols, so a better estimate has to use the interactions between rules, as the
chain length of \citet{berglund2023formalizing}, their Remark~5, does for the lookahead constant. Under the ordering
condition the rest of this subsection computes both right radii exactly from those interactions, by chains of rules. On
a byte alphabet the chain family stops at \(m = 255\). The bound (ii) is loose in general and says only that a finite
radius is bounded in the size of the dictionary, without the ordering condition. On \((a, b)\) it is exact, at one,
where \(\binom{n - 1}{2}\) is six. The chain length of the source paper's Remark~5 is the longest sequence of rules of
decreasing priority that every dictionary with the same tokenizer contains, a property that bounds the lookahead. The
chains below are the ones a particular input realizes, each weighted by its left operands, and their largest weight is
the right span radius. The first step, Proposition~\ref{prop:startspan} in Appendix~\ref{app:windows}, holds for every
dictionary. For a string \(x\) and a symbol \(b\), the \emph{last common boundary} \(c\) is the end of the longest
common prefix of complete tokens of \(\T^D(x)\) and \(\T^D(xb)\), the \emph{changed suffix} is \(s = x[c{:}]\), and
\(t\) and \(t'\) are the first tokens of \(\T^D(s)\) and \(\T^D(sb)\).

Under the ordering condition the changed suffix is built from a chain of rules. For a string \(s\) and a symbol \(b\)
write \(f_r\) for the end of the longest common prefix of complete tokens of \(N_r(s)\) and \(N_r(sb)\), the
\emph{frontier} after stage \(r\), so that \(f_0 = |s|\). The next lemma is proved in Appendix~\ref{app:windows}.


\begin{lemma}[No absorption]\label{lem:noabsorb}
Assume the ordering condition, and let \(s\) be a string and \(b\) a symbol. The frontier never moves right. It moves at
stage \(r\) exactly when the pass of the rule \((u, v)\) of rank \(r\) merges an occurrence crossing \(f_{r-1}\), which
happens in at most one of the two runs, since the token ending at \(f_{r-1}\), the same in both, is \(u\), and the token
beginning there is \(v\) in that run and not in the other. Then \(f_r = f_{r-1} - |u|\), rule \(r\) is useful, and after
the pass the tokens beginning at \(f_r\) are \(uv\) in that run and \(u\) in the other. When no crossing occurrence is
merged, \(f_r = f_{r-1}\), and two different tokens beginning there before the pass leave two different tokens beginning
there after it.
\end{lemma}

\begin{theorem}[Chain form]\label{thm:chainform}
Assume the ordering condition, let \((x, b)\) have positive right span ripple, and let \(s\) be its changed suffix. Then
\(f_{|D|} = 0\). Let \(r_1 < \cdots < r_k\) be the stages at which the frontier of \((s, b)\) moves. Then
\(s = u_{r_k} \cdots u_{r_1}\), each rule \(r_i\) is useful, \(v_{r_1} = b\), and after stage \(r_i\) the tokens
beginning at the frontier are \(u_{r_i}\) in one run and \(u_{r_i} v_{r_i}\) in the other and never again coincide. The
right span ripple of \((x, b)\) is \(|s|\), the chain's left operands summed, and the right start ripple is
\(|s| - \min(|t|, |t'|)\) with \(\min(|t|, |t'|) \ge |u_{r_k}|\). In particular \(Q_{\mathrm{span}}(D) \le Q(D)\), and
the right start ripple of an input is at most \(Q(D) - |u_{r_k}|\) for the last rule \(r_k\) of its chain.
\end{theorem}

\begin{proof}
By Proposition~\ref{prop:startspan} \(\T^D(s)\) and \(\T^D(sb)\) differ in their first token, so \(f_{|D|} = 0\), while
\(f_0 = |s|\). By Lemma~\ref{lem:noabsorb} the frontier moves only at the stages \(r_i\), by \(|u_{r_i}|\) each, so the
moves sum to \(|s|\), and at the move of stage \(r_i\) the token \(u_{r_i}\) ends the common prefix, occupying
\([f_{r_i}, f_{r_{i-1}})\), since the frontier stands still between moves; so \(s = u_{r_k} \cdots u_{r_1}\), and each
\(r_i\) is useful by the lemma. Before the first move the frontier is \(|s|\) and the token beginning there in the
second run is \(b\), so the first move merges \(u_{r_1}\) with \(b\), and \(v_{r_1} = b\). After the move at \(r_i\) the
tokens beginning at the frontier are \(u_{r_i} v_{r_i}\) and \(u_{r_i}\), and the lemma's last sentence, applied at
every stage until the next move, keeps them different. Proposition~\ref{prop:startspan} gives both ripples. After stage
\(r_k\) the token beginning at position \(0\) is \(u_{r_k}\) or \(u_{r_k} v_{r_k}\), and a later merge can only grow it,
since a merge with it as right operand would need a token before position \(0\); so both final first tokens have at
least \(|u_{r_k}|\) symbols. The chain's rules are distinct and useful, so \(|s| \le Q(D)\).
\end{proof}

The theorem turns the search for an attaining input into a search over chains. Call \((s, b)\), a string and a symbol, a
\emph{state} when \(\T^D(s)\) and \(\T^D(sb)\) differ in their first token, and give it the rank \(r_k\) of its last
moving stage. Under the ordering condition the merges of a run come in nondecreasing rank, so one run of
Definition~\ref{def:sp} on \(s\) that records how the token at position \(0\) grows, the \emph{first-token history} of
\(s\), gives the first token of every \(N_r(s)\).


\begin{theorem}[Chain search]\label{thm:chainsearch}
Assume the ordering condition. The states with one moving stage are exactly the pairs \((u_r, v_r)\) for the useful
rules \(r\) whose right operand is a symbol, each of rank \(r\). Let \((s, b)\) be a state of rank \(r_k\), let
\(r' > r_k\) be useful with \(u = u_{r'}\), and let \(t\) and \(t'\) be the first tokens of \(N_{r'-1}(s)\) and
\(N_{r'-1}(sb)\). Then \((us, b)\) is a state of rank \(r'\) whose moving stages are those of \((s, b)\) and \(r'\)
exactly when \(v_{r'}\) is one of \(t\) and \(t'\) and, with \(w\) the other, \(N_{r'-1}(uw) = u \mid w\). Every state
is reached from a one-stage state by a unique sequence of such extensions. Hence the search that starts from the
one-stage states and extends each state by every rule that passes the test visits every state exactly once, and
\(Q_{\mathrm{span}}(D)\) is the largest \(|s|\) and \(Q_{\mathrm{start}}(D)\) the largest \(|s| - \min(|t|, |t'|)\) over
the states visited, where the first tokens of \(\T^D(s)\) and \(\T^D(sb)\) take the place of \(t\) and \(t'\). Each test
costs one lookup in the first-token histories of \(s\) and \(sb\) and one tokenization of \(uw\) with the rules of rank
below \(r'\).
\end{theorem}

The proof is in Appendix~\ref{app:windows}. A graph on the rules bounds the search and prunes it, by
Proposition~\ref{prop:rulegraph} in Appendix~\ref{app:windows}, which bounds \(Q_{\mathrm{span}}(D)\) and
\(Q_{\mathrm{start}}(D)\) by its heaviest paths \(C(D)\) and \(C_{\mathrm{start}}(D)\).

The lookahead constant of the source paper's Definition~6, for a dictionary proper in the conference text's sense, is
the least nonnegative integer \(l(D)\) such that a tokenization \(\phi\tau\) of a string with
\(|\yield(\tau)| \ge l(D)\) keeps \(\phi\) under every appended string. The condition reads for every dictionary, and
the corollary below gives its least value.

\begin{corollary}[Lookahead constant]\label{cor:lookconst}
For every dictionary \(l(D) = \min(Q_{\mathrm{span}}(D), A(D) + 1)\), where \(A(D)\) is the largest \(|s| - |t|\) over
the pairs \((s, b)\) of a string and a symbol whose tokenizations \(\T^D(s)\) and \(\T^D(sb)\) differ in their first
token, \(t\) the first token of \(\T^D(s)\), and \(A(D) = 0\) when there is no such pair. If there is such a pair,
\(l(D) = A(D) + 1 \le \min(Q_{\mathrm{span}}(D), Q_{\mathrm{start}}(D) + 1)\), and if there is none,
\(l(D) = Q_{\mathrm{span}}(D) = 0\).

With \(l_b(D)\) one more than the largest \(|s| - |t|\) over the pairs \((s, b)\) with the symbol \(b\), and zero when
there is none, the least constant of the source paper's Definition~6 for the appended strings over a set \(\Sigma'\) of
symbols is \(\max(\{l_b(D) : b \in \Sigma'\} \cup \{0\})\), and \(l_b(D)\) is the mirrored constant of the reversed
overlap graph under \(b\), so that \(l(D)\) is computed with an attaining input in polynomial time.

For every dictionary, \(l_b(D) = 0\) exactly when no canonical word of two symbols or more ends in \(b\), exactly when
no rule that fires has the right operand \(b\). The symbols \(b\) with \(l_b(D) = 0\) thus form the largest set \(Z\) of
symbols with \(\T^D(xz) = \T^D(x) \mid z_1 \mid \cdots \mid z_n\) for every string \(x\) and every string
\(z = z_1 \cdots z_n\) over \(Z\), and when \(l_b(D) > 0\) the shortest strings \(s\) of a pair \((s, b)\) are exactly
the shortest left operands \(u\) of the rules \((u, b)\) that fire. Read from the left, the origin radius under a symbol
\(a\) is zero exactly when no rule that fires has the left operand \(a\), and otherwise the shortest inputs \(y\) with
\(\rho(a, y) > 0\) are exactly the shortest right operands \(v\) of the rules \((a, v)\) that fire. A symbol \(c\) that
is an operand of no rule that fires, on either side, is a universal delimiter,
\(\T^D(xcy) = \T^D(x) \mid c \mid \T^D(y)\) for all strings \(x\) and \(y\), and no other symbol is, so the universal
delimiters are the symbols of \(Z\) whose origin radius is zero, which are the symbols that occur in no canonical word
of two symbols or more.

When the rules that fire satisfy the ordering condition, \(A(D)\) is the largest \(|s| - |t|\) over the states of the
chain search.
\end{corollary}

\begin{proof}
A pair \((s, b)\) with \(\T^D(s) = t\psi\) gives a tokenization of \(s\) with \(|\yield(\psi)| = |s| - |t|\) whose first
token appending \(b\) removes, so \(l(D) > |s| - |t|\). Conversely let \(\T^D(x) = \phi\tau\), with
\(|\yield(\tau)| \ge A(D) + 1\) when a pair exists, and suppose an appended string removes \(\phi\). Take a shortest
one, \(zb\), as in the proof of Lemma~\ref{lem:delay}. Then \(\T^D(xz)\) begins with \(\phi\) and \(\T^D(xzb)\) does
not, so the longest common prefix of complete tokens of the two is a proper prefix of \(\phi\), ending at some \(c\) and
followed in \(\T^D(xz)\) by a token \(t\) of \(\phi\), with \(c + |t| \le |\yield(\phi)|\). By Lemma~\ref{lem:proj} on
the interval from \(c\), with \(s = (xz)[c{:}]\), the rests of the two tokenizations are \(\T^D(s)\) and \(\T^D(sb)\),
which differ in their first token, and \(|s| - |t| \ge |xz| - |\yield(\phi)| \ge |\yield(\tau)|\). If a pair exists,
this exceeds \(A(D)\), and if none does, \((s, b)\) would be one, so in either case no appended string removes \(\phi\),
and \(l(D) \le A(D) + 1\), or \(l(D) = 0\) when no pair exists. A pair \((s, b)\) has last common boundary zero, so by
Proposition~\ref{prop:startspan} its right span ripple is \(|s| \ge |s| - |t| + 1\) and its right start ripple
\(|s| - \min(|t|, |t'|) \ge |s| - |t|\), \(t'\) the first token of \(\T^D(sb)\), which gives the two bounds. A positive
right span ripple gives a pair by the same proposition, so with none \(Q_{\mathrm{span}}(D) = 0\).

For a set \(\Sigma'\) of symbols the same argument holds with the appended strings drawn from \(\Sigma'^*\). A pair
\((s, b)\) with \(b \in \Sigma'\) gives a tokenization whose first token the appended \(b\) removes, and when a string
over \(\Sigma'\) removes \(\phi\), a shortest one, \(zb\), has \(b \in \Sigma'\) and gives a pair \((s, b)\) with
\(|s| - |t| \ge |\yield(\tau)|\). The least constant is thus the largest \(l_b(D)\) over \(b \in \Sigma'\), or zero, and
with \(\Sigma' = \Sigma\) it is \(l(D)\). A changed token \([c, f)\) of \(\T^D(x)\) under \(b\) begins at \(|x| - f\) in
the reversed string, so the mirrored constant of the reversed overlap graph of Definition~\ref{def:revoverlap} under
\(b\) is one more than the largest \(|x| - f\) over the first changed tokens, and zero when no token changes. The first
changed token begins at the last common boundary \(c\), and by Lemma~\ref{lem:proj}, \(s = x[c{:}]\) gives a pair
\((s, b)\) whose first token \(t\) ends at \(f\), with \(|x| - f = |s| - |t|\), while every pair is the case \(x = s\)
and \(c = 0\). The mirrored constant is thus \(l_b(D)\), and Theorem~\ref{thm:overlap} computes it on the reversed
graph, with an attaining path, in polynomial time.

For the zero case, a canonical word \(w = w'b\) of two symbols or more gives the pair \((w', b)\), since \(\T^D(w')\)
begins with a token of \(w'\) and \(\T^D(w'b) = w\). Conversely the last token of \(\T^D(sb)\) for a pair \((s, b)\)
ends in \(b\), and it is not the symbol \(b\) alone, since then \(|s|\) would be a boundary of \(\T^D(sb)\) and
Lemma~\ref{lem:proj} would give \(\T^D(sb) = \T^D(s) \mid b\), with the same first token. By Lemma~\ref{lem:gtrees} a
canonical word of two symbols or more ending in \(b\) has a rule that fires with the right operand \(b\), while the
product of such a rule is such a word. Over \(Z\) the least constant is zero, so an appended string over \(Z\) keeps all
of \(\T^D(x)\), and every token after it is a canonical word ending in a symbol of \(Z\), which is that symbol alone,
while a symbol outside \(Z\) has a pair \((s, b)\) and changes \(\T^D(s)\). For the shortest pair, a rule \((u, b)\)
that fires gives the pair \((u, b)\), since its operand \(u\) is canonical and its product is one token. Conversely the
last token of \(\T^D(sb)\) is a canonical word \(w'b\) with \(w'\) a suffix of \(s\), and by Lemma~\ref{lem:gtrees} it
has a rule \((u, b)\) that fires whose product \(ub\) is a suffix of \(w'b\), so that \(u\) is a suffix of \(w'\). For a
shortest \(s\), \(|u| \ge |s|\) since \((u, b)\) is a pair, so \(u = s\). From the left, by Lemma~\ref{lem:reset},
\(\rho(a, y) > 0\) exactly when the first token of \(\T^D(ay)\) is a canonical word \(av\) with \(v\) a nonempty prefix
of \(y\), and by Lemma~\ref{lem:gtrees} it has a rule \((a, v')\) that fires with \(av'\) a prefix of \(av\), while such
a rule gives \(\rho(a, v') > 0\), as \(\T^D(v') = v'\) and \(\T^D(av') = av'\), and for a shortest \(y\), \(v' = y\) in
the same way. For the delimiters, in the run on \(xcy\) the symbol \(c\) starts as a token of its own, and a merge that
consumed it would apply a rule with the operand \(c\) on one side, while a rule that is applied fires. The token \(c\)
thus survives to the end, both its boundaries are boundaries of every step, no merge crosses either, and
Lemma~\ref{lem:proj} on the intervals \(x\) and \(y\) makes the merges inside them the whole runs on \(x\) and on \(y\)
alone. Conversely a rule \((u, c)\) that fires has \(\T^D(uc) = uc \ne \T^D(u) \mid c\), and a rule \((c, v)\) that
fires has \(\T^D(cv) = cv \ne c \mid \T^D(v)\). By the two zero criteria above, no rule that fires has the right operand
\(c\) exactly when \(c \in Z\), and none has the left operand \(c\) exactly when the radius under \(c\) is zero. A
symbol in a canonical word \(w\) of two symbols or more is a leaf of the derivation of \(w\) (Lemma~\ref{lem:gtrees}),
and the rule of its parent is a producer, which fires, with that symbol as an operand. Conversely the product of a rule
that fires with the operand \(c\) is a canonical word of two symbols or more that contains \(c\).

The useful sublist of Lemma~\ref{lem:gtrees} has the tokenizer of \(D\), and under the ordering condition the chain
search visits every pair, its states, exactly once (Theorem~\ref{thm:chainsearch}).
\end{proof}

The bound by the start radius can be strict, as on \((a, a), (b, a)\) and on \((a, b), (b, a)\), where \(l(D) = 1\) and
\(Q_{\mathrm{start}}(D) + 1 = 2\). Without the ordering condition the chain search's values need not be radii,
Proposition~\ref{prop:chainsfail} in Appendix~\ref{app:windows}.

The width \(R + q + 1\) of Theorem~\ref{thm:pinned} is the least that pins byte \(R\) in every window exactly under a
condition on the attaining inputs, and the condition can fail. Appendix~\ref{app:windows} gives the condition,
Proposition~\ref{prop:exactwidth}, the least width, Proposition~\ref{prop:leastwidth}, and a family on which the least
width falls without bound below the width of Theorem~\ref{thm:pinned}, Proposition~\ref{prop:slack}.

In practice a planner has three routes. With a finite radius, Theorem~\ref{thm:pinned} is a worst-case guarantee with a
constructive cut; on an actual text, a window's cut can be decided as it occurs, by Lemma~\ref{lem:localref}, for every
dictionary, infinite radius included; and an exact pre-tokenizer, where the pipeline has one, supplies cuts of its own,
certified for the deployed patterns by \citet{nidhogg2026pretokenizer}. A sequential fallback stands under all three.
Two caveats go with the last two. GPT-2's encoder applies the merges to each match of its pattern separately, so a
certificate on the raw table certifies the byte-pair stage alone, not normalization, special tokens or the composed
pipeline; and a pre-tokenizer's pieces can be arbitrarily long, so pre-tokenization alone bounds no chunk.

\subsection{Circularity}\label{sec:circularity}

With the right lookahead in hand, the cycle of Theorem~\ref{thm:overlap} loses its conditions. Whether the radius is
infinite does not depend on sources, exits or marks, only on whether one periodic word has two canonical tilings that
never share a boundary. The name is taken from codes, where a code is circular exactly when every periodic bi-infinite
word has at most one decomposition \citep[Proposition~2.9]{devolder1992finitary}. The canonical tokens of a dictionary
need not form a code, and its tilings are constrained by \(E\), so the theorem below is not that one. Its second form is
Proposition~3 of \citet{beal2024unambiguously} for the spelling automaton, which is unambiguous.


\begin{definition}[Circularity certificate]\label{def:circular}
A \emph{circular token list} is a nonempty list \(a_1, \dots, a_h\) of tokens of \(V\) with every consecutive pair in
\(E\), the pair \((a_h, a_1)\) included. A \emph{circularity certificate} is two circular lists, a common multiple \(N\)
of the lengths of their yields, and a shift \(0 < k < N\). Write \(U\) and \(Y\) for the two yields repeated to length
\(N\). The certificate requires \(Y = U[k{:}]\,U[{:}k]\), and that no token start of the first list repeated to length
\(N\) is congruent modulo \(N\) to a token start of the second plus \(k\).
\end{definition}

\begin{theorem}[Circularity]\label{thm:circular}
For every dictionary, \(\ripple(D)\) is infinite exactly when \(D\) has a circularity certificate. Equivalently, take
the pairs \(((s, i), (t, j))\) of a canonical token and a position inside it, \(0 \le i < |s|\) and \(0 \le j < |t|\),
the position being the number of the token's symbols already read, with \((i, j) \ne (0, 0)\), and an edge when
\(s[i] = t[j]\) and each side either advances by one position or, at its last symbol, passes to position zero of a token
that follows it in \(E\); the radius is then infinite exactly when this graph has a cycle.
\end{theorem}

\begin{proof}
Let a certificate be given and put \(p = U[k{:}]\). Then \(p\,U^n = Y^n\,p\). By Theorem~\ref{thm:bigram}, \(\T^D(U^n)\)
is the first list repeated and \(\T^D(Y^n)\) the second. By Theorem~\ref{thm:lookahead}, \(\T^D(Y^n p)\) shares with
\(\T^D(Y^n)\) a complete token prefix of at least \(nN - Q(D)\) symbols. In the coordinates of \(U^n\), the tokens of
that prefix start at the second list's starts shifted by \(k\), which the certificate keeps apart from the first list's.
No symbol of \(U^n\) in that range then has the same covering span in \(\T^D(U^n)\) and in \(\T^D(p\,U^n)\), and
prepending \(p\) has ripple at least \(nN - Q(D) - |p|\), unbounded in \(n\). Lemma~\ref{lem:D} turns the word \(p\)
into a symbol.

Conversely let the radius be infinite. By Lemma~\ref{lem:reset} an input of ripple \(r\) has no common boundary before
\(r\). Read the two tokenizations along the input and record at each position the two covering tokens and the offsets
inside them. Before \(r\) the offsets are never both zero. There are finitely many such records, so for \(r\) large one
repeats, and the stretch between is a cycle of the graph in the statement. Along the cycle each side passes a token
boundary, since an offset that only grew could not return. Rotate the cycle to a token start of the first side, and let
\(k\) be the position of a token start of the second. The tokens each side completes around the cycle are two circular
lists, their consecutive pairs in \(E\) because they are consecutive in a tokenization. They spell the same cyclic word,
shifted by \(k\), and they share no start because the cycle has no pair with both offsets zero. That is a certificate. A
cycle of the graph gives one the same way, which is the second statement.
\end{proof}

The graph has no sources, no exits and no marks, which Theorem~\ref{thm:overlap} needed in order to compute the value,
but for the verdict alone they were never necessary. The graph is the square of the spelling automaton with the common
boundaries removed, the test \citet[Propositions~3 and~14]{beal2024unambiguously} use for two bi-infinite paths with one
label, unconstrained and on a sofic shift. For the dictionary of Proposition~\ref{prop:noselfpair} the two lists are the
single tokens \(abaa\) and \(aaba\), at shift three. With its fifth rule changed to \((aaba, a)\) one list serves twice.
The list \(aabaa \mid aba\) has yield \((aaba)^2\) and starts tokens at \(0\) and \(5\) modulo eight, against \(4\) and
\(1\) after a shift of four. Certificates are not of bounded size, by Proposition~\ref{prop:certsize} in
Appendix~\ref{app:windows}.

\section{A rule criterion under the ordering condition}\label{sec:criterion}

The source paper asked for an easily decidable property of the dictionaries with infinite ripple. Under the ordering
condition there is one, read off the rules.


\begin{definition}[Effective self-pair]\label{def:selfpair}
A rule \((t, t)\) of rank \(i\) is effective when \(\T^{D_{<i}}(tt) = t \mid t\), where \(D_{<i}\) is the dictionary of
the rules before rank \(i\).
\end{definition}

Effectiveness is a property of the rule at its own stage. A rule \((t, t)\) whose \(t\) is never a token at stage \(i\)
is not effective, nor is one whose two copies of \(t\) are never adjacent there. A rule of a different rank with the
same product must not be mistaken for it. Under the ordering condition the stage at which a token exists is well
defined, since Lemma~\ref{lem:stab} makes the run one pass per rule in rank order.

\begin{theorem}[The self-pair criterion]\label{thm:selfpair}
Under the ordering condition, \(\ripple(D)\) is infinite exactly when \(D\) has an effective self-pair. Clause (a) of
the revised Definition~5 is not assumed.
\end{theorem}

\begin{proof}
By Lemma~\ref{lem:stab} the run is one left-to-right pass per rule in rank order. By Lemma~\ref{lem:D} prepended strings
of any length may be considered.

Suppose rule \(i\) is an effective self-pair \((t, t)\). Theorem~\ref{thm:bigram} applied to \(D_{<i}\) gives
\(\T^{D_{<i}}(t^r) = t \mid \cdots \mid t\) for every \(r\). Take \(x = t^{2h}\) and prepend \(t\). At stage \(i\) the
pass on \(x\) pairs the copies from the left, with tokens starting at \(0, 2|t|, 4|t|, \dots\). The pass on \(tx\) pairs
them from its own left end, with tokens starting at \(-|t|, |t|, 3|t|, \dots\) in the coordinates of \(x\), the last
copy left alone. Before the end of \(x\) the two sets of boundaries are disjoint. Every later stage only removes
boundaries, so no position of \(x\) has the same covering span in the two runs, since a common span would start at a
common boundary at or before it. The origin ripple is \(2h|t|\), unbounded in \(h\).

Suppose no self-pair is effective. If a self-pair rule \((t, t)\) of rank \(i\) ever fires, its input at stage \(i\)
holds two adjacent tokens \(t\). Theorem~\ref{thm:bigram} applied to \(D_{<i}\) then gives
\(\T^{D_{<i}}(tt) = t \mid t\), so the rule is effective. Hence every self-pair rule is inactive on every input, and the
run consists of the passes of rules \((u, v)\) with \(u \ne v\). Such a pass is local. Its occurrences are pairwise
disjoint, since an occurrence starting at a token \(v\) would need \(v = u\). A merge creates no new occurrence, \(uv\)
being neither \(u\) nor \(v\). The pass merges every occurrence present at its start, in whatever order. Now suppose
that before such a pass the two runs, aligned on \(x\), agree from a common boundary \(b\) to the end. An occurrence
inside the common suffix is merged in both. An occurrence crossing \(b\) is \(u\) before \(b\) and \(v\) after it, in
one run or both. It consumes the suffix's first token and nothing more, since that token is not \(u\). After the pass
the two runs then agree from a common boundary at most \(|v|\) further right. Before any pass they agree from position
\(0\). Composing over the passes bounds the origin ripple by the sum of the \(|v_i|\) over the rules with
\(u_i \ne v_i\), for every input and every prepended string.
\end{proof}

With Theorem~\ref{thm:overlap}, the criterion is equivalent under the ordering condition to a cycle in \(C\). With
Corollary~\ref{cor:agree} it is equivalent to Theorem~\ref{thm:origin}'s cycle. The census dictionary
\((a, b), (b, a), (ba, ba)\) and its image under exchanging \(a\) and \(b\) carry a rule \((t, t)\) that is inactive,
and both are bounded at radius two. On the first the prefix dictionary tokenizes \(baba\) as \(b \mid ab \mid a\), and
on its image \(abab\) as \(a \mid ba \mid b\). The criterion needs the stage and not the syntax.
Table~\ref{tab:exec-criterion} lists the executions of this section. One half of the criterion needs no hypothesis at
all.


\begin{theorem}[Effective self-pairs, every dictionary]\label{thm:selfgeneral}
For every dictionary, an effective self-pair makes \(\ripple(D)\) infinite. Precisely, let rule \(i\) be \((t, t)\) with
\(\T^{D_{<i}}(tt) = t \mid t\), let \(M\) be the length of the longest product among the first \(i\) rules, and put
\(K = \lfloor M/|t| \rfloor\). Then for every \(n > K\), prepending \(t\) to \(t^n\) changes the covering span of each
of the first \((n - K)|t|\) symbols.
\end{theorem}

\begin{proof}
Theorem~\ref{thm:bigram} for \(D_{<i}\) gives \(\T^{D_{<i}}(t^n) = t \mid \cdots \mid t\). The run on \(t^n\) uses only
rules before \(i\) until none applies, so it reaches that state. From there, while a rule of the first \(i\) applies,
every token is a power of \(t\). Count in units of \(t\). The rules that can apply are those of the first \(i\) whose
operands are powers of \(t\). Among them rule \(i\), which merges two units, has the lowest priority, and it is the only
one with both operands a unit, since \(t \mid t\) was final for \(D_{<i}\). No token exceeds \(K\) units. Let \(F(n)\)
be the token sequence when no rule of the first \(i\) applies, and \(B_n\) its set of boundaries, in units. The full run
passes through \(F(n)\), and whatever it does afterwards only removes boundaries.

Every state of this phase is a run of tokens of two or more units followed by a run of units. A merge by rule \(i\)
takes the first two units and creates a token of two units, the active token. A rule before \(i\) has an operand of two
or more units, so it applies only inside the run of larger tokens or at its last token and the first unit. No such rule
applied just before the merge by rule \(i\), so every merge by a rule before \(i\) that follows contains the active
token, which absorbs the larger token before it or the unit after it. When none applies and two units remain, rule \(i\)
takes them. Hence \(F(n)\) has no token of one unit except possibly its last.

Run the same phase on \(n\) and on \(n + 1\) units. While two or more common units remain, the two runs have the same
larger tokens and the same choices, and make the same merge. They first differ in one of two ways. One common unit
remains and no rule before \(i\) applies, and the longer run merges that unit with its extra one. Or no common unit
remains, and the extra unit lets the active token absorb it in the longer run, which can outrank a merge the shorter run
still makes. After that every merge of either run contains its own active token, which ends at the run's right end and
has at most \(K\) units. Hence no boundary at or before \(n - K\) differs, that is,
\(B_n \cap [0, n - K] = B_{n+1} \cap [0, n - K]\).

Suppose \(j \le n - K - 1\) were a boundary of \(F(n)\) and \(j + 1\) a boundary of \(F(n + 1)\), which is \(j\) after
the shift by the prepended unit. Then \(j\) and \(j + 1\) would both be boundaries of \(F(n)\), a token of one unit that
is not its last, since \(M \ge 2|t|\) gives \(K \ge 2\). The boundaries of \(F(n)\) and the shifted boundaries of
\(F(n + 1)\) are then disjoint on \([0, n - K - 1]\), and so are those of the two final tokenizations, which are
subsets. A symbol among the first \((n - K)|t|\) of \(t^n\) is covered in the base run by a token starting at a boundary
in that range, and no token of the prepended run starts there. Its covering span differs. Lemma~\ref{lem:D} turns the
prepended word \(t\) into a symbol.
\end{proof}

The proof asks nothing of the rules after \(i\) and nothing about producers. Priorities may alternate afterwards as they
like, since a merge can remove a boundary and never place one. The alternation inside the phase is real. With units for
powers of \(t\), the rules \((2, 1)\), \((3, 2)\), \((1, 1)\) in that order give \(F(5) = 5\) and \(F(6) = 3 \mid 3\).
The two runs part at \(3 \mid 2\) against \(3 \mid 2 \mid 1\), where the shorter still merges \((3, 2)\) and the longer
takes \((2, 1)\) first. The longer run is thus not the shorter one continued, and the comparison is between their
boundaries away from the right end.

The effective self-pair is therefore a sufficient condition for every dictionary, decided in polynomial time, by
tokenizing \(tt\) under each \(D_{<i}\), and by Theorem~\ref{thm:selfpair} a necessary one under the ordering condition.
Without the ordering condition it is not necessary, as the next proposition shows.


\begin{proposition}[The criterion needs the ordering condition]\label{prop:noselfpair}
The dictionary \((a,b)\), \((b,a)\), \((a,ba)\), \((a,aba)\), \((aba,a)\), \((ab,a)\) satisfies clause (a), fails the
ordering condition and has no rule \((t, t)\). For every \(h \ge 1\), \(\T^D((abaa)^h)\) is \(h\) copies of \(abaa\),
and \(\T^D(a(abaa)^h)\) is \(h\) copies of \(aaba\) followed by \(a\), giving origin ripple \(4h\) and label distance
\(h + 1\).
\end{proposition}

\begin{proof}
Rule one merges every \(ab\). After that no token \(b\) or \(ba\) remains, so rules one to three never apply again. From
then on the run proceeds block by block. Rule six creates \(aba\) at the leftmost block \(ab \mid a\), and rule four or
five consumes that \(aba\) before rule six is chosen again, since both outrank it. In the run on \((abaa)^h\) the token
before the new \(aba\) is a completed \(abaa\) or nothing, so rule four cannot fire, and rule five merges the \(aba\)
with the block's second \(a\), completing \(abaa\). In the run on \(a(abaa)^h\) a free \(a\) precedes the new \(aba\),
the prepended one at the first block and the block's own second \(a\) thereafter. Rule four outranks rule five and
merges it, leaving the block's second \(a\) free for the next block. The final tokens \(abaa\), \(aaba\) and \(a\) match
no rule. The token starts of the two runs are incongruent modulo four before the final \(a\), so the origin ripple is
\(4h\). No label agrees, so the distance is at least \(h + 1\), and \(h\) substitutions with one insertion attain it.
\end{proof}

The mechanism is that of Proposition~\ref{prop:sixrule} with the self-pair removed, a lower rule creating an operand of
a higher one, which the ordering condition forbids. The criterion is thus sharp in its hypothesis. For dictionaries with
clause (a) alone, the only characterization this paper has is the cycle of Theorem~\ref{thm:overlap}, a decision
procedure and not a property read off the rules.

\subsection{Extremal inputs under the ordering condition}\label{sec:extremal}


\begin{theorem}[Extremal inputs of few tokens]\label{thm:witness}
Let the rules of \(D\) that fire satisfy the ordering condition, and let \(\ripple(D)\) be finite. Let \(x\) have origin
ripple \(r > 0\) under the prepended symbol \(a\). Then the first \(r\) symbols of \(x\) have the same origin ripple and
the same mirrored quantity under \(a\). Their two tokenizations have at most \(k + 1\) tokens together, \(k\) the number
of rules that fire, so at most \(|D| + 1\).
\end{theorem}

\begin{proof}
By Lemma~\ref{lem:gtrees} the useful sublist has the tokenizer of \(D\), so let every rule fire. The ordering condition
then holds, and by Theorem~\ref{thm:selfpair} no self-pair fires, so the \(k\) rules have unequal operands and by
Lemma~\ref{lem:stab} the run is their passes. Follow the two runs, aligned on \(x\), as two affected prefixes and a
common suffix. Initially the prefixes are empty and \(a\), before the common suffix \(x\) as symbols. In a pass, merges
inside the prefixes and inside the suffix keep the picture. A merge across the \emph{agreement point}, the first
position from which the two runs agree, consumes the suffix's first token, which is then \(v\), in one run or both, and
by the locality in the proof of Theorem~\ref{thm:selfpair} nothing more. When that happens, move that token into both
prefixes. At most one token is absorbed per pass, so at most \(k\) in all. The absorbed tokens are disjoint factors of
an initial segment of \(x\), each a whole token in both runs when absorbed and never split afterwards, so every boundary
of either final tokenization inside that segment is an \emph{absorbed end}, the end of an absorbed token. By
Lemma~\ref{lem:reset}, \(r\) is the first common boundary, and the cut at \(r\) preserves both tokenizations before it,
hence both quantities. The two cut tokenizations are nonempty. Their boundaries inside \((0, r]\) are absorbed ends that
coincide only at \(r\). The prepended run has no boundary at \(0\), which would be a common boundary before \(r\). The
union of their boundaries has \(n_0 + n_1 - 1\) elements among at most \(k\) absorbed ends, and \(n_0 + n_1 \le k + 1\).
\end{proof}

\begin{corollary}[The label distance of a finite ordered dictionary]\label{cor:labelm}
Let the rules of \(D\) that fire satisfy the ordering condition, with \(\ripple(D)\) finite. Every input under every
prepended symbol has label distance at most \(\max(1, k)\), \(k\) the rules that fire as in Theorem~\ref{thm:witness},
so at most \(\max(1, |D|)\). For every \(m \ge 1\) the dictionary \((ab^{i-1}, b)_{i = 1, \dots, m}\) attains \(m\),
with origin radius and mirrored constant \(m\) as well.
\end{corollary}

\begin{proof}
With ripple \(r > 0\) the two tokenizations agree after \(r\). Before it they have \(n_0 + n_1 \le k + 1\) tokens with
\(n_0, n_1 \ge 1\), so each has at most \(k\). Substituting the shorter prefix's tokens and inserting or deleting the
rest costs at most the longer, at most \(k\). With ripple zero the second tokenization is the first with the prepended
symbol as its own token, distance one. The family satisfies both clauses and has unequal operands. On \(x = b^m\),
prepending \(a\) turns \(m\) tokens \(b\) into the one token \(ab^m\), giving distance \(m\), origin ripple \(m\) and
mirrored quantity \(m\). The bound of Theorem~\ref{thm:selfpair}'s proof, the sum of the \(|v_i|\), is \(m\) too.
\end{proof}

The distance bound is for a prepended symbol. Theorem~\ref{thm:label} bounds a symbol edit anywhere by \(F_{2|D|+2}\).
For a prepend to a finite ordered dictionary the corollary is the sharper of the two, and Theorem~\ref{thm:label} covers
every edit and every ordered dictionary. The family shows that the prepend distance can grow linearly with the rules. A
weighted recurrence on the rules alone does not give the exact radius. Take \((a,b), (ab,a), (b,a), (a,ab)\), ordered
and with every rule useful. The recurrence that extends a disturbance by \(|v_i|\) along each rule whose left operand a
previous rule produced predicts four. The radius is three, since the pass of \((b, a)\) changes the token at the
agreement point without moving it and blocks the last rule.


The attaining input cut at its ripple has a shape. Fix an ordered dictionary with finite radius and a symbol \(a\), and
for an input \(x\) with ripple \(r > 0\) call a rank \(i\) \emph{moving} when the pass of rule \(i\) moves the agreement
point in the proof of Theorem~\ref{thm:witness}.

\begin{corollary}[Absorbed tokens]\label{cor:chunks}
The first \(r\) symbols of \(x\) are \(v_{i_1} v_{i_2} \cdots v_{i_k}\) for the moving ranks \(i_1 < \cdots < i_k\),
\(k \le |D|\), each pass moving the agreement point by exactly the length of its right operand. Consequently the radius
under \(a\) is the largest ripple of a word \(v_{i_1} \cdots v_{i_k}\) with \(i_1 < \cdots < i_k\), and \(\ripple(D)\)
the maximum over \(a\). If the moving ranks of \(v_{i_1} \cdots v_{i_k}\) are \(i_1, \dots, i_k\) themselves, those of
\(v_{i_1} \cdots v_{i_{k-1}}\) are \(i_1, \dots, i_{k-1}\).
\end{corollary}

\begin{proof}
In the proof of Theorem~\ref{thm:witness} a pass moves the agreement point only by absorbing the suffix's first token,
which is then \(v\), so each moving pass moves it by \(|v_i|\) exactly, and the token absorbed spells \(v_i\). The
moving ranks increase with the passes, and the agreement point after the last pass is \(r\), by Lemma~\ref{lem:reset}.
Every input has ripple at most its length, with equality on the attaining input cut at its ripple, which gives the
maximum. For the closure, cut the word at the agreement point before the pass of \(i_k\). No merge crosses that position
before that pass in either run, so by Lemma~\ref{lem:proj} the runs on the shorter word and on it with \(a\) prepended
coincide with the longer ones there up to that pass, the agreement point is then at the word's end, and later passes
cannot move it past the end.
\end{proof}

The corollary reads the radius off \(2^{|D|}\) absorbed strings \(v_{i_1} \cdots v_{i_k}\), each evaluated in polynomial
time.

\section{The label measure}\label{sec:labels}

Lemma~\ref{lem:E} in Appendix~\ref{app:labels} shows that one pass moves one edit to at most three, and
Lemma~\ref{lem:script} gives the edits it makes of a substitution, an indel and a prepended token.


\begin{theorem}[The label measure is Lipschitz under the ordering condition]\label{thm:label}
Let \(D\) satisfy the ordering condition, with \(m = |D|\) and \(F_j\) the Fibonacci numbers, \(F_1 = F_2 = 1\). One
substituted symbol moves the tokenization by at most \(F_{2m+1}\) tokens, one inserted or deleted symbol by at most
\(F_{2m+2}\), and for \(m \ge 1\) one prepended symbol by at most \(F_{2m}\). Hence for all inputs \(x, x'\),
\(d(\T^D(x), \T^D(x')) \le F_{2m+2}\, d(x, x')\), \(d\) on the right the edit distance on symbols.
\end{theorem}

\begin{proof}
By Lemma~\ref{lem:stab}, \(\T^D = f_m \circ \cdots \circ f_1\), the passes in priority order, each the map \(f\) of
Lemma~\ref{lem:E} for its rule. Let \(\sigma_r\) and \(\iota_r\) be the largest distances between the images, under any
\(r\) passes, of two token sequences one substitution apart and one indel apart. Then \(\sigma_0 = \iota_0 = 1\). Given
\(r + 1\) passes, apply the first. By Lemma~\ref{lem:script} its two images are joined by a chain of single edits
through token sequences, and the other \(r\) passes move the ends of each link at most \(\sigma_r\) or \(\iota_r\)
apart. Hence \(\sigma_{r+1} \le \max(2\sigma_r,\, \sigma_r + \iota_r)\) and
\(\iota_{r+1} \le \max(\sigma_r + 2\iota_r,\, 2\sigma_r + \iota_r)\). By induction \(\sigma_r \le F_{2r+1}\) and
\(\iota_r \le F_{2r+2}\). All four terms are then at most \(F_{2r+1} + F_{2r+2} = F_{2r+3}\) or
\(F_{2r+1} + 2F_{2r+2} = F_{2r+4}\) respectively. A symbol edit is a token edit before the first pass. A prepended
symbol is one token edit after the first pass, so at most \(\iota_{m-1}\) after the last. The last claim follows along a
chain of single symbol edits.
\end{proof}

Lemma~\ref{lem:E} alone gives \(3^{|D|}\); counting the two kinds of edit apart gives \(F_{2m+2}\), which grows as
\(2.62^m\). At one rule it is three, and attained. The bound is not uniform in the dictionary, since on the doubling
dictionary of \(k\) rules one prepend moves the tokenization by \(k\) tokens, Proposition~\ref{prop:doubling} in
Appendix~\ref{app:labels}.


\begin{definition}[The amplification constant]\label{def:amp}
\(C_m\) is the largest label distance \(d(\T^D(x), \T^D(x'))\) over dictionaries \(D\) of \(m\) rules satisfying the
ordering condition and inputs \(x\), \(x'\) one symbol insertion, deletion or substitution apart, which
Theorem~\ref{thm:label} bounds by \(F_{2m+2}\).
\end{definition}

The substitution constant and the prepend constant are the same suprema with \(x'\) one symbol substitution from \(x\)
and with \(x' = ax\).

\begin{theorem}[The amplification constant exceeds twice the rule count]\label{thm:amp}
For every \(m \ge 1\), \(C_m \ge 2m + 1\). On the doubling dictionary of \(m\) rules, the input \(x = a^{(m+1)2^m - 1}\)
and the \(x'\) obtained from it by inserting one \(b\) after \(a^{2^m - 1}\) satisfy \(d(\T^D(x), \T^D(x')) = 2m + 1\).
\end{theorem}

The proof is in Appendix~\ref{app:labels}. Table~\ref{tab:exec-labels} lists the executions of Theorems~\ref{thm:label}
and~\ref{thm:amp}. The bounds on the label distance under one edit form a hierarchy by hypothesis. Under the ordering
condition alone it is at most \(F_{2m+2}\) and at most \((m + 1)(m + 2)/2\) (Theorems~\ref{thm:label}
and~\ref{thm:quadratic}); when no self-pair is effective it is at most \(1 + \lceil 3k/2 \rceil\) in the \(k\) rules
with unequal operands that fire, and that is attained (Theorem~\ref{thm:region}); and when the origin radius is finite a
prepended symbol moves the tokenization by at most \(\max(1, |D|)\) tokens (Corollary~\ref{cor:labelm}).
Table~\ref{tab:amp} gathers them with the lower bounds.

Thus \(2m + 1 \le C_m \le F_{2m+2}\). The executed tiers of Table~\ref{tab:exec-labels}, one per rule count, suggest the
lower end, \(C_m = 2m + 1\), and the same executions suggest \(m + 1\) for a prepended symbol from two rules on. Both
fail, the first from six rules on and the second from five. Proposition~\ref{prop:counterfamily} in
Appendix~\ref{app:labels} refutes the linear bounds, with a family of \(m\) rules that reaches \(3m - 4\) under one edit
and \(2m - 3\) under a prepended symbol.

The constant is quadratic, in both directions. A family of dictionaries carries a staircase of lone tokens, one per
level of a doubling hierarchy, down from each of several junctions. A decomposition of the two runs into common blocks
and differing clusters shows that nothing grows faster. Theorem~\ref{thm:staircase} and
Proposition~\ref{prop:staircaseinsert} in Appendix~\ref{app:labels} build the staircases, which give
\(\lceil (m^2 + 5m)/10 \rceil\) under a prepended symbol and one more under a substituted symbol for \(m \ge 5\),
\(\lfloor (m^2 + 11m + 20)/10 \rfloor\) under one insertion for \(m \ge 7\), and under a substituted symbol \(4m - 21\)
for \(9 \le m \le 11\), the value for \(7 \le m \le 11\) passing the staircase from nine on, and two below the
insertion's value, \(\lfloor (m^2 + 11m)/10 \rfloor\), for \(m \ge 12\).

For the upper bound, write the two token lists of the runs at any stage as
\[ X = B_0 A_0 B_1 A_1 \cdots A_{c-1} B_c, \qquad Y = B_0 A'_0 B_1 A'_1 \cdots A'_{c-1} B_c, \]
with \emph{blocks} \(B_j\) common to both, each inner block \(B_1, \dots, B_{c-1}\) nonempty and a power of one token,
and \emph{clusters} \(A_j, A'_j\) arbitrary. The weight of the decomposition is \(w = \sum_j \max(|A_j|, |A'_j|)\), and
matching the blocks and editing each cluster on its own gives \(d(X, Y) \le w\). If an inner block empties, its two
clusters join, which does not raise the weight, since \(\max(a + b, a' + b') \le \max(a, a') + \max(b, b')\). Before the
first pass the symbols of the two inputs form one cluster of weight one. Lemmas~\ref{lem:passunequal}
and~\ref{lem:passself} in Appendix~\ref{app:labels} bound what one pass does to a decomposition with \(c\) clusters and
weight \(w\), at most \(c\) clusters and weight \(w + c + 1\) after a rule with unequal operands, and at most \(c + 1\)
clusters and the same weight bound after a rule \((t, t)\).


\begin{theorem}[The constant is quadratic]\label{thm:quadratic}
Under the ordering condition one symbol insertion, deletion or substitution moves the tokenization by at most
\((m + 1)(m + 2)/2\) tokens, and for \(m \ge 1\) a prepended symbol by at most \(m(m + 1)/2\), with \(m = |D|\). For
\(m \ge 1\) a substituted symbol moves it by at most \(m(m + 3)/2\). With the rules \((t, t)\) at ranks
\(j_1, \dots, j_r\), counted from one, the distance after one insertion, deletion or substitution is at most
\(2m + 1 + \sum_i (m - j_i) \le 2m + 1 + rm - r(r + 1)/2\), after a substituted symbol, for \(m \ge 1\), at most
\(2m + \sum_i (m - j_i)\), and after a prepended symbol at most \(2m - 1 + \sum_{j_i \ge 2} (m - j_i)\). Consequently
\(C_m\), the substitution constant and the prepend constant are all \(\Theta(m^2)\).
\end{theorem}

\begin{table}[t]
\centering
\small
\setlength{\tabcolsep}{4pt}
\begin{tabular}{@{}>{\raggedright\arraybackslash}p{0.13\textwidth}lll>{\raggedright\arraybackslash}p{0.15\textwidth}@{}}
\toprule
bound & one edit, \(C_m\) & substitution & prepended symbol & statement \\
\midrule
upper & \(F_{2m+2}\) & \(F_{2m+1}\) & \(F_{2m}\) & Theorem~\ref{thm:label} \\
\addlinespace
upper & \((m + 1)(m + 2)/2\) & \(m(m + 3)/2\) & \(m(m + 1)/2\) & Theorem~\ref{thm:quadratic} \\
\addlinespace
exact, no effective self-pair & \(1 + \lceil 3k/2 \rceil\) & \(1 + \lfloor 3k/2 \rfloor\) & \(\max(1, k)\)
  & Theorem~\ref{thm:region} \\
\addlinespace
lower & \(2m + 1\) & & \(m\) & Theorem~\ref{thm:amp}, Proposition~\ref{prop:doubling} \\
\addlinespace
lower, \(m \ge 5\) & \(3m - 4\) & & \(2m - 3\) & Proposition~\ref{prop:counterfamily} \\
\addlinespace
lower, \(m \ge 5\) & \(\lceil (m^2 + 5m)/10 \rceil + 1\) & \(\lceil (m^2 + 5m)/10 \rceil + 1\)
  & \(\lceil (m^2 + 5m)/10 \rceil\) & Theorem~\ref{thm:staircase} \\
\addlinespace
lower, \(m \ge 7\) & \(\lfloor (m^2 + 11m + 20)/10 \rfloor\) & \(\lfloor (m^2 + 11m)/10 \rfloor\), \(m \ge 12\)
  & & Proposition~\ref{prop:staircaseinsert} \\
\addlinespace
lower, \(9 \le m \le 11\) & & \(4m - 21\) & & Proposition~\ref{prop:staircaseinsert} \\
\bottomrule
\end{tabular}
\caption{The label distance under one symbol edit over the ordered dictionaries of \(m\) rules, with \(m \ge 1\) for
the substitution and the prepend: the upper bounds, the exact values when no self-pair is effective, \(k\) the rules
with unequal operands that fire, and the lower bounds the families attain.}
\label{tab:amp}
\end{table}


The proof is in Appendix~\ref{app:labels}. The ordering condition need hold only on the rules that fire, since deleting
the others changes no run (Section~\ref{sec:semantics}), and the theorem holds whenever the rules that fire satisfy it,
with \(m\) their number and the ranks counted among them, and so does Theorem~\ref{thm:region}. On
\((b, a), (ab, a), (a, b)\), unordered, whose middle rule never fires, the two give a substitution at most \(4\), and
the exact supremum is \(2\), equal to the largest distance over every substitution in every input of one to ten symbols.

Theorem~\ref{thm:label} is nowhere the sharper bound. The two agree for a substituted symbol at \(m \le 2\), for an
insertion or deletion at \(m = 1\) and for a prepended symbol at \(m \le 2\), and the last two are attained. Between
\(m^2/10\) and \(m^2/2\) the exact leading constant is open.

\section{The label measure decided}\label{sec:labelgraph}

Section~\ref{sec:overlap} decided the origin measure for every dictionary and left the label measure open where the
origin radius is infinite and the ordering condition fails. That it is decidable follows from
Corollary~\ref{cor:rational}. Under a prepended symbol \(a\) the label distance is the distance between the rational
functions \(x \mapsto \T^D(x)\) and \(x \mapsto \T^D(ax)\), and under one edit it is the diameter of a rational
relation, and both are computable for the edit distance on tokens by \citet[Theorem~3.25 and
Corollary~3.12]{aiswarya2024edit}. This section decides it directly on the overlap graph, in time polynomial in that
graph. Cut the overlap graph at the edges on which both runs complete a token. The label distance is bounded exactly
when every strongly connected component of what remains is a single cycle along which the two runs complete their tokens
with a fixed shift, and then the supremum is computed exactly. The same construction with one edit edge decides the
supremum of the label distance under one symbol edit anywhere, so the amplification constant of Section~\ref{sec:labels}
is computed exactly on small universes instead of read off a horizon.


Call an edge of the overlap graph a \emph{common-boundary edge} when both copies complete a token on it. The
\emph{disagreement graph} \(G_a\) is the set of live pairs reachable from the sources of \(a\) through edges that are
not common-boundary edges, with those edges. An \emph{exit} is a common-boundary edge leaving a pair of \(G_a\), and its
target is jointly final. For a walk \(\pi\) let \(\Lambda(\pi)\) and \(\mathrm{P}(\pi)\) be its \emph{token words}, the
tokens the first and the second copy complete along it, in order. Let \(C_L(\pi)\) and \(C_R(\pi)\) be its
\emph{completion sets}, the indices of the edges on which each copy completes a token. A closed walk is \emph{conjugate}
when its two token words are rotations of one another. A strongly connected component of \(G_a\) is \emph{nontrivial}
when it contains an edge. It is \emph{phase-consistent} when there are a primitive word \(w\) and a map \(\varphi\) from
its pairs to \(\mathbb{Z}/|w|\) such that every edge from \(\tau\) reads the symbol \(w[\varphi(\tau)]\) and enters a
pair of phase \(\varphi(\tau) + 1\).

\begin{theorem}[The label measure]\label{thm:labelgraph}
Fix a dictionary and a symbol \(a\). Every phase-consistent nontrivial component of \(G_a\) is a single simple cycle.
\(\sup_x \lambda(a, x)\) is finite exactly when every nontrivial component is phase-consistent and its cycle \(c\) is
conjugate; with \(n = |c|\) and \(w\) the primitive root of the word \(c\) spells, that is when some
\(\delta \in |w|\mathbb{Z}\) with \(0 < \delta < n\) has \(C_R(c) = C_L(c) + \delta \pmod n\). When the supremum is
infinite there are a closed walk \(c\) of \(G_a\), a stem \(s\) from a source and an exit path \(e\) with
\(\lambda(a, s c^k e) \to \infty\). When it is finite it is at most \(2(M + 3)|G_a| + 3\), with \(M\) the longest token,
and it is computed exactly, with an attaining input. The condition is decided in time polynomial in the overlap graph.
\end{theorem}

On the one-rule dictionary \((a, a)\), \(G_a\) has one cycle, of length two, spelling \(aa\), with completion sets
\(\{0\}\) and \(\{1\}\), \(w = a\) and \(\delta = 1\), so the label distance is bounded, with supremum one. On the
six-rule dictionary of Proposition~\ref{prop:sixrule} under \(b\) there is a cycle spelling \(abba\) with completion
sets \(\{3\}\) and \(\{2\}\), and \(abba\) is primitive, so no \(\delta\) exists, and the label distance is unbounded,
with the certificate \(a\,(abba)^k\,abb\), whose label distance is \(2, 6, 10, 18, 34\) at \(k = 0, 4, 8, 16, 32\).

The proof rests on four lemmas, namely cancellation of a common token, conjugacy of the words whose powers stay close,
truncation at the ripple, and rigidity of a component.

\begin{lemma}[Cancellation]\label{lem:cancel}
For token sequences \(u, v\) and a token \(c\), \(d(cu, cv) = d(u, v) = d(uc, vc)\).
\end{lemma}

The proof is in Appendix~\ref{app:labels}. The first statement of the next lemma is, for two words, the loop criterion
of \citet[Lemma~4.1 and Claim~4.9]{aiswarya2024edit}, and the proof is given for the explicit bound in the second.

\begin{lemma}[Conjugacy]\label{lem:conjugacy}
For words \(u, v\) over any alphabet, \(\sup_k d(u^k, v^k)\) is finite exactly when \(u\) and \(v\) are conjugate. If
\(u\) and \(v\) are not conjugate, then \(d(u^k, v^k) \ge k\) for every \(k \ge 1\).
\end{lemma}

\begin{lemma}[Truncation]\label{lem:truncate}
For every \(x\) let \(r = \rho(a, x)\). Then \(\T^D(x) = \alpha\beta\) and \(\T^D(ax) = \alpha'\beta\) with
\(\alpha = \T^D(x[0{:}r])\) and \(\alpha' = \T^D(a\,x[0{:}r])\), and
\(\lambda(a, x) = d(\alpha, \alpha') = \lambda(a, x[0{:}r])\). When \(r > 0\), the accepting product path of
\(x[0{:}r]\) is a path of \(G_a\) followed by one exit, and every such path is the accepting path of the input it
spells. Hence \(\sup_x \lambda(a, x)\) is the supremum of \(d(\Lambda(\pi), \mathrm{P}(\pi))\) over the paths \(\pi\)
from a source of \(a\) through \(G_a\) to an exit, the second copy's token completed at the source, if any, counted in
\(\mathrm{P}(\pi)\).
\end{lemma}

The proofs of Lemmas~\ref{lem:conjugacy}, \ref{lem:truncate} and~\ref{lem:rigid} are in Appendix~\ref{app:labels}. A
source whose second copy has already completed the one-symbol token \(a\) has a common boundary at position zero. From
it both copies tokenize \(x\) identically, by Lemma~\ref{lem:proj}, so those paths contribute \(\lambda = 1\) and no
cycle.

\begin{lemma}[Rigidity]\label{lem:rigid}
Let \(S\) be a nontrivial strongly connected component of \(G_a\). If \(S\) is not phase-consistent, it contains a
closed walk that is not conjugate. If \(S\) is phase-consistent, every pair of \(S\) has exactly one successor inside
\(S\), so \(S\) is a simple cycle.
\end{lemma}

\begin{proof}[Proof of Theorem~\ref{thm:labelgraph}]
The first claim is Lemma~\ref{lem:rigid}. Suppose some nontrivial component fails the condition. Lemma~\ref{lem:rigid},
or the failed shift test on a phase-consistent cycle, gives a closed walk \(c\) at a pair \(\sigma\) of \(G_a\) whose
token words are not conjugate. With a stem \(s\) from a source to \(\sigma\) and an exit path \(e\) from \(\sigma\), put
\(x_k = s c^k e\), the words spelled. By Lemma~\ref{lem:truncate} its accepting path is the concatenation, so
\(\T^D(x_k) = A\,\Lambda(c)^k B\) and \(\T^D(a x_k) = A'\,\mathrm{P}(c)^k B'\), with \(A, A', B, B'\) the tokens
completed on the stem and on the exit. By the triangle inequality
\(\lambda(a, x_k) \ge d(\Lambda(c)^k, \mathrm{P}(c)^k) - |A| - |A'| - |B| - |B'|\), unbounded by
Lemma~\ref{lem:conjugacy}.

Suppose every nontrivial component is a simple cycle with a conjugate shift. By Lemma~\ref{lem:truncate} it suffices to
bound \(d(\Lambda(\pi), \mathrm{P}(\pi))\) over paths \(\pi\) of \(G_a\) from a source to an exit. The condensation of
\(G_a\) is acyclic, so \(\pi\) meets each cycle component \(S\) in one contiguous segment, an arc of the cycle plus
\(n_S \ge 0\) full turns. Removing all full turns leaves a simple path of at most \(|G_a|\) edges. Let \(S\) have cycle
length \(N\) and shift \(\delta\), and let \(d_S \in \{\delta, \delta - N\}\) have \(|d_S| \le N/2\). Along the segment
inside \(S\) the completion sets are those of the cycle repeated, so \(C_R = C_L + d_S\) there, and the symbols read are
\(|w|\)-periodic with \(|w|\) dividing \(d_S\). Match each token of the first copy completed inside the segment and
covering \([p, p + \ell)\) of the input to the second copy's token covering \([p + d_S, p + \ell + d_S)\) whenever that
token is also completed inside the segment. They spell the same word, by the periodicity, and the pairs increase on both
sides. A token of the segment left unmatched lies within \(|d_S| + M\) symbols of one end of the segment, so at most
\(2(N/2 + M + 1) = N + 2M + 2\) per side are unmatched. Tokens completed outside all segments number at most
\(|G_a| + 1\) per side, plus one on the second side for a token completed at the source. Matches in different components
are ordered as the components are, so the union is a monotone alignment whose cost is at most the number of unmatched
tokens. With \(\sum_S N_S \le |G_a|\) and at most \(|G_a|/2\) cycle components, since a self-loop of the product is a
common-boundary edge and a cycle of \(G_a\) has at least two pairs,
\(\lambda \le 2(2|G_a| + (M + 1)|G_a| + 1) + 1 = 2(M + 3)|G_a| + 3\).

\(G_a\), its components, a shortest closed walk, the phase search, the successor count and the shift test are computed
in time polynomial in the overlap graph. The exact value is computed by the profiles below.
\end{proof}

Under the ordering condition Theorem~\ref{thm:label} makes the supremum finite for every dictionary, and the executed
decision finds every ordered product bounded, as it must.

\paragraph{The exact supremum.} Fix a cap \(k \ge 1\). Along a path \(\pi\) of \(G_a\) the product state is known, so an
alignment of \(\Lambda(\pi)\) and \(\mathrm{P}(\pi)\) is built by a nondeterministic \emph{alignment automaton}. Its
state is a buffer of pending tokens of one side, read in order of completion, the first copy's token before the second's
on the same edge. A completed token is resolved in one of two ways. When the buffer is empty or holds its own side, it
is pushed as pending. When the buffer holds the other side, it is declared unmatched, at cost one, or front tokens are
popped as unmatched at cost one each until one is matched to it, at cost one if the labels differ and zero otherwise, or
the buffer empties and the token is pushed. A front token may be popped as unmatched at any time, so a token of the
buffer's side is declared unmatched once it reaches the front, at once when it was pushed on an empty buffer, and every
buffer is a suffix of the tokens completed so far on its side. At an exit the value is the cost plus the number of
pending tokens. A run whose cost exceeds \(k\) is discarded, and so is one whose buffer holds more than
\(\lfloor k/2 \rfloor M + (k \bmod 2)\) tokens once both completions of an edge are processed. A source whose second
copy completed the token \(a\) starts with it pending or unmatched.

\begin{lemma}[Profiles]\label{lem:profiles}
Let \(\pi\) be a path of \(G_a\) from a source to an exit, and \(\eta\) the length of the second input less that of the
first. Every run on \(\pi\) yields a monotone alignment of the two token words whose cost is the run's value. If
\(d(\Lambda(\pi), \mathrm{P}(\pi)) \le k\), some run has exactly that cost, and along every prefix of \(\pi\) its buffer
holds at most \(\Delta = \lfloor k/2 \rfloor M + (k \bmod 2)|\eta|\) tokens once both completions of an edge are
processed, and one more between them.
\end{lemma}

The proof is in Appendix~\ref{app:labels}. The \emph{profile} of a path is the map from buffers to the least cost of a
run reaching them, restricted to costs at most \(k\). Every buffer is a suffix of the tokens completed so far on its
side, so a profile is determined by the last \(\Delta\) tokens completed on each side and, for each of the
\(2\Delta + 1\) suffix lengths, a least cost or an absence, which gives at most
\[
P' = \Bigl(\sum_{j \le \Delta} v^j\Bigr)^2 (k + 2)^{2\Delta + 1}
\]
profiles along any path, \(v = |V|\). The profile of \(\pi e\) is a function of the profile of \(\pi\) and the edge
\(e\). A breadth-first search over pairs of a product state and a profile from the sources, at most \(|G_a| P'\) of
them, evaluating every exit, computes \(\max_\pi \min(d(\Lambda(\pi), \mathrm{P}(\pi)), k + 1)\) and records an
attaining path, of at most \(|G_a| P'\) symbols since the search keeps a shortest path to every pair. A path on which
every run exceeds the cap contributes \(k + 1\). Iterating \(k = 1, 2, \dots\) until the maximum is at most \(k\)
returns \(\sup_x \lambda(a, x)\) whenever Theorem~\ref{thm:labelgraph} has declared it finite, and each iteration is
finite. \(P'\) is at most \(((v + 1)(k + 2))^{2\Delta + 1}\), exponential in \(\Delta\) and so in \(kM\), so the value,
unlike the decision, is not computed in polynomial time here.

\paragraph{One symbol edit anywhere.} The \emph{edit product} has states \((\text{side}, q, q')\) with \(q, q'\) states
of the spelling automaton. On side zero both copies read the same symbol. The edit edges lead to side one. An insertion
moves the second copy on a symbol and leaves the first in place, a deletion the reverse, and a substitution moves both
on different symbols. On side one both copies read the same symbol again. The diagonal sources are \((0, q, q)\) with
\(q\) the start or a token-end state, and the \emph{edit graph} \(G_e\) is the part that the diagonal sources reach
through edges that are not common-boundary edges. An exit is a common-boundary edge into a jointly final side-one pair.
For an edit at the end of the input it is any edge of \(G_e\) into a jointly final side-one pair, since an appended
symbol or a deleted last symbol ends the interval at the edit itself. Let \(x, x'\) be one edit apart at position \(p\).
Let \(b \le p\) be the largest position that starts a token in both tokenizations, and \(r\) the first position after
the edit at which both have a boundary. By Lemma~\ref{lem:proj} the two tokenizations agree before \(b\), with both runs
in the state of the last token before \(b\), and from \(r\) on. Canceling the common prefix and suffix,
\(d(\T^D(x), \T^D(x'))\) is the distance of the tokens in the \emph{edit interval} \([b, r)\), and the interval is a
path of \(G_e\) from a diagonal source to an exit. Conversely take a path of \(G_e\) from \((0, q, q)\) with \(q\) the
end state of a token \(t\), and prefix it by the run on \(t\) in both copies. The result is the pair of accepting runs
of \(t x_w\) and \(t x'_w\), the two words the path spells, which are one edit apart. By cancellation their distance is
that of the path. \(G_e\) has no common-boundary edge. A closed walk lies on one side, and pumping it pumps both inputs
by the same word, keeping them one edit apart. Each copy's run is the unique accepting run of its own input.
Lemmas~\ref{lem:rigid} and~\ref{lem:profiles} and the proof of Theorem~\ref{thm:labelgraph} thus apply to \(G_e\)
unchanged, and \(\sup d(\T^D(x), \T^D(x'))\) over all pairs one symbol edit apart is decided and, when finite, computed
exactly, with an attaining pair whose two inputs have at most \(|G_e| P'\) symbols each, since the search keeps a
shortest path and, both tokenizations having a boundary after \(t\), the two words a path spells are themselves at its
distance. For a substitution \(\eta = 0\), and Lemma~\ref{lem:profiles} gives buffers of at most
\(\lfloor k/2 \rfloor M\) tokens after an edge and one more between its completions. Table~\ref{tab:exec-labelgraph}
lists the executions of the results of this section. Boundedness needs only the prepended symbols, since zero delay
reduces an edit anywhere to prepends.

\begin{corollary}[Prepend bounds decide edit bounds]\label{cor:prependedit}
For a dictionary on a finite alphabet let \(C_a\) be the supremum of \(\lambda(a, x)\) over the inputs \(x\), \(\Csym\)
the largest \(C_a\), \(\Eedit\) the supremum of \(d(\T^D(x), \T^D(x'))\) over the pairs one symbol edit apart, and
\(\hat{q} = Q_{\mathrm{span}}(D)\). Then \(\Csym \le \Eedit\). Let a factor \(r\) be replaced by a factor \(s \ne r\)
between a prefix \(u\) and a suffix \(v\), with \(u = \yield(\Stable(u))\,w\). Then
\[
d(\T^D(urv), \T^D(usv)) \le 2\sum_{c \in w} C_c + \sum_{c \in r} C_c + \sum_{c \in s} C_c,
\]
the sums running over the positions of each word, a bound that is finite whenever the symbols of \(w\), \(r\) and \(s\)
have finite constants, whatever the constants of the other symbols, and that is attained for every \(m \ge 1\) by an
inserted factor of \(m\) symbols on a dictionary with \(\Csym\) infinite.

For a set \(\Gamma\) of symbols let \(Q_\Gamma\) be the largest \(\sum_{c \in w} C_c\) over the words \(w\) over
\(\Gamma\) with \(\Stable(w)\) empty, which are the held suffixes over \(\Gamma\), the empty word among them with the
sum zero and the constants taken in the nonnegative integers with infinity, so that \(Q_\Gamma\) is infinite exactly
when a held suffix over \(\Gamma\) holds a symbol of infinite constant. When the constants on \(\Gamma\) are finite,
with \(C_\Gamma\) the largest of them, zero when \(\Gamma\) is empty, \(Q_\Gamma \le \hat{q}C_\Gamma\), and \(Q_\Gamma\)
is the largest weight of a path over \(\Gamma\) in the product of two copies of the reversed spelling automaton, from a
pair of the start and another state to a pair of final states whose tokens differ, each edge weighing the constant of
its symbol, and zero when there is no such path. When \(w\), \(r\) and \(s\) are over \(\Gamma\), the distance is at
most \(2Q_\Gamma + \sum_{c \in r} C_c + \sum_{c \in s} C_c\), at most \((2\hat{q} + |r| + |s|)C_\Gamma\) for finite
constants.

For any set \(\Gamma\) of symbols, finite constants or not, no rule that fires has a left operand that is a single
symbol of \(\Gamma\), that is, no canonical word of two symbols or more begins with a symbol of \(\Gamma\), exactly when
every symbol of \(\Gamma\) has origin radius zero, by the zero criterion of Corollary~\ref{cor:lookconst}, exactly when
\(\T^D(cx) = c \mid \T^D(x)\) for every \(c \in \Gamma\) and every \(x\). Then every symbol of \(\Gamma\) has constant
one and \(Q_\Gamma = 0\), every prefix \(u\) whose held suffix \(w\) is over \(\Gamma\) has \(w\) empty, and for such a
\(u\) and factors \(r\) and \(s\) over \(\Gamma\), \(d(\T^D(urv), \T^D(usv))\) is the edit distance of \(r\) and \(s\)
as strings of symbols, for every suffix \(v\). When the test fails, some symbol \(c\) of \(\Gamma\) has \(\Stable(c)\)
empty and \(Q_\Gamma \ge 1\), infinite exactly when a constant on \(\Gamma\) is.

When \(\Csym\) is finite and \(\Gamma\) is the whole alphabet, \(C_\Gamma = \Csym\) and \(Q_\Gamma\) is the largest
weight \(Q_C\) of a held suffix, so insertions and deletions move the tokenization by at most
\(2Q_C + \Csym \le (2\hat{q} + 1)\Csym\), substitutions by at most \(2Q_C + 2\Csym \le 2(\hat{q} + 1)\Csym\), and an
inserted or deleted factor of \(m\) symbols by at most \(2Q_C + m\Csym \le (2\hat{q} + m)\Csym\), which is attained for
every \(m \ge 1\). Hence the label distance is bounded under every prepended symbol exactly when it is bounded under one
edit anywhere.
\end{corollary}

\begin{proof}
A prepend is an insertion, so \(\Csym \le \Eedit\). By the proof of Theorem~\ref{thm:zerodelay}, with
\(P = \Stable(u)\), \(\T^D(uz) = P\,\T^D(wz)\) for every \(z\) and \(|w| \le \hat{q}\). Canceling \(P\) by
Lemma~\ref{lem:cancel} gives \(d(\T^D(urv), \T^D(usv)) = d(\T^D(wrv), \T^D(wsv))\). Prepending the symbols of \(wr\) to
\(v\) one at a time, from the last, moves the tokenization by at most \(C_c\) for each symbol \(c\), so
\(d(\T^D(v), \T^D(wrv))\) is at most the sum of \(C_c\) over the positions of \(wr\), and likewise for \(ws\). The
triangle inequality through \(\T^D(v)\) gives the pointwise bound, which uses only the constants of the symbols of
\(w\), \(r\) and \(s\). A held suffix has empty stable output. With \(P = \Stable(u)\) and \(u = \yield(P)\,w\),
\(\T^D(uz) = P\,\T^D(wz)\) for every \(z\), so \(\Stable(u)\) is \(P\) followed by \(\Stable(w)\), which is therefore
empty, and a word with empty stable output is its own held suffix, so it has at most \(\hat{q}\) symbols. When \(w\),
\(r\) and \(s\) are over \(\Gamma\), \(w\) is a held suffix over \(\Gamma\), so the sum over \(w\) is at most
\(Q_\Gamma\), which gives the uniform bound, and \(C_c \le C_\Gamma\) on \(\Gamma\) with \(|w| \le \hat{q}\) gives
\(Q_\Gamma \le \hat{q}C_\Gamma\). An insertion has \(r\) empty and \(|s| = 1\), a deletion the reverse, and a
substitution \(|r| = |s| = 1\). The computation of \(Q_\Gamma\) is proved in Appendix~\ref{app:labels}.

For the zero test, every constant is at least one, since \(\T^D(x)\) and \(\T^D(cx)\) have different yields, so they are
different token sequences and \(\lambda(c, x) \ge 1\) for every \(x\). If a rule \((c, v)\) with \(c \in \Gamma\) fires,
then by Lemma~\ref{lem:gtrees} it is the producer of \(cv\), so \(cv \in V\) and \(\T^D(cv) = cv\), while
\(\T^D(c) = c\). The two first tokens differ, so \(\Stable(c)\) is empty, \(c\) is a held suffix over \(\Gamma\), and
\(Q_\Gamma \ge C_c \ge 1\). Suppose instead that no rule that fires has a single symbol of \(\Gamma\) as its left
operand, and let \(c \in \Gamma\). For every \(x\) the first token of \(\T^D(cx)\) is a canonical word beginning with
\(c\). If it had two symbols or more, the spine clause of Lemma~\ref{lem:gtrees} would give a rule that fires with the
left operand \(c\) alone. The first token is thus \(c\), the boundary after it is a boundary of the result, no merge
crossed it (a merge removes a boundary and never restores one), and Lemma~\ref{lem:proj} on the interval \(x\) gives
\(\T^D(cx) = c \mid \T^D(x)\). Hence \(\lambda(c, x) = 1\) for every \(x\) and \(C_c = 1\). For a nonempty word
\(w = cw'\) over \(\Gamma\), \(\T^D(wz) = c \mid \T^D(w'z)\) for every \(z\), so \(c\) is a common first token and
\(\Stable(w)\) is nonempty. The only word over \(\Gamma\) with empty stable output is then the empty one, of weight
zero, and \(Q_\Gamma = 0\). The two directions give the equivalence, and the clause of Corollary~\ref{cor:lookconst}
that the origin radius under \(a\) is zero exactly when no rule that fires has the left operand \(a\) gives the second
form. Conversely, the prepend identity for every \(x\) makes the first token of \(\T^D(cv)\) the symbol \(c\) for every
rule \((c, v)\), so no such rule fires, which gives the third form. The canonical form is the spine clause read at the
first symbol: the parent of the first leaf of a canonical word of two symbols or more beginning with \(c\) is a rule
that fires with the left operand \(c\), and the product of such a rule is such a word. Neither direction uses that the
constants on \(\Gamma\) are finite, and a symbol \(c\) of infinite constant fails the identity, since passing it gives
\(C_c = 1\), so the held suffix \(c\) has infinite weight and \(Q_\Gamma\) is infinite exactly when a constant on
\(\Gamma\) is.

For the identity, let \(Q_\Gamma = 0\). The held suffix \(w\) of \(u\) has \(\Stable(w)\) empty, as shown above, so when
\(w\) is over \(\Gamma\) it is empty, and \(\T^D(uz) = P\,\T^D(z)\) for every \(z\), with \(P = \Stable(u)\). With
\(r = r_1 \cdots r_k\) over \(\Gamma\), the prepend identity \(\T^D(cx) = c \mid \T^D(x)\) applied \(k\) times from the
right gives \(\T^D(rv) = r_1 \mid \cdots \mid r_k \mid \T^D(v)\), and likewise for \(s\). Lemma~\ref{lem:cancel} cancels
the common prefix \(P\) and the common suffix \(\T^D(v)\) one token at a time, and what remains is the distance of
\(r_1 \mid \cdots \mid r_k\) and \(s_1 \mid \cdots \mid s_l\), sequences of one-symbol tokens, two of which are equal
exactly when they are the same symbol. That is the edit distance of \(r\) and \(s\) as strings.

On \((a, a)\) over \(\{a, b\}\), \(\hat{q} = \Csym = Q_C = 1\), and inserting \(b^m\) into \(aa\) turns the one token
\(aa\) into \(a \mid b \mid \cdots \mid b \mid a\), with \(m + 2\) tokens, none of them \(aa\). The lengths need
\(m + 1\) insertions, and the token \(aa\) one more edit, so the distance is
\(m + 2 = 2Q_C + m\Csym = (2\hat{q} + m)\Csym\), and one substitution with \(m + 1\) insertions attains it. On
\((c, ab), (ab, c), (a, b), (d, d)\) over \(\{a, b, c, d, e\}\) the run on \((abc)^n\) joins each \(ab\) with the \(c\)
after it, and the run on \(c\,(abc)^n\) joins it with the \(c\) before it, which \((c, ab)\) prefers, leaving
\((cab)^n c\), which shares no token with \((abc)^n\), so that \(C_c\) and \(\Csym\) are infinite. The symbol \(e\) is
in no rule, so prepending it puts the one token \(e\) in front, and \(d\) is only in \((d, d)\), so by
Lemma~\ref{lem:proj} a maximal run of \(d\) tokenizes on its own, as \(dd\) from the left, and prepending \(d\) changes
or adds one of its tokens; hence \(C_e = C_d = 1\). Inserting \(e^m\) into \(dd\), where \(u = w = d\) and \(v = d\),
gives \(d \mid e \mid \cdots \mid e \mid d\) against \(dd\), at distance \(m + 2 = 2C_d + mC_e\), the pointwise bound.
\end{proof}

At \(m = 1\) this is the insertion bound, attained by \(aa\) against \(aba\). The held suffix \(w\) is paid for once for
a whole factor, so a factor of \(m\) symbols costs at most \((2\hat{q} + m)\Csym\), where iterating the bound for one
insertion gives \(m(2\hat{q} + 1)\Csym\). The corollary decides boundedness and leaves the exact \(\Eedit\) to the edit
product. On \((a, b), (ab, c)\) over \(\{a, b, c\}\), where \(\hat{q} = \Csym = 2\) and \(Q_C = 3\), the bounds \(10\)
and \(12\) for an insertion and a substitution fall to \(8\) and \(10\). On \((c, ab), (ab, c), (a, b)\) over
\(\{a, b, c, d, e\}\), with \(\Gamma = \{b, d, e\}\), the left operands \(c\), \(ab\) and \(a\) of the rules include no
symbol of \(\Gamma\), so the zero test gives \(Q_\Gamma = 0\) and \(C_b = C_d = C_e = 1\), and an inserted symbol of
\(\Gamma\) after a prefix whose held suffix is over \(\Gamma\) moves the tokenization by exactly one, where the held
suffixes of all prefixes, the symbols outside \(\Gamma\) weighing nothing, weigh up to one.

\subsection{The amplification constant computed}\label{sec:ampcomputed}

Section~\ref{sec:labels} showed \(C_m = \Theta(m^2)\), between \(\lfloor (m^2 + 11m + 20)/10 \rfloor\), for \(m \ge 7\),
and \((m + 1)(m + 2)/2\). The one-edit supremum of a fixed dictionary is now computable, so \(C_m\) restricted to a
finite universe of dictionaries is computed exactly, every value attained by a replayed pair of inputs. Over the ordered
proper dictionaries of \(m\) rules on two symbols the constant is \AmpExactOne{}, \AmpExactTwo{}, \AmpExactThree{} and
\AmpExactFour{} at \(m = 1, 2, 3, 4\). On three symbols it is \AmpExactThreeSymOne{} and \AmpExactThreeSymTwo{} at
\(m = 1, 2\). On the one-symbol ordered dictionaries of five rules with distinct products and a foreign symbol it is
\AmpExactUnaryFive{}, on the doubling dictionary alone, by the grid of Proposition~\ref{prop:unary} in
Appendix~\ref{app:labels}, which equals the edit product on every such dictionary of one to four rules. Over all
\FiveEditTables{} ordered proper dictionaries of five rules on two symbols it is \FiveEditMax{}, attained by
\FiveEditAttaining{} of them, the doubling dictionary among them. Its computation took hours and up to 42 GB of memory,
so it is a recorded campaign outside the archived run of Appendix~\ref{app:exec}. That run checks its record,
\code{closure/campaign/full-five-edit.json}, for consistency and replays its \FiveEditAttaining{} attaining pairs of
inputs. On these universes \(C_m = 2m + 1\) is thus a computed fact, and by Proposition~\ref{prop:counterfamily} it
stops holding at six rules. The extremal dictionaries at three and four rules include mixed ones beside the doubling
family, namely \((a,a), (aa,b), (aab,aa)\) at \(m = 3\), on \(aababaaabaa\) against \(aabaaaabaa\), and
\((a,a), (aa,b), (aab,aa), (aabaa,aa)\) at \(m = 4\).

On one symbol the one-edit supremum is finite and computed exactly in polynomial time by a finite grid, with or without
the ordering condition, Proposition~\ref{prop:unary} in Appendix~\ref{app:labels}. On \((aa, a), (a, a)\), which fails
the ordering condition and is equivalent to no ordered dictionary, the supremum is three, attained by \(aa\) against
\(aba\). A dictionary outside the condition can thus keep the label measure bounded under every edit, as the unbounded
six-rule dictionary of Proposition~\ref{prop:sixrule} does not.

Where the ordering condition holds and no self-pair is effective, the bound is proved, by following the tokens left
between the common prefix and suffix of two lists, their \emph{region}.

\begin{theorem}[The region bound]\label{thm:region}
Let \(D\) satisfy the ordering condition and have no effective self-pair, and let \(k\) be the number of its rules with
\(u_i \ne v_i\) that fire on some input. Then \(d(\T^D(x), \T^D(x')) \le 1 + \lceil 3k/2 \rceil\) for all \(x, x'\) one
symbol insertion, deletion or substitution apart, \(d(\T^D(x), \T^D(x')) \le 1 + \lfloor 3k/2 \rfloor\) for a
substitution, and \(\lambda(a, x) \le \max(1, k)\) for every prepend. When \(x\) and \(x'\) are over two symbols and one
insertion or deletion apart, \(d(\T^D(x), \T^D(x')) \le \max(1, \lfloor 3k/2 \rfloor)\), so that
\(1 + \lfloor 3k/2 \rfloor\) bounds any one edit over two symbols. Each bound is attained at every \(k \ge 0\) by a
dictionary of \(k\) rules on two symbols, the one-edit bound by the insertion of a third symbol and the other three on
inputs over the two symbols.
\end{theorem}

The proof is in Appendix~\ref{app:labels}. The bounds are the maxima over the class, dictionaries and inputs together,
which a single dictionary need not reach. At \(k = 3\) the dictionary \((a, b), (b, ab), (ab, bab)\) tokenizes
\(abbababbab\) as \(abbab \mid abbab\) and \(abbacbabbab\), with one \(c\) inserted, as
\(ab \mid b \mid a \mid c \mid bab \mid bab\), at distance six, and the proof grows this pair, and the pair \(babbab\),
\(baabab\) of \((a, b), (b, ab)\) at distance four, by two rules at a time. Coded onto two symbols, the pair \(ab\),
\(acb\) of \((a, b)\) becomes \(bab\), \(baab\) on \((a, b), (b, ab)\), at distance three, the two-symbol insertion
value at \(k = 2\). The one-edit bound exceeds \(1 + \lfloor 3k/2 \rfloor\) at odd \(k\) only, and the ordered
dictionaries of one to four rules on two symbols, whose inputs are on two symbols too, attain the latter at every \(k\)
from one to four: over those without an effective self-pair the largest one-edit suprema are \NoselfEditOne{},
\NoselfEditTwo{}, \NoselfEditThree{} and \NoselfEditFour{}, and the substitution suprema, computed on the edit product
with substitutions alone, never exceed \(1 + \lfloor 3k/2 \rfloor\) and are largest at \NoselfSubOne{}, \NoselfSubTwo{},
\NoselfSubThree{} and \NoselfSubFour{}. On two symbols the insertion or deletion maximum is one below the substitution
maximum for every \(k \ge 1\), so the largest two-symbol value is a substitution's, and the insertion suprema, computed
on the edit product with insertions alone, never exceed \(\max(1, \lfloor 3k/2 \rfloor)\) and are largest at
\NoselfInsOne{}, \NoselfInsTwo{}, \NoselfInsThree{} and \NoselfInsFour{}. The prepend bound is attained at every
\(k \ge 1\). On \((a, b), (ab, b), \dots, (ab^{k-1}, b)\), ordered with \(k\) rules of unequal operands that fire,
\(b^k\) is \(k\) single tokens and \(ab^k\) one token, at distance \(k\); with no rule a prepend is at distance one. The
ordered dictionaries of one to four rules on two symbols attain it at every \(k\) from one to four as well. For
\(m \ge 2\) every dictionary attaining \(C_m\) contains an effective self-pair, since
\(C_m \ge 2m + 1 > 1 + \lceil 3m/2 \rceil\) (Theorems~\ref{thm:amp} and~\ref{thm:region}), while at \(m = 1\) the rule
\((a, b)\) attains \(C_1 = 3\) on \(ab\) against \(acb\). A self-pair pass is where the region argument stops. It
regroups a run \(u^j\) into pairs from the left. On \(P_0 u^p u^q S_0\) against \(P_0 u^p\, x\, u^q S_0\) with \(x\) a
distinct symbol and \(p, q\) odd, the pass of \((u, u)\) leaves \(T^{(p+q)/2}\) against
\(T^{(p-1)/2} \mid u \mid x \mid T^{(q-1)/2} \mid u\) between \(P_0\) and \(S_0\), \(T = uu\), so the maximal common
prefix and suffix leave regions of \((q + 1)/2\) and \((q + 5)/2\) tokens, unbounded in \(q\), while the distance is
three, so the region is not preserved by self-pair passes. The distance is not additive over passes either, and a pass
can add more than two to it with no self-pair in sight. On \((a,b), (b,a), (ba,b), (ab,ba)\), ordered, proper and of
finite radius, the inputs \(abbabbba\) and \(abbaabba\) are at distances \ObstacleDistances{} after zero to four rules.
Neither the region nor a per-pass increment thus bounds the distance in general, and
Proposition~\ref{prop:counterfamily} shows that no bound of \(2m + 1\) exists to be proved. Theorem~\ref{thm:quadratic}
settles the growth by a different invariant, the decomposition into blocks and clusters.

The prepend constant is the largest \(\lambda(a, x)\) over ordered dictionaries of \(m\) rules, symbols and inputs. It
is at least \(m + 1\) for \(m \ge 2\), and at least \(2m - 3\) for \(m \ge 5\) by Proposition~\ref{prop:counterfamily}.
Take the dictionary \((a,a), (aa, c_2), (aac_2, c_3), \dots\), whose \(i\)th rule merges the previous product with
\(c_i = b\) for even and \(a\) for odd \(i\). It is ordered, since its products have distinct lengths. On
\(x_m = aa\,c_2 \cdots c_m\) it tokenizes \(x_m\) as one token. On \(a x_m\) the pass of \((a, a)\) pairs the first two
symbols and leaves \(aa\) before \(a\) rather than \(b\), so no later rule fires, and the tokenization is
\(aa \mid a \mid c_2 \mid \cdots \mid c_m\). The two sequences share no token, so the distance is \(m + 1\). Executed,
the exact prepend supremum over every ordered proper dictionary of one to four rules on two symbols is
\PrependExactOne{}, \PrependExactTwo{}, \PrependExactThree{} and \PrependExactFour{}, so the constant is \(m + 1\) on
these universes for \(m = 2, 3, 4\). At five rules it is \PrependExactFive{}, computed over all \PrependFiveTables{}
ordered proper dictionaries on two symbols and attained by \PrependFiveAttaining{} of them, the dictionary of
Proposition~\ref{prop:counterfamily} among them. That is \(2m - 3\) and not \(m + 1\). Theorem~\ref{thm:quadratic} shows
it is \(\Theta(m^2)\), and its exact value is open.

\section{A production vocabulary}\label{sec:production}

\begin{table}[t]
\centering
\small
\begin{tabular}{@{}lp{0.62\textwidth}@{}}
\toprule
merges & measure \\
\midrule
4, 8, 12 & origin radius \RealRadiusFour{}, \RealRadiusEight{}, \RealRadiusTwelve{}; mirrored constant
\RealMirroredFour{}, \RealMirroredEight{}, \RealMirroredTwelve{} \\
\addlinespace
32 & origin radius \RealRadiusThirtyTwo{}; token machine of \RealStates{} states at a tail of \RealTail{} \\
\addlinespace
1 to \GptRadiusEarlyLast{} & origin radius \GptRadiusEarly{} \\
\addlinespace
to \GptFinitePrefix{} & origin radius \GptRadiusLate{}; mirrored constant \GptMirroredFortyOne{} at
\GptFinitePrefix{} \\
\addlinespace
100 & infinite under \GptInfiniteHundred{} of \GptSymbolsHundred{} symbols, at most \GptMaxFiniteHundred{} under the
rest \\
\addlinespace
200 & infinite under \GptInfiniteTwoHundred{} of \GptSymbolsTwoHundred{}, at most \GptMaxFiniteTwoHundred{} \\
\addlinespace
400 & infinite under \GptInfiniteFourHundred{} of \GptSymbolsFourHundred{}, at most \GptMaxFiniteFourHundred{} \\
\addlinespace
800 & infinite under \GptInfiniteEightHundred{} of \GptSymbolsEightHundred{}, at most \GptMaxFiniteEightHundred{} \\
\addlinespace
\GptRules{} & \GptSelfPairs{} effective self-pairs, the least at merge \GptFinitePrefix{};
\(Q_{\mathrm{span}} = \ChainGptQspan\), \(Q_{\mathrm{start}} = \ChainGptQstart\), \(Q(D) = \ChainGptQD\),
\(l(D) = \LookGptConstant\); class span radii \ClassRadiusLetters{} for letters, \ClassRadiusDigits{} for digits,
\ClassRadiusSymbols{} for symbols, \ClassRadiusWhitespace{} for whitespace and \ClassRadiusContractions{} for
contractions \\
\bottomrule
\end{tabular}
\caption{GPT-2's merge table and its prefixes: the origin radius and mirrored constant where they are finite, the
verdict per prepended symbol beyond, and on the whole table the self-pairs, the right radii, the lookahead constant and
the span radii inside the classes of the pre-tokenizer, merges counted from zero for the self-pair.}
\label{tab:gpt2}
\end{table}


Table~\ref{tab:gpt2} gathers what this section measures on GPT-2. We take the first \(N\) merges of GPT-2 at the six
fixed sizes \(N = 4, 8, 12, 24, 28, 32\). Each satisfies the ordering condition and gives a bounded origin radius,
\RealRadiusFour{} at four merges and \RealRadiusThirtyTwo{} at eight and beyond. At thirty-two merges the machine has
\RealStates{} states over an alphabet of \RealAlphabet{} symbols, with a tail of \RealTail{}. The mirrored constant is
\RealMirroredFour{} at four merges, \RealMirroredEight{} at eight and \RealMirroredTwelve{} at twelve, against origin
radii of \RealRadiusFour{}, \RealRadiusEight{} and \RealRadiusTwelve{}. That is where the token machine stops. The
criterion and the overlap graph go on.


The whole merge table has \GptRules{} rules, read from the \code{merges.txt} of the published model, held to its digest.
It satisfies clause (a) and the ordering condition on every rule, so Theorem~\ref{thm:selfpair} applies to it and to
each of its prefixes. It carries \GptSelfPairs{} rules \((t, t)\), every one effective at its stage. The least of them
is the rule \((l, l)\) at merge \GptFinitePrefix{}, counted from zero. GPT-2's origin radius is thus infinite, as is
that of every prefix of more than \GptFinitePrefix{} merges, and it is finite for every prefix of at most
\GptFinitePrefix{}. The overlap graph, in a C++ implementation held to the Python one on \CppDictionaries{} dictionaries
and \CppVerdicts{} verdicts, computes the finite prefixes in a fraction of a second. The radius is \GptRadiusEarly{} on
every prefix of 1 to \GptRadiusEarlyLast{} merges and \GptRadiusLate{} on every longer one to \GptFinitePrefix{} merges,
where the mirrored constant is \GptMirroredFortyOne{}. Beyond that the graph gives the verdict per prepended symbol,
where the criterion gives only the maximum. At 100 merges over \GptSymbolsHundred{} symbols the radius is infinite under
\GptInfiniteHundred{} of them and at most \GptMaxFiniteHundred{} under the rest. At 200 it is infinite under
\GptInfiniteTwoHundred{} of \GptSymbolsTwoHundred{} and at most \GptMaxFiniteTwoHundred{}. At 400, under
\GptInfiniteFourHundred{} of \GptSymbolsFourHundred{} and at most \GptMaxFiniteFourHundred{}. At 800, under
\GptInfiniteEightHundred{} of \GptSymbolsEightHundred{} and at most \GptMaxFiniteEightHundred{}, the largest product
explored holding \GptNodesEightHundred{} nodes. The source paper recorded that from random testing the ripple seems to
often be finite. On GPT-2's table it is infinite from the forty-second merge on, and on the three tables below from the
first. The pre-tokenizer is not modeled. The question asked is about the merge table as a dictionary under
Definition~\ref{def:sp}, and a pre-tokenizer that bounds the length of its pieces bounds every ripple by that length.

Three more production tables have no finite nonempty prefix. The rank files of cl100k\_base and o200k\_base, the tables
of OpenAI's \code{tiktoken} library, and of Llama~3, from Meta's public \code{llama-models} repository, list tokens as
byte strings with ranks. They are held to their SHA-256 digests. From each we derive the \emph{rank-derived list}. For
every token of two bytes or more, in rank order, the function \code{byte\_pair\_merge} of \code{tiktoken}, the
\emph{rank merge} below, runs on its bytes with only the ranks below it, and the two parts it leaves are the merge; a
token left in three parts or more gives none. Table~\ref{tab:vocabs} sets the four tables side by side. On cl100k\_base
and o200k\_base every token derives, clause (a) and the ordering condition hold on every merge, and the derived list
under Definition~\ref{def:sp} tokenizes every token of the rank file as the rank merge with all ranks does. On Llama~3,
whose first \LlamaShared{} merges are cl100k\_base's, \LlamaUnreachable{} tokens ranked after them are left in three
parts or more by the rank merge with the ranks below them. Clause (a) first fails at merge \LlamaClauseA{}, one of whose
parts is such a token, while the ordering condition holds on every merge. The derived list also disagrees with the rank
merge on \LlamaRankPairDisagree{} of the \LlamaVocabTokens{} tokens of two bytes or more, because a merge of smaller
rank can become applicable only after a token of larger rank exists. On \code{.:.:}, the least of them by rank, the rank
merge forms \code{:.}, then \code{.:.}, and then \code{.:.:}, whose rank is below that of \code{.:.}, while the derived
list leaves \code{.:.} and \code{:}, a pair it cannot hold, since \code{.:.} ranks after \code{.:.:}. Llama~3's
rank-derived list is thus not its published pair list.

\begin{table}[t]
\centering
\setlength{\tabcolsep}{5pt}
\begin{tabular}{@{}lrrrrl@{}}
\toprule
table & merges & ordered prefix & rules \((t, t)\) & effective & least effective \\
\midrule
GPT-2 & \GptRules{} & \GptRules{} & \GptSelfPairs{} & \GptEffective{} & merge \GptFinitePrefix{}, \((l, l)\) \\
cl100k\_base & \ClMerges{} & \ClAdmissible{} & \ClSelfPairs{} & \ClEffective{} & merge \ClLeastMerge{}, two spaces \\
o200k\_base & \OMerges{} & \OAdmissible{} & \OSelfPairs{} & \OEffective{} & merge \OLeastMerge{}, two spaces \\
Llama~3 (rank-derived) & \LlamaMerges{} & \LlamaAdmissible{} & \LlamaSelfPairs{} & \LlamaEffective{} & merge
\LlamaLeastMerge{}, two spaces \\
\bottomrule
\end{tabular}
\caption{The self-pair criterion on four production merge tables. The columns give the merges, read for GPT-2 from
its \code{merges.txt} and derived from the rank file for the other three, the longest ordered prefix, its rules
\((t, t)\), the effective ones, and the least effective one, counted from zero. Llama~3's derived list is not its
published pair list. The origin radius is finite exactly on the prefixes before it, the first \GptFinitePrefix{} merges
of GPT-2 and the empty prefix of the other three.}
\label{tab:vocabs}
\end{table}

The published list of \LlamaPublishedMerges{} merges is every split of a token of two bytes or more into two tokens,
ordered by the product's rank and then by the left part's. Its first consumer violating either clause is its merge
\LlamaPublishedFail{}, which joins a space to three spaces, a token first produced at merge \LlamaPublishedOrdered{}, so
its longest ordered prefix has \LlamaPublishedOrdered{} merges, with \LlamaPublishedSelfPairs{} rules \((t, t)\), all
effective. The criterion of Theorem~\ref{thm:selfpair} reads the whole of each derived list, which is ordered.

Under Definition~\ref{def:sp} the published list equals the rank merge with all ranks on every string, and so does the
shorter \emph{self-split list}, the last join of the rank merge on each token that it leaves whole, in rank order, which
is \LlamaSelfSplit{} merges for Llama~3 and the derived list itself for the other two tables. Take a run of the rank
merge and two adjacent parts \(u\) and \(v\) whose concatenation \(t\) is a token. No merge so far has crossed the ends
of \(uv\), and a merge chosen inside it has the least rank there and is the leftmost of that rank, so, as in
Lemma~\ref{lem:proj}, the merges inside are a prefix of the run on \(t\) alone, which reaches \(u \mid v\) and joins it
last. Three facts follow. Every pair the rank merge can join at a partition it reaches is the last join of its product,
one pair for each product, so no other split of a product is ever a candidate there, and the order of the splits within
one product's rank does not matter. Each list holds every such last join, in the order of the product's rank, so the
rules that can apply at a reachable partition keep the priorities of the ranks. And Definition~\ref{def:sp} takes the
leftmost occurrence of the rule of highest priority, as the rank merge takes the leftmost pair of least rank. By
induction the two procedures make the same merge at every step. The restriction is the source paper's Lemma~2(i), its
Corollary~1 says that a rule is useful exactly when it fires on its own product, and \citet[Appendices~A.2
and~C]{jiang2026incremental} keep only such last joins of a merge list; the argument here adapts them to a table of
ranks. Run on the self-split list, the zero-delay tokenizers of Theorems~\ref{thm:zerodelay} and~\ref{thm:zerodelayauto}
therefore emit the stable output of \code{tiktoken}'s rank merge for every rank table, Llama~3's included, and in
polynomial time, since each rule of the self-split list spells its token once, so that the list written out is no longer
than the table. The agreement is between merge procedures. On Llama~3, \LlamaNotWhole{} tokens of the rank file, the
least at rank \LlamaNotWholeLeast{}, are left in three parts or more by the rank merge with all ranks, and so by both
lists. \code{tiktoken} and the published tokenizer return such a token whole only because they look a whole piece up in
the vocabulary before merging, \code{tiktoken} in every encoding and the published tokenizer by its
\code{ignore\_merges} flag. Executed, the two lists equal the rank merge on every token of two bytes or more and on
\VocabSeededStrings{} seeded strings per table.

Looking a piece up first keeps the zero delay, at the cost of waiting while the piece read so far could still become a
vocabulary entry that the merge would split. Let a tokenizer \(g\) return each string \(v\) of a finite vocabulary
\(\mathcal{V}\) as the one token \(v\) and every other string as \(\T^D(v)\), let \(\mathcal{B}\) be the strings of
\(\mathcal{V}\) that \(\T^D\) does not leave whole, and let
\(M_{\mathcal{B}} = \max(\{|b| : b \in \mathcal{B}\} \cup \{0\})\). With the rank file as the vocabulary and the
self-split list as \(D\), \(\mathcal{B}\) is empty for cl100k\_base and o200k\_base and holds the \LlamaNotWhole{}
tokens above for Llama~3.


\begin{corollary}[Zero delay through whole-piece lookup]\label{cor:wholepiece}
For every string \(p\), the longest common prefix of \(g(pq)\) over all strings \(q\) is empty when \(p\) is a prefix of
a string of \(\mathcal{B}\), and \(\Stable(p)\) otherwise. A streaming tokenizer that keeps the prefixes of
\(\mathcal{B}\) as a trie, holds the input while it is one of them, and otherwise runs the tokenizer of
Theorem~\ref{thm:zerodelay} emits exactly that prefix after each symbol and has emitted \(g(w)\) at the end of the input
\(w\). Its largest held suffix over the inputs read is \(\max(M_{\mathcal{B}}, Q_{\mathrm{span}}(D))\) symbols.
\end{corollary}

\begin{proof}
If \(p\) is no prefix of a string of \(\mathcal{B}\), neither is any \(pq\), and a string of \(\mathcal{V}\) outside
\(\mathcal{B}\) is one token under \(\T^D\) too, so \(g(pq) = \T^D(pq)\) for every \(q\) and the common prefix is
\(\Stable(p)\). Let \(p\) be a prefix of \(b \in \mathcal{B}\). The continuation that completes \(b\) gives the one
token \(b\), so a nonempty common prefix would begin with \(b\), and the empty continuation would make \(p = b\). The
vocabulary is finite, so some \(bq\) lies outside it, and \(\T^D(bq)\) would begin with the token \(b\), so that by
Lemma~\ref{lem:proj} on the interval of \(b\), \(\T^D(b) = b\), against \(b \in \mathcal{B}\). The common prefix is thus
empty. The streaming tokenizer emits nothing while the input is a prefix of a string of \(\mathcal{B}\), holding at most
\(M_{\mathcal{B}}\) symbols. Once the input is none, no extension is one either, and the tokenizer of
Theorem~\ref{thm:zerodelay}, run on the symbols held and then on the rest, emits \(\Stable\) of the input read with a
held suffix of at most \(Q_{\mathrm{span}}(D)\). At the end of the input the tokenizer flushes. If it is still in the
trie, the whole input \(w\) is a prefix of a string of \(\mathcal{B}\), and it emits \(g(w)\), the one token \(w\) when
\(w\) lies in \(\mathcal{V}\) and \(\T^D(w)\) otherwise. If it has left the trie, \(w\) is no prefix of a string of
\(\mathcal{B}\), so \(g(w) = \T^D(w)\), and the tokenizer of Theorem~\ref{thm:zerodelay} emits \(\T^D\) of its held
suffix, which completes \(\T^D(w)\), whether or not that held suffix is the whole input. When
\(M_{\mathcal{B}} \ge Q_{\mathrm{span}}(D)\) and \(\mathcal{B}\) is nonempty, a longest string of \(\mathcal{B}\) has an
empty common prefix and so a held suffix of \(M_{\mathcal{B}}\) symbols. When
\(Q_{\mathrm{span}}(D) > M_{\mathcal{B}}\), a string \(x\) of right span ripple \(Q_{\mathrm{span}}(D)\) under some
symbol has, by Lemma~\ref{lem:delay} and its proof, a held suffix of \(Q_{\mathrm{span}}(D)\) symbols under \(\T^D\),
and \(|x| \ge Q_{\mathrm{span}}(D) > M_{\mathcal{B}}\) puts it outside the prefixes of strings of \(\mathcal{B}\), where
the two common prefixes agree. When both are zero, every held suffix is empty.
\end{proof}

Executed on \WholeDictionaries{} dictionaries, with the symbols and the products as the vocabulary, \WholeWithB{} of
them with a product the tokenization splits, the common prefix by brute force over the continuations is as the corollary
says on \WholePrefixes{} prefixes, \WholeInTrie{} of them prefixes of a string of \(\mathcal{B}\), and the streaming
tokenizer emits it after every symbol of \WholeInputs{} inputs and flushes \(g\) of each. The bound on the held suffix
is attained on every one of the \WholeDictionaries{} dictionaries, by a longest string of \(\mathcal{B}\) on \WholeByB{}
and by a string attaining \(Q_{\mathrm{span}}(D)\) on the other \WholeBySpan{}, each replayed by the brute force over
the continuations.

Among the lists over the same bytes whose tokenizer under Definition~\ref{def:sp} is the rank merge with all ranks, the
self-split list uses the fewest rules and the least total length of operands. It has one rule for each token of two
bytes or more that the merge leaves whole, the last join of that token, which fires on it, so it is its own useful
sublist, and Lemma~\ref{lem:gtrees} gives both minima. Deleting any of its rules changes the tokenization of that rule's
product. The rank-derived list is a subsequence of it, since when the rank merge with the ranks below a token leaves two
parts, the merge with all ranks runs the same way on that token and joins those two parts last.

In all three rank files the lowest-ranked token of two bytes is two spaces, so the first merge of each derived list, and
of Llama~3's published list, is the rule that joins two spaces. A first rule \((a, a)\), with \(a\) a symbol, makes the
origin radius infinite whatever the rules after it. On \(a^n\) the rule has the highest priority, so the run pairs the
symbols from the left until no two single \(a\) are adjacent, leaving its tokens starting at the even positions of
\(x = a^n\) and, in the coordinates of \(x\), at the odd positions \(-1, 1, 3, \dots\) for \(ax\). Every later merge
only removes starts, so every position of \(x\) is covered by a token starting at an even position in one run and at an
odd one in the other, every position changes its covering span, and the origin ripple of \(a\) on \(a^n\) is \(n\) for
every \(n\). Every prefix of one merge or more of the derived lists and of the published list thus has infinite origin
radius, which Theorem~\ref{thm:selfgeneral} gives as well; this is the linear worst case on whitespace that the interval
encoder of the \code{bpe} crate notes \citep{vanantwerpen2024bpe}. A space prepended to \(n\) spaces moves the covering
token of every one of the \(n\) bytes, replayed by Definition~\ref{def:sp} for \(n\) up to \(64\) under the first merge,
under the whole derived list and, for Llama~3, under the published list, and the overlap graph on the first merge alone
is infinite under the space. At 100 merges it is infinite under \ClInfiniteHundred{} of the \ClSymbolsHundred{} symbols
of cl100k\_base and under \OInfiniteHundred{} of the \OSymbolsHundred{} of o200k\_base, and at most
\ClMaxFiniteHundred{} and \OMaxFiniteHundred{} under the rest. The first 100 merges of Llama~3's derived list are
cl100k\_base's. As for GPT-2, the question is about the merge table as a dictionary under Definition~\ref{def:sp}, not
about the pre-tokenizer, normalization, special tokens or the composed pipeline.


The right radii of GPT-2's whole table come from the rules. Every prefix of the table is ordered with every rule useful,
so Theorem~\ref{thm:chainsearch} and Proposition~\ref{prop:rulegraph} apply, and the pruned chain search computes the
exact right span radius and right start radius of the whole table in about a second, giving
\(Q_{\mathrm{span}} = \ChainGptQspan\) and \(Q_{\mathrm{start}} = \ChainGptQstart\), where \(Q(D) = \ChainGptQD\) and
the rule graph, of \ChainGptEdges{} edges on the \GptRules{} rules with \ChainGptSources{} sources, gives
\(C(D) = \ChainGptC\) and \(C_{\mathrm{start}}(D) = \ChainGptCstart\). The search visits \ChainGptStates{} states and
prunes \ChainGptPruned{}. The span radius is attained by a string of \ChainGptQspan{} symbols with \(e\) appended, along
a chain of \ChainGptSpanRules{} rules from \((r, e)\) to the rule that merges a space-led \code{neighbourhood} with
\(s\), and the start radius, a start ripple of \ChainGptQstart{} on a string of \ChainGptStartString{} symbols, along
\ChainGptStartRules{} rules; both are replayed by definition through \code{gpt2\_full.py}'s tokenizer.
Table~\ref{tab:chains} gives the prefixes. The exactness rests on three things, that the attaining strings replayed by
definition are lower bounds, that the rule graph's bounds are upper bounds, and that the two coincide at every size but
200 merges, where \(C_{\mathrm{start}}(D) = \ChainGptTwoHundredCstart\) exceeds the start radius
\ChainGptTwoHundredQstart{} and the search still returns the latter, held to the append product there. The product,
which explores the pair graph of two spelling runs over the canonical tokens, agrees to 400 merges and exhausts 12 GB of
address space at 800, the last prefix the C++ overlap graph completes. The whole table thus has an exact right
lookahead, in that after a prefix \(x\) of a pre-tokenized piece, every token of \(\T^D(x)\) ending at or before
\(|x| - \ChainGptQspan\) is a token of \(\T^D(xz)\) for every continuation \(z\), and every token start before
\(|x| - \ChainGptQstart\) is a start of \(\T^D(xz)\), by Lemmas~\ref{lem:delay} and~\ref{lem:rightstart}. The first
figure is the delay of \citet{mamouras2026bpe} for this table under Definition~\ref{def:sp}, by Lemma~\ref{lem:delay},
which their Theorem~10 bounds by the left operands of all rules summed, \(Q(D) = \ChainGptQD\) here since every rule of
the table is useful; the second is the finer start version. The origin side stays infinite from the forty-second merge
on, so the pinned-byte width of Section~\ref{sec:density} is not available on the whole table.

\begin{table}[t]
\centering
\begin{tabular}{@{}rrrrrrr@{}}
\toprule
merges & \(Q(D)\) & \(C(D)\) & \(C_{\mathrm{start}}(D)\) & \(Q_{\mathrm{span}}\) & \(Q_{\mathrm{start}}\) & states \\
\midrule
\ChainGptRows
\bottomrule
\end{tabular}
\caption{The right radii of GPT-2's merge prefixes by the pruned chain search, with the lookahead bound
\(Q(D)\), the rule graph's bounds and the states the search visits; the append product agrees to 400 merges and the
plain search to 1,600.}
\label{tab:chains}
\end{table}


Inside one piece of GPT-2's pre-tokenizer the delay depends on the kind of piece. The pattern of the released encoder,
as \citet{nidhogg2026pretokenizer} states it, cuts the input into contractions, runs of letters, of digits and of other
symbols, each run with at most one leading space, and runs of whitespace, and the byte-pair stage tokenizes each piece
alone. On ASCII input its classes are the runs of the \ClassBytesLetters{} letters, of the \ClassBytesDigits{} digits
and of the \ClassBytesSymbols{} bytes that are none of these and none of the \ClassBytesWhitespace{} whitespace bytes,
each with an optional leading space, the runs of those whitespace bytes, and the \ClassContractionForms{} contractions,
as the pre-tokenizer of the \code{tokenizers} library confirms on every byte. For such a class \(K\), the class span
radius \(Q^K_{\mathrm{span}}\) is the supremum of the right span ripple of \((x, b)\) over the strings with \(x\) and
\(xb\) both pieces of \(K\). Every string that extends a piece of a class and is a prefix of a longer piece of it is a
piece too, so the proof of Lemma~\ref{lem:delay} goes through inside \(K\). After a piece \(x\) of \(K\), every token of
\(\T^D(x)\) ending at or before \(|x| - Q^K_{\mathrm{span}}\) is a token of \(\T^D(xz)\) for every \(z\) with \(xz\) a
piece of \(K\), and the bound is attained. The chain search computes \(Q^K_{\mathrm{span}}\) when it keeps only the
states \((s, b)\) with \(sb\) a factor of a piece of \(K\) and counts only those with \(s\) and \(sb\) pieces, since a
state extends its parent on the left and a string with a suffix outside the factors of \(K\) is outside them too. On the
whole table it gives \ClassRadiusLetters{} for letters, \ClassRadiusDigits{} for digits, \ClassRadiusSymbols{} for
symbols, \ClassRadiusWhitespace{} for whitespace and \ClassRadiusContractions{} for contractions, none of which extends
another, with every attaining string replayed by definition. The letters value is the whole table's, on the same string,
so the figure \ChainGptQspan{} above is the delay of a run of letters, while within a run of digits a streaming
tokenizer withholds at most \ClassRadiusDigits{} symbols, within a run of symbols \ClassRadiusSymbols{}, and within a
run of whitespace \ClassRadiusWhitespace{}.

The source paper's Theorem~2 bounds the lookahead constant by \(|D| \max \{|uv|\}\) over the rules, and GPT-2's table,
proper since it satisfies clause (a) on every rule, has \GptRules{} rules and a longest product of \LookGptLongest{}
symbols, which gives \LookGptBound{}. On GPT-2's merge table \(l(D) = Q_{\mathrm{start}}(D) + 1 = \LookGptConstant\).
There is a string of \LookGptString{} symbols whose first token, of \LookGptFirst{} symbols and followed by
\LookGptAfter{} more, is removed when \(e\) is appended, so \(l(D) > \LookGptAfter\), while
\(Q_{\mathrm{start}}(D) = \LookGptAfter\) and Corollary~\ref{cor:lookconst} bounds \(l(D)\) by
\(Q_{\mathrm{start}}(D) + 1\).

By the corollary \(A(D) + 1 = l(D) = \LookGptConstant\), so \(A(D) = \LookGptAfter\) on the whole table, and the string,
replayed by definition, is a state of the chain search with the largest \(|s| - |t|\), \(t\) the first token of
\(\T^D(s)\). Its first token is a space and \(G\), followed by the two bytes of a UTF-8 \"o, so it lies outside the
ASCII classes above. The source paper's Definition~6 quantifies over every string with no pre-tokenizer, and \(l(D)\) is
one number for the whole table.

\paragraph{The decisions in the libraries.} Release 3.0.0 of \code{munch}~\citep{munch} computes \(\ripple(D)\) and
\(L(D)\) of a merge table in \code{ripple\_radius()} by the overlap graph of Theorem~\ref{thm:overlap}, with an
attaining input, and reports an infinite radius by the first effective self-pair of Theorem~\ref{thm:selfpair}. It takes
the tables that satisfy clause (a) with distinct products, which satisfy the ordering condition, gives the largest
values over the prepended symbols, and counts in bytes, so that over an alphabet of characters its values are in bytes
where this paper's are in symbols. On byte-level tables of that kind its \code{rule\_ceilings()} computes
\(Q_{\mathrm{span}}(D)\) and \(Q_{\mathrm{start}}(D)\) by the chain search of Theorems~\ref{thm:chainform}
and~\ref{thm:chainsearch}, with the ceiling of Proposition~\ref{prop:radiibounds}(iii), and \code{window\_density()}
decides within a budget whether the certified windows are dense, as in Proposition~\ref{prop:lastbyte}, with
Theorems~\ref{thm:pinned} and~\ref{thm:density} and Corollary~\ref{cor:lastbyte}. Over a table's token automaton,
\code{miscovering()} decides a window certificate by Theorem~\ref{thm:bigram}, within the bound of
Lemma~\ref{lem:localref}, and for byte tables \code{chunk\_plan()} plans its cuts with the window of
Theorem~\ref{thm:pinned}, the two sides of each cut tokenized alone as the whole input tokenizes them by
Theorem~\ref{thm:bigram}. The library's \code{encoder\_divergence()} decides whether two tables tokenize alike by their
canonical words and pairs, as in Corollary~\ref{cor:equivalence}, with a separating string of at most \(2M - 1\) bytes,
\(M\) the longest canonical word of either table, a shortest one when the canonical words agree. The streaming encoder
\code{Stream\_encoder} of release 1.0.0 of \code{matisse}~\citep{matisse} applies Definition~\ref{def:stable},
Lemma~\ref{lem:finitecont} and Theorem~\ref{thm:zerodelay} inside each piece of the pre-tokenizer, with the whole-piece
lookup of Corollary~\ref{cor:wholepiece}, on tables that satisfy the ordering condition. The record of \code{munch}'s
release 3.0.0 is \doi{10.5281/zenodo.23422563}, under the concept identifier \doi{10.5281/zenodo.21752996}, and that of
\code{matisse}'s release 1.0.0 is \doi{10.5281/zenodo.23782292}, under \doi{10.5281/zenodo.23782291}.

\section{Related work}\label{sec:related}

The question answered here is the third future-work item of \citet{berglund2023formalizing}, Section~5, carried
unchanged in the revision. Everything this paper inherits from that paper is named where it is used. The neighbors below
are credited for what this paper shares with them, and they mark where it stops.

The right side of this question is occupied by the source paper's bounded lookahead \citep{berglund2023formalizing} and,
with Martens, its authors' construction of a tokenization DFA and of a subsequential transducer
\citep[Theorem~4]{berglund2024dfa}, with Jonsson a compact form of the token DFA \citep{berglund2025compact}, the last
two for proper dictionaries tokenized by one pass per rule in priority order, Algorithm~1 of \citet{berglund2024dfa},
which is Definition~\ref{def:sp} under the ordering condition and not in general. Streaming byte-pair encoding
\citep{mamouras2026bpe} and incremental byte-pair encoding \citep{jiang2026incremental} occupy it as well, and the exact
token-neighbor distance of maximal-munch tokenization grammars with its complexity \citep{li2026streaming} occupies it
for lexers. The first of these defines the delay of a ruleset and bounds it by the sum of the left operands (their
Theorem~10), for trained rulesets where the tokenizer is one pass per rule. It then tightens the bound by analyzing
batches of rules, on the 200,000-rule table it studies from roughly 700,000 to 29,872, and to 2,969 after about twelve
hours of analysis that composes sixteen neighboring batches (their Section~3.3), and builds its streaming tokenizer on
the tightened bound. Section~\ref{sec:lookahead} proves the first bound for every dictionary and shows its worst case
attained. The delay of \citet{mamouras2026bpe} is the right span radius of Section~\ref{sec:lookahead}
(Lemma~\ref{lem:delay}), computed in Section~\ref{sec:density} exactly from the rules under the ordering condition. It
is \ChainGptQspan{} on GPT-2's whole table, where their Theorem~10 gives \ChainGptQD{}, and neither paper computes the
batched bound on that table. Their monolithic transducer, whose delay is exact, stops at 923 rules of the 200,000-rule
table they analyze, in their Section~3.2. Online tokenization with a bounded delay \citep{dalirrooyfard2026online}
measures how far an algorithm's output lags behind the longest prefix that no continuation changes, and leaves open
whether byte-pair encoding alone admits no lag at all (their Remark~1.4). The right span radius is how far that prefix
stays behind the input, and Theorems~\ref{thm:zerodelay} and~\ref{thm:zerodelayauto} emit that prefix exactly, in
polynomial time, for their lists and for every dictionary, not in their near-linear time and not for the composed
pipeline, which their Theorem~1.1 shows NP-hard. Emitting that prefix from a compiled subsequential transducer is
classical \citep{mohri1994compact}, and Section~\ref{sec:lookahead} sets the two theorems beside it and beside the
streaming of \citet{hayase2025bytesampler} and \code{tiktoken}. None of the byte-pair works that follow the source paper
yields a result for prepending. The source paper only exhibits it.

Incremental retokenization of byte-pair encoding has also been argued informally.
\citet[Section~4.1]{hellsten2024incremental} stops retokenizing at the first unchanged token and concedes that a change
can propagate through the entire text, which is the infinite radius seen from an implementation. The interval encoder of
GitHub's \code{bpe} crate re-encodes a slice from its left end until its tokens rejoin the whole text's, and notes a
linear worst case on whitespace, where the encoding changes when the starting position changes
\citep{vanantwerpen2024bpe}.

Section~\ref{sec:density} puts the two sides together. The lookahead constant of the source paper's Definition~6,
bounded in its Theorem~2 by the number of rules times the longest product, the delay of \citet{mamouras2026bpe}, bounded
in their Theorem~10 by the sum of the left operands, and the eager output of \citet{jiang2026incremental}, which emits a
token once its boundary is settled, each fix a prefix of the output against appended input. None combines that with a
lookbehind, and none asks for a cut that holds in every context. Practice has asked for one at width two. Gigatoken
finds cuts before selected punctuation in SentencePiece input where no vocabulary token contains the two symbols at the
cut in that order, arguing in a source comment that a merge's result is always a vocabulary token, so no merge can span
the boundary \citep{rod2026gigatoken}; the runtime of SentencePiece Lite states the same principle
\citep{sentencepiecelite2026}, and \citet{xu2026parallel} checks a merge table for cuts across which no merge joins.
Each is a sufficient condition for a cut \((W, 1, 1)\) with \(|W| = 2\), and none needs a finite radius.
Section~\ref{sec:density} adds a width at which every window has a cut, the least width at which every window pins its
byte \(\ripple(D)\), and the equivalence of a finite radius with dense cuts. The window certificate is that of
\citet{nidhogg2026finitelyverifiable}, decided there over a finite-state verifier, and for maximal munch
\citet[Theorem~3]{nidhogg2026splitting} decides the counterpart of density, whether the anchor-free span is bounded. The
chain length of the source paper's Remark~5, which looks for a sharper lookahead bound through chains of interacting
rules, is a property of the tokenizer up to equivalence; the chains of Section~\ref{sec:density} are the ones an input
realizes, and their weight is the exact right radius.

The bigram characterization of canonical token strings is Theorem~1 in Appendix~B of \citet{vieira2025canonical}, for
merge lists that need not be proper, used there for membership tests in language-model decoding. It was stated before,
for a ranked list of tokens and with a paragraph's argument, in the documentation of GitHub's \code{bpe} crate
\citep{vanantwerpen2024bpe}, and it follows, for both of the source paper's definitions, from that paper's Corollary~2
by induction on the number of tokens. Section~\ref{sec:overlap} writes its proof out with the projection lemma explicit
and builds on it the automaton whose product answers the question, which neither asks. The characterization is also what
makes a window certificate's refutation local here.

Tokenization as finite-state transduction \citep{cognetta2025transduction} and the classical theory of subsequential
transducers \citep{mohri1997transducers} decide properties of a compiled tokenizer, and two of the properties studied
here have been settled that way. \citet[Theorem~4]{berglund2024dfa} obtain the subsequential transducer of a proper
dictionary, under one pass per rule, from bounded variation. With Corollary~\ref{cor:rational}, the edit distance of
functional transducers \citep[Theorem~3.25 and Corollary~3.12]{aiswarya2024edit} decides whether the label distance is
bounded and computes its supremum. The robustness analysis of \citet[Definition~2.1 and
Theorem~4.4]{samanta2013robustness} takes a transducer, a bound on the number of edits and a proposed factor, and
decides whether inputs at most that many edits apart have outputs at most the factor times as far apart, in space
exponential in the number of edits for the Levenshtein distance. For one edit that is the question whether the one-edit
label distance is at most the factor. Section~\ref{sec:labelgraph} adds to the second a decision in polynomial time on
the overlap graph, which needs no proposed bound, the shape of a bounded component and a bound on the supremum.

For codes the shape of two of this paper's results is classical. A code is circular exactly when every periodic
bi-infinite word has at most one decomposition \citep[Proposition~2.9]{devolder1992finitary}, and a finite code has
finite synchronization delay exactly when it is very pure, which is circular \citep{restivo1975synchronization}; in this
paper the two statements are Theorems~\ref{thm:circular} and~\ref{thm:density}. The method is classical too. The
synchronization delay of a local automaton is computed on the product of two of its runs, whose part on distinct states
is acyclic, as the height of that part plus one \citep[Section~1]{beal1998synchronization}, and an unambiguous automaton
has two bi-infinite paths with one label exactly when its square has a cycle off its diagonal
\citep[Proposition~3]{beal2024unambiguously}. The overlap graph of Theorem~\ref{thm:overlap} is that product for the
spelling automaton. The spelling automaton is unambiguous, since every state is reached from the start and reaches a
final state and an input has one accepting run, and its bi-infinite paths are the canonical tilings. Proposition~3 of
\citet{beal2024unambiguously} thus applies to it as it stands, and it is the second form of Theorem~\ref{thm:circular},
that two canonical tilings of one bi-infinite word exist exactly when the square has a cycle off its diagonal. Their
Proposition~14 reaches the same tilings as the paths of the tokens' automaton constrained by the canonical pairs.
Theorem~\ref{thm:circular} adds that such a pair of tilings is an infinite radius, which needs the right lookahead, and
the certificate that exhibits it. The canonical tokens need not form a code, so the code theorems of
\citet{devolder1992finitary} and \citet{restivo1975synchronization} do not apply as they stand.

The question is one instance of a general one: at which positions does the past of a deterministic left-to-right
computation stop mattering, and how is that certified in a form one can execute? Its other instances so far are a
lexer's split points and split windows \citep{nidhogg2026splitpoints,nidhogg2026splitwindows}, recovery after a lexical
error \citep{nidhogg2026panicmode}, splitting and edit bounds for maximal munch \citep{nidhogg2026splitting}, and
retokenization under appended input, which Section~\ref{sec:lookahead} and Corollary~\ref{cor:lookconst} treat for
byte-pair encoding. Stating it once across them is separate work.

\section{Limitations}\label{sec:limitations}

The execution behind Lemma~\ref{lem:max} reaches the tail lengths the construction reports and inputs to length eight,
while the lemma is stated for every tail length at which the exploration completes and every input. Past that horizon
only the proof carries it. The exploration completes for every dictionary (Corollary~\ref{cor:completion}), at a tail
that can be exponential in the number of rules.

The lookahead bound \(Q(D)\) is attained by the doubling dictionary and lies far above the right span radius on GPT-2's
table, \ChainGptQspan{} against \ChainGptQD{}. The right span radius is computed on the reversed overlap graph, at
polynomial cost, and under the ordering condition by the chain search, whose cost is discussed below. The width
\(\ripple(D) + Q_{\mathrm{start}}(D) + 1\) of Theorem~\ref{thm:pinned} is the least for every dictionary of the pool of
Table~\ref{tab:exec-overlap} and not in general. The least width is \(\max(\ripple(D) + 1, M + 1)\) by
Proposition~\ref{prop:leastwidth}, where \(M\), the length defined there, is computed by the append product or by
enumeration and not from the rules, and it can fall arbitrarily far below the theorem's width
(Proposition~\ref{prop:slack}). A criterion read off the rules for dense certified windows, in the way the self-pair
criterion reads off whether the radius is finite, is open. Proposition~\ref{prop:lastbyte} decides it per dictionary,
and Corollary~\ref{cor:lastbyte} gives it at start radius zero.

Under the ordering condition both right radii are computed exactly from the rules by the chain search, whose plain form
enumerates every state, exponentially many in the span radius in the worst case, and whose pruned form has no bound
proved; it answered in a second on every GPT-2 prefix, where the rule graph that prunes it is exact at every size of
Table~\ref{tab:chains} but one, the start radius at 200 merges. Without the condition the search's values are lower
bounds once replayed (Proposition~\ref{prop:chainsfail}), and the exact computation stays with the append product, which
explores the whole pair graph, was run on the GPT-2 prefixes to 400 merges, and exhausts 12 GB at 800. A search over
episodes rather than rules was not attempted. The rule graph is loose for the span radius on \ChainGraphSlackSpan{}
dictionaries of that pool and for the start radius on \ChainGraphSlackStart{}, and for the start radius at 200 merges,
so an exact graph would need an edge condition that reads the forced intermediate merges off the rules. The C++ overlap
graph explores the whole reachable product before deciding, and runs out of memory on GPT-2 prefixes beyond 800 merges.

The exact label supremum is computed by profiles exponential in the cap, where the decision is polynomial. Over ordered
dictionaries of \(m \ge 1\) rules the amplification constant is \(\Theta(m^2)\), between
\(\lfloor (m^2 + 11m + 20)/10 \rfloor\) and \((m + 1)(m + 2)/2\), the substitution constant between
\(\lfloor (m^2 + 11m)/10 \rfloor\), for \(m \ge 12\), and \(m(m + 3)/2\), and the prepend constant between
\(\lceil (m^2 + 5m)/10 \rceil\) and \(m(m + 1)/2\); the leading constants lie between \(1/10\) and \(1/2\), and their
exact values are open. The constant is \(2m + 1\) on the universes of Section~\ref{sec:ampcomputed}, computed exactly to
four rules in the archived run of Appendix~\ref{app:exec} and at five rules by a recorded campaign. The prepend constant
is \(2m - 3\) at five rules on two symbols and quadratic in general. An input attaining the supremum of a fixed
dictionary under a prepended symbol has at most \(|G_a| P'\) symbols, and each input of an attaining pair under one edit
at most \(|G_e| P'\), by the profile search of Section~\ref{sec:labelgraph}, exponential in the value and the longest
token, and no bound polynomial in the dictionary is proved. For the constants over all ordered dictionaries of \(m\)
rules, whether an extremal pair of inputs of length polynomial in \(m\) exists is open.

The origin and label censuses run over two symbols, while the four dictionaries outside the census, the \BeyondProper{}
and \BeyondUnproper{} of the criterion's check and the \CppDictionaries{} of the C++ graph's reach three. The token
machine's production vocabulary is cut at thirty-two merges, and the overlap graph's at 800; the criterion reads the
whole table, and for cl100k\_base, o200k\_base and Llama~3 the whole rank-derived list. For Llama~3 that list is not its
published pair list, whose infinite radius rests on its first rule alone. The zero-delay tokenizers are polynomial in
the dictionary written out and not in the number of rules. Both are executed here on small dictionaries, and that of
Theorem~\ref{thm:zerodelay} also inside each piece on production tables by \code{matisse}.

\section{Conclusion}

The question \citet{berglund2023formalizing} left open has an answer for every dictionary. The origin radius, how far
into the input a prepend can change covering spans and so token origins, is decided and computed exactly on a product of
two runs of the automaton that guesses the final tokens, with radius certificates in both directions, and it is infinite
exactly when one periodic word has two canonical tilings with no common boundary. Under the ordering condition the
radius is infinite exactly when some rule \((t, t)\) is effective at its own stage, a property read off the rules. After
every symbol a streaming tokenizer emits exactly the tokens that no continuation changes, in polynomial time, which
answers the question \citet{dalirrooyfard2026online} leave open for byte-pair encoding. Boundedness of the label
distance is decided for every dictionary in polynomial time, and under the ordering condition the largest label distance
one edit causes over dictionaries of \(m\) rules grows as \(m^2\). A finite radius makes every window of width
\(\ripple(D) + Q_{\mathrm{start}}(D) + 1\) have a cut, and the radius is finite exactly when cuts are dense at some
width. These costs are in the dictionary written out. Measured by the number of rules, with the dictionary given as a
straight-line program, the decision and the exact radius become questions about compressed words, which the compact
companion takes up.

The production tables sit on the infinite side. GPT-2's merge table, read as a dictionary with the pre-tokenizer not
modeled, has infinite origin radius from the forty-second merge on, and those of cl100k, o200k and Llama~3 from the
first, while a window's cut can still be decided as it occurs in a text, by Lemma~\ref{lem:localref}. Every measured
number is its program's output, and Section~\ref{sec:limitations} names what is proved and not executed.

\section*{Tools}

Large language models assisted with drafting, with deriving proofs, with checking citations against primary records and
with reviewing the implementation; the author verified each suggestion by derivation, against primary sources, or
against the implementation, its tests and the archived artifacts, and is responsible for all content.

\appendix

\section{The token machine and the tail product}\label{app:machine}

This appendix proves the lemmas of Sections~\ref{sec:machine} and~\ref{sec:product}, Theorem~\ref{thm:origin}, and the
counts of Corollary~\ref{cor:tail}, stated as Lemma~\ref{lem:tailcounts}, with the coarsening of its relations.

\begin{proof}[Proof of Lemma~\ref{lem:A}]
Completion means that on every reachable transition from a truncated suffix the backup found its shared token inside the
suffix. A suffix never truncated is the whole tokenization, on which the backup may run to the first token harmlessly.
The source paper's Algorithm~1 on the full token list \(E(p)\,S(p)\) followed by the byte takes exactly the same steps,
since each step tokenizes a suffix of the current tokens and the byte, reading from the end backwards. The algorithm
thus halts at the same shared token inside \(S(p)\). By the source paper's Theorem~1 its output is \(\T^D(pc)\). That
output agrees with \(E(p)\,S(p)\) followed by \(c\) on everything before the shared token, so \(E(p)\) is a prefix of
\(\T^D(pc)\). Induction over the bytes of \(z\) gives the claim. Nothing here uses the coincidence of the two
definitions. The source paper's Corollary~2, the gluing corollary, and its Remark~3 hold for every dictionary.
\end{proof}

\begin{proof}[Proof of Lemma~\ref{lem:B}]
Let \(f\) be the first token of \(S\). Keep beside \(\T^D(p) = ES\) the invariant that whenever \(E\) is nonempty,
\(Ef\) is a prefix of \(\T^D(pz)\) for every \(z\). It holds at each truncation, where \(S\) becomes the last \(B\)
tokens of the current tokenization, so \(f\) is token \(n - m\) and the stability statement of Lemma~\ref{lem:stab}
applies. The invariant is preserved between truncations, where merges may shorten \(S\) below \(B\) without moving
\(f\). By the invariant, \(\T^D(pc) = Ef\psi'\) for some \(\psi'\). The source paper's Remark~3 says a suffix of a
correct tokenization that begins at a token boundary is correct, so \(\T^D(\yield(f\psi')) = f\psi'\). Since
\(\yield(f\psi') = \yield(S)c\), the isolated tokenization of the retained suffix followed by \(c\) begins with \(f\),
so the backup halts at \(f\) at the latest. The new retained suffix begins with \(f\) unless it exceeds \(B\) tokens, in
which case it is truncated and the invariant is established afresh. When \(E\) is empty the suffix is the whole
tokenization, and the backup may run to its front harmlessly. The construction flags a never-truncated suffix that has
reached \(B\) tokens as if truncated. This is safe because such a suffix has \(m + 1 > m\) tokens, so the stability
statement pins its first token and the backup halts there. Each reachable suffix has at most \(B\) tokens from the
finite set of tokens the dictionary can form. The reachable suffixes are thus finitely many, and the exploration
completes.
\end{proof}


The bound \(|D| + 1\) is the least tail this construction accepts on the source paper's Example~4 family for
\(n \le 4\), with \ExampleFourStates{} states at \(n = 4\). Table~\ref{tab:growth} reports the machine's growth on the
census against the tail-count bound \(\sum_{k=0}^{|D|+1} (|\Sigma| + |D|)^k\). At three rules the measured maximum comes
within a factor of \GrowthSlackThree{} of that bound.


\begin{table}[t]
\centering
\begin{tabular}{@{}lrrrr@{}}
\toprule
rules & states, median & states, max & tail-count bound & product nodes / states\(^2\), worst \\
\midrule
1 & \GrowthMedianOne & \GrowthMaxOne & \GrowthBoundOne & \multirow{3}{*}{\GrowthRatio} \\
2 & \GrowthMedianTwo & \GrowthMaxTwo & \GrowthBoundTwo & \\
3 & \GrowthMedianThree & \GrowthMaxThree & \GrowthBoundThree & \\
\bottomrule
\end{tabular}
\caption{Machine growth over the census of proper dictionaries on two symbols; the product of
Section~\ref{sec:product} holds at most \PendingMax{} pending tokens per run on this census.}
\label{tab:growth}
\end{table}

\begin{proof}[Proof of Lemma~\ref{lem:finite}]
A retained suffix has at most \(B\) tokens from the finite set of tokens the dictionary can form. The pending tokens of
a run cover the bytes between the lesser freeze point and its own, which the other run still retains. These tokens lie
within at most \(BM\) bytes, with \(M\) the longest token, plus one token straddling the lesser freeze point, at most
\(BM + 1\) tokens in all. The freeze points, as offsets from the node's position, are bounded by the tails' lengths in
bytes. Each part of a node thus takes finitely many values.
\end{proof}

\begin{proof}[Proof of Lemma~\ref{lem:max}]
By Definition~\ref{def:origin} the ripple compares, for each position of the input, the covering span in the two final
tokenizations. A token that has left a tail is final by Lemma~\ref{lem:A}. Each position is compared exactly once, and
with final tokens. A position the lesser freeze point passes is compared at that step, where the tokens covering it have
left both tails and the edge records the last disagreement that step reveals. A position the lesser freeze point has not
passed when the input ends is compared in the terminal contribution of the node the input ends in, where the input
having ended makes every token final. Every comparison is then between final tokens, and every position is compared.
Every disagreeing position of the input lies in the range of a recorded event or of the terminal contribution, each of
which names the largest disagreeing position in its own range. The ripple, one more than the largest disagreeing
position, is then the largest of their values. Position \(-1\), which only the prepended run covers, is no position of
the input, and a contribution naming it has value zero at the node that reads it. The clamp at zero is
Definition~\ref{def:origin}'s convention that the ripple is zero when no position differs. For the last claim, an event
names a disagreement at a position the lesser freeze point passes at its step, and the terminal contribution names one
at or after the lesser freeze point at the last node. Freeze points only advance, so every event of the path lies before
the terminal position. The terminal value \(|x| + e\) is at least zero, a contribution at position \(-1\) counting zero,
so it is the largest value the path reads.
\end{proof}

\begin{proof}[Proof of Lemma~\ref{lem:dead}]
The two runs are then in the same state with the same uncompared region. Both read the same bytes from then on, and the
transition is a function of the suffix and the byte (Section~\ref{sec:machine}), so their retained suffixes and their
frozen tokens agree at every later node. Every later comparison agrees, on the edges and at the nodes alike. The values
already recorded on the path to the node are properties of earlier edges, and the pruning leaves them untouched.
\end{proof}

\begin{lemma}[The suffix-pair counts]\label{lem:tailcounts}
In the setting of Corollary~\ref{cor:tail}, let \(N\) be the reachable suffixes, the empty one included, \(\mathcal{Y}\)
the yields of the nonempty ones, and \(C_{\mathrm{suf}}\) the number of pairs \((u, v)\) of words of \(\mathcal{Y}\)
with \(u\) a proper suffix of \(v\). A finite radius under a symbol is at most
\(C_{\mathrm{suf}} \le \min\bigl(\binom{N - 1}{2}, \sum_{u \in \mathcal{Y}} (|u| - 1)\bigr)\). With \(K\) the classes of
the quotient machine, \(K_c\) the number of those whose output ends in the symbol \(c\), and \(C^{Q}_{\mathrm{suf}}\)
the number of pairs of two distinct classes holding two nonempty suffixes whose yields are one a proper suffix of the
other, a finite radius is at most
\(C^{Q}_{\mathrm{suf}} \le \min\bigl(C_{\mathrm{suf}}, \sum_c \binom{K_c}{2}\bigr) \le \binom{K - 1}{2}\), and the
product of the quotient on unordered pairs, with the pairs of equal classes merged into one dead end, has at most
\(C^{Q}_{\mathrm{suf}} + 2\) reachable states. The count \(C^{Q}_{\mathrm{suf}}\) does not depend on \(B\), and the
pairs it counts are the pairs of distinct classes \(\{q_0 u, qu\}\) over the classes \(q\) and the nonempty words \(u\),
which are the closure of the pairs \(\{q_0 c, qc\}\) over the symbols \(c\) under transitions on a common symbol,
computed from the quotient alone in time \(O(|\Sigma|(K + C^{Q}_{\mathrm{suf}}))\).
\end{lemma}

\begin{proof}[Proof of Corollary~\ref{cor:tail}, the counts, and of Lemma~\ref{lem:tailcounts}]
The machine keeps \(\T^D(p) = E(p)\,S(p)\), and \(S(p)\) is nonempty for nonempty \(p\), so the last token of \(S(p)\)
is that of \(\T^D(p)\). The pair reached on a nonempty \(y\) is thus marked exactly when \(\T^D(y)\) and \(\T^D(ay)\)
end in different tokens, that is when \(\rho(a, y) = |y|\), and by the identity of the last tokens of
Section~\ref{sec:product} the radius under \(a\) is the supremum of \(|y|\) over the marked \(y\). Two runs at one
suffix stay together, so a pair on the diagonal reaches no mark. If a cycle reaches a mark, every \(u v^k z\) ends at a
marked pair, and its ripple is its length, which grows without bound. Otherwise the pairs that reach a mark form a
finite acyclic graph. If no mark is reachable, the radius is zero, attained by the empty input. If a mark is reachable,
a longest path in that graph from the start pair to a marked pair spells an input \(y\) with \(\rho(a, y) = |y|\) and no
longer marked input. Its pairs after the start are distinct, none on the diagonal, and none holds the empty suffix,
which only the start pair does. Exchanging the two runs maps the product to itself and keeps the marks, so a pair
followed later on the path by its exchange would close a cycle through pairs that reach a mark, the segment between them
followed by the same segment with the runs exchanged, each of whose pairs reaches the mark; the starting pairs need not
be symmetric for this. The pairs after the start are thus distinct sets of two nonempty suffixes, at most
\(\binom{N - 1}{2}\) of them, one per symbol of \(y\). A retained suffix begins at a boundary of \(\T^D(p)\), so by
Lemma~\ref{lem:proj} it is the tokenization of its yield, and distinct suffixes have distinct yields. The yields of the
two suffixes of the pair reached on a prefix \(p\) of \(y\) are suffixes of \(p\) and of \(ap\), so the shorter is a
proper suffix of the longer, and sending each pair after the start to its two yields, the shorter first, is injective on
those sets. That gives \(C_{\mathrm{suf}}\), and a word \(u\) of \(\mathcal{Y}\) has at most \(|u| - 1\) nonempty proper
suffixes. For the quotient, give each suffix its last token as output and the empty suffix an output of its own. The
equivalence that identifies two suffixes with the same output after every continuation is respected by the transitions,
so the quotient's runs on \(y\) end at the classes of the machine's runs, and the pair of classes reached on a nonempty
\(y\) has different outputs exactly when the pair of suffixes is marked. The two products thus have the same marked
inputs, so the same verdict, value and attaining inputs, and a lasso of the quotient product is a certificate as above.
A pair of equal classes reaches no mark, the empty class is reached only at the start, and exchanging the runs maps the
quotient product to itself, so the count above, made on the classes, gives \(\binom{K - 1}{2}\).

The count gives more. The two suffixes reached on a prefix \(p\) of \(y\) have yields one a proper suffix of the other,
so the pairs of classes on the path after the start are distinct pairs that \(C^{Q}_{\mathrm{suf}}\) counts. The two
yields of a pair that \(C^{Q}_{\mathrm{suf}}\) counts end in one symbol \(c\), and so do the last tokens, the outputs of
its two classes, so the pair is one of the \(\binom{K_c}{2}\) pairs of classes whose outputs end in \(c\). Choosing for
each pair that \(C^{Q}_{\mathrm{suf}}\) counts one pair of suffixes in its two classes with comparable yields is
injective into the pairs that \(C_{\mathrm{suf}}\) counts, and the \(K_c\) sum to \(K - 1\), which gives the chain.
Every pair off the diagonal reached after the start is counted there by the same argument, so the unordered product has
the start, one merged diagonal state and at most \(C^{Q}_{\mathrm{suf}}\) more. Its marked inputs are those of the
quotient product, since a mark does not depend on the order of the pair, and so are its verdict and value. Two distinct
classes differ in output after some continuation \(z\), possibly empty, so a pair of them reached on a nonempty input
reaches a marked pair along \(z\), and the pairs after the start that reach a mark are exactly those off the diagonal. A
cycle reaching a mark is then any reachable cycle off the diagonal. When there is none, the last pair of a longest path
from the start pair through pairs off the diagonal has only diagonal successors, which reach no mark, so it is marked
itself, or the path is empty and the radius zero.

The output of the machine after \(p\) is the last token of \(\T^D(p)\), with an output of its own after the empty input,
and every suffix is reached. Its quotient by equal outputs after every continuation is thus the minimal Moore machine of
that map, whose classes are those of the inputs \(p\) and \(p'\) such that \(\T^D(pz)\) and \(\T^D(p'z)\) end in the
same token for every \(z\), whatever \(B\). If \(\ripple(D)\) is finite, then by Lemma~\ref{lem:D} the origin ripple of
every prepended string is at most \(\ripple(D)\), so for every \(p\), every \(w\) of at least \(\ripple(D) + 1\) symbols
and every \(z\) the last position of \(wz\) keeps its span, and \(\T^D(pwz)\) and \(\T^D(wz)\) end in the same token, so
that \(w\) sends the class of every \(p\) to the class of \(w\). Conversely, if every word of length \(\ell\) sends all
classes to one, then for every symbol \(a\) and every \(y\) of at least \(\ell\) symbols the classes of \(ay\) and \(y\)
agree, so do their outputs, \(\rho(a, y) < |y|\), and the radius under \(a\) is below \(\ell\) by the identity of the
last tokens of Section~\ref{sec:product}. An input of \(\ripple(D)\) symbols attaining the radius under some \(a\), or
the empty word when the radius is zero, the empty class standing apart, does not send all classes to one.

For the pairs, every class other than the empty one holds the retained suffix \(S(p)\) of some nonempty input \(p\), the
class that \(u\) leads to from it is the class of \(S(pu)\), and \(u\) leads from the empty class to the class of
\(S(u)\). The yields of \(S(u)\) and \(S(pu)\) are suffixes of \(u\) and of \(pu\), so the shorter is a suffix of the
longer, a proper one when the classes differ, since distinct suffixes have distinct yields; so a pair \(\{q_0 u, qu\}\)
of distinct classes is one that \(C^{Q}_{\mathrm{suf}}\) counts. Conversely let two distinct classes hold \(S(p)\) and
\(S(p')\) with \(\yield(S(p))\) a proper suffix of \(\yield(S(p')) = w\,\yield(S(p))\), \(w\) nonempty. Every emitted
token is final (Lemma~\ref{lem:A}), so \(\T^D(pz)\) begins with \(E(p)\) for every \(z\), and Lemma~\ref{lem:proj} on
the interval after it gives \(\T^D(pz) = E(p)\,\T^D(\yield(S(p))\,z)\); the inputs \(p\) and \(\yield(S(p))\) thus have
the same last token after every continuation and lie in one class, the class of \(S(p)\). With \(u = \yield(S(p))\), the
class of \(S(p)\) is the one that \(u\) leads to from the empty class and the class of \(S(p')\) the one that \(wu\)
leads to, that is, the one that \(u\) leads to from the class \(q\) of \(w\), and the pair is \(\{q_0 u, qu\}\). Writing
\(u = cu'\), that pair is reached from \(\{q_0 c, qc\}\) by \(u'\) through transitions on common symbols, and a pair of
equal classes leads only to pairs of equal classes, so the pairs of distinct classes are the closure, which a search
from the at most \(|\Sigma| K\) starting pairs visits with \(|\Sigma|\) transitions per pair found.
\end{proof}

\begin{proof}[Proof of Corollary~\ref{cor:tail}, the coarsening]
A step of the recurrence for the relations sorts the \(K\) tuples of successor blocks stably, one symbol at a time, by
counting over the at most \(K\) block numbers, in time \(O(|\Sigma| K)\) with no hashing, and there are at most
\(K - 1\) steps. Merging instead keeps for each class the tuple of the blocks of its successors in a hash table, with
the list of the transitions into each class, joins two blocks when the tuples of two of their classes collide, relabels
the classes of the smaller block, rewrites the changed entry in the stored tuple of each class with a transition into a
relabeled class and rehashes it, records each join with its round, and makes the joins that the collisions of round
\(n\) reveal at round \(n + 1\). A class is relabeled at most \(\log_2 K\) times, since its block at least doubles each
time, and the transitions into all classes number \(|\Sigma| K\), so at most \(|\Sigma| K \log_2 K\) entries are
rewritten in all, each with one rehash of a tuple of \(|\Sigma|\) entries, and the time is \(O(|\Sigma|^2 K \log K)\) in
expectation over the hash table. Keeping each class's tuple of successor blocks as a balanced binary tree over the
\(|\Sigma|\) positions, of one fixed shape, whose nodes are interned by the identifiers of their children, makes equal
tuples equal roots. A rewritten entry copies one path, \(O(\log(|\Sigma| + 1))\) nodes, each interned by one lookup that
takes constant expected time, so the \(|\Sigma| K \log_2 K\) rewrites cost \(O(|\Sigma| K \log K \log(|\Sigma| + 1))\)
in expectation, after \(O(|\Sigma| K)\) to build the trees, and a collision is a repeated root. The executed merging
keeps the tuples in a hash table, in expected time \(O(|\Sigma|^2 K \log K)\). After the joins of round \(n\) the blocks
are those of \(\sim_n\), so the round at which two classes first share a block, the largest round on the path between
them in the forest of joins, is the least \(n\) at which they are equivalent. From the forest, the verdict and the
radius under a symbol \(a\) are one query of the round at which \(q_0\) and \(q_a\) first share a block, \(|\Sigma|\)
queries for all the symbols. An attaining input is read off the relations one symbol at a time, and at each of its at
most \(K - 2\) symbols up to \(|\Sigma|\) candidate symbols are tried, each with one such query, so all \(|\Sigma|\)
attaining inputs cost \(O(|\Sigma|^2 K)\) queries. A query is the largest round on the path between two classes in the
forest of joins, \(O(K)\) as the program reads it. Recorded instead as a binary forest, whose leaves are the classes and
whose internal nodes are the joins, each labeled by its round with the two blocks it joined as its subtrees, the round
at which two classes first share a block is the label of their lowest common ancestor, since the rounds do not decrease
toward a root, and the query takes \(O(1)\) after linear preprocessing.
\end{proof}

\begin{proof}[Proof of Theorem~\ref{thm:origin}]
By Lemma~\ref{lem:max} the ripple of an input is the maximum of the values its path reads, so \(\ripple(D)\) is the
supremum of that maximum over the paths from the start nodes. Lemma~\ref{lem:dead} lets the construction prune the nodes
whose two runs have become identical, from which no contribution is reachable. For part 2, suppose a node from which a
contribution is reachable lies on a cycle reachable from its start node. Take a stem to it, the loop, and a shortest
exit path to a node that carries a contribution. Suppose first that this contribution is the node's terminal one, of
offset \(e\). The input \(\mathrm{stem}\,\mathrm{loop}^k\,\mathrm{exit}\) reaches the node after
\(|\mathrm{stem}| + k|\mathrm{loop}| + |\mathrm{exit}|\) bytes and reads it there. A contribution is a property of the
node, so its value there is that count plus \(e\). Suppose instead that the node's only contribution is an event on an
outgoing edge. Then that edge is appended to the exit path, so that the pumped input traverses it and records the event.
The event's value where it is read is again the bytes read plus its offset. Either way Lemma~\ref{lem:max} makes that
value a lower bound on the ripple. The value grows by \(|\mathrm{loop}|\) at every repetition, so the radius is
infinite. Where the contribution read is a terminal one, the bound is an equality, whatever the sign of \(e\), by the
last claim of Lemma~\ref{lem:max}. For part 3, suppose no such node lies on a reachable cycle. Then the subgraph of
nodes from which a contribution is reachable is acyclic, and finite by Lemma~\ref{lem:finite}. Take the longest weighted
path from its start node. At a node it takes the best of three values, namely its own terminal offset, one plus an event
offset on an outgoing edge, and one plus the best of a successor in the subgraph. By Lemma~\ref{lem:max} this path gives
the supremum, clamped at zero. The supremum is attained, and the choices along the path spell an input, which the
construction replays against Definition~\ref{def:origin} before asserting the value. Part 1 is the two together, the
radius being infinite when such a node lies on a reachable cycle and finite when none does. The cycle test and the
longest path terminate on the finite product, and the product is built whenever the exploration completes.
\end{proof}

\section{Right lookahead and zero delay}\label{app:lookahead}

This appendix holds the two lemmas behind Theorem~\ref{thm:lookahead}, the proof of Theorem~\ref{thm:zerodelayauto} and
Corollary~\ref{cor:completion}.

\begin{lemma}[Episodes]\label{lem:episodes}
Run \(A_r\) on an \((r-1)\)-normal sequence. The run is a sequence of episodes. Each begins with a merge by rule \(r\),
its seed, at the leftmost occurrence of rule \(r\). Every later merge of the episode is by a rule before \(r\) and has
the token the previous merge created as an operand, the active token. Each seed lies strictly to the right of the
previous episode's final active token. Every merge produces a token born at \(r\), and within an episode no rule is used
twice.
\end{lemma}

\begin{proof}
The first merge is by rule \(r\), the input being \((r-1)\)-normal. After a merge, every adjacent pair that does not
contain the token it created was adjacent before it, so a rule before \(r\) that now applies does so at a pair
containing that token. By induction along the episode, every merge by a rule before \(r\) contains the active token, and
every other pair is one that stood in the \((r-1)\)-normal state before the seed. The active token is never an operand
of rule \(r\), since the seed is longer than either operand of rule \(r\) and the active token only grows. When no rule
before \(r\) applies, the next merge by rule \(r\) is then disjoint from the active token. It is not to its left, since
a pair to the left has stood unchanged since before the seed, which was the leftmost occurrence of rule \(r\) then. It
thus lies strictly to the right, and starts the next episode. A merged interval is one token of a run with the first
\(r\) rules, so by Lemma~\ref{lem:proj} its word is canonical for them. It spans two or more tokens of the
\((r-1)\)-normal input, whose boundaries the run with \(r - 1\) rules never crossed, so by the same lemma that run on
the interval's word alone ends in those tokens, and the word is not canonical for \(r - 1\) rules. It is thus born at
\(r\). After a rule \((u, v)\) is used, the active token is longer than \(u\) and than \(v\) and only grows, so the rule
is not used again in the episode.
\end{proof}

\begin{lemma}[One stage]\label{lem:onestage}
Let \(U\) and \(V\) be \((r-1)\)-normal sequences that share a complete token prefix of \(c\) symbols. Then \(A_r(U)\)
and \(A_r(V)\) share a complete token prefix of at least \(c - Q_r\) symbols.
\end{lemma}

\begin{proof}
Take \(c\) as large as it can be, and follow the two runs while they make the same merges inside the shared prefix. By
Lemma~\ref{lem:episodes} a merge by a rule before \(r\) happens only at the active token, and seeds are taken left to
right. The first merge that differs between the runs, or that one run makes and the other does not, then involves the
part at or after \(c\). An active token grows to the left only by a merge in which it is the right operand, and that
moves its start by the left operand's length, while a merge in which it is the left operand does not move its start.
There are two cases. In the first, a seed touches the part at or after \(c\). If it crosses \(c\), its left operand
\(u_r\) ends at \(c\) and its product starts at \(c - |u_r|\); otherwise it starts at or after \(c\). In the second, an
active token \(t\) common to both runs ends at \(c\), and a merge by a rule \((t, v)\) with the token after it is made
in one run and not, or not then, in the other. Charge \(|t|\) to that rule, since the active token starts at
\(c - |t|\). The rule cannot be charged again for a growth to the left. That would need the active token to be \(v\),
and \(v\) is a token of the \((r-1)\)-normal input, born before \(r\), while every active token is born at \(r\). In
both cases the later growths to the left in the episode use distinct useful rules born at \(r\), other than the one
charged, so the active token's start stays at or after \(c - Q_r\), and its end at or after \(c\). Every later seed
starts to the right of that token, hence at or after \(c\), and its episode moves left by at most \(Q_r\) as well. Hence
no merge that differs between the runs touches a symbol before \(c - Q_r\). The tokens of the two results that touch a
symbol before \(c - Q_r\) therefore coincide, ends included, and they form a complete prefix of at least \(c - Q_r\)
symbols.
\end{proof}

\begin{proof}[Proof of Theorem~\ref{thm:zerodelayauto}]
The automaton has no parallel edges, every state is reached from the start, and every state reaches a final state,
\((t, j)\) through the rest of \(t\). By Theorem~\ref{thm:bigram} and the determinism of Definition~\ref{def:sp} every
input has exactly one accepting run. Two distinct paths from the start on \(p\) ending at one state, followed by one
path from it to a final state with label \(y\), would be two accepting runs on \(py\). The paths on \(p\) are then at
most one per state, and the prefix of the accepting run on \(p\) is one of them. Keep them as a map from state to
pending output, with the emitted list \(P\), and the invariant that the completed output of each active path is \(P\)
followed by its pending list, \(P\) being \(\Stable(p)\).

Write \(\mathrm{out}(\pi)\) for the completed output of a path \(\pi\) and \(H\) for the longest common prefix of
\(\mathrm{out}(\pi)\) over the active paths. For any continuation \(q\), the accepting run on \(pq\) begins with an
active path \(\pi\), so \(\mathrm{out}(\pi)\), and with it \(H\), is a prefix of \(\T^D(pq)\), and \(H\) is a prefix of
\(\Stable(p)\). Conversely let \(\pi\) end at a state from which \(\theta\) leads to a final state with label \(y\).
Then \(\pi\theta\) is the accepting run on \(py\), and \(\T^D(py) = \mathrm{out}(\pi)\,\mathrm{out}(\theta)\), of which
\(\Stable(p)\) is a prefix. Each token of \(\mathrm{out}(\theta)\) is completed on a symbol of \(y\), so it ends beyond
\(|p|\), while \(\Stable(p)\), a prefix of \(\T^D(p)\), spells at most \(|p|\) symbols. It follows that \(\Stable(p)\)
is a prefix of \(\mathrm{out}(\pi)\) for every active \(\pi\), and that \(H = \Stable(p)\).

On a symbol \(a\), extend each active path along its edges reading \(a\), appending the edge's token, if any, to the
pending list; paths without such an edge die, and the accepting run's prefix survives. No two extended active paths end
at one state, by the first paragraph. Every new completed output still begins with \(\Stable(p)\), and \(\Stable(p)\) is
a prefix of \(\Stable(pa)\), since every \(\T^D(paq)\) is a \(\T^D(pq')\). The longest common prefix of the new pending
lists is thus \(\Stable(pa)\) less \(\Stable(p)\). Emit it and remove it from every pending list, and the invariant
persists. At the end of the input exactly one active state is final, since two would be two accepting runs, and its
pending list, which is \(\T^D(p)\) less \(\Stable(p)\), is flushed.

As in the proof of Theorem~\ref{thm:zerodelay}, \(\Stable(p)\) spells at least \(|p| - Q_{\mathrm{span}}(D)\) symbols,
and each completed output spells at most \(|p|\), so each pending list holds at most \(Q_{\mathrm{span}}(D)\) symbols
and tokens after emission and at most \(Q_{\mathrm{span}}(D) + 1\) before. The active path ending at a final state is
the accepting run on \(p\), whose pending list is \(\T^D(p)\) less \(\Stable(p)\), so the largest pending yield equals
the held suffix of Theorem~\ref{thm:zerodelay}, and \(Q_{\mathrm{span}}(D)\) is attained by it as well. Per symbol the
edges of the active states are scanned once, at most \(E_s\) looks, and at most \(n\) edges survive, one per target.
Copying their pending lists, the longest common prefix of at most \(n\) lists of at most \(Q_{\mathrm{span}}(D) + 1\)
identifiers and its removal cost \(O(n\,(Q_{\mathrm{span}}(D) + 1))\). The working space is the automaton and the \(n\)
pending lists. The preprocessing is \(V\), \(E\), the automaton and the identifiers.
\end{proof}


\begin{corollary}[Completion for every dictionary]\label{cor:completion}
For every dictionary the exploration of Section~\ref{sec:machine} completes at \(B = Q(D) + 1\), and \(B \le 2^{s}\)
with \(s\) the number of useful rules.
\end{corollary}

\begin{proof}
Follow the proof of Lemma~\ref{lem:B} with \(Q(D)\) in place of \(m\). At a truncation \(S\) is the last \(B\) tokens of
\(\T^D(p)\), so the \(Q(D)\) tokens after its first token \(f\) hold at least \(Q(D)\) symbols, and \(f\) ends at or
before \(|p| - Q(D)\). By Theorem~\ref{thm:lookahead}, \(\T^D(p)\) and \(\T^D(pz)\) share a complete token prefix of at
least \(|p| - Q(D)\) symbols for every \(z\), and the end of \(f\) is a boundary of \(\T^D(p)\) inside it, so \(Ef\) is
a prefix of \(\T^D(pz)\). That is the invariant of Lemma~\ref{lem:B}, which the ordering condition supplied there, and
the rest of that proof holds as written. A never-truncated suffix flagged at \(B\) tokens has its first token pinned by
the same count. The bound on \(B\) is that of Theorem~\ref{thm:lookahead}.
\end{proof}

Theorem~\ref{thm:origin} therefore holds for every dictionary, at a tail that can be exponential in the number of rules,
and the overlap graph remains the construction to run. Executed, the exploration at \(B = Q(D) + 1\) completes on
\CompletionUnordered{} dictionaries that fail the ordering condition, with \(B\) up to five.

\section{Pinned bytes, cuts and right radii}\label{app:windows}

This appendix holds the propositions of Sections~\ref{sec:density} and~\ref{sec:circularity} that compare, bound and
sharpen the pinned bytes, the cuts and the right radii, and the longer proofs of those sections.


\begin{proposition}[The two sides]\label{prop:sides}
For every dictionary, (i) for every symbol \(a\) and string \(x\) the left start ripple of \(a\) on \(x\) equals the
origin ripple \(\rho(a, x)\), so the left start radius is \(\ripple(D)\), and (ii)
\(Q_{\mathrm{start}}(D) \le Q_{\mathrm{span}}(D) - 1\) when \(Q_{\mathrm{span}}(D) \ge 1\), and both are zero otherwise.
\end{proposition}

\begin{proof}
(i) A changed start is a changed span, so the left start ripple is at most \(\rho(a, x)\). Let \(e = \rho(a, x) \ge 1\),
so that position \(e - 1\) is the last whose span changes. If \(e < |x|\), position \(e\) keeps its span, and that span
begins at \(e\), since a span beginning before \(e\) would cover \(e - 1\) in both runs and \(e - 1\) would keep it. The
token covering \(e - 1\) then ends at \(e\) in both runs, and if \(e = |x|\) it ends at \(|x|\) in both. Its span is
\([s, e)\) in the first run and \([s', e)\) in the second, shifted, and the two differ, so \(s \ne s'\) and position
\(e - 1\) changes its start. (ii) Fix \(x\) and \(b\) with right span ripple \(r \ge 1\), and let \(\ell\) be the
smallest position of \(x\) whose span changes. The span of \(\ell\) in \(\T^D(x)\) is \([s, e)\) with \(s \le \ell\). If
\(s < \ell\), position \(s\) has that same span in \(\T^D(x)\), which is not a span of \(\T^D(xb)\), so \(s\) is a
smaller changed position. Hence \(s = \ell\), and \(\ell\) begins a token of \(\T^D(x)\). In \(\T^D(xb)\) position
\(\ell\) is covered by \([s', e')\) with \(s' \le \ell\). If \(s' < \ell\), position \(s'\) is covered in \(\T^D(x)\) by
a token ending at or before \(\ell\), a different span, so \(s'\) is a smaller changed position. Hence \(s' = \ell\).
The first changed position keeps its start, every position before it keeps its span, and so every changed start lies
beyond \(\ell\), which makes the right start ripple of \((x, b)\) at most \(r - 1\). Taking suprema gives (ii), and when
\(Q_{\mathrm{span}}(D) = 0\) no start changes either.
\end{proof}

Neither proof uses anything of byte-pair encoding. Theorem~\ref{thm:pinned} and Proposition~\ref{prop:sides} hold for
any deterministic segmentation of every string into nonempty spans with a finite origin radius and a finite right start
radius; what byte-pair encoding supplies is the finite right lookahead of Theorem~\ref{thm:lookahead} and the adjacency
theorem behind the converse below. The literal converse of the density bound, that a finite radius makes the certified
windows of some width dense, is false.


\begin{proposition}[The literal converse fails]\label{prop:converse}
The dictionary \((a, b), (a, ab)\) over \(\{a, b\}\) has origin radius \(2\), and no factor of \(a^n\) is certified at
any origin, for any \(n\). Hence the certified windows of no width are dense.
\end{proposition}

\begin{proof}
The canonical tokens are \(a\), \(b\), \(ab\) and \(aab\). Each of \(b\), \(ab\) and \(aab\) is followed by every token,
and \(a\) by \(a\) and \(aab\) only, by tokenizing the pairs. No rule applies to \(a^n\), so \(\T^D(a^n)\) is \(n\)
single tokens. In \(a^k b\) with \(k \ge 2\) the only applicable rule is \((a, b)\), at the last two symbols, after
which \((a, ab)\) fires once and nothing else applies, so \(\T^D(a^k b) = a^{k-2} \mid aab\). Take \(W = a^k\) with
\(k \ge 2\). At the occurrence of \(W\) in \(W\) itself the token covering its last byte is \(a\), beginning at offset
\(k - 1\); at the occurrence at \(0\) in \(Wb\) it is \(aab\), beginning at offset \(k - 2\). Two offsets, so no origin
certifies \(W\). For \(W = a\), the occurrence at \(0\) in \(a\) is covered by a token beginning at offset \(0\), and
the occurrence at \(1\) in \(aab\) by one beginning at offset \(-1\). The strings \(a^n\) are arbitrarily long and
contain no certified window.

For the radius, prepending \(b\) changes no span, since \(b\) is followed by every token, so \(b \mid \T^D(y)\) is
canonical and by Theorem~\ref{thm:bigram} it is \(\T^D(by)\). Prepend \(a\) to \(x = a^k w\) with \(w\) empty or
beginning with \(b\). If \(w\) is empty, \(\T^D(ax)\) is single tokens and no span changes. Otherwise \(w = bz\), and
since \(ab\) and \(aab\) are followed by every token, \(\T^D(x) = a^{k-2} \mid aab \mid \T^D(z)\) and
\(\T^D(ax) = a^{k-1} \mid aab \mid \T^D(z)\) for \(k \ge 2\), where the token \(aab\) covers the same positions of \(x\)
in both runs and no span changes; \(\T^D(x) = ab \mid \T^D(z)\) against \(\T^D(ax) = aab \mid \T^D(z)\) for \(k = 1\),
where positions \(0\) and \(1\) change and the ripple is \(2\); and \(\T^D(x) = b \mid \T^D(z)\) against
\(ab \mid \T^D(z)\) for \(k = 0\), ripple \(1\). The radius is thus \(2\), attained on \(x = ab\).
\end{proof}

Two rules refute it. The dictionary \((a, b), (ab, c), (ab, abc)\) over three symbols does the same with radius
\DensityBRadius{}, its avoiding strings \((ab)^n\) tokenized into two-symbol tokens, and the chain decision refutes
every factor of \((ab)^{12}\) at every origin.

\begin{proof}[Proof of Proposition~\ref{prop:radiibounds}]
(i) For \(L \ge 2\) take a longest canonical token \(t\) and prepend \(t[0]\) to \(t[1{:}]\). Then \(\T^D(t) = t\),
whose shifted span begins at \(-1\), while every span of \(\T^D(t[1{:}])\) begins at \(0\) or later, so all \(L - 1\)
positions change. (ii) Every canonical token is a symbol or a product, so \(n \le n_0\). Fix \(a\) and a path of
Theorem~\ref{thm:overlap} attaining the radius \(R\) under \(a\), made of \(d(q)\) edges inside \(C\) from a source to
\(q\), then the marked edge, \(R\) edges in all. The \(R\) pairs after the source have no component at the start state,
which no transition enters, and the two components of each were entered by the symbol both copies read. None lies on the
diagonal. A pair \((p, p)\) before the marked edge reaches a live marked edge, whose two targets differ since an
annotation is fixed by its target, so two different accepting runs read one word from \(p\), and the run of the first
copy from the start state to \(p\) makes them two accepting runs on one input, against Theorem~\ref{thm:bigram}; the
last pair is the target of a marked edge. Nor are the two components at one offset \(j\) of two tokens. The two tokens
would then begin at one position of an accepting completion, a boundary of both runs, and by Lemma~\ref{lem:proj} both
runs would tokenize the rest of it alike, so the tokens would be equal and the pair diagonal. No two of the \(R\) pairs
have the same components. In the same order they would close a cycle in \(C\). In opposite orders, \((p, p')\) and later
\((p', p)\), exchanging the copies, which maps the overlap graph to itself and keeps the marks, closes a walk through
pairs that are live and reach the marked edge, as in the proof of Corollary~\ref{cor:tail}, so it lies in \(C\). Either
way the radius is infinite, so the \(R\) pairs are distinct sets of two states other than the start state, entered by
one symbol and at different offsets, which is the count \(B\). Each of them is live, so both copies read the rests of
their tokens next, and those rests, the unread suffixes, are one a prefix of the other, which is the count \(B'\). The
consumed prefixes, the parts of their tokens the two copies have read, are suffixes of the input read so far, the second
copy's with the prepended symbol, so one is a suffix of the other, which is the count \(B''\). At most one of the \(R\)
pairs has both states ending their tokens. Such a pair puts the ends of both tokens at one position of an accepting
completion, a boundary of both runs, and by Lemma~\ref{lem:proj} both runs tokenize the rest alike, so every later pair
would be diagonal. No diagonal pair precedes the marked edge, so only the last pair, the marked edge's target, can be
one, and at least \(F - 1\) of the \(F\) such sets, whose unread suffixes are empty and which \(B'\) counts too, are not
among the \(R\), and at least \(F'' - 1\) of the \(F''\) that \(B''\) counts. The \(F - F''\) jointly final sets of
\(B'\) that \(B''\) leaves out are among those it drops, so \(B'' \le B' - (F - F'')\), which gives
\(B'' - \max(0, F'' - 1) \le B' - \max(0, F - 1)\), and \(B' \le B\) gives the next inequality. (iii) By
Proposition~\ref{prop:sides}(ii) and Lemma~\ref{lem:rightstart}(i),
\(Q_{\mathrm{start}}(D) \le Q_{\mathrm{span}}(D) - 1 \le Q(D) - 1 \le S - 1\) when \(Q_{\mathrm{span}}(D) \ge 1\), and
\(Q_{\mathrm{start}}(D) = 0\) otherwise. (iv) By Corollary~\ref{cor:chunks} an input of ripple \(r\) begins with the
right operands of its moving ranks, distinct rules each of which fires and so is useful, so \(\ripple(D)\) is at most
the sum of the useful right operands; and \(Q_{\mathrm{start}}(D) \le Q(D) - 1\) as in the proof of (iii), \(Q(D)\)
being the sum of the useful left operands, at least \(1\) when a rule is useful. With no useful rule every string is
single symbols and both radii are \(0\). (v) Every operand of \(K_m\) is a symbol and no product is an operand, so a run
merges disjoint pairs of symbols in priority order, and whether a pair of adjacent symbols of an input merges depends on
nothing but the input. Prepend \(c\) to \(x\). The pair \((c, x[0])\) is a rule only when \(c = a_{i-1}\) and
\(x[0] = a_i\), of lower priority than \((a_i, a_{i+1})\), the one rule that can take \(x[0]\) from inside \(x\), so the
pairs inside \(x\) merge as before and only the span of position \(0\) can change, so the ripple is at most \(1\),
attained by \(a_0\) on \(a_1\). The products have two symbols, so \(L = 2\). Append \(a_m\) to
\(x = a_0 a_1 \cdots a_{m-1}\) with \(m \ge 2\). In \(\T^D(x)\) the pairs
\((a_{m-2}, a_{m-1}), (a_{m-4}, a_{m-3}), \dots\) merge, from the right, and in \(\T^D(x a_m)\) the pairs
\((a_{m-1}, a_m), (a_{m-3}, a_{m-2}), \dots\), so every position of \(x\) changes its span and every position but \(0\)
changes its start. Hence \(Q_{\mathrm{span}}(K_m) \ge m\) and \(Q_{\mathrm{start}}(K_m) \ge m - 1\), while
\(Q_{\mathrm{span}}(K_m) \le Q(K_m) = m\) and (iii) gives the rest. For \(m = 1\), \(Q_{\mathrm{start}} = 0\) by (iii)
and \(Q_{\mathrm{span}} = 1\) on \(a_0\) with \(a_1\) appended.
\end{proof}


\begin{proposition}[Start from span]\label{prop:startspan}
For every dictionary, let \((x, b)\) have positive right span ripple. Then \(s\) is nonempty, \(t \ne t'\), and
\(\T^D(s)\) and \(\T^D(sb)\) share no boundary in \((0, |s|]\). The right span ripple of \((x, b)\) is \(|s|\), the
positions whose start changes are exactly those at or beyond \(c + \min(|t|, |t'|)\), and the right start ripple is
\(|s| - \min(|t|, |t'|)\). Consequently \(Q_{\mathrm{span}}(D)\) is the largest \(|s|\), and \(Q_{\mathrm{start}}(D)\)
the largest \(|s| - \min(|t|, |t'|)\), over the pairs \((s, b)\) of a string and a symbol whose tokenizations
\(\T^D(s)\) and \(\T^D(sb)\) differ in their first token.
\end{proposition}

\begin{proof}
Write \(\T^D(x) = t_1 \cdots t_k\) and \(\T^D(xb) = t'_1 \cdots t'_l\), and let \(j\) be the number of leading tokens
they share, so that \(c = |t_1 \cdots t_j|\). If \(j = k\), then \(\T^D(xb)\) is \(\T^D(x)\) followed by the tokens
covering \(b\) alone, the single token \(b\), and no span of \(x\) changes, against the hypothesis. Thus \(j < k\),
\(s\) is nonempty, and \(t = t_{j+1}\) and \(t' = t'_{j+1}\) differ. Since \(c\) is a boundary of both final
tokenizations, no merge crossed it in either run, and Lemma~\ref{lem:proj} on the interval from \(c\) gives
\(\T^D(s) = t_{j+1} \cdots t_k\) and \(\T^D(sb) = t'_{j+1} \cdots t'_l\). A boundary \(p\) of both with \(0 < p < |s|\)
would, by the same lemma, make both tokenizations begin with \(\T^D(s[{:}p])\) and so with the same token, against the
maximality of \(j\); a boundary of \(\T^D(sb)\) at \(|s|\) would make \(\T^D(sb) = \T^D(s) \mid b\), the same
contradiction. A span \([a, e)\) of \(\T^D(x)\) with \(a \ge c\) is a span of \(\T^D(xb)\) only if \(e - c\), which lies
in \((0, |s|]\), is a shared boundary, so every position at or beyond \(c\) changes its span, while the positions before
\(c\) lie in shared tokens, so the span ripple is \(|x| - c = |s|\). For the starts, a position in
\([c, c + \min(|t|, |t'|))\) has start \(c\) in both runs. Two different tokens beginning at \(c\) have different
lengths, so let \(p \ge c + \min(|t|, |t'|)\). In the run with the shorter first token the start of \(p\) is a boundary
of that run in \((c, p]\), and in the other run it is \(c\) or a boundary of that run in \((c, p]\). The two are not
equal, since \((c, p]\) holds no shared boundary. The changed starts are then exactly the positions at or beyond
\(c + \min(|t|, |t'|)\), and the start ripple is \(|s| - \min(|t|, |t'|)\). The last sentence follows by taking suprema,
since every \((x, b)\) of positive span ripple yields the pair \((s, b)\), with differing first tokens and the same
ripples, and every such pair is its own changed suffix.
\end{proof}

\begin{proof}[Proof of Lemma~\ref{lem:noabsorb}]
Write \(N_{r-1}(s) = P A\) and \(N_{r-1}(sb) = P B\), with \(P\) the common prefix of complete tokens ending at
\(f = f_{r-1}\). \(B\) is nonempty. When \(A\) is nonempty its first token \(a\) and the first token \(b'\) of \(B\)
differ, by maximality; when \(A\) is empty, \(f = |s|\) and \(B\) is the single token \(b\). By Lemma~\ref{lem:stab},
applied to the rules of rank at most \(r\), stage \(r\) is one pass of rule \(r\), and since a merge creates no new
occurrence of its own rule, the pass is the greedy left-to-right matching of the occurrences present when it starts. The
occurrences inside \(P\) are the same in both lists and are decided first, identically. An occurrence crossing \(f\) is
the last token of \(P\) with \(a\), or with \(b'\), present exactly when that token is \(u\) and the token after it is
\(v\). Since \(a\) and \(b'\) differ, at most one list has one, and whether the token \(u\) is still free when the pass
reaches it is decided inside \(P\), the same in both lists.

Suppose no crossing occurrence is merged. Then \(P\) is processed identically, \(A\) and \(B\) separately, and the
common prefix after the pass is the processed \(P\) extended by the leading tokens the processed \(A\) and \(B\) share,
which the claim below shows to be none when \(A\) is nonempty; when \(A\) is empty, \(B\) stays the single token \(b\).
Hence \(f_r = f\). Suppose instead that the crossing occurrence of one list, say the first, is merged. It merges \(u\)
with \(a = v\) into \(uv\), a token beginning at \(f - |u|\). In the second list \(u\) stays, since its right neighbor
is not \(v\) and can only grow, and an occurrence with \(u\) as right operand lies inside \(P\) and was decided
identically in both lists. The parts before \(u\) are processed identically, so the common prefix after the pass ends at
\(f - |u|\), where the tokens \(uv\) and \(u\) begin; rule \(r\) fired, so it is useful.

The claim is that if the first tokens of \(A\) and \(B\) differ before a pass that merges no crossing occurrence, then
the first tokens of the processed \(A\) and \(B\) differ after it. Two different tokens beginning at \(f\) have
different lengths. In the pass each of \(a\) and \(b'\) is merged at most once and only as a left operand, since a merge
with it as right operand would cross \(f\). The first tokens after the pass are therefore \(a\) or \(av\) and \(b'\) or
\(b'v\), and they are equal only if exactly one grew, say \(av = b'\). Then the word \(b'\) is produced in the first
list at stage \(r\) by rule \(r\) from the adjacent tokens \(a \mid v\), inside an interval whose endpoints no merge has
crossed, and by Lemma~\ref{lem:proj} the run on \(b'\) alone ends with that merge, by rule \(r\); but \(b'\) is a token
of \(N_{r-1}(sb)\), so by the same lemma the run on \(b'\) alone is completed by rules of rank below \(r\) and ends with
one of them. A run has one last merge, which is the contradiction.
\end{proof}

\begin{proof}[Proof of Theorem~\ref{thm:chainsearch}]
A state with one moving stage \(r\) has \(s = u_r\) and \(b = v_r\) by Theorem~\ref{thm:chainform}, with \(r\) useful.
Conversely let \(r\) be useful with \(v_r\) a symbol. Rule \(r\) fires at its pass on two adjacent tokens
\(u_r \mid v_r\) of an \((r-1)\)-normal sequence, so by Lemma~\ref{lem:proj} \(N_{r-1}(u_r) = u_r\), a single token,
which stays one under more rules, and \(N_{r-1}(u_r v_r) = u_r \mid v_r\), after which rule \(r\) merges the two. It
follows that \(\T^D(u_r) = u_r\) and \(\T^D(u_r v_r) = u_r v_r\), a single token, and that \((u_r, v_r)\) is a state
whose frontier is \(|u_r|\) through stage \(r - 1\) and \(0\) from stage \(r\), of rank \(r\).

Now let \((s, b)\) be a state of rank \(r_k\), \(r' > r_k\) useful, and suppose \(v_{r'}\) is one of \(t\) and \(t'\)
with \(N_{r'-1}(uw) = u \mid w\) for the other. Since rule \(r'\) is useful it fires at its pass on adjacent tokens
\(u \mid v_{r'}\) of an \((r'-1)\)-normal sequence, so under \(D_{<r'}\) the words \(u\) and \(v_{r'}\) are canonical
and \((u, v_{r'})\) is a canonical pair, by Lemma~\ref{lem:proj}. Theorem~\ref{thm:bigram} for \(D_{<r'}\) then makes
\(u\, N_{r'-1}(s)\) the tokenization of \(us\) under \(D_{<r'}\), the pair \((u, t)\) being canonical by the above when
\(t = v_{r'}\) and by hypothesis when \(t = w\), and likewise \(u\, N_{r'-1}(sb)\) that of \(usb\). The boundary \(|u|\)
stands in both runs through stage \(r' - 1\), and by Lemma~\ref{lem:proj} the runs on \(us\) and \(usb\) to the right of
it are the runs on \(s\) and \(sb\) through that stage. Hence the frontier of \((us, b)\) through stage \(r' - 1\) is
\(|u|\) more than that of \((s, b)\), which is \(0\) from stage \(r_k\) on. At stage \(r'\) the token ending at \(|u|\)
is \(u\) in both lists, the first token of each, and the token beginning there is \(v_{r'}\) in exactly one; no
occurrence precedes the crossing one, so the pass merges it, and by Lemma~\ref{lem:noabsorb} the frontier moves to
\(0\), where it stays. This makes \((us, b)\) a state of rank \(r'\) whose moving stages are those of \((s, b)\) and
\(r'\).

Conversely let \((s', b)\) be a state with moving stages \(r_1 < \cdots < r_k\), \(k \ge 2\), so that \(s' = u_{r_k} s\)
with \(s = u_{r_{k-1}} \cdots u_{r_1}\). The frontier of \((s', b)\) is at least \(|u_{r_k}|\) before stage \(r_k\) and
equal to it from stage \(r_{k-1}\), so the boundary \(|u_{r_k}|\) stands in both runs through stage \(r_k - 1\), and by
Lemma~\ref{lem:proj} the runs on \(s\) and \(sb\) are the runs on \(s'\) and \(s'b\) to the right of it through that
stage. The frontier of \((s, b)\) moves at \(r_1, \dots, r_{k-1}\) and is \(0\) from stage \(r_{k-1}\), where it stays
by Lemma~\ref{lem:noabsorb}, so \((s, b)\) is a state of rank \(r_{k-1}\). The move at \(r_k\) is one of the lemma, so
\(v_{r_k}\) is one of \(t\) and \(t'\), and \(N_{r_k - 1}(s') = u_{r_k} N_{r_k - 1}(s)\) is a tokenization under
\(D_{<r_k}\), so by Theorem~\ref{thm:bigram} the pair \((u_{r_k}, w)\) is canonical there for the other first token
\(w\), that is, \(N_{r_k - 1}(u_{r_k} w) = u_{r_k} \mid w\). The moving stages are determined by \((s', b)\), so the
sequence of extensions is unique, and induction on \(k\) reaches every state once. The maxima are those of
Proposition~\ref{prop:startspan}, every attaining pair being a state. For the cost, the first token of \(N_{r'-1}(s)\)
is the last entry of the first-token history of \(s\) with rank at most \(r' - 1\), since the merges come in
nondecreasing rank.
\end{proof}

A graph on the rules bounds the search and prunes it. Let \(G(D)\) have the useful rules as nodes, each weighted by
\(|u_r|\), the rules whose right operand is a symbol as sources, and an edge \(r \to r'\) whenever \(r < r'\), \(u_r\)
lies on the left spine of the derivation tree of \(v_{r'}\), and the rule above it on that spine, if any, has rank at
least \(r\). The left spine of a canonical word is the word, then the left spine of its producer's left operand, by
Lemma~\ref{lem:gtrees}.


\begin{proposition}[Rule graph]\label{prop:rulegraph}
Assume the ordering condition. The chain of every state is a path of \(G(D)\) from a source, so that
\(Q_{\mathrm{span}}(D) \le C(D) \le Q(D)\), with \(C(D)\) the heaviest path from a source, and
\(Q_{\mathrm{start}}(D) \le C_{\mathrm{start}}(D) \le \max(0, Q(D) - 1)\), with \(C_{\mathrm{start}}(D)\) the heaviest
path counted without its last node. Both are longest paths in a directed acyclic graph built from the rules and the
derivation trees alone, computed in time linear in its size. The chain search may drop a state \((s, b)\) with last rule
\(r\) whenever \(|s| - |u_r| + g(r)\) is at most the largest \(|s|\) found so far and \(|s| - |u_r| + f(r)\) at most the
largest start value found so far, \(g(r)\) and \(f(r)\) the heaviest paths from \(r\) with and without their last node;
its maxima are those of the full search.
\end{proposition}

\begin{proof}
Let \(r_i \to r_{i+1}\) be consecutive moving stages of a state, and consider the run in which the token beginning at
the frontier \(f_{r_i}\) just before stage \(r_{i+1}\) is \(v_{r_{i+1}}\). After stage \(r_i\) that token was \(w\), one
of \(u_{r_i}\) and \(u_{r_i} v_{r_i}\), by Theorem~\ref{thm:chainform}. Between the two stages the frontier stands, so
the token beginning at \(f_{r_i}\) is never a right operand, and it grows only as a left operand, keeping its start. By
Lemma~\ref{lem:proj} the merges inside the interval of \(v_{r_{i+1}}\) form the run on that word alone, whose last merge
is its producer and whose left operand was the token beginning at \(f_{r_i}\) just before. Inductively the tokens that
began at \(f_{r_i}\) during that run are the nodes of the left spine of \(v_{r_{i+1}}\), and \(w\) is one of them, with
\(u_{r_i}\) either \(w\) itself or the node below it, since \(u_{r_i} v_{r_i}\) is produced by rule \(r_i\). The rule
above \(w\) on the spine is the merge that first grew \(w\), which came after the pass of \(r_i\), since in that pass
\(u_{r_i} v_{r_i}\) cannot be a left operand, being longer than \(u_{r_i}\), and \(u_{r_i}\) in the other run is
followed by a token that is not \(v_{r_i}\). The rule above \(u_{r_i}\) on the spine, if any, is then \(r_i\) itself or
has rank above \(r_i\), and \(r_i \to r_{i+1}\) is an edge. The first rule of a chain has \(v_{r_1} = b\), a symbol, so
the path starts at a source. Its weight is \(|s|\) by Theorem~\ref{thm:chainform}, and the start value is at most
\(|s| - |u_{r_k}|\), the weight without the last node. Ranks increase along edges, so the graph is acyclic and a path
visits distinct useful rules, which gives \(C(D) \le Q(D)\) and \(C_{\mathrm{start}}(D) \le Q(D) - 1\) when a rule is
useful; with no useful rule all four values are \(0\). For the pruning, every extension of a state is an edge of
\(G(D)\) from its last rule \(r\), so a state \((s', b)\) reached from \((s, b)\) has \(|s'| \le |s| - |u_r| + g(r)\)
and start value at most \(|s'| - |u_{r_z}| \le |s| - |u_r| + f(r)\), \(r_z\) its last rule, and \((s, b)\) itself
satisfies both bounds, since \(g(r) \ge |u_r|\) and \(f(r) \ge 0\). A dropped state and its descendants raise neither
maximum.
\end{proof}


\begin{proposition}[Chains need the hypothesis]\label{prop:chainsfail}
Without the ordering condition the chain search's values need not be radii. On the unordered dictionary \((ac, a)\),
\((acaaca, a)\), \((caca, c)\), \((c, aca)\), \((acaaca, b)\), \((c, c)\), \((a, c)\), \((aca, aca)\) the search visits
\ChainNonstatePairs{} pairs, among them \((cacaaca, a)\), whose two tokenizations \(caca \mid aca\) and
\(caca \mid aca \mid a\) share their first token, and reports a start value of \ChainNonstateStart{} where the start
radius is \ChainNonstateProductStart{}. Without the hypothesis the search's maxima are lower bounds on the radii once
each is replayed by definition, and the exact values stay with the append product.
\end{proposition}

\begin{proof}
The dictionary fails the ordering condition, and every value stated is a finite computation that
\code{check\_chain\_families.py} carries out, namely the chain search on it and the ripples of each maximum by
Definition~\ref{def:rightripple}, which make the replayed maxima lower bounds. The extension test of
Theorem~\ref{thm:chainsearch} is proved sufficient under the ordering condition alone, and the eight-rule dictionary
shows it is not sufficient without it.
\end{proof}


\begin{proposition}[Exact width]\label{prop:exactwidth}
Let \(R = \ripple(D)\) be finite and \(q = Q_{\mathrm{start}}(D) \ge 1\). The following are equivalent: (a) some window
of width \(R + q\) fails to pin its byte \(R\), so that the width \(R + q + 1\) of Theorem~\ref{thm:pinned} is the least
at which every window pins its byte \(R\); (b) some \((x, b)\) attains \(q\) with its first changed position
\(c + \min(|t|, |t'|)\) at least \(R\); (c) some pair \((s, b)\) attaining \(q\) with differing first tokens \(t\) and
\(t'\) admits a string \(y\), possibly empty, with \(|y| + \min(|t|, |t'|) \ge R\) such that \(\T^D(y)\,\T^D(s)\) and
\(\T^D(y)\,\T^D(sb)\) are \(\T^D(ys)\) and \(\T^D(ysb)\).
\end{proposition}

\begin{proof}
(b) implies (a). Let \(p = c + \min(|t|, |t'|) \ge R\) be the first changed position. The window \(x[p - R{:}]\) has
width \(R + q\), its byte \(R\) is the position \(p\) of \(x\), and the contexts \((x[{:}p - R], \varepsilon)\) and
\((x[{:}p - R], b)\) cover that byte from two starts. (a) implies (b). Let \(W\) of width \(R + q\) fail at byte \(R\)
under two contexts. By Lemma~\ref{lem:rightstart}(iii) on each left context and the position \(R\) of \(Wv\), the start
of byte \(R\) under a context \((u, v)\) is its start in \(\T^D(Wv)\) shifted by \(|u|\), so the two contexts differ in
their right parts and one of them differs from the empty one, so some string \(v\) changes the start of byte \(R\) from
its value in \(\T^D(W)\). Take a shortest such \(v\), and write \(v = v'b\). If \(v'\) is nonempty it does not change
that start, and position \(R\) of \(Wv'\) has \(|Wv'| - R > q\), so by Lemma~\ref{lem:rightstart}(ii) appending \(b\)
leaves its start, which is then its start in \(\T^D(W)\), a contradiction. Hence a single symbol \(b\) changes the start
of byte \(R\) of \(W\); the start ripple of \((W, b)\) is at least \(|W| - R = q\), hence \(q\), and by
Proposition~\ref{prop:startspan} the changed starts are exactly the positions at or beyond \(c + \min(|t|, |t'|)\), so
that position is \(R\). (b) and (c) are the same statement. Given (b), write \(x = ys\) with \(s\) the changed suffix,
so that \((s, b)\) attains \(q\). Then \(c = |y|\) is a boundary of both runs, so Lemma~\ref{lem:proj} gives
\(\T^D(x) = \T^D(y)\,\T^D(s)\) and \(\T^D(xb) = \T^D(y)\,\T^D(sb)\), and \(|y| + \min(|t|, |t'|) \ge R\). Given (c), the
common prefix of complete tokens of \(\T^D(ys)\) and \(\T^D(ysb)\) is \(\T^D(y)\), so \((ys, b)\) has changed suffix
\(s\), attains \(q\), and has first changed position \(|y| + \min(|t|, |t'|) \ge R\).
\end{proof}

By Theorem~\ref{thm:bigram} the lists in (c) are those tokenizations exactly when \(y\) is empty or the last token of
\(\T^D(y)\) precedes both \(t\) and \(t'\) in \(E\). Under the ordering condition the chain search decides (c) from its
attaining states, by a capped longest-path relaxation over the canonical pair graph for the strings \(y\); for every
dictionary the start version of the append product reports whether (b) holds, as the longest path before a marked edge
whose completions attain \(q\). The two agree on every ordered dictionary of the pool of Table~\ref{tab:exec-overlap}.
When some symbol of the alphabet occurs in no operand, (c) holds with \(y\) a power of that symbol, which is a token of
its own that may precede and follow every token, so the width is the least for every such dictionary, \DensityInert{} of
that pool's among them. When the condition fails, the least width is below the theorem's, and the next proposition says
what it is.


\begin{proposition}[Least width]\label{prop:leastwidth}
Let \(R = \ripple(D)\) be finite and \(q = Q_{\mathrm{start}}(D)\). Let \(M\) be the largest \(|x|\) over the inputs
\((x, b)\) of positive right start ripple whose first changed position lies at or before \(R\), and \(0\) when there is
none. For \(h > R\), every window of width \(h\) pins its byte \(R\) exactly when \(h > M\), so the least width \(h^*\)
at which every window pins its byte \(R\) is \(\max(R + 1, M + 1)\), and
\(\max(R + 1, q + m + 1) \le h^* \le R + q + 1\) for every pair \((s, b)\) attaining \(q\) with differing first tokens
whose shorter first token has \(m \le R\) symbols. For \(q \ge 1\) the upper end is attained exactly under
Proposition~\ref{prop:exactwidth}, and not in general, since the ordered dictionary \((b, a)\), \((a, b)\), \((ba, a)\),
\((ba, ab)\), \((baa, ab)\), \((a, ab)\), \((baa, aab)\), \((ab, baab)\) over \(\{a, b\}\) has \(R = \LeastWidthAR\) and
\(q = \LeastWidthAQstart\), and every window of width \(\LeastWidthALeast = R + q\) pins its byte \(R\).
\end{proposition}

\begin{proof}
Let \(W\) have width \(h > R\). By Lemma~\ref{lem:rightstart}(iii) the start of byte \(R\) of \(W\) under a context
\((u, v)\) is its start in \(\T^D(Wv)\) shifted by \(|u|\), so \(W\) fails to pin byte \(R\) exactly when some string
\(v\) changes that start from its value in \(\T^D(W)\). Suppose \(h > M\) and some \(v\) does, and take a shortest one,
\(v = v'b\). Then \(v'\) leaves the start and \(b\) changes it, so the input \(x = Wv'\) with \(b\) appended has a
changed start at position \(R\), hence a first changed position at most \(R\), and \(|x| = h + |v'| > M\), against the
choice of \(M\). Conversely let \(R < h \le M\), and let \((x, b)\) attain \(M\) with first changed position
\(p \le R\); by Proposition~\ref{prop:startspan} every position at or beyond \(p\) changes its start, position \(R < M\)
among them, so the window \(x\) itself, of width \(M\), fails at byte \(R\) under the contexts
\((\varepsilon, \varepsilon)\) and \((\varepsilon, b)\), and so does its prefix of width \(h\), since extending a window
keeps its pinned bytes. Widths at most \(R\) hold no byte \(R\), which gives \(h^* = \max(R + 1, M + 1)\). An input with
first changed position \(p \le R\) has \(|x| = p + \text{ripple} \le R + q\), so \(h^* \le R + q + 1\). A pair
\((s, b)\) as stated has first changed position \(m\) and \(|s| = m + q\); if \(m + q > R\) it counts toward \(M\), so
\(h^* \ge m + q + 1\), and otherwise \(h^* \ge R + 1 \ge m + q + 1\). For \(q \ge 1\) the upper end is attained exactly
when some window of width \(R + q\) fails, which is Proposition~\ref{prop:exactwidth}(a). The dictionary named is worked
below.
\end{proof}

Worked, the dictionary of the proposition has the canonical tokens \(a\), \(b\), \(ba\), \(ab\), \(baa\), \(baab\),
\(baaab\), \(aab\), \(baaaab\) and \(abbaab\), so \(L = \LeastWidthAL\), and \(Q(D) = \LeastWidthAQD\). Its origin
radius is \LeastWidthAR{}, attained by \(a\) prepended to \(abbaab\), whose single token becomes \(aab \mid baab\). Its
span radius is \LeastWidthAQspan{} and its start radius \LeastWidthAQstart{}, since appending \(a\) to \(abbaab\)
retokenizes the single token as \(ab \mid baa \mid ba\), so the starts change from position \LeastWidthAAttaining{} on,
the first changed position. No canonical token precedes \(ab\) or \(abbaab\) in \(E\), since every canonical token
followed by either merges across the join, so every input attaining the start radius has the empty prefix. The chain
search, which visits every state, finds \((abbaab, a)\) the only state attaining it, so (c) fails,
\LeastWidthAAttaining{} being below \(R\). The least width is \LeastWidthASlack{} below the theorem's, with
\(M = \LeastWidthAM\) attained by the eight inputs \(y\,baaaab\) with \(a\) appended, \(y\) any three symbols, where
\(baaaab\) is a token of \(\T^D(y\,baaaab)\) and becomes \(baa \mid a \mid a \mid ba\), so that the window's byte \(6\),
the fourth symbol of \(baaaab\), has its start moved from the window's position \(3\) to its position \(6\). Those eight
are exactly the windows of width \(9\) that fail, \LeastWidthAFailNine{} of \LeastWidthAWindowsNine{}, under the
contexts \((\varepsilon, \varepsilon)\) and \((\varepsilon, a)\), and at width \(10\) none of \LeastWidthAWindowsTen{}
fails. This gives \(h^* = R + q = \LeastWidthALeast\) here, between the proposition's bounds
\(\max(R + 1, q + m + 1) = 7\), with \(m = 2\), and \(R + q + 1 = 11\). Table~\ref{tab:exec-overlap} lists the
executions of Propositions~\ref{prop:leastwidth} and~\ref{prop:slack}.


\begin{proposition}[Unbounded slack]\label{prop:slack}
Let \(A\) be the dictionary of Proposition~\ref{prop:leastwidth}, \(X = abbaab\), \(Y = baaaab\), \(B = baa\) and
\(C_i = Y B^i\). For \(n \ge 1\) the dictionary \(A_n\), the rules of \(A\) followed by
\((C_0, B), (C_1, B), \dots, (C_{n-1}, B)\) and \((X, C_n)\), has \(n + 9\) rules over \(\{a, b\}\), satisfies the
ordering condition, and has \(\ripple(A_n) = 12 + 3n\), \(Q_{\mathrm{span}}(A_n) = 15 + 3n\),
\(Q_{\mathrm{start}}(A_n) = 9 + 3n\) and least width \(h^* = \ripple(A_n) + 7\). Thus \(h^*\) lies \(3n + 3\) below the
width \(\ripple(A_n) + Q_{\mathrm{start}}(A_n) + 1\) of Theorem~\ref{thm:pinned}.
\end{proposition}

\begin{proof}
Under \(A\) the string \(Y B^i\) tokenizes as \(Y \mid B \mid \cdots \mid B\) and \(X Y B^n\) as
\(X \mid Y \mid B \mid \cdots \mid B\), so the counter rules build \(C_1, \dots, C_n\) from the left and the last rule
joins \(X\) and \(C_n\) into \(F = X C_n\). The canonical tokens are the ten of \(A\), the \(C_i\) and \(F\), every
operand is a symbol or an earlier product, and no rule has equal operands, so the ordering condition holds and the
radius is finite by Theorem~\ref{thm:selfpair}.

Prepending \(a\) to \(F\) gives \(aab \mid baab \mid C_n\) against the single token \(F\), since \(a X\) tokenizes as
\(aab \mid baab\) under \(A\) and \(baab \mid Y\), \(Y \mid B\) and \(B \mid B\) are canonical pairs there, so every
position of \(F\) changes and the ripple is \(|F| = 12 + 3n\). For the upper bound follow the agreement point of
Corollary~\ref{cor:chunks}. After the passes of \(A\) it is at most \(\ripple(A) = 6\). A counter pass moves it by
\(|B| = 3\) when it moves it. The tokens ending at the new agreement point are \(C_{i+1}\) and \(B\), and later counter
passes can replace them only by counter products, never by \(X\), so once a counter pass has moved the agreement point
the last rule cannot. The ripple is thus at most \(6 + 3n\), or \(6 + |C_n| = 12 + 3n\) when the last pass moves the
agreement point and no counter pass did.

For the right radii run the chain search of Theorem~\ref{thm:chainsearch}. Its fourteen states over \(A\) are fixed, and
seven of them, the set \(Z\) of the states \((z, b')\) whose two first tokens include \(B\), are the ones a counter rule
extends, to \((C_i z, b')\) with first tokens \(C_i\) and \(C_{i+1}\) and start value \(|z| \le 6\). No counter state
extends by another counter rule, whose right operand is \(B\), and the last rule extends exactly the seven states made
by \((C_{n-1}, B)\), to \((X C_{n-1} z, b')\) with first tokens \(X\) and \(F\). The new right operands are longer than
one symbol, so there are no new sources, and the search has \(21 + 7n\) states, the longest \(15 + 3n\) symbols and the
largest start value \(|C_{n-1}| + 6 = 9 + 3n\).

For the least width write \(R = \ripple(A_n)\). No canonical token precedes \(X\). For the tokens of \(A\) this is the
computation of Proposition~\ref{prop:leastwidth}, and every \(C_i\) and \(F\) ends in \(B\) under \(A\), which merges
across a following \(X\). A state with first tokens \(X\) and \(F\) thus has the empty prefix only and length at most
\(R + 3\), while every other state has start value at most \(6\), so an input whose first changed position is at most
\(R\) has at most \(R + 6\) symbols, and \(M \le R + 6\). The state \((YY, a)\) has start value \(6\), and \(a\)
precedes \(a\), \(Y\) and \(C_1\) in \(E\), so \(a^{R - 6} YY\) with \(a\) appended has first changed position \(R\) and
length \(R + 6\). Hence \(M = R + 6\) and \(h^* = R + 7\) by Proposition~\ref{prop:leastwidth}.
\end{proof}


\begin{proposition}[Certificates are not of bounded size]\label{prop:certsize}
For \(n \ge 2\) let \(t_j = ab\,a^j\) and \(T = ab\,a^n b\), and let \(D\) be the rules \((a, b)\), \((b, a)\),
\((a, ba)\), \((t_1, a), \dots, (t_{n-2}, a)\), \((t_{n-1}, ab)\), \((ab, a)\) in this order, \(n + 3\) rules. It
satisfies clause (a), has no rule \((t, t)\) and has infinite radius, and each list of each of its circularity
certificates has at least \(n\) tokens.
\end{proposition}

\begin{proof}
The composite operands \(ba\), \(t_1 = a \cdot ba\), \(t_2, \dots, t_{n-1}\) and \(ab\) are products of earlier rules.
Every token is \(a\), \(b\), \(ba\), a \(t_j\) with \(0 \le j \le n - 1\), or \(T\). Take a word in which every \(b\)
follows an \(a\). The first pass merges every \(ab\) and leaves no token \(b\), so the second and third rules never
apply. The last rule then makes \(t_1\) at the leftmost \(ab\) that an \(a\) follows. The rules \((t_j, a)\) grow that
token while an \(a\) follows it, and \((t_{n-1}, ab)\) closes it into \(T\) when \(ab\) follows, all before the last
rule is used again. It follows that \(\T^D(t_j) = t_j\), \(\T^D(T) = T\), \(\T^D(Ta) = T \mid a\) and
\(\T^D(aT) = a \mid T\), and that the list of \(n - 1\) tokens \(a\) followed by \(T\) is circular. Its yield
\((a^n b)^2\) is its own rotation by \(n + 1\), and its starts \(0, \dots, n - 1\) are disjoint from
\(n + 1, \dots, 2n\). That is a certificate, and the radius is infinite by Theorem~\ref{thm:circular}.

Now take any certificate. Its lists, repeated, are two tilings of one periodic word by canonical tokens, adjacent tokens
forming canonical pairs, with no common boundary. No boundary lies between an \(a\) and a \(b\) that follows it, since
by Theorem~\ref{thm:bigram} the two tokens around it would be the tokenization of their word, whose first pass merges
that pair. A token \(b\) or \(ba\) then follows a \(b\). No token contains \(bb\), so the position between those two
\(b\) is a boundary of both tilings, which is excluded. Hence the word has no \(bb\), and each of its \(b\) is in each
tiling the second symbol of a \(t_j\) or of \(T\), or the last symbol of \(T\). It is not the same in both tilings,
since the two tokens would start, or end, at one place. Hence every \(b\) is the last symbol of a \(T\) in one tiling,
the \(b\) before it is \(n + 1\) positions back, and the word is \(a^n b\) repeated. Let a \(b\) be the second symbol of
a token \(t_j\). If \(j \le n - 2\), two or more \(a\) follow the token, so the next token is \(a\), but the word of
that pair is \(t_{j+1}\), one token. If \(j = n - 1\), one \(a\) follows before the next \(b\), so the next token starts
with \(ab\), but the run on that pair's word grows the first \(ab\) to \(t_{n-1}\) and merges it with the second \(ab\),
across the boundary. Neither pair is canonical, so every \(b\) that is second in its token is in a \(T\). Each tiling is
then the tokens \(T\), each from one \(b\) to the next, separated by the \(n - 1\) symbols \(a\) left between them, each
of which is a token, every other token having a \(b\). That sequence has least period \(n\) in tokens, so each list has
a multiple of \(n\) tokens.
\end{proof}

\section{The label measure}\label{app:labels}

This appendix holds the lemmas, the constructions and the longer proofs of Sections~\ref{sec:labels}
and~\ref{sec:labelgraph}.


\begin{lemma}[One pass moves one edit to at most three]\label{lem:E}
Let \(f\) replace, in one left-to-right pass, every adjacent pair \(u, v\) by \(uv\). If \(s'\) is obtained from \(s\)
by one token edit, then \(d(f(s), f(s')) \le 3\), and three is attained.
\end{lemma}

\begin{proof}
Write \(s = LxR\) and \(s' = LyR\) with \(x, y\) single tokens or empty, not both empty. If \(u \ne v\), occurrences of
the pair cannot overlap, since the shared token would be both \(v\) and \(u\), so the pass replaces every occurrence
wherever it lies. Inside \(L\) only the last token can be treated differently than in \(L\) alone, when it is \(u\) and
the next token is \(v\). Hence \(f(s)\) begins with \(f(L)\) less at most its last token and, by the mirror argument,
ends with \(f(R)\) less at most its first. Every occurrence inside \(L\) and \(R\) is merged in \(s\) too, so
\(|f(s)| \le |f(L)| + 1 + |f(R)|\), leaving a middle of at most three tokens. The same holds for \(s'\) with the same
outer parts. If \(u = v\), the pass replaces each maximal run of \(k\) tokens \(u\) by \(g(k)\), \(\lfloor k/2 \rfloor\)
tokens \(uu\) followed by one \(u\) when \(k\) is odd. Write \(L = L_0 u^p\) and \(R = u^q R_0\), with \(L_0\) not
ending and \(R_0\) not starting in \(u\), and \(m = p + q\). The runs inside \(L_0\) and \(R_0\) are maximal in both
sequences, so \(f\) factors through them and the distance is that between the images of \(u^p x u^q\) and \(u^p y u^q\).
If \(x, y\) are both tokens other than \(u\), one substitution. If one is \(u\) and the other empty, the images are
\(g(m+1)\) and \(g(m)\), one edit apart. If \(x\) is \(u\) or empty and \(y\) is a token \(w\) other than \(u\), the
images are \(g(m + \delta)\) with \(\delta \in \{0, 1\}\) and \(g(p)\,w\,g(q)\). When \(p\) is even, deleting \(w\)
gives \(g(m)\). When \(p\) is odd, replacing the lone \(u\) that ends \(g(p)\) by \(uu\) and deleting \(w\) gives
\(g(m+1)\). In either case \(g(m+\delta)\) is at most one further edit away. Three is attained by \([a, b]\) against
\([a, c, b]\) under \((a, b)\) and by \([a, a]\) against \([a, b, a]\) under \((a, a)\).
\end{proof}


\begin{lemma}[The script of one pass]\label{lem:script}
Let \(f\) be as in Lemma~\ref{lem:E}, and call a token edit a substitution or an indel, an insertion or a deletion. If
\(s'\) is obtained from \(s\) by one substitution, then \(f(s')\) is obtained from \(f(s)\) by at most two
substitutions, or by at most one substitution and one indel. If \(s'\) is obtained by one indel, then \(f(s')\) is
obtained by at most one substitution and two indels, or by at most two substitutions and one indel. If \(s'\) is \(s\)
with one token prepended, by at most one edit.
\end{lemma}

\begin{proof}
Let \(u \ne v\). Occurrences of the pair do not overlap, so the pass merges every occurrence, and an occurrence that
does not contain an edited token is merged in both sequences. Insert \(c\) between two tokens. If they are \(u\) and
\(v\), the images have \(uv\) there against \(u, c, v\) when \(c\) is neither, one substitution and two insertions, and
against \(u, uv\) when \(c = u\) and against \(uv, v\) when \(c = v\), one insertion. Otherwise \(c\) completes at most
one occurrence, since it is not both \(u\) and \(v\), and its partner was unmerged in \(s\). The image has \(uv\) in
place of the partner, one substitution, or it gains \(c\), one insertion. Substitute \(c'\) for \(c\). At most one
occurrence of \(s\) contains the position, and at most one of \(s'\). With neither, one substitution. With one, a merged
token stands against two tokens, one substitution and one indel. With both, the two lie on opposite sides of the
position, since on the same side \(c' = c\), and the images have \(uv, v\) against \(u, uv\) or the reverse, two
substitutions. A prepended token merges at most with the first token of \(s\), which is then \(v\) and began no
occurrence, at the cost of one substitution, and otherwise of one insertion.

Let \(u = v\), with \(g\) as in the proof of Lemma~\ref{lem:E} and \(T = uu\). Inserting or deleting a \(u\) changes the
image of its run from \(g(k)\) to \(g(k \pm 1)\), one indel or one substitution. Exchanging two tokens other than \(u\)
is one substitution. There remain a token \(w\) other than \(u\) inserted into a run or substituted for one of its
\(u\), and the inverses, which use the inverse scripts. Let \(p\) and \(q\) count the \(u\) of the run before and after
the position. Inserted, the images are \(g(p + q)\) and \(g(p)\,w\,g(q)\). For \(p\) even the first is \(g(p)\,g(q)\),
one insertion, and for \(p\) odd drop the common \(g(p - 1)\). With \(q\) even, \(T^{q/2}\,u\) stands against
\(u\,w\,T^{q/2}\), where the first \(T\) becomes \(u\), \(w\) is inserted and the last \(u\) becomes \(T\), two
substitutions and one insertion, and one insertion when \(q = 0\). With \(q\) odd, \(T^{(q+1)/2}\) stands against
\(u\,w\,T^{(q-1)/2}\,u\), one substitution and two insertions. Substituted, the images are \(g(p + 1 + q)\) and
\(g(p)\,w\,g(q)\). For \(p\) even drop \(g(p)\), leaving \(T^{q/2}\,u\) against \(w\,T^{q/2}\) when \(q\) is even, at
most two substitutions, and \(T^{(q+1)/2}\) against \(w\,T^{(q-1)/2}\,u\) when \(q\) is odd, one substitution and one
insertion. For \(p\) odd drop \(g(p - 1)\), leaving \(T\,g(q)\) against \(u\,w\,g(q)\), one substitution and one
insertion. A prepended \(u\) changes \(g(k)\) to \(g(k + 1)\), and any other prepended token is one insertion.
\end{proof}


\begin{proposition}[The bound is not uniform in the dictionary]\label{prop:doubling}
The doubling dictionary \((a,a)\), \((aa,aa)\), \(\dots\), \((a^{2^{k-1}}, a^{2^{k-1}})\) is ordered; prepending \(a\)
to \(a^{2^k - 1}\) gives label distance exactly \(k\), and inserting \(b\) after the first symbol of \(a^{2^k}\) gives
\(k + 2\).
\end{proposition}

\begin{proof}
After pass \(i\) the run \(a^N\) is \((a^{2^i})^{\lfloor N/2^i \rfloor}\) followed by the digits of \(N\) below \(2^i\),
one token \(a^{2^j}\) for each set bit \(j < i\) in decreasing order. This holds because pass \(i+1\) pairs the leading
blocks left to right and leaves one over exactly when bit \(i\) of \(N\) is set. In particular \(\T^D(a^{2^k - 1})\) is
the \(k\) tokens \(a^{2^{k-1}}, \dots, a\), and \(\T^D(a^{2^k})\) is the single token \(a^{2^k}\). No label is shared,
so the distance is the longer length. Inserting \(b\) after the first symbol of \(a^{2^k}\) gives \(a\), \(b\) and those
\(k\) tokens against the single token.
\end{proof}

\begin{proof}[Proof of Theorem~\ref{thm:amp}]
Write \(B = a^{2^m}\) and \(d_j = a^{2^{m-j}}\) for \(1 \le j \le m\). These \(m + 1\) tokens are pairwise distinct, and
none of them is \(b\). By the pass computation of Proposition~\ref{prop:doubling} a run \(a^N\) tokenizes as
\(\lfloor N/2^m \rfloor\) copies of \(B\) followed by one token for each set bit of \(N\) below \(2^m\), in decreasing
order. Since \((m+1)2^m - 1 = m2^m + (2^m - 1)\), \(\T^D(x) = B^m d_1 \cdots d_m\). Every operand of every rule is a
power of \(a\), so no merge crosses the inserted \(b\) and \(\T^D(x') = d_1 \cdots d_m\, b\, B^m\). Its left part is the
tokenization of \(a^{2^m - 1}\), and its right part that of \(a^{m2^m}\). Deleting \(d_1 \cdots d_m\) from the first
sequence and inserting \(d_1, \dots, d_m, b\) at its front gives the second, so the distance is at most \(2m + 1\).

For the lower bound, read an alignment as a strictly increasing partial injection \(\varphi\) from the \(2m\) positions
of the first sequence to the \(2m + 1\) of the second. Each pair in \(\varphi\) is a match or a substitution and each
position outside it a deletion or an insertion, so the cost is \(2m + (2m + 1) - |\varphi| - k\). Here \(k\) counts the
pairs whose two tokens are equal. Such a pair pairs two copies of \(B\) or two copies of one \(d_j\), since the tokens
are distinct otherwise. The copies of \(B\) precede the digits in the first sequence and follow them in the second. A
matched pair of blocks together with a matched pair of digits would invert the order, so \(k\) counts blocks only or
digits only. Suppose the \(k\) equal pairs are blocks. Their positions are at most \(m\) in the first sequence and at
least \(m + 2\) in the second. The largest of them sits at position at most \(m\) on the left and at least \(m + 1 + k\)
on the right. The pairs of \(\varphi\) up to and including it number at most \(m\). The pairs beyond it are sent to the
at most \(m - k\) positions of the second sequence beyond position \(m + 1 + k\), so they number at most \(m - k\). The
two counts give \(|\varphi| \le m + (m - k)\). Suppose instead they are digits. Their positions are at least \(m + 1\)
in the first sequence and at most \(m\) in the second. The smallest of them sits at position at least \(m + 1\) on the
left and at most \(m - k + 1\) on the right. The pairs before it are sent to the at most \(m - k\) positions of the
second sequence before that one, so they number at most \(m - k\). The pairs from it on sit at left positions \(m + 1\)
to \(2m\), so they number at most \(m\), and again \(|\varphi| \le (m - k) + m\). With \(k = 0\) the injection is
bounded by \(2m\), so \(|\varphi| + k \le 2m\) in every case and the cost is at least \(2m + 1\).
\end{proof}


\begin{proposition}[The linear bounds fail]\label{prop:counterfamily}
Write \(A = aa\), \(T = aaaa\), \(U = aab\), \(V = aabaa\) and \(P_j = T^j V\). For \(h \ge 1\) let \(D_h\) be the rules
\((a,a)\), \((A,b)\), \((U,A)\), \((A,A)\) followed by \((T, P_{j-1})\) for \(j = 1, \dots, h\), a dictionary of
\(m = h + 4\) rules that satisfies both clauses of the revised Definition~5. Then prepending \(a\) to \(P_h^2\) gives
label distance \(2m - 3\). Inserting \(b\) before the last symbol of the first \(P_h\) in \(P_h^3\) gives label distance
\(3m - 4\). Hence the prepend constant is at least \(2m - 3\) and \(C_m \ge 3m - 4\) for every \(m \ge 5\).
\end{proposition}

\begin{proof}
Every composite operand is the product of an earlier rule, and the products \(A\), \(U\), \(V\), \(T\),
\(P_1, \dots, P_h\) are distinct, so no rule produces an operand of an earlier one. Both clauses hold, and the run is
one pass per rule (Lemma~\ref{lem:stab}). On \(P_h^2 = a^{4h+2}\,b\,a^{4h+4}\,b\,a^2\) the first pass gives
\(A^{2h+1}\,b\,A^{2h+2}\,b\,A\), the second merges each \(A\) before a \(b\), the third each \(U\) with the \(A\) after
it, giving \((A^{2h}\,V)^2\), the fourth \((T^h\,V)^2\), and the remaining passes fold each block into \(P_h\), so the
tokenization is \(P_h \mid P_h\). The same passes give \(P_h \mid P_h \mid P_h\) on \(P_h^3\). With \(a\) prepended the
first run of \(a\) has odd length, so the first pass leaves \(A^{2h+1}\,a\,b\,A^{2h+2}\,b\,A\). The first \(b\) now
follows \(a\) and not \(A\), so the second pass merges only at the second \(b\), and the third pass gives
\(A^{2h+1}\,a\,b\,A^{2h+1}\,V\). The fourth gives \(T^h\,A\,a\,b\,T^h\,A\,V\). Now \(V\) follows \(A\) and not \(T\), so
no \(P_j\) is ever made and the remaining rules find nothing. That is \(2h + 5 = 2m - 3\) tokens, none of them \(P_h\),
against two tokens \(P_h\), so the distance is \(2m - 3\). For the insertion, the edited input is
\(a^{4h+2}\,b\,a\,b\,a^{4h+3}\,b\,a^{4h+4}\,b\,a^2\). The passes give in turn
\(A^{2h+1}\,b\,a\,b\,A^{2h+1}\,a\,b\,A^{2h+2}\,b\,A\), then \(A^{2h}\,U\,a\,b\,A^{2h+1}\,a\,b\,A^{2h+1}\,U\,A\), then
the same with \(V\) for the final \(U\,A\), then \(T^h\,U\,a\,b\,T^h\,A\,a\,b\,T^h\,A\,V\), where again \(V\) follows
\(A\) and nothing more happens. That is \(3h + 8 = 3m - 4\) tokens, none of them \(P_h\), against three tokens \(P_h\).
\end{proof}

At five rules the family gives a prepend distance of seven where \(m + 1\) is six, and at six rules a one-insertion
distance of fourteen where \(2m + 1\) is thirteen. The dictionaries have an effective self-pair, so their origin radius
is infinite and Corollary~\ref{cor:labelm}, which bounds the prepend distance of a finite ordered dictionary by
\(\max(1, |D|)\), is untouched. The tiers' values are correct for their universes. Section~\ref{sec:ampcomputed}
computes the constant exactly to five rules on two symbols, the fifth by a recorded campaign whose attaining pairs the
archived run of Appendix~\ref{app:exec} replays, and it is \(2m + 1\) there. The family equals \(2m + 1\) at five rules
and passes it at six, so no bound linear in \(m\) with these constants holds. The mechanism of the doubling family, one
token per set bit on each side of the edit, is not the extremal one. Here an edit desynchronizes a block of \(h\) tokens
\(T\) from the token that would have absorbed it, three times over.


\begin{theorem}[A staircase]\label{thm:staircase}
For \(1 \le J \le L - 1\) let \(U_i = a^{2^i}\) and let \(D(L, J)\) be the rules \((a, a)\); then for
\(i = 1, \dots, J\) the rules \((U_{i-1}, b)\), \((U_{i-1}b, U_i)\) and \((U_i, U_i)\); then \((U_i, U_i)\) for
\(i = J + 1, \dots, L - 1\). It has \(m = L + 2J\) rules and satisfies both clauses. Let
\(x = a^{2^L - 1}(b\,a^{2^L})^J\). Then \(\T^D(ax) = U_L\,(b\,U_L)^J\), and \(\T^D(x)\) has \(JL - J(J+1)/2 + L\)
tokens, none of them a token of \(\T^D(ax)\). The inputs \(ax\) and \(bx\) are one substitution apart,
\(\T^D(bx) = b \mid \T^D(x)\), and the two lists are at distance \(JL - J(J + 1)/2 + L + 1\), the longer length,
although they share the label \(b\). Hence for \(m \ge 5\) the prepend constant is at least
\(\lceil (m^2 + 5m)/10 \rceil\), and the substitution constant, and so \(C_m\), at least
\(\lceil (m^2 + 5m)/10 \rceil + 1\), on two symbols.
\end{theorem}

\begin{proof}
The products are distinct words, and no product is an operand of an earlier rule, so both clauses hold and the run is
one pass per rule (Lemma~\ref{lem:stab}). Call the maximal blocks of \(a\) runs. In \(ax\) every run has \(2^L\)
symbols. Each pass \((U_{i-1}, U_{i-1})\) halves every run with nothing left over, so no lone \(U_{i-1}\) stands before
a \(b\) and the two junction rules of level \(i\) find nothing. This gives \(\T^D(ax) = U_L\,(b\,U_L)^J\). In \(x\) the
first run has \(2^L - 1\) symbols. After stage \(i\), the passes of level \(i\), each run \(r \le i\) ends in a lone
token at each level from \(i\) down to \(r\), followed by \(K_r = U_{r-1}\,b\,U_r\), which has absorbed the \(b\) after
the run and the first \(U_r\) of the next run, and run \(i + 1\) has lost that \(U_i\), so it is one short of a power of
two. The induction is that one lone \(U_{i-1}\) stands before a \(b\), at the end of run \(i\), so \((U_{i-1}, b)\)
merges once, and the product meets the first \(U_i\) of run \(i + 1\). After stage \(J\) the remaining halvings leave
run \(r \le J\) as \(U_{L-1} \cdots U_r\,K_r\) and run \(J + 1\) as \(U_{L-1} \cdots U_J\). That is
\(\sum_{r=1}^{J}(L - r + 1) + L - J = JL - J(J+1)/2 + L\) tokens, all of them \(U_i\) with \(i < L\) or \(K_r\). The two
lists share no label, so the distance is the longer length. With \(m = 5J + \rho\), \(0 \le \rho < 5\), and
\(L = m - 2J\) the count is \((m^2 + 5m + \rho(5 - \rho))/10\), and since \(\rho(5 - \rho) \le 6\) that is
\(\lceil (m^2 + 5m)/10 \rceil\). No rule has the left operand \(b\), and \(b\) is a right operand only after a power of
\(a\), so the leading \(b\) of \(bx\) is merged by nothing, the boundary after it is a boundary of the result, and
Lemma~\ref{lem:proj} gives \(\T^D(bx) = b \mid \T^D(x)\). The tokens of \(\T^D(x)\) are none of \(\T^D(ax)\), so the two
lists share the label \(b\) alone, which the longer list holds once, at its front. An alignment with \(h\) unmatched
tokens of the shorter list and \(t\) matched pairs of equal labels costs the longer length plus \(h - t\), and here
\(t \le 1\), with \(t = 1\) matching the front \(b\) of the longer list to a \(b\) of the shorter, before which the
first token \(U_L\) of the shorter is left unmatched, so \(h \ge t\) and the distance is the longer length,
\(JL - J(J + 1)/2 + L + 1\) against \(2J + 1\), the first being larger as \(L \ge J + 1\). The inputs \(ax\) and \(bx\)
differ in their first symbol only, so the dictionary of \(m\) rules on two symbols, on inputs over the two, gives the
count above plus one under a substitution.
\end{proof}

The staircase passes \(3m - 4\) from \(m = 24\) on and the prepend family \(2m - 3\) from \(m = 13\) on, and at
\(m = 100\) the staircase gives \StaircaseHundred{}. Inserting one \(a\) into the first run gives the same two lists, so
the one-edit constant grows at least as fast.


\begin{proposition}[A staircase for one insertion or substitution]\label{prop:staircaseinsert}
On \(D(L, J)\), with \(K_r = U_{r-1}\,b\,U_r\) as there, let \(N = 2^L\), \(x = a^{2N}(b\,a^N)^J\), and let \(x'\)
insert one \(b\) after the first \(N - 1\) symbols of \(x\). Then \(\T^D(x) = U_L\,U_L\,(b\,U_L)^J\) and
\[ \T^D(x') = (U_{L-1} \cdots U_1\,K_1)^2 \Bigl(\prod_{r=2}^{J} U_{L-1} \cdots U_r\,K_r\Bigr)\, U_{L-1} \cdots U_J, \]
at distance \((J + 2)L - J(J + 1)/2\). Hence \(C_m \ge \lfloor (m^2 + 11m + 20)/10 \rfloor\) for \(m \ge 7\). For
\(J \ge 2\) let \(x''\) instead replace the symbol of \(x\) at position \(N - 2\) by \(b\). Then
\[ \T^D(x'') = U_{L-1} \cdots U_2\,K_2\; U_{L-1} \cdots U_2\,K_1 \Bigl(\prod_{r=2}^{J} U_{L-1} \cdots U_r\,K_r\Bigr)\,
  U_{L-1} \cdots U_J, \]
at distance \((J + 2)L - J(J + 1)/2 - 2\) from \(\T^D(x)\), the inputs over the two symbols. Hence the substitution
constant is at least \(\lfloor (m^2 + 11m)/10 \rfloor\) for \(m \ge 12\), two below the insertion's value, and
\(4m - 21\) for \(7 \le m \le 11\). These bounds, and the prepend's in Theorem~\ref{thm:staircase}, are the largest
counts of their kind over the \(L, J\) with \(L + 2J = m\).
\end{proposition}

\begin{proof}
Every run of \(x\) has a multiple of \(N\) symbols, so as in the proof of Theorem~\ref{thm:staircase} no junction rule
fires, and \(\T^D(x)\) is as stated. In \(x'\) the first two runs have \(N - 1\) and \(N + 1\) symbols. The pass of
\((a, a)\) leaves each of them ending in a lone \(a\) before a \(b\), so \((a, b)\) merges at both junctions, and
\((ab, U_1)\) absorbs the first \(U_1\) after each. The second run had one \(U_1\) to spare and the third is left one
\(U_1\) short. The remaining halvings make each of the first two runs \(U_{L-1} \cdots U_1\,K_1\). From level two on,
the induction of Theorem~\ref{thm:staircase} runs on the third run and those after it, and gives the rest of the list.
Its length is \(2L + \sum_{r=2}^{J}(L - r + 1) + L - J = (J + 2)L - J(J + 1)/2\), at least the \(2J + 2\) tokens of
\(\T^D(x)\), since \(L \ge J + 1\). No token of either list is a token of the other, so the distance is the longer
length.

In \(x''\) the first two runs have \(N - 2\) and \(N + 1\) symbols, so the first junction acts at level two and the
second at level one. The pass of \((a, a)\) leaves the second ending in a lone \(a\) before the second \(b\) and the
first with none, so \((a, b)\) merges at the second junction only, \((ab, U_1)\) absorbs the first \(U_1\) of the third
run, and the third run is left one \(U_1\) short, as in \(x'\). The first run holds \(2^{L-1} - 1\) tokens \(U_1\), an
odd number, so the pass of \((U_1, U_1)\) leaves it ending in a lone \(U_1\) before the first \(b\), while the second
run, its lone \(a\) gone, holds \(2^{L-1}\) of them and is halved with nothing left over. At level two, which exists
since \(J \ge 2\), \((U_1, b)\) merges that lone \(U_1\) with the first \(b\) and \((U_1 b, U_2)\) absorbs the first
\(U_2\) of the second run, giving \(K_2\), which holds a \(b\) and is no operand of a later rule. Each of the first two
runs now holds \(2^{L-2} - 1\) tokens \(U_2\), one short of a power of two, followed by \(K_2\) or \(K_1\), so its
remaining halvings leave \(U_{L-1} \cdots U_2\). From the third run on, the induction of Theorem~\ref{thm:staircase}
runs from level two as it does in \(x'\). The list has two tokens fewer than \(\T^D(x')\),
\((J + 2)L - J(J + 1)/2 - 2\), at least \(2J + 2\) since \(L \ge J + 1\) and \(J \ge 2\), and no token of it is \(b\) or
\(U_L\), so the distance is its length.

It remains to choose \(J\). Put \(L = m - 2J\), so that \(L \ge J + 1\) reads \(J \le (m - 1)/3\). The counts of the
insertion and of the substitution both change by \(m - 5J - 7\) from \(J\) to \(J + 1\), which falls as \(J\) grows.
With \(q = \lfloor (m - 2)/5 \rfloor\) and \(m = 5q + 2 + \theta\), \(0 \le \theta < 5\), the change is \(\theta \ge 0\)
from \(q - 1\) to \(q\) and \(\theta - 5 < 0\) from \(q\) to \(q + 1\), so \(q\), which is at most \((m - 1)/3\), gives
the largest count. There the substitution's count is \((m^2 + 11m - (\theta - 2)(\theta - 3))/10\), and since
\((\theta - 2)(\theta - 3) \le 6\) it is \(\lfloor (m^2 + 11m)/10 \rfloor\), and the insertion's, two more, is
\(\lfloor (m^2 + 11m + 20)/10 \rfloor\). This settles the insertion for \(m \ge 7\), where \(q \ge 1\), and the
substitution for \(m \ge 12\), where \(q \ge 2\). For \(7 \le m \le 11\) the substitution needs \(J \ge 2 > q\), every
change from \(J = 2\) on is negative, and \(J = 2\) gives \(4m - 21\). The count of Theorem~\ref{thm:staircase} changes
by \(m - 5J - 5\), and the same argument makes \(J = \lfloor m/5 \rfloor\) its best for \(m \ge 5\).
\end{proof}

The insertion's value passes \(3m - 4\) from \(m = 16\) on, and at \(m = 100\) the proposition gives
\StaircaseInsertHundred{} under the insertion and \StaircaseSubHundred{} under the substitution, against the
\StaircaseHundred{} of the prepend. The substitution passes the bound of Theorem~\ref{thm:staircase} from \(m = 9\) on,
with \StaircaseSubNine{}, \StaircaseSubTen{} and \StaircaseSubEleven{} at \(m = 9, 10, 11\).


\begin{lemma}[A pass with unequal operands]\label{lem:passunequal}
After the pass of \((u, v)\), \(u \ne v\), there is a decomposition with at most \(c\) clusters and weight at most
\(w + c + 1\).
\end{lemma}

\begin{proof}
Occurrences of the pair are disjoint and all merged. Mark the first token of a block when it is the right operand of an
occurrence straddling into the block, in \(X\) or in \(Y\), and its last token when it is the left operand of one
straddling out. The first block can lose only its last token and the last block only its first. An inner block, a power
of one token \(\tau\), loses its first token only if \(\tau = v\) and its last only if \(\tau = u\), so at most one
token of each block is marked, \(c + 1\) in all. Move the marked tokens into the adjacent clusters, on both sides. Each
raises the larger side of its cluster by one. Every occurrence now lies inside one part, the pass acts on each part
separately and on a block identically in \(X\) and \(Y\), and a pass never adds a token. An inner block stays a power of
one token, and if it becomes empty its two clusters join.
\end{proof}

\begin{lemma}[A self-pair pass]\label{lem:passself}
After the pass of \((t, t)\) there is a decomposition with at most \(c + 1\) clusters and weight at most \(w + c + 1\).
\end{lemma}

\begin{proof}
The pass regroups each maximal run of \(t\) from its left end into pairs \(T = tt\) and a lone \(t\) when the run is
odd. Give each new token to the part of both its constituents, a pair across a block and a cluster to the cluster, and a
pair of two block tokens across an empty cluster to that cluster. A cluster's own tokens do not grow in number, since
each token it owns from its own pairs consumes at least one of its tokens. An inner block \(\tau^l\) with \(\tau \ne t\)
is untouched. An inner block \(t^l\) lies in one run in each list, and its first \(t\) may pair backward and its last
forward, on each side independently. On each side it owns \(T^k\), \(T^{k-1}t\) or \(T^{k-1}\) when \(l = 2k\), and
\(T^k t\) or \(T^k\) when \(l = 2k + 1\). Keep the longest power of \(T\) common to both sides and move the rest into
the following cluster. That rest is at most one token per side, a \(T\) or a \(t\). When the following cluster is empty
and owns a pair across it, the block has given its last \(t\) to that pair, so its rest on that side is empty. In the
first block the pairs agree except that its last \(t\) may pair forward on one side only, again a rest of at most one
token per side. The larger side of each cluster grows by at most one. In the last block only the run at its start can be
paired at different parities. The common power of \(T\) there becomes a new inner block, followed by a new cluster of at
most one token per side, or that one token joins the last cluster when the common power is empty. Hence the weight grows
by at most \(c + 1\) and the clusters by at most one.
\end{proof}

\begin{proof}[Proof of Theorem~\ref{thm:quadratic}]
By Lemma~\ref{lem:stab} the runs are the passes. Let \(s_i\) count the rules \((t, t)\) among the first \(i\). By the
two lemmas there are at most \(1 + s_{i-1} \le i\) clusters before pass \(i\), and the weight grows by at most
\(2 + s_{i-1} \le i + 1\) at pass \(i\). After \(m\) passes it is then at most
\(1 + \sum_{i=1}^{m}(i + 1) = (m + 1)(m + 2)/2\), and at most \(1 + 2m + \sum_{i=1}^{m} s_{i-1}\), in which a rule
\((t, t)\) at rank \(j\) counts once for each of the \(m - j\) passes after it, which is \(2m + 1 + \sum_i (m - j_i)\),
largest when the ranks are \(1, \dots, r\).

A substitution gains one less at the first pass. Before it the lists are \(B_0\,c\,B_1\) and \(B_0\,c'\,B_1\) with
symbols \(c \ne c'\). If the first rule is \((u, v)\) with \(u \ne v\), a token of \(B_0\) is marked only as a \(u\)
before a \(v\) in the cluster on some side, and it then merges with that \(v\), so that side of the cluster does not
grow, and likewise at \(B_1\); the two marks come from different sides, since the cluster's symbol on one side is not
both \(u\) and \(v\), so each side grows by at most one. If the first rule is \((t, t)\) and neither \(c\) nor \(c'\) is
\(t\), the pass leaves the cluster alone. Otherwise one of them is \(t\), say \(c\), and the maximal runs of \(t\)
through the edit read \(t^{p + q + 1}\) on one side and \(t^p\,c'\,t^q\) on the other, the rest alike. After the pass
both begin with \(T^{\lfloor p/2 \rfloor}\), \(T = tt\), and what follows is \(T^{\lfloor (e + q + 1)/2 \rfloor}\), with
a lone \(t\) when \(e + q + 1\) is odd, against \(t^e\,c'\,T^{\lfloor q/2 \rfloor}\), with a lone \(t\) when \(q\) is
odd, where \(e\) is the parity of \(p\). In each of the four cases a common power of \(T\) stands between at most two
clusters of weight at most two in all. The weight after the first pass is at most two, the later passes add at most
\(2 + s_{i - 1} \le i + 1\), and the total is at most \(2 + \sum_{i=2}^{m}(i + 1) = m(m + 3)/2\). By the ranks it is at
most \(2 + \sum_{i=2}^{m} (2 + s_{i-1}) = 2m + \sum_i (m - j_i)\), since \(s_1, \dots, s_{m-1}\) count a rule \((t, t)\)
at rank \(j\) once for each \(i\) from \(j\) to \(m - 1\).

For a prepended symbol the first pass leaves the two lists one substitution or one insertion apart. A pass of
\((u, v)\), \(u \ne v\), merges the prepended symbol with the first symbol of the input or leaves it alone, and that
first symbol is then \(v\), which cannot begin another occurrence of this rule. A pass of \((t, t)\) turns \(T^k\) into
\(T^k t\), or \(T^k t\) into \(T^{k+1}\), and leaves the rest alike. Hence one cluster of weight one remains, there are
at most \(1 + s_{i-1} - s_1 \le i - 1\) clusters before pass \(i\) for \(i \ge 2\), and the other \(m - 1\) passes add
at most \(2, 3, \dots, m\), in all \(m(m + 1)/2\), and by the ranks at most
\(1 + 2(m - 1) + \sum_{j_i \ge 2} (m - j_i)\). The lower bound is Theorem~\ref{thm:staircase}.
\end{proof}

\begin{proof}[Proof of Lemma~\ref{lem:cancel}]
For the upper bound, extend an optimal alignment of \(u\) and \(v\) by matching the two copies of \(c\). For the lower
bound, take an optimal alignment of \(cu\) and \(cv\), a strictly increasing partial injection between positions whose
pairs are matches or substitutions, its cost the unpaired positions plus the substitutions. If the two copies of \(c\)
are paired, the rest aligns \(u\) with \(v\). If neither is paired, pairing them lowers the cost by two. If the left
\(c\) is paired with a later position \(j\) of \(cv\), the first position of \(cv\) is unpaired by monotonicity. Pair
the two copies of \(c\) and leave \(j\) unpaired, at no greater cost. The right \(c\) is symmetric. The suffix case is
the mirror image.
\end{proof}

\begin{proof}[Proof of Lemma~\ref{lem:conjugacy}]
If \(u = pq\) and \(v = qp\), then \(d(u^k, v^k) \le 2\min(|p|, |q|) \le |u|\), by deleting the leading \(p\) of \(u^k\)
and appending it, or by inserting \(q\) at the front and deleting the trailing \(q\). If \(|u| \ne |v|\), then
\(d(u^k, v^k) \ge k\,\bigl||u| - |v|\bigr|\). Let \(|u| = |v| = \ell\), suppose \(u\) and \(v\) are not conjugate, and
fix an optimal alignment of cost \(K\). Its matched pairs, those with equal labels, fall into maximal \emph{stretches}
of pairs each continuing the last on both sides. Between two consecutive stretches some position is unpaired or
substituted, and it lies between them on both sides, so different gaps charge different units of cost and there are at
most \(K + 1\) stretches. A stretch of \(\ell\) pairs from \((i, j)\) would make the rotation of \(u\) by
\(i \bmod \ell\) equal the rotation of \(v\) by \(j \bmod \ell\), so each stretch has at most \(\ell - 1\) pairs. At
most \(K\) left positions are unmatched, so \(k\ell - K \le (K + 1)(\ell - 1)\), that is \(k\ell \le (K + 1)\ell - 1\),
and \(k \le K\).
\end{proof}

\begin{proof}[Proof of Lemma~\ref{lem:truncate}]
By Lemma~\ref{lem:reset}, \(r\) is the first common boundary and the covering spans agree from \(r\) on, so the tokens
after \(r\) coincide, and they are \(\beta\). The boundary \(r\) is never crossed in either run, so Lemma~\ref{lem:proj}
gives \(\alpha\) and \(\alpha'\) as stated, and Lemma~\ref{lem:cancel} cancels \(\beta\). A common boundary at a
position \(p < r\) would be a common boundary before \(r\), so the accepting path of \(x[0{:}r]\) crosses none before
its last edge. A path of \(G_a\) with an exit is an accepting product path, and each copy has one accepting run per
word.
\end{proof}

\begin{proof}[Proof of Lemma~\ref{lem:rigid}]
No edge of \(G_a\) is a common-boundary edge, so every closed walk in \(G_a\) has disjoint completion sets, and both are
nonempty, since a closed walk of the spelling automaton passes a token end. The token words of a closed walk are the
factors of the cyclic word \(z\) it spells between consecutive cuts. They are thus conjugate exactly when some rotation
by \(\delta\) carries \(C_L\) onto \(C_R\) and fixes \(z\). With disjoint completion sets
\(\delta \not\equiv 0 \pmod{|z|}\), so \(z\) is a proper power \(w^j\), \(j \ge 2\), of its primitive root, and
\(\delta \in |w|\mathbb{Z}\). In particular a closed walk spelling a primitive word is not conjugate.

Suppose \(S\) is not phase-consistent. Fix \(\sigma \in S\) and a shortest closed walk \(c_1\) at \(\sigma\) inside
\(S\), spelling \(z_1\). If \(z_1\) is primitive, \(c_1\) is the walk sought. Otherwise \(z_1 = w^j\) with \(w\)
primitive and \(j \ge 2\). Assign phases from \(\sigma\) by breadth-first search inside \(S\), setting
\(\varphi(\sigma) = 0\) and, along an edge from \(\tau\) reading \(b\) into \(\tau'\), requiring
\(b = w[\varphi(\tau)]\) and setting, or checking, \(\varphi(\tau') = \varphi(\tau) + 1\). Suppose every closed walk at
\(\sigma\) inside \(S\) spelled a power of \(w\). Then for two walks \(p, p'\) from \(\sigma\) to \(\tau\) and a walk
\(q\) back, \(pq\) and \(p'q\) would spell powers of \(w\). That forces \(|p| \equiv |p'| \pmod{|w|}\) and both to spell
prefixes of \(w^\infty\), and the search would find no inconsistency. An inconsistency then yields a closed walk \(c_2\)
at \(\sigma\) spelling a word \(z_2\) that is not a power of \(w\). If \(z_2\) is primitive, \(c_2\) is not conjugate.
Otherwise \(z_2 = w'^{j'}\) with \(w'\) primitive, \(w' \ne w\) and \(j' \ge 2\), and \(c_1 c_2\) spells
\(w^j w'^{j'}\). Were \(c_1 c_2\) conjugate, this word would be a proper power. The theorem of Lyndon and
Sch\"utzenberger \citep{lyndon1962equation} says that \(x^a y^b = z^c\) with \(a, b, c \ge 2\) makes \(x\), \(y\) and
\(z\) powers of one word, so \(w\) and \(w'\) would be powers of a common word, hence equal. The walk \(c_1 c_2\) is
thus not conjugate.

Suppose \(S\) is phase-consistent, with word \(w\), and let \(w'\) be its rotation to the phase of \(\sigma \in S\).
Suppose \(\sigma\) had two distinct successors \(\tau_1 \ne \tau_2\) inside \(S\). Choose closed walks \(c_1\) through
\(\tau_1\) and \(c_2\) through \(\tau_2\) at \(\sigma\), of \(j_1\) and \(j_2\) blocks of \(w'\). Let \(s\) be a walk
from a source of \(a\) to \(\sigma\) and \(e\) an exit path from \(\sigma\). The walks \(s c_1 c_2 e\) and
\(s c_2 c_1 e\) are accepting product paths that spell the same word, since \(c_1 c_2\) and \(c_2 c_1\) both spell
\(w'^{j_1 + j_2}\). Each copy has exactly one accepting run on that word, by Theorem~\ref{thm:bigram} and the
determinism of Definition~\ref{def:sp}, so the two product paths coincide edge by edge, and their first edges after
\(s\) enter \(\tau_1\) and \(\tau_2\). Hence \(\tau_1 = \tau_2\). A strongly connected graph in which every state has
one successor is a simple cycle.
\end{proof}

\begin{proof}[Proof of Lemma~\ref{lem:profiles}]
A pair joins the front pending token of one side to the arriving token of the other. Two pairs made in this order
increase on both sides, since the later arriving token is later in its sequence, and the later front token was pending
behind, or pushed after the emptying of, the earlier one. Unmatched declarations and substitutions cost one each.
Conversely, take an optimal alignment of cost at most \(k\) and process the tokens in completion order, pushing every
token of the buffer's side, those whose partner has not yet arrived as pending and those the alignment leaves unmatched
behind them, and popping a front token as unmatched as soon as the alignment leaves it unmatched. Pending tokens of both
sides at once would give a crossing pair, so one side is pending at a time, and each token is resolved as the automaton
allows, at the alignment's cost and pairs. The run this defines keeps its cost within \(k\) on every prefix of \(\pi\),
since the cost only grows, and the search keeps it provided its buffer stays within the cap, which the rest of the proof
shows. Since \(\pi\) runs from a source to an exit, the two token words spell the two whole inputs. Let \(h\) be the
total length of the tokens up to and including two paired tokens on the second side less that on the first, where
\(\eta\) is one for a prepended symbol and at most one in absolute value for an edit. The steps between the completions
of the two paired tokens number \(|h|\) or \(|h - \eta|\), the copy that the edit leaves in place being one step behind
from the edit on. Pairs of equal labels contribute equally, so \(|h|\) is at most the total length of the unmatched
tokens and the length differences of the substituted pairs up to and including that pair, at most \(M\) for each unit of
cost spent there, and \(|\eta - h|\) is at most \(M\) for each unit spent after it. With \(i\) the units spent up to and
including the pair and \(j\) those after it, \(i + j \le k\), \(|h| \le iM\) and \(|h - \eta| \le jM\), and the two
differ by at most \(|\eta|\). For \(k = 2r + 1\) one of \(i\) and \(j\) is at most \(r\), so one of \(|h|\) and
\(|h - \eta|\) is at most \(rM\) and the other at most \(rM + |\eta|\). For \(k = 2r\) either \(i\) and \(j\) are both
at most \(r\), so that both gaps are at most \(rM\), or one of them is at most \(r - 1\), and then both gaps are at most
\((r - 1)M + |\eta| \le rM\), as \(|\eta| \le 1 \le M\). Both \(|h|\) and \(|h - \eta|\) are thus at most \(\Delta\),
which is at most \(\lfloor (kM + 1)/2 \rfloor\), the bound that adding the two inequalities gives, and below it when
\(k\) is odd and \(M \ge 3\). Once both completions of an edge are processed, the front of the buffer is pending, since
unmatched front tokens are popped at once, and its partner completes on a later edge. The tokens of the buffer are of
one side, so they complete on different edges, between the front's, at step \(s\), and the current one, which comes
before the step \(t\) of the front's partner. There are thus at most \(t - s \le \Delta\) of them. Between the two
completions of an edge, the second may be the front's partner, which allows one more.
\end{proof}

\begin{proof}[Proof of Corollary~\ref{cor:prependedit}, the computation of \(Q_\Gamma\)]
For the computation, a nonempty \(w\) has empty stable output exactly when some \(\T^D(wy)\) begins with another token
than \(\T^D(w)\). The token of \(\T^D(wy)\) covering the last symbol of \(w\) then ends inside \(y\), since a boundary
at \(|w|\) would give \(\T^D(wy) = \T^D(w)\,\T^D(y)\) by Lemma~\ref{lem:proj}, and by the same lemma \(y\) may be cut
after that token. Read from the right, \(\T^D(w)\) is an accepting run on \(w^{\mathrm{rev}}\) from the start of the
reversed spelling automaton, and the cut \(\T^D(wy)\) is a run on \(w^{\mathrm{rev}}\) from the state its last token
reaches inside \(y\), ending in a final state; conversely, by Theorem~\ref{thm:bigram} read from the right, every such
pair of runs ending in different tokens is a pair of reversed tokenizations of \(w\) and of some \(wy\) with different
first tokens. The words over \(\Gamma\) with empty stable output are thus the words whose reversals label the paths over
\(\Gamma\) from a pair \((\mathrm{start}, r)\), \(r\) any other state, to a pair of final states with different tokens,
together with the empty word, of weight zero. A cycle on such a path would give arbitrarily long such words, against
\(|w| \le \hat{q}\), so the part of the product that reaches such a pair is acyclic, and a longest-path computation
gives \(Q_\Gamma\) with an attaining word. When \(\Gamma\) is the whole alphabet the sources of the reversed overlap
graph of Definition~\ref{def:revoverlap} suffice, by Lemma~\ref{lem:delay} with weights, which reduces the held suffixes
to the symbols of a string \(x\) after the longest common prefix of complete tokens of \(\T^D(x)\) and \(\T^D(xb)\),
read reversed by an accepting path from a source of \(b\) up to its last marked edge.
\end{proof}


\begin{proposition}[One-symbol dictionaries]\label{prop:unary}
Let every operand of \(D\) be a power of one symbol \(a\), with or without the ordering condition, and let inputs also
carry symbols no rule consumes. Then the supremum of \(d(\T^D(x), \T^D(x'))\) over pairs one symbol edit apart is
finite, and it is computed, with an attaining pair, in polynomial time.
\end{proposition}

\begin{proof}
Write \(U(n) = \T^D(a^n)\). By Theorem~\ref{thm:bigram} each \(U(n)\) is the unique walk of canonical pairs on the
canonical tokens spelling \(a^n\). Two distinct cycles of that graph, or one through two tokens, would give two such
walks of one length, so there is exactly one cycle, a loop at a token \(T\). No edge enters \(T\) from another token
\(t\), since then \(U(|t| + k|T|) = t\,T^k\) for every \(k\), while \(U(j|T|) = T^j\), and the two inputs, one a prefix
of the other, would differ in their first token at every length, against Theorem~\ref{thm:lookahead}. It follows that
\(U(n) = T^A\tau\) with \(\tau\) a walk avoiding \(T\), of at most \(H\) symbols and \(s\) tokens, \(H\) and \(s\) the
longest such walk in symbols and in tokens. Write \(\ell = |T|\). The other symbols are tokens of their own and split an
input into runs of \(a\) tokenized independently, so by Lemma~\ref{lem:cancel} every edit comes to \(U(n)\) against
\(U(n + 1)\), or to \(U(p + q + \delta)\) against \(U(p)\,b\,U(q)\) with \(\delta \in \{0, 1\}\), or costs one. The
\(T\)-counts on the two sides differ by an integer of absolute value at most \((2H + 1)/\ell\), so by at most
\(J = \lfloor (2H + 1)/\ell \rfloor\), and aligning the powers of \(T\) bounds every distance by \(C = 2s + J + 1\). If
\(n > H\), then \(U(n)\) holds a \(T\) and so begins with it, \(U(n) = T\,U(n - \ell)\), and likewise for \(p > H\),
where \(U(p + q + \delta)\) begins with \(T\) too; removing that \(T\) from both sides keeps the distance, by
Lemma~\ref{lem:cancel}, so \(n, p \le H\) suffice. An alignment in which no token of the right power \(T^F\) of \(U(q)\)
is matched at no cost costs at least \(F + 1\), the one more for \(b\), which is not a power of \(a\). When \(F \ge C\)
an optimal alignment thus matches some \(T\) of that power to a \(T\) of the other side, which lies in its leading
power, since the tails avoid \(T\). If \(q > H + \ell C\), then \(F > C\). Deleting that column gives an alignment of
the pair at \(q - \ell\) of no greater cost, and the pair at \(q - \ell\), whose right power has \(F - 1 \ge C\) tokens,
has such a column to duplicate, so the two distances are equal. The supremum is then the largest distance over
\(n, p \le H\) and \(q \le H + \ell C\), each an actual edit. With \(L\) the longest canonical token these bounds are
\(O(|D|L)\), and the canonical graph, its longest walks, the runs and the edit-distance tables are of polynomial size.
\end{proof}

\begin{proof}[Proof of Theorem~\ref{thm:region}]
For token sequences \(\tau = PAS\) and \(\tau' = PA'S\) with a common prefix \(P\) and suffix \(S\), call \(A, A'\) the
regions. Let \(f\) be the pass of a rule \((u, v)\) with \(u \ne v\). Its occurrences are pairwise disjoint, a shared
token being both \(v\) and \(u\), and a merge creates none, so the pass merges exactly the occurrences present. An
occurrence inside \(P\) is one in \(PAS\) and conversely, and the last token of \(P\) belongs to an occurrence crossing
the end of \(P\) only as its \(u\), with the first token of \(A\), or of \(S\) when \(A\) is empty, as its \(v\). Then
\(f(PAS)\) begins with \(f(P)\) less at most its last token, which is the last token of \(P\), and by the mirror
argument ends with \(f(S)\) less at most its first token. Between them lie the last token of \(P\), the tokens of \(A\)
and the first token of \(S\), merged among themselves, at most \(|A| + 2\) tokens, and at most \(|A| + 1\) when \(P\) is
empty. Likewise with \(A'\). Now \(\T^D\) is the composition of the passes of the \(k\) rules that fire, by
Lemma~\ref{lem:stab} and Theorem~\ref{thm:selfpair}'s proof, the inactive self-pairs firing on nothing. Initially
\(x = PAS\) and \(x' = PA'S\) as symbol sequences with \(|A|, |A'| \le 1\), and each pass keeps the form.

Call the last token of \(P\) and the first token of \(S\) the ends, and let \(E\) be the ends that some side merges into
its region in the pass, at most two. After the pass the new common prefix is \(f(P)\) less its last token when that
token is in \(E\), and likewise the suffix, so the new region of a side is its old region together with the ends of
\(E\), less one token for each merge made among those tokens on that side. An end merges only across the region, the
last token of \(P\) as the \(u\) of an occurrence whose \(v\) is the first token of the region or of \(S\), so the ends
of \(E\) are counted once each and the rest of \(P\) and \(S\) is merged the same way on both sides. Three facts bound
the growth. A side whose region is empty absorbs both ends with one merge, the occurrence formed by the two ends, or
neither, and after absorbing them its new region is the one token \(uv\). A side whose region is one token \(t\) absorbs
at most one end, since absorbing the left end makes \(t\) the \(v\) of the rule and absorbing the right end makes it the
\(u\), and \(u \ne v\). A side whose region has two tokens or more absorbs each end with a merge of its own, the two
merges being disjoint occurrences, and after absorbing both its new region begins and ends with \(uv\), the merges
inside it leaving those two tokens alone. A region thus grows by at most two in a pass, and by two only when its own
side makes no merge and the other side absorbs both ends, whose new region then begins and ends with the same token
\(uv\), the one token \(uv\) when it was empty. Nothing changes between passes, so at the next pass of a rule that
fires, a rule \((u', v')\) with \(u' \ne v'\), that region absorbs both ends only if its first token \(uv\) is \(v'\)
and its last token \(uv\) is \(u'\), which cannot be, and the first side does not grow by two in two consecutive passes
of rules that fire. Each region has at most one token at the start, so after the passes of the \(k\) rules that fire a
region has at most \(1 + k + \lceil k/2 \rceil = 1 + \lceil 3k/2 \rceil\) tokens, having grown by two in at most
\(\lceil k/2 \rceil\) of the passes and by at most one in each of the others. Two sequences with a common prefix and
suffix are at distance at most the larger region, which gives the one-edit bound. From a substitution both regions have
one token, so neither side grows by two in the first pass of a rule that fires, the growths by two lie in the last
\(k - 1\) such passes, at most \(\lfloor k/2 \rfloor\) of them, and a region has at most \(1 + \lfloor 3k/2 \rfloor\)
tokens. From an insertion over two symbols the first pass of a rule that fires leaves the two lists one token insertion
or one token substitution apart. Its rule has two distinct single symbols \(u\) and \(v\) as operands, since no earlier
rule has produced a token, and it merges every occurrence of \(uv\) present, which are disjoint. The inserted symbol
\(w\) lies in at most one occurrence, as the \(u\) of one or the \(v\) of one. If the insertion breaks an occurrence of
the other input, that input ends \(u\) before the point of insertion and begins \(v\) after it, so its two symbols are
\(u\) and \(v\), \(w\) is one of them, and the pass turns \(uuv\) or \(uvv\) into \(u \mid uv\) or \(uv \mid v\) against
the one token \(uv\), every other occurrence being common to the two inputs. If it breaks none, every occurrence of the
other input is one of the edited input, and \(w\) either stays alone, one token inserted, or merges with its neighbor at
that point, which is alone in the other input, one token substituted. The regions of the lists after the pass, taken
afresh, have at most one token each, and the remaining \(k - 1\) such passes leave a region of at most
\(1 + \lceil 3(k - 1)/2 \rceil = \lfloor 3k/2 \rfloor\) tokens for \(k \ge 1\), with no such pass the lists are one
token apart, and a deletion is an insertion read from the other input. With the substitution bound this gives
\(1 + \lfloor 3k/2 \rfloor\) for any one edit over two symbols. For a prepend \(P\) is empty and stays empty. The first
pass of a rule that fires merges the prepended symbol with the first token of the input, which is its \(v\) and so
cannot begin another occurrence of this rule, or leaves it alone, so the regions have at most one token each after it.
Each later pass raises the larger region by at most one, so \(\lambda(a, x) \le \max(1, k)\).

The dictionaries attaining the prepend bound are given after the theorem. For the others, write tokenizations as lists.
With no rule, the empty input against \(c\) and \(a\) against \(b\) are at distance one. On \((a, b)\), \(ab\) against
\(acb\) gives \([ab]\) against \([a, c, b]\), at distance three, and \(ab\) against \(aa\) gives \([ab]\) against
\([a, a]\), at distance two. On \((a, b), (a, ab)\), \(aab\) against \(aacb\) gives \([aab]\) against \([a, a, c, b]\),
at distance four. On \((a, b), (b, ab), (ab, bab)\), \(abbababbab\) against \(abbacbabbab\) gives \([abbab, abbab]\)
against \([ab, b, a, c, bab, bab]\), at distance six, and on \((a, b), (b, ab)\), \(babbab\) against \(baabab\) gives
\([bab, bab]\) against \([b, a, ab, ab]\), at distance four. Each of these dictionaries is ordered with every rule
firing and no self-pair, which covers \(k \le 2\) for one edit and \(k \le 1\) for a substitution, and the last two
pairs seed an induction that adds two rules at a time. Suppose that \(D\), on the symbols \(a\) and \(b\), is ordered
with its \(k\) rules firing and none a self-pair, that \(\T^D(X) = [T, T]\) and \(\T^D(Y) = B\) with \(B\) of
\(n \ge 1\) tokens none of which is \(T\), where \(X\) and \(Y\) are one insertion apart or one substitution apart.
Suppose further that \(T\) is a longest token of \(D\), that \(Z\) is a token of \(D\) with \(|Z| < |T|\), and that
before the pass of each rule \((u, v)\) of \(D\) the run on none of the four strings \(T\), \(Z\), \(X\), \(Y\) alone
ends in the token \(u\). The seeds satisfy this with \(Z = bab\), \(T = abbab\) and with \(Z = ab\), \(T = bab\), the
four strings ending in \(b\) before \((a, b)\), in \(ab\) before \((b, ab)\) and in \(bab\) before \((ab, bab)\). By
induction over the passes, a concatenation of the four strings is tokenized string by string, since an occurrence
crossing a boundary between them would begin at the last token of one of them. Let \(X' = TZXZT\) and \(Y' = TZYZT\),
one edit apart of the same kind as \(X\) and \(Y\), and append the rules \((Z, T)\) and \((T, ZT)\). Both are ordered
after \(D\), their operands produced earlier and their products longer than every earlier token, neither is a self-pair,
and both fire, as the runs on \(ZT\) and \(TZT\) show. The run of \(D\) gives \([T, Z, T, T, Z, T]\) on \(X'\) and
\([T, Z] B [Z, T]\) on \(Y'\). The pass of \((Z, T)\) gives \([T, ZT, T, ZT]\) and \([T, Z] B [ZT]\), since the first
\(Z\) of \(Y'\) precedes a token of \(B\), which is not \(T\), and canceling the common \(T\) and \(ZT\) by
Lemma~\ref{lem:cancel} leaves \([ZT, T]\) against \([Z] B\), which share no label, \(ZT\) being longer than every token
of \(B\) and \(T\) neither \(Z\) nor a token of \(B\), so the distance is \(n + 1\), with \(k + 1\) rules. The pass of
\((T, ZT)\) then gives \([TZT, TZT]\) against \([T, Z] B [ZT]\) unchanged, since its \(T\) precedes \(Z\) and its \(ZT\)
follows a token of \(B\), so the distance is \(n + 3\), with \(k + 2\) rules, and the hypothesis holds again with
\(TZT\) for \(T\), \(ZT\) for \(Z\) and \(X'\), \(Y'\) for \(X\), \(Y\): the four new strings end in the old string
\(T\), so before every old pass they end as \(T\) does, before the pass of \((Z, T)\) all four end in \(T \ne Z\), and
before the pass of \((T, ZT)\) all four end in \(ZT \ne T\). From the seeds at \(k = 3\) and \(k = 2\), at distances six
and four, each step adds one and then two, which is how \(1 + \lceil 3k/2 \rceil\) grows from an odd \(k\) and
\(1 + \lfloor 3k/2 \rfloor\) from an even one, and the inputs stay over \(a\), \(b\) and the one \(c\) of the insertion
seed, or over \(a\) and \(b\) for the substitution.

For an insertion over two symbols, code the pair at \(k - 1\) rules above, whose one \(c\) stands immediately before a
\(b\) in \(acb\), \(aacb\), the seed \(abbacbabbab\) and every \(TZYZT\), replacing each \(a\) by \(b\), each \(b\) by
\(ab\) and the \(c\) by \(a\), in the inputs and in the operands of every rule, and putting the rule \((a, b)\) first.
The codes \(b\) and \(ab\) begin with different symbols, so the coding of words over the two old symbols is injective,
every nonempty code ends in \(b\), and the inserted \(a\) is followed by the \(a\) of a code \(ab\), so the pass of
\((a, b)\) merges exactly the pairs inside the codes \(ab\) and leaves each old symbol as its code and the \(c\) as the
token \(a\), which no code equals. The coded dictionary is ordered with every rule firing and no self-pair, as the old
one is, since the coding keeps products and operands distinct or equal as they were, the new rule produces the code
\(ab\), which only later rules consume, no coded product is \(a\) or \(ab\), and each coded rule fires on its coded
product, whose run after the first pass is the old run on the product relabeled. The old run on each input is relabeled
the same way pass by pass, the labels present being coded injectively, so the two coded inputs, over \(a\) and \(b\) and
one insertion of \(a\) apart, are at the old distance \(1 + \lceil 3(k - 1)/2 \rceil = \lfloor 3k/2 \rfloor\) under
\(k\) rules, for every \(k \ge 2\), and at \(k \le 1\) the empty input against \(a\) is at distance one. On
\((a, b), (b, ab)\) the coded pair is \(bab\) against \(baab\), tokenized \(bab\) and \(b \mid a \mid ab\), at distance
three.
\end{proof}

\section{Executions}\label{app:exec}

Each table lists the executions that check the results of one section, one sentence per row, beside the result and the
programs that run it. Every number is written from the programs' output by \code{paper/emit\_stats.py}, and the comments
of the source name the stages that produce it. The programs and their outputs are archived together as a standalone
record at \doi{10.5281/zenodo.23988871}.

{\small
\begin{longtable}{@{}>{\raggedright\arraybackslash}p{0.31\textwidth}>{\raggedright\arraybackslash}p{0.65\textwidth}@{}}
\caption{The executions of Section~\ref{sec:product}.}\label{tab:exec-product} \\
\toprule
result and program & execution \\
\midrule
\endfirsthead
\toprule
result and program & execution \\
\midrule
\endhead
\bottomrule
\endlastfoot
Lemma~\ref{lem:max} \newline{\footnotesize\code{path\_maximum.py}} & Executed, the maximum is the ripple on every
nonempty input to length eight of every product of the census, of the \HardCases{} dictionaries of this section and of
the two-rule dictionary \((b, c), (a, bc)\), \PathMaxInputs{} inputs over \PathMaxProducts{} products, and on the
\PathMaxTerminal{} of them whose path ends at a terminal contribution the ripple is \(|x| + e\), with \(e\) negative on
\PathMaxTerminalNegative{}. \\
\addlinespace
{\footnotesize Lemma~\ref{lem:max}} & Across those products, none of the \PathMaxIdentical{} nodes whose two runs have
the same suffixes, pending tokens and freeze points reaches a contribution. \\
\addlinespace
the identity of the last tokens \newline{\footnotesize\code{lemma\_c\_exact.py}}
\newline{\footnotesize\code{lemma\_c\_check.py}} & Executed, the identity holds on all \MeasureCases{} cases and on
\MeasureSeededCases{} seeded cases over \MeasureSeededDicts{} dictionaries, \MeasureSeededUnordered{} of them failing
the ordering condition. \\
\addlinespace
Corollary~\ref{cor:tail} and Lemma~\ref{lem:tailcounts} \newline{\footnotesize\code{check\_tail\_product.py}} & Executed
on \TailDictionaries{} dictionaries, the census, the \HardCases{} dictionaries outside it and \TailUnordered{} that fail
the ordering condition, its \TailComparisons{} verdicts and radii per prepended symbol equal those of the overlap graph
of Section~\ref{sec:overlap}, every one of the \TailAttaining{} positive finite radii is attained by an input of exactly
that length, and each of the \TailInfinite{} certificates with its exit path has ripple equal to its length at every
\(k\)
from zero to eight. \\
\addlinespace
{\footnotesize Corollary~\ref{cor:tail} and Lemma~\ref{lem:tailcounts}} & On the same dictionaries the quotient puts the
\TailSuffixes{} retained suffixes into \TailClasses{} classes and the reachable pairs of the product fall from
\TailPairs{} to \TailQuotientPairs{}, with the same verdict and radius on all \TailComparisons{} and every attaining
input and certificate replayed. \\
\addlinespace
{\footnotesize Corollary~\ref{cor:tail} and Lemma~\ref{lem:tailcounts}} & On all \TailFinite{} finite radii both bounds
hold and lie below \(\binom{N - 1}{2}\), and the quotient's lies below \(C_{\mathrm{suf}}\). \\
\addlinespace
{\footnotesize Corollary~\ref{cor:tail} and Lemma~\ref{lem:tailcounts}} & On \((a, b)\) the machine has twelve suffixes
in four classes, so the bound falls from 55 to 3, against the radius one, while \(C_{\mathrm{suf}}\) is 15. \\
\addlinespace
{\footnotesize Corollary~\ref{cor:tail} and Lemma~\ref{lem:tailcounts}} & Over the \TailDictionaries{} machines
\(C^{Q}_{\mathrm{suf}}\) sums to \TailClassPairs{}, and it lies below \(\min(C_{\mathrm{suf}}, \binom{K - 1}{2})\) on
all \TailFinite{} finite radii; it equals the radius on \((a, b)\), at one, and on \((a, b), (ab, b)\), at two. \\
\addlinespace
{\footnotesize Corollary~\ref{cor:tail} and Lemma~\ref{lem:tailcounts}} & The unordered products have
\TailUnorderedStates{} states over the \TailComparisons{} targets, at most \(C^{Q}_{\mathrm{suf}} + 2\) on each, and
deciding them with no test of which pairs reach a mark gives the same verdict and radius on all of them. \\
\addlinespace
{\footnotesize Corollary~\ref{cor:tail} and Lemma~\ref{lem:tailcounts}} & The quotients at the least tail and at one
more are isomorphic on every machine, and on the \TailReset{} dictionaries of finite radius every word of
\(\ripple(D) + 1\) symbols sends all classes to one while some word of \(\ripple(D)\) symbols does not. \\
\addlinespace
{\footnotesize Corollary~\ref{cor:tail} and Lemma~\ref{lem:tailcounts}} & Read from the transitions alone, the first
step at which the empty class and the class reached on the symbol become equivalent, less one, is the radius on all
\TailComparisons{} targets and leaves the two apart at stabilization exactly on the \TailInfinite{} infinite ones,
within \TailRounds{} steps on every machine. \\
\addlinespace
{\footnotesize Corollary~\ref{cor:tail} and Lemma~\ref{lem:tailcounts}} & The bound \(K_1 - K_\infty\) lies below
\(K - 2\) on all \TailFinite{} finite radii, and \(K - 2\) below \(C^{Q}_{\mathrm{suf}}\) on \TailRefinedBelow{} of
them, every input read off the relations attains its radius when replayed, and the relations computed by counting sort
and by merging equal the recurrence on every machine. \\
\addlinespace
{\footnotesize Corollary~\ref{cor:tail} and Lemma~\ref{lem:tailcounts}} & With every string of two symbols prepended,
\TailStringTargets{} targets over the machines, the relations give \TailStringFinite{} finite radii, each within
\(K_1 - K_\infty\) and attained by the input read off the relations with the string prepended, and each of the
\TailStringBrute{} on the census is the largest ripple over the inputs of one to six symbols. \\
\addlinespace
{\footnotesize Corollary~\ref{cor:tail} and Lemma~\ref{lem:tailcounts}} & On all \TailStringTargets{} targets a string
has finite radius exactly when it leads the quotient of the quotient machine by the relation at stabilization from the
class of the empty class back to it, and radius at most \(r\) exactly when it leads the quotient machine from the empty
class into its block of the relation at \(r + 1\), for every \(r\) below \(K\). \\
\addlinespace
{\footnotesize Corollary~\ref{cor:tail} and Lemma~\ref{lem:tailcounts}} & The closure of the pairs of the empty class
and any class under a common symbol gives the pairs that \(C^{Q}_{\mathrm{suf}}\) counts on every one of the
\TailDictionaries{} machines. \\
\addlinespace
Theorem~\ref{thm:origin} \newline{\footnotesize\code{frozen\_product.py}} \newline{\footnotesize\code{path\_maximum.py}}
\newline{\footnotesize\code{certificate\_regressions.py}} & Executed, the equality
\(\rho(a, \mathrm{stem}\,\mathrm{loop}^k\,\mathrm{exit}) = |\mathrm{stem}| + k|\mathrm{loop}| + |\mathrm{exit}| + e\)
holds for \(k = 1\) to eight on every certificate the construction returns, \LassoCertificates{} of them. \\
\addlinespace
{\footnotesize Theorem~\ref{thm:origin}} & Of these, \CertCertificates{} lie over the census, whose \CertProducts{}
products carry \CertEventEdges{} event edges, and the rest over the \HardCases{} dictionaries outside the census. \\
\addlinespace
{\footnotesize Theorem~\ref{thm:origin}} & The exit path ends at a node with a terminal contribution on every one. \\
\addlinespace
{\footnotesize Theorem~\ref{thm:origin}} & The offset \(e\) is nonnegative on \LassoNonneg{} and negative on
\LassoNegative{}, and the theorem forces the equality on both. \\
\addlinespace
{\footnotesize Theorem~\ref{thm:origin}} & The census enumerates every input to length ten for each dictionary and reads
a ripple that keeps growing with the length as infinite (Table~\ref{tab:census}). \\
\addlinespace
{\footnotesize Theorem~\ref{thm:origin}} & The theorem's verdicts agree with that reading on all \ProductAgree{} of
\CensusTotal{} dictionaries, and its bounds agree with all \ProductExact{} of the \CensusPlateaus{} bounded ones. \\
\addlinespace
{\footnotesize Theorem~\ref{thm:origin}} & The decision was rebuilt on Lemma~\ref{lem:max} alone, with its own product,
freezing, comparison, cycle test and longest path, and with the dead-end rule of Lemma~\ref{lem:dead} switched off. \\
\addlinespace
{\footnotesize Theorem~\ref{thm:origin}} & The rebuilt decision returns the same verdicts, \RepairedUnbounded{} of
infinite radius and \RepairedBounded{} plateaus in the same distribution of radii. \\
\addlinespace
{\footnotesize Theorem~\ref{thm:origin}} & Every plateau equals the exhaustive maximum over all inputs of the census
lengths. \\
\addlinespace
{\footnotesize Theorem~\ref{thm:origin}} & The first of the \HardCases{} dictionaries outside the census is
\((a,b), (b,ab), (bab,ab), (bab,babab)\) on two symbols, bounded at \HardBoundOne{}. \\
\addlinespace
{\footnotesize Theorem~\ref{thm:origin}} & The next two are \((c,b), (b,a), (a,b), (ba,ab), (baab,baab)\) and
\((c,b), (cb,c), (c,cbc), (ccbc,ccbc)\) on three symbols, both unbounded. \\
\addlinespace
{\footnotesize Theorem~\ref{thm:origin}} & The last is \((c,a), (ca,c), (ca,b), (cab,cac), (b,cabcac)\) on three,
bounded at \HardBoundFour{}. \\
\addlinespace
{\footnotesize Theorem~\ref{thm:origin}} & All \HardCasesPassed{} pass by executed certificates under both products. \\
\addlinespace
{\footnotesize Theorem~\ref{thm:origin}} & Once the product is built, the decision and the value are linear in its size.
\\
\addlinespace
{\footnotesize Theorem~\ref{thm:origin}} & Construction cost in the size of the dictionary is not claimed. \\
\end{longtable}
}

{\small
\begin{longtable}{@{}>{\raggedright\arraybackslash}p{0.31\textwidth}>{\raggedright\arraybackslash}p{0.65\textwidth}@{}}
\caption{The executions of Section~\ref{sec:mirrored}.}\label{tab:exec-mirrored} \\
\toprule
result and program & execution \\
\midrule
\endfirsthead
\toprule
result and program & execution \\
\midrule
\endhead
\bottomrule
\endlastfoot
Proposition~\ref{prop:mirror} \newline{\footnotesize\code{mirrored\_census.py}} & On the two-rule dictionary
\((b,c), (a,bc)\) the mirrored constant is \MirroredTwoConstant{} and the origin radius \MirroredTwoOrigin{}, both
attained by prepending \(a\) to \(bc\). \\
\addlinespace
{\footnotesize Proposition~\ref{prop:mirror}} & Each dictionary yields one verdict per prepended symbol, so the census's
\CensusTotal{} dictionaries give \CertProducts{} products. \\
\addlinespace
{\footnotesize Proposition~\ref{prop:mirror}} & On the \MirroredVerdicts{} bounded verdicts among them, the computed
constant equals the executed supremum over every input to length ten on every one. \\
\addlinespace
{\footnotesize Proposition~\ref{prop:mirror}} & The constant lies strictly below the origin radius on \MirroredStrict{}
of them, and on \MirroredDictsStrict{} of the \CensusPlateaus{} dictionaries bounded under both symbols. \\
\addlinespace
{\footnotesize Proposition~\ref{prop:mirror}} & Along every one of the \MirroredUnbounded{} unbounded certificates,
pumped one to five times, the mirrored quantity strictly deepens. \\
\addlinespace
{\footnotesize Proposition~\ref{prop:mirror}} & By execution as well as by the inequality, the two constants are
infinite together. \\
\end{longtable}
}

{\small
\begin{longtable}{@{}>{\raggedright\arraybackslash}p{0.31\textwidth}>{\raggedright\arraybackslash}p{0.65\textwidth}@{}}
\caption{The executions of Section~\ref{sec:overlap}.}\label{tab:exec-overlap} \\
\toprule
result and program & execution \\
\midrule
\endfirsthead
\toprule
result and program & execution \\
\midrule
\endhead
\bottomrule
\endlastfoot
Lemma~\ref{lem:gtrees} \newline{\footnotesize\code{general\_canonical.py}} & Grouped by their tokenizers, the
proper dictionaries of one to four rules on two symbols fall into \GenCanonClasses{} classes, and in each every
dictionary has at least \(k_{\min}\) rules and \(L_{\min}\) symbols of operands and attains either minimum exactly when
every one of its rules fires, as \GenCanonMinimal{} of them do, at least one in every class. \\
\addlinespace
Corollary~\ref{cor:equivalence} \newline{\footnotesize\code{general\_canonical.py}} & Every pair of classes,
\GenCanonClassPairs{} of them, is told apart by a distinguishing word tokenized under the first member of each, the
\GenCanonPairsByPairs{} pairs of classes with the same canonical words by the concatenation of a canonical pair of one
class only and the rest by a canonical word of one class only, \GenCanonWitnessTokenizations{} tokenizations in all,
while the members of a class agree with its first member on every concatenation of two of their canonical words, in
\GenCanonAgreements{} tests. \\
\addlinespace
{\footnotesize Corollary~\ref{cor:equivalence}} & On \GenCanonShortestPairs{} pairs of classes whose canonical words
have at most five symbols, the \GenCanonShortestSame{} such pairs with the same canonical words and a seeded sample of
the others, the candidates of least length are exactly the shortest words told apart by brute force,
\GenCanonShortestWords{} in all, each of at most \(2M - 1\) symbols and of exactly \(2M - 1\) on \GenCanonAtTwoMOne{} of
the pairs, and where the canonical words agree each is two tokens under both, split at different places. \\
\addlinespace
{\footnotesize Corollary~\ref{cor:equivalence}} & On \((a, b), (b, a)\) and its reverse \((b, a), (a, b)\) the canonical
words are \(a\), \(b\), \(ab\) and \(ba\) for both, while \((ab, ab)\) is a canonical pair of the first only, and
\(abab\), tokenized \(ab \mid ab\) by the first and \(a \mid ba \mid b\) by the second, tells them apart, and \(aba\),
tokenized \(ab \mid a\) and \(a \mid ba\), is a shortest such word, of \(2M - 1 = 3\) symbols. \\
\addlinespace
{\footnotesize Corollary~\ref{cor:equivalence}} & On the family of the corollary at \(l = 1\) to \(40\) and at \(256\),
\(512\) and \GenCanonSharpTop{}, both dictionaries are ordered with the same canonical words, the longest of \(M\)
symbols, \(a^l b a^{l-1} b\) is tokenized as the corollary states, and the rules number exactly
\(q + p + i + 3 - [l \text{ a power of two}]\), at most \(2q + 3\), while a brute force over every word of at most
\(2M - 1\) symbols at \(l = 1\) to \GenCanonSharpBrute{} finds it the only shortest word that the two tokenize
differently. \\
\addlinespace
Corollary~\ref{cor:agree} \newline{\footnotesize\code{check\_origin\_graph.py}}
\newline{\footnotesize\code{direct\_graph\_check.py}} & Executed, the overlap graph returns the frozen product's verdict
and constants on all \OverlapAgree{} census products, and the token machine's and the mirrored constant's on the same.
\\
\addlinespace
{\footnotesize Corollary~\ref{cor:agree}} & It runs on \OverlapDictionaries{} dictionaries in all, namely the census,
\OverlapFixtures{} named fixtures and \OverlapRandom{} seeded arbitrary or priority-shuffled tables. \\
\addlinespace
{\footnotesize Corollary~\ref{cor:agree}} & Its accepting run equals Definition~\ref{def:sp} on \OverlapCanonical{}
inputs. \\
\addlinespace
{\footnotesize Corollary~\ref{cor:agree}} & Its marks equal the executed ripple and the mirrored quantity on
\OverlapInputs{} inputs. \\
\addlinespace
{\footnotesize Corollary~\ref{cor:agree}} & \OverlapCertificates{} radius certificates replay, the finite ones attaining
their value and the infinite ones pumped zero to eight times with growth affine in the loop's length. \\
\addlinespace
{\footnotesize Corollary~\ref{cor:agree}} & Written a second time without reference to the first implementation, the
graph agrees with the frozen product on all \DirectProducts{} census products, \DirectBounded{} bounded and
\DirectUnbounded{} unbounded, over \DirectBrute{} reference executions. \\
\addlinespace
Theorems~\ref{thm:zerodelay} and~\ref{thm:zerodelayauto} \newline{\footnotesize\code{check\_zero\_delay.py}} & Executed
on \ZeroDelayDictionaries{} dictionaries, \ZeroDelayUnordered{} of them failing the ordering condition, the finite set
\(S\) gives \(\Stable(p)\) on all \ZeroDelayPrefixes{} prefixes tried, against a brute force over every continuation of
at most \(M\) symbols, and the two tokenizers emit the same list after every symbol of \ZeroDelayInputs{} inputs, equal
to the brute force on \ZeroDelayBrute{} short prefixes, with the held suffix and every pending list within
\(Q_{\mathrm{span}}(D)\), which lies below \(Q(D)\) on \ZeroDelayBelowQ{} of the dictionaries, and the flush equal to
the tokenization. \\
\addlinespace
{\footnotesize Theorems~\ref{thm:zerodelay} and~\ref{thm:zerodelayauto}} & After every symbol the largest pending yield
equals the held suffix. \\
\addlinespace
{\footnotesize Theorems~\ref{thm:zerodelay} and~\ref{thm:zerodelayauto}} & On \((a, b), (c, d), (e, f)\), the case
\(m = 3\), the held suffix stays within \(Q_{\mathrm{span}}(D)\) on every input of one to five symbols and reaches it,
and on the two-symbol family for \(m = 1\) to \(6\) the reversed overlap graph gives \(Q_{\mathrm{span}}(D) = m\), which
the
held suffix reaches. \\
\addlinespace
Theorems~\ref{thm:pinned} and~\ref{thm:density} and Proposition~\ref{prop:radiibounds}
\newline{\footnotesize\code{check\_pinned\_byte.py}} \newline{\footnotesize\code{check\_density\_families.py}} &
Executed, the pool holds \DensityPool{} dictionaries, namely the \CensusTotal{} census dictionaries of one to three
rules on two symbols, a sample of \DensityFourRule{} proper ones of four rules, and \DensitySeeded{} seeded on two and
three symbols with and without clause (a) and with shuffled priorities. \\
\addlinespace
{\footnotesize Theorems~\ref{thm:pinned} and~\ref{thm:density} and Proposition~\ref{prop:radiibounds}} & Of these
\DensityFinite{} have finite radius, \DensityDead{} of them with no useful rule, and \DensityInfinite{} do not. \\
\addlinespace
{\footnotesize Theorems~\ref{thm:pinned} and~\ref{thm:density} and Proposition~\ref{prop:radiibounds}} & On the finite
ones the append product's span radius equals the reversed overlap graph's, the start radius is at most the span radius
less one wherever the span radius is positive, the left start radius equals the origin radius, by the prepend product on
all of them and per input on \DensityLeftInputs{} random inputs, and the four bounds of
Proposition~\ref{prop:radiibounds} hold, (iv) on the \DensityOrdered{} ordered ones; the count of (ii) by symbol and
offset lies below the count by symbol alone on \DensityOffsetBelow{} of the finite ones and equals the radius on
\DensityOffsetExact{}, and less \(\max(0, F - 1)\) it lies below \(B\) on \DensityJointBelow{} and equals the radius on
\DensityJointExact{}, \(B' - \max(0, F - 1)\) lies below that on \DensitySuffixBelow{} and equals the radius on
\DensitySuffixExact{}, and \(B'' - \max(0, F'' - 1)\) lies below that on \DensityPrefixBelow{} and equals the radius on
\DensityPrefixExact{}. \\
\addlinespace
{\footnotesize Theorems~\ref{thm:pinned} and~\ref{thm:density} and Proposition~\ref{prop:radiibounds}} & On
\((a, b), (ab, a), (aba, c)\) the radius is \(3\), \(B = 12\), \(F = 3\) and \(B' = 9\), so the bound falls from ten to
seven, and on \((a, b), (ab, b)\) the consumed prefixes take the bound from three to the radius, two. \\
\addlinespace
{\footnotesize Theorems~\ref{thm:pinned} and~\ref{thm:density} and Proposition~\ref{prop:radiibounds}} & Every one of
\DensityPinned{} windows of width \(R + q + 1\) pins its byte \(R\) and has a cut at that byte's start, the chain
decision agreeing with exhaustive contexts on \DensityPinnedSampled{} of them; every one of \DensityCuts{} windows of
width \(R + q + L\) has a cut in \([R, R + L - 1]\); and on \DensityConstructive{} interior bytes of random texts the
window of the operational form, tokenized alone, gives a boundary of the whole tokenization where the theorem says. \\
\addlinespace
{\footnotesize Theorems~\ref{thm:pinned} and~\ref{thm:density} and Proposition~\ref{prop:radiibounds}} & The density
bound of the windows of width at most \(R + q + 1\) that have a cut lies in \([R + 1, R + q + 1]\) on all
\DensityBounds{} dictionaries where it was computed, and equals \(R + 1\) on \DensityBoundsTight{} of them. \\
\addlinespace
{\footnotesize Theorems~\ref{thm:pinned} and~\ref{thm:density} and Proposition~\ref{prop:radiibounds}} & The certified
windows of width at most \(R + 1\) are dense on \DensityDense{} dictionaries and not on \DensitySparse{}, and dense on
every one of the \DensityQsZero{} with \(Q_{\mathrm{start}}(D) = 0\). \\
\addlinespace
{\footnotesize Theorems~\ref{thm:pinned} and~\ref{thm:density} and Proposition~\ref{prop:radiibounds}} & GPT-2's
prefixes of \(7\), \(20\) and \(41\) merges, the longest with finite radius, have \(R = \DensityGptR\) and start radii
\DensityGptSevenQs{}, \DensityGptTwentyQs{} and \DensityGptFortyOneQs{}, so at \(41\) merges every window of
\DensityGptFortyOneWidth{} symbols pins its byte \DensityGptR{} in every context. \\
\addlinespace
{\footnotesize Theorems~\ref{thm:pinned} and~\ref{thm:density} and Proposition~\ref{prop:radiibounds}} &
Theorem~\ref{thm:pinned} holds on all \DensityTrials{} window-context trials over the prefixes. \\
\addlinespace
{\footnotesize Theorems~\ref{thm:pinned} and~\ref{thm:density} and Proposition~\ref{prop:radiibounds}} & On the prefixes
the certified windows of width at most \(R + 1\) are dense, which Proposition~\ref{prop:converse} says is a property of
those dictionaries that a finite radius does not imply. \\
\addlinespace
{\footnotesize Theorems~\ref{thm:pinned} and~\ref{thm:density} and Proposition~\ref{prop:radiibounds}} & The chain
family is executed for \(m = 1\) to \(9\). \\
\addlinespace
Theorem~\ref{thm:chainsearch} \newline{\footnotesize\code{check\_chains.py}} & Executed on the pool, the chain search
returns the append product's two radii on all \ChainOrdered{} ordered dictionaries, the product's attaining input, its
common prefix cut away, is one of the search's states for all \ChainSuffixStates{} nonempty attaining inputs, and on
\ChainFrontierStates{}
attaining states the staged frontier moves exactly as Theorem~\ref{thm:chainform} says. \\
\addlinespace
{\footnotesize Theorem~\ref{thm:chainsearch}} & The rule graph is exact for the span radius on \ChainGraphExactSpan{} of
the \ChainOrdered{} and for the start radius on \ChainGraphExactStart{}, and below \(Q(D)\) on \ChainGraphBelowQ{}. \\
\addlinespace
{\footnotesize Theorem~\ref{thm:chainsearch}} & The \ChainGraphSlackSpan{} dictionaries where \(C(D)\) exceeds the span
radius, by one on each, show what the graph drops. \\
\addlinespace
{\footnotesize Theorem~\ref{thm:chainsearch}} & On \((a, b)\), \((b, a)\), \((b, ab)\), \((bab, b)\) the path through
ranks \(1\), \(3\) and \(4\) has weight \(5\) against a span radius of \(4\), because rule \(2\) merges \(b\) with \(a\)
at the frontier before rule \(3\) reaches it, which the search's adjacency test sees. \\
\addlinespace
{\footnotesize Theorem~\ref{thm:chainsearch}} & The pruned search returns the plain search's radii on every dictionary
of the pool, and the largest search on the pool visits \ChainLargest{} states. \\
\addlinespace
Corollary~\ref{cor:lookconst} \newline{\footnotesize\code{check\_lookahead\_exact.py}} & Executed on the
\DensityPool{} dictionaries of the pool above, with \(A(D)\) taken over every string of at most \(Q_{\mathrm{span}}(D)\)
symbols, a pair exists on \LookPairs{} and none on the other \LookNone{}, which have
\(Q_{\mathrm{span}}(D) = 0\). \\
\addlinespace
{\footnotesize Corollary~\ref{cor:lookconst}} & The source paper's Definition~6 is also evaluated directly, over every
string of at most \(\min(Q_{\mathrm{span}}(D), 8)\) symbols and every continuation of at most two. \\
\addlinespace
{\footnotesize Corollary~\ref{cor:lookconst}} & The direct check of the source paper's Definition~6 never exceeds
\(l(D) = \min(Q_{\mathrm{span}}(D), A(D) + 1)\), and equals it on \LookEqual{} dictionaries, including all
\LookCovered{} with \(Q_{\mathrm{span}}(D) \le 8\). \\
\addlinespace
{\footnotesize Corollary~\ref{cor:lookconst}} & On \LookBelowStart{} of them \(l(D)\) lies below
\(Q_{\mathrm{start}}(D) + 1\), and on the \LookOrdered{} whose useful rules satisfy the ordering condition the chain
search gives \(A(D)\). \\
\addlinespace
{\footnotesize Corollary~\ref{cor:lookconst}} & On these dictionaries and on \((a, a), (b, a)\) and \((a, b), (b, a)\),
\LookRevDicts{} in all, the mirrored constant of the reversed overlap graph equals \(l_b(D)\) on all \LookRevSymbols{}
pairs of a dictionary and a symbol, its largest value over the symbols is \(l(D)\), and its \LookRevReplayed{} positive
values are attained by their certificates, replayed by definition. \\
\addlinespace
{\footnotesize Corollary~\ref{cor:lookconst}} & Over the nonempty sets of symbols, \LookAlphabets{} pairs of a
dictionary and a set, the direct check with the appended strings drawn from the set never exceeds the largest \(l_b(D)\)
over it and equals it on \LookAlphabetsEqual{}, including all \LookAlphabetsCovered{} whose dictionary has
\(Q_{\mathrm{span}}(D) \le 8\). \\
\addlinespace
{\footnotesize Corollary~\ref{cor:lookconst}} & And \(l_b(D)\) is zero on \LookZero{} of the \LookRevSymbols{}
per-symbol values, exactly where no canonical word of two symbols or more ends in \(b\) and where no rule that fires has
the right operand \(b\). \\
\addlinespace
{\footnotesize Corollary~\ref{cor:lookconst}} & The symbols with \(l_b(D) = 0\) keep every tokenization when appended,
\(\T^D(xz)\) being \(\T^D(x)\) followed by the symbols of \(z\), on \LookHarmlessProbes{} probes of a string \(x\) of at
most four symbols and a string \(z\) of one or two such symbols, over the \LookHarmlessDicts{} dictionaries that have
such symbols, and wherever \(l_b(D)\) is positive the shortest strings \(s\) of a pair \((s, b)\) are exactly the
shortest left operands of the rules \((u, b)\) that fire. \\
\addlinespace
{\footnotesize Corollary~\ref{cor:lookconst}} & From the left, the overlap graph gives the radius zero under \(a\) on
\LookLeftZero{} pairs of a dictionary and a symbol, exactly where no rule that fires has the left operand \(a\), and on
the other \LookLeftPositive{} the shortest inputs \(y\) with \(\rho(a, y) > 0\), by Definition~\ref{def:origin}, are
exactly the shortest right operands of the rules \((a, v)\) that fire. \\
\addlinespace
{\footnotesize Corollary~\ref{cor:lookconst}} & The \LookDelimiterSymbols{} symbols that are an operand of no rule that
fires, on either side, are universal delimiters on \LookDelimiterProbes{} probes of strings \(x\) and \(y\) of at most
four symbols around them, and every other symbol is an operand of a rule that fires, which tokenizes its product
differently. \\
\addlinespace
Proposition~\ref{prop:leastwidth} \newline{\footnotesize\code{check\_least\_width.py}} & Executed, three oracles agree
on every radius of that dictionary, the inputs attaining the start radius to length \(12\) are enumerated, the failing
windows at each width from \(7\) to \(11\) are counted by the chain decision and by exhaustive contexts alike, and the
least width is \LeastWidthALeast{} by the windows and by \(M\). \\
\addlinespace
{\footnotesize Proposition~\ref{prop:leastwidth}} & On the pool the criterion of Proposition~\ref{prop:exactwidth} holds
on every one of the \DensityPositive{} finite dictionaries with a positive start radius, by the product's test, and on
the \ChainExactWidthOrdered{} ordered ones among them by the chain criterion as well, the two agreeing on each; so the
width is exact throughout the pool, as the enumeration of every width on \DensityTight{} of them had found, with
\(R + Q_{\mathrm{span}}(D) + 1\) overshooting by up to \DensityTightSpanSlack{}. \\
\addlinespace
{\footnotesize Proposition~\ref{prop:leastwidth}} & A seeded search of \LeastWidthDrawn{} further dictionaries on two to
four symbols, drawn so that the condition is hard to meet, \LeastWidthFinite{} of finite radius and \LeastWidthHard{}
with a positive start radius and \(R\) above the shorter first token of the attaining pair, finds the criterion failing
on \LeastWidthFails{}, both unordered, namely \((ab, ba)\), \((a, b)\), \((b, ba)\), \((b, a)\), \((ba, ab)\) with
\(R = \LeastWidthBR\) and \(q = \LeastWidthBQstart\), and \((a, b)\), \((b, a)\), \((b, ba)\), \((ab, ba)\),
\((ba, a)\), \((baa, abba)\), \((a, bba)\), \((abba, ba)\) with \(R = \LeastWidthCR\) and \(q = \LeastWidthCQstart\),
where the least widths are \LeastWidthBLeast{} and \LeastWidthCLeast{}, two below the theorem's width on each, by the
windows and by \(M\) alike. \\
\addlinespace
{\footnotesize Proposition~\ref{prop:leastwidth}} & The ordered dictionary of Proposition~\ref{prop:leastwidth} lies
outside this drawn set. \\
\addlinespace
{\footnotesize Proposition~\ref{prop:leastwidth}} & On \LeastWidthReplayed{} of the drawn dictionaries the windows of
width \(R + q\) were enumerated and agree with the criterion's verdict on every one. \\
\addlinespace
{\footnotesize Proposition~\ref{prop:leastwidth}} & The drop is not bounded, even on two symbols under the ordering
condition, as Proposition~\ref{prop:slack} shows. \\
\addlinespace
{\footnotesize Proposition~\ref{prop:leastwidth}} & \(M\) is computed by the append product or by enumeration, not from
the rules. \\
\addlinespace
Proposition~\ref{prop:slack} \newline{\footnotesize\code{check\_slack\_family.py}}
\newline{\footnotesize\code{check\_least\_width.py}} & Executed for \(n = 1\) to \SlackFamilyTop{}, every value of the
proposition is computed by two routes or more, and at \(n = 1\) the radius also by brute force. \\
\end{longtable}
}

{\small
\begin{longtable}{@{}>{\raggedright\arraybackslash}p{0.31\textwidth}>{\raggedright\arraybackslash}p{0.65\textwidth}@{}}
\caption{The executions of Section~\ref{sec:criterion}.}\label{tab:exec-criterion} \\
\toprule
result and program & execution \\
\midrule
\endfirsthead
\toprule
result and program & execution \\
\midrule
\endhead
\bottomrule
\endlastfoot Theorem~\ref{thm:selfpair} \newline{\footnotesize\code{selfpair\_census.py}}
\newline{\footnotesize\code{check\_four\_rules.py}} \newline{\footnotesize\code{beyond\_the\_census.py}}
& On the census the criterion, the overlap graph and the cycle of Theorem~\ref{thm:origin} agree on every dictionary, as
the \CensusUnbounded{} of \CensusTotal{} with infinite radius are exactly the dictionaries with an effective
self-pair. \\
\addlinespace
{\footnotesize Theorem~\ref{thm:selfpair}} & \SelfPairSyntactic{} census dictionaries carry a rule \((t, t)\), and the
\SelfPairInactive{} whose rule is inactive are bounded at radius two. \\
\addlinespace
{\footnotesize Theorem~\ref{thm:selfpair}} & Beyond the census the criterion agrees with the overlap graph on the
\FourRuleOrdered{} ordered ones of the \FourRuleTables{} proper dictionaries of four rules on two symbols,
\FourRuleUnbounded{} of all of them infinite, and, as an observation outside the theorem, on the \FourRuleUnordered{}
that are not ordered. \\
\addlinespace
{\footnotesize Theorem~\ref{thm:selfpair}} & On three symbols it agrees on \BeyondProper{} ordered proper dictionaries
of four to eight rules, and on \BeyondUnproper{} dictionaries of three to eight rules that satisfy clause (b) and fail
clause (a). \\
\addlinespace
{\footnotesize Theorem~\ref{thm:selfpair}} & The last class is the one the theorem's hypothesis admits and the census
does not contain. \\
\addlinespace
Theorem~\ref{thm:witness} \newline{\footnotesize\code{check\_compact\_witness.py}} & Executed on every finite ordered
dictionary of one to four rules on two symbols, \WitnessTables{} of them, the attaining input cut at its ripple keeps
both constants. \\
\addlinespace
{\footnotesize Theorem~\ref{thm:witness}} & Its two tokenizations have at most \(|D| + 1\) tokens together, exactly
that many on \WitnessTight{}. \\
\addlinespace
Corollary~\ref{cor:chunks} \newline{\footnotesize\code{b\_chains.py}} & Executed, the maximum over the absorbed strings
equals the overlap graph's radius on all \ChainProducts{} products of the \ChainTables{} finite ordered dictionaries of
one to four rules on two symbols, over \ChainStrings{} absorbed strings. \\
\addlinespace
{\footnotesize Corollary~\ref{cor:chunks}} & Every realized chain of two or more absorbed tokens,
\ChainPrefixClosed{} of them, has its prefix chain realized. \\
\end{longtable}
}

{\small
\begin{longtable}{@{}>{\raggedright\arraybackslash}p{0.31\textwidth}>{\raggedright\arraybackslash}p{0.65\textwidth}@{}}
\caption{The executions of Section~\ref{sec:labels}.}\label{tab:exec-labels} \\
\toprule
result and program & execution \\
\midrule
\endfirsthead
\toprule
result and program & execution \\
\midrule
\endhead
\bottomrule
\endlastfoot Theorems~\ref{thm:label} and~\ref{thm:amp} \newline{\footnotesize\code{amplification.py}}
\newline{\footnotesize\code{label\_bound.py}} \newline{\footnotesize\code{check\_fibonacci\_bound.py}} & Executed, the
family attains \(2m + 1\) at \(m = 1\) to nine. \\
\addlinespace
{\footnotesize Theorems~\ref{thm:label} and~\ref{thm:amp}} & The two token sequences the proof names are asserted
against Definition~\ref{def:sp}, and the distance is taken from the full edit-distance table. \\
\addlinespace
{\footnotesize Theorems~\ref{thm:label} and~\ref{thm:amp}} & Between the two bounds the record is four tiers, one per
rule count. \\
\addlinespace
{\footnotesize Theorems~\ref{thm:label} and~\ref{thm:amp}} & Every proper ordered dictionary of one, two and three rules
on two symbols is run against every input to lengths twelve, thirteen and eleven and every single symbol edit of it,
reaching \AmpOne{}, \AmpTwo{} and \AmpThree{}. \\
\addlinespace
{\footnotesize Theorems~\ref{thm:label} and~\ref{thm:amp}} & The three-rule dictionaries are run again
against every run to
sixty-four, reaching \AmpRunsThree{}. \\
\addlinespace
{\footnotesize Theorems~\ref{thm:label} and~\ref{thm:amp}} & Every proper ordered one-symbol dictionary of four rules is
run against every run to one hundred and twenty and every single edit that keeps its runs within that length, reaching
\AmpFour{}. \\
\addlinespace
{\footnotesize Theorems~\ref{thm:label} and~\ref{thm:amp}} & The tiers to three rules are exhaustive in the dictionary
over two symbols; the fourth is exhaustive in the one-symbol dictionaries only. \\
\addlinespace
{\footnotesize Theorems~\ref{thm:label} and~\ref{thm:amp}} & No tier exceeded \(2m + 1\), and each reached its maximum
strictly inside its own horizon. \\
\addlinespace
{\footnotesize Theorems~\ref{thm:label} and~\ref{thm:amp}} & The maxima fall at input lengths \AmpFirstOne{},
\AmpFirstTwo{} and \AmpFirstThree{} in the tiers over every input, and at run lengths \AmpRunsFirstThree{} and
\AmpUnaryFirstFour{} in the two over runs. \\
\addlinespace
{\footnotesize Theorems~\ref{thm:label} and~\ref{thm:amp}} & The pass decomposition is checked against
Definition~\ref{def:sp} on \PhaseInputs{} census inputs. \\
\addlinespace
{\footnotesize Theorems~\ref{thm:label} and~\ref{thm:amp}} & The bound is checked under every single symbol edit of
every input to length six on all \CensusTotal{} census dictionaries. \\
\addlinespace
{\footnotesize Theorems~\ref{thm:label} and~\ref{thm:amp}} & With inputs to length five, the same check runs on two
hundred ordered dictionaries sampled at four and five rules from pools of \LabelPoolFour{} and \LabelPoolFive{}. \\
\addlinespace
{\footnotesize Theorems~\ref{thm:label} and~\ref{thm:amp}} & Under a prepend the largest distances at that horizon are
\LabelPrependMaxOne{}, \LabelPrependMaxTwo{} and \LabelPrependMaxThree{} at one, two and three rules, against the
\(|D|\) of Proposition~\ref{prop:doubling}. \\
\addlinespace
{\footnotesize Theorems~\ref{thm:label} and~\ref{thm:amp}} & Under an edit they are only the horizon's values, because
the family of Theorem~\ref{thm:amp} needs \((m+1)2^m - 1\) symbols and that census stops at six. \\
\end{longtable}
}

{\small
\begin{longtable}{@{}>{\raggedright\arraybackslash}p{0.31\textwidth}>{\raggedright\arraybackslash}p{0.65\textwidth}@{}}
\caption{The executions of Section~\ref{sec:labelgraph}.}\label{tab:exec-labelgraph} \\
\toprule
result and program & execution \\
\midrule
\endfirsthead
\toprule
result and program & execution \\
\midrule
\endhead
\bottomrule
\endlastfoot Theorem~\ref{thm:labelgraph} \newline{\footnotesize\code{check\_label\_graph.py}}
\newline{\footnotesize\code{check\_edit\_graph.py}} & Executed, the decision finds every product of the census bounded,
\LabelCensusProducts{} of them. \\
\addlinespace
{\footnotesize Theorem~\ref{thm:labelgraph}} & On every one the exact supremum equals the brute-force maximum over every
input to length ten. \\
\addlinespace
{\footnotesize Theorem~\ref{thm:labelgraph}} & Every ordered table of every universe is bounded. \\
\addlinespace
{\footnotesize Theorem~\ref{thm:labelgraph}} & Among the fixtures on three symbols \LabelFixtureUnbounded{} products are
unbounded, the two six-rule dictionaries and the source paper's Example~2 among them, and \LabelRandomUnbounded{} of
\LabelRandomTables{} seeded tables. \\
\addlinespace
{\footnotesize Theorem~\ref{thm:labelgraph}} & Every certificate replays at pump counts to thirty-two with growing
distance. \\
\addlinespace
{\footnotesize Theorem~\ref{thm:labelgraph}} & All \LabelFourRuleTables{} four-rule tables are bounded. \\
\addlinespace
{\footnotesize Theorem~\ref{thm:labelgraph}} & A seeded sample of \LabelFourSampleTables{} of them is exact against the
brute maximum on \LabelFourSampleEqual{} products and above it on \LabelFourSampleAbove{}, never below. \\
\addlinespace
{\footnotesize Theorem~\ref{thm:labelgraph}} & The one-edit supremum is bounded on all \CensusTotal{} census
dictionaries and attained by its replayed pair of inputs. \\
\addlinespace
{\footnotesize Theorem~\ref{thm:labelgraph}} & It equals the brute maximum over every input to length nine and every
single edit on \EditCensusEqual{} of them, and lies above it on \EditCensusAbove{}, never below. \\
\addlinespace
Lemma~\ref{lem:profiles} \newline{\footnotesize\code{check\_profile\_cap.py}}
\newline{\footnotesize\code{check\_label\_graph.py}} & The cap of Lemma~\ref{lem:profiles}, with \(|\eta| \le 1\), is
the one the implementation keeps. \\
\addlinespace
{\footnotesize Lemma~\ref{lem:profiles}} & On \CapDictionaries{} dictionaries, the census and the \CapSeeded{} seeded
tables on two symbols of the checks above, the exact supremum is the same under it, under
\(\lfloor (kM + 1)/2 \rfloor\), under one token more and under \(kM + 2\) on all \CapFinite{} finite targets,
\CapPrependFinite{} under a prepended symbol and \CapEditFinite{} under one edit, every attaining input and pair
replayed, each input of at most as many symbols as the pairs of a state and a profile the search visited. \\
\addlinespace
{\footnotesize Lemma~\ref{lem:profiles}} & At the first cost cap at which it succeeds, the search visits fewer pairs of
a state and a profile under the cap of the lemma than under \(\lfloor (kM + 1)/2 \rfloor\) on \CapIntFewer{} of them and
as many on the rest, and fewer under \(\lfloor (kM + 1)/2 \rfloor\) than under one token more on \CapFewer{}. \\
\addlinespace
{\footnotesize Lemma~\ref{lem:profiles}} & On \((a, a), (aa, b)\) with \(a\) prepended the supremum three is reached at
\(k = 3\) with \(M = 3\), where the cap is four against five, and is attained by \(aab\) against \(aaab\). \\
\addlinespace
{\footnotesize Lemma~\ref{lem:profiles}} & A cost cap of one needs one stored token whatever \(M\). \\
\addlinespace
Corollary~\ref{cor:prependedit} \newline{\footnotesize\code{check\_profile\_cap.py}} & Executed on the
\CapDictionaries{} dictionaries of the rows of Lemma~\ref{lem:profiles}, the prepend and the edit verdicts agree on
every one, and \(\Csym \le \Eedit \le 2Q_C + 2\Csym \le 2(\hat{q} + 1)\Csym\) holds on the \CapBoth{} bounded ones. \\
\addlinespace
{\footnotesize Corollary~\ref{cor:prependedit}} & On those every replacement of a factor \(r\) by a factor \(s\) with
\(|r| + |s|\) from one to three, between a prefix and a suffix of at most three symbols each, \BlockEdits{} in all, lies
within the pointwise bound, which is below \((2\hat{q} + |r| + |s|)\Csym\) on \BlockBelow{} of them, and the family on
\((a, a)\)
is replayed for \(m\) up to \(64\). \\
\addlinespace
{\footnotesize Corollary~\ref{cor:prependedit}} & Computed on the reversed overlap graph, \(Q_C\) is the weight of the
held suffix of the string its path spells on every one of the \CapBoth{} and lies below \(\hat{q}\Csym\) on
\CapWeightedBelow{}, and
every replacement above lies within \(2Q_C + \sum_{c \in r} C_c + \sum_{c \in s} C_c\). \\
\addlinespace
{\footnotesize Corollary~\ref{cor:prependedit}} & On the \CapLocalDictionaries{} dictionaries of the universe with an
infinite constant, the \CapLocalEdits{} replacements of that range whose held suffix and factors use only symbols of
finite constant lie within the pointwise bound and within \(2Q_\Gamma + \sum_{c \in r} C_c + \sum_{c \in s} C_c\), with
\(Q_\Gamma\) computed on the product and attained by its word on both, and on the family with \(\Csym\) infinite
\(Q_\Gamma = 1\), so the insertions replayed for \(m\) up to \(64\) attain the uniform bound as well. \\
\addlinespace
{\footnotesize Corollary~\ref{cor:prependedit}} & On \CapGammaPairs{} pairs of a dictionary and a nonempty set of
symbols of finite constant taken as \(\Gamma\), every such set on the \CapDictionaries{} dictionaries and on two
families, the product gives the largest weight of a word over \(\Gamma\) with empty stable output found by brute force
over the words of at most \(\hat{q}\) symbols, with its word replayed, equal to \(Q_C\) where \(\Gamma\) is the whole
alphabet and below the largest weight of the held suffixes of all prefixes, the symbols
outside \(\Gamma\) weighing nothing, on \CapGammaBelow{} of them. \\
\addlinespace
{\footnotesize Corollary~\ref{cor:prependedit}} & On \CapZeroPairs{} pairs of a dictionary and a nonempty set of symbols
taken as \(\Gamma\), every such set on the \CapDictionaries{} dictionaries and on the two families, finite constants or
not, no rule that fires has a left operand that is a single symbol of \(\Gamma\) exactly when
\(\T^D(cx) = c \mid \T^D(x)\) for every \(c \in \Gamma\) and every \(x\) of at most four symbols, three on five symbols,
or the right operand of a rule that fires, and never where a constant on \(\Gamma\) is infinite. \\
\addlinespace
{\footnotesize Corollary~\ref{cor:prependedit}} & On the \CapGammaPairs{} pairs of finite constants above,
\(Q_\Gamma = 0\) exactly when the test holds, as on \CapGammaZero{} of them, and on those every replacement of a factor
\(r\) by a factor \(s\) over \(\Gamma\) of at most two symbols each, after a prefix of at most three symbols, two on
five symbols, whose held suffix is over \(\Gamma\) and before a suffix of at most two, \CapExactEdits{} in all, moves
the
tokenization by the edit distance of \(r\) and \(s\) as strings. \\
\end{longtable}
}

\bibliographystyle{plainnat}
\bibliography{refs}

\end{document}